\pdfoutput=1
\documentclass[11pt]{article}
\usepackage[T1]{fontenc}
\usepackage[utf8]{inputenc}
\usepackage{lmodern}
\usepackage[a4paper,margin=25mm,headheight=14pt]{geometry}
\usepackage{amsmath,amssymb,amsthm,mathtools,bm}
\usepackage{microtype,booktabs,longtable,array,needspace}
\usepackage{tabularx,xltabular}
\usepackage{graphicx}
\usepackage{caption}
\usepackage{float}
\usepackage{placeins}
\usepackage{tikz}
\usetikzlibrary{arrows.meta,decorations.markings,decorations.pathreplacing,calc}
\tikzset{cutp/.style={line width=.6pt,draw=black!85},
  cutmark/.style={postaction={decorate},decoration={markings,mark=at position .5 with {\draw[line width=.8pt] (-2pt,-3.6pt)--(-.2pt,3.6pt); \draw[line width=.8pt] (.7pt,-3.6pt)--(2.5pt,3.6pt);}}},
  cutI/.style={line width=1.6pt,draw=orange!85!black,cutmark},cutO/.style={line width=1.6pt,draw=violet!80!black,cutmark},
  hfprop/.style={line width=.7pt,draw=black!85,postaction={decorate},decoration={markings,mark=at position .58 with {\arrow{Stealth[length=2mm,width=1.5mm]}}}},
  hfheat/.style={line width=1.6pt,draw=black,cutmark},
  hfout/.style={line width=.7pt,draw=teal!55!black,-{Stealth[length=2mm,width=1.5mm]}},
  hfv/.style={circle,fill=black,inner sep=1.6pt},hfo/.style={rectangle,draw,fill=white,inner sep=2.6pt,line width=.6pt},
  cutv/.style={circle,fill=black,inner sep=1.4pt},cuto/.style={rectangle,draw,fill=white,inner sep=2.3pt}}
\newcommand{\bbtri}[4]{\begin{scope}[xshift=#1]
  \coordinate (a) at (-1.0,1.5); \coordinate (b) at (1.0,1.5); \coordinate (o) at (0,0);
  \draw[#2] (a)--(b); \draw[#3] (a)--(o); \draw[#4] (b)--(o);
  \draw[hfout] (a)--++(-.42,.6); \draw[hfout] (b)--++(.42,.6);
  \node[hfv] at (a) {}; \node[hfv] at (b) {}; \node[hfo] at (o) {};
  \node[font=\scriptsize,teal!45!black] at (-1.55,2.33) {$b$}; \node[font=\scriptsize,teal!45!black] at (1.55,2.33) {$c$};
  \node[font=\scriptsize,anchor=east] at (-1.1,1.4) {$x_1$}; \node[font=\scriptsize,anchor=west] at (1.1,1.4) {$x_2$};
  \node[font=\scriptsize] at (0,-.38) {$bb\,(0)$};
  \node[font=\tiny,black!60] at (-.62,1.68) {$c$}; \node[font=\tiny,black!60] at (.62,1.68) {$b$};
  \node[font=\tiny,black!60] at (-1.0,1.05) {$c$}; \node[font=\tiny,black!60] at (-.42,.2) {$b$};
  \node[font=\tiny,black!60] at (1.0,1.05) {$c$}; \node[font=\tiny,black!60] at (.42,.2) {$b$};
  \end{scope}}
\newcommand{\cutone}[4]{\begin{scope}[xshift=#1]
  \coordinate (a) at (-.6,1.05); \coordinate (b) at (.6,1.05); \coordinate (o) at (0,0);
  \draw[#2] (a)--(b); \draw[#3] (a)--(o); \draw[#4] (b)--(o);
  \node[cutv] at (a) {}; \node[cutv] at (b) {}; \node[cuto] at (o) {};
  \node[above] at (a) {\scriptsize$1$}; \node[above] at (b) {\scriptsize$2$}; \node[below] at (o) {\scriptsize$\mathcal O$};
  \end{scope}}
\newcommand{\cuttwo}[6]{\begin{scope}[xshift=#1]
  \coordinate (a) at (-.9,.95); \coordinate (b) at (0,1.2); \coordinate (c) at (.9,.95); \coordinate (o) at (0,0);
  \draw[#2] (a)--(b); \draw[#3] (b)--(c); \draw[#4] (a)--(o); \draw[#5] (b)--(o); \draw[#6] (c)--(o);
  \node[cutv] at (a) {}; \node[cutv] at (b) {}; \node[cutv] at (c) {}; \node[cuto] at (o) {};
  \node[above] at (a) {\scriptsize$1$}; \node[above] at (b) {\scriptsize$2$}; \node[above] at (c) {\scriptsize$3$}; \node[below] at (o) {\scriptsize$\mathcal O$};
  \end{scope}}
\usepackage[most]{tcolorbox}
\tikzset{
  pict/.style={baseline={(0,-.55ex)},line width=.5pt,font=\scriptsize},
  pR/.style={rectangle,draw,fill=white,minimum width=4.4mm,minimum height=1.5mm,inner sep=0pt,sloped,line width=.5pt},
  pdot/.style={circle,fill=black,inner sep=1.05pt},
  parr/.style={-{Stealth[length=1.3mm,width=1mm]},line width=.45pt},
  pcur/.style={postaction={decorate},decoration={markings,mark=at position .84 with {\arrow{Stealth[length=1.3mm,width=1mm]}}}},
}
\newcommand{\pictResistor}{\tikz[pict]{\draw (0,0)--(1.1,0); \node[pR] at (.55,0) {}; \node at (.55,.22) {$t_e$};}}
\newcommand{\pictCurrent}{\tikz[pict]{\draw[pcur] (0,0)--(1.25,0); \node[pR] at (.48,0) {}; \node at (1.02,.2) {$y_e$};}}
\newcommand{\pictSource}{\tikz[pict]{\node[pdot] (v) at (0,0) {}; \draw (v)--(-.42,-.2) (v)--(.42,-.2); \draw[parr] (0,.46)--(v); \node[anchor=west] at (.03,.34) {$\lambda_v$}; \node[anchor=north] at (0,-.04) {$v$};}}
\newcommand{\pictKirchhoff}{\tikz[pict]{\node[pdot] (v) at (0,0) {}; \draw[parr] (0,.46)--(v); \node[anchor=west] at (.03,.34) {$\lambda_v$};
  \draw[pcur] (v)--(.55,-.22); \draw[pcur] (v)--(-.55,-.22); \draw[pcur] (-.6,.16)--(v);}}
\newcommand{\pictLoop}{\tikz[pict]{\coordinate (a) at (0,-.3); \coordinate (b) at (1.3,-.3); \coordinate (c) at (.65,.5);
  \draw (a)--node[pR,minimum width=3.4mm,minimum height=1.3mm]{}(b)--node[pR,minimum width=3.4mm,minimum height=1.3mm]{}(c)--node[pR,minimum width=3.4mm,minimum height=1.3mm]{}(a);
  \draw[parr] (.79,.1) arc (0:290:.13);}}
\newcommand{\pictConductance}{\tikz[pict]{\node[pdot] (u) at (0,0) {}; \node[pdot] (w) at (1.1,0) {}; \draw (u)--node[pR]{}(w); \node at (.55,.24) {$1/t_e$};}}
\newcommand{\pictInsert}{\tikz[pict]{\draw (0,0)--(1.3,0); \node[pR] at (.42,0) {};
  \draw[blue!65!black,line width=.9pt,-{Stealth[length=1.5mm,width=1.2mm]}] (.78,.12)--(1.24,.12); \node[blue!65!black] at (1.0,.31) {$y_{e,+\dot a}$};}}
\newcommand{\pictLaman}{\tikz[pict]{\draw[dashed,fill=black!6] (.65,.08) ellipse (.95 and .5);
  \node[pdot] (a) at (0,-.18) {}; \node[pdot] (b) at (1.3,-.18) {}; \node[pdot] (c) at (.65,.42) {};
  \draw (a)--node[pR,minimum width=3.4mm,minimum height=1.3mm]{}(b)--node[pR,minimum width=3.4mm,minimum height=1.3mm]{}(c)--node[pR,minimum width=3.4mm,minimum height=1.3mm]{}(a);
  \node at (1.72,.42) {$S$};}}
\newcommand{\pictVolume}{\tikz[pict]{\draw[parr] (0,-.25)--(1.05,-.25); \draw[parr] (0,-.25)--(0,.48);
  \fill[blue!12,draw=blue!65!black,line width=.45pt] (.2,-.14)--(.86,-.08) to[bend right=25] (.34,.38)--cycle;
  \draw[parr,blue!65!black] (.6,.06) arc (0:280:.12);}}
\usepackage[hyphens]{url}
\usepackage[numbers,sort&compress]{natbib}
\usepackage[hidelinks]{hyperref}
\numberwithin{equation}{section}
\numberwithin{figure}{section}
\numberwithin{table}{section}
\newcounter{rthmc}[section]
\renewcommand{\therthmc}{\thesection.\arabic{rthmc}}
\newcommand{\rthm}[3]{\par\needspace{3\baselineskip}\addvspace{\medskipamount}\noindent\refstepcounter{rthmc}\label{#2}\textbf{#1~\therthmc}\if\relax\detokenize{#3}\relax\else\ (#3)\fi\textbf{.}}
\newtcolorbox{proofbox}{breakable,enhanced,colback=white,colframe=black!28,boxrule=0.4pt,arc=1pt,
  left=5pt,right=5pt,top=3pt,bottom=3pt,before skip=6pt,after skip=8pt,parbox=false}
\newcommand{\HQ}{|H(Q_0)|}
\newcommand{\HQC}{|H(Q_0)|_{\mathrm{Cartan}}}

\allowdisplaybreaks[3]
\hypersetup{pdftitle={Holomorphic Feynman diagrams in four dimensions and the quantum BPS ring},pdfauthor={Botao Li, Kangning Liu, Chi-Ming Chang},pdfsubject={Manuscript revision 20261123-180653}}
\title{Holomorphic Feynman diagrams in four dimensions\\ and the quantum BPS ring}
\author{Botao Li$^{1}$, Kangning Liu$^{2,3,4}$ and Chi-Ming Chang$^{2,5}$\\[6pt]
{\small $^{1}$Peking University}\\[3pt]
{\small $^{2}$Yau Mathematical Sciences Center (YMSC), Tsinghua University,}\\
{\small Beijing, China}\\[3pt]
{\small $^{3}$Department of Mathematical Sciences, Tsinghua University,}\\
{\small Beijing, China}\\[3pt]
{\small $^{4}$Jefferson Physical Laboratory, Harvard University,}\\
{\small Cambridge, Massachusetts 02138, USA}\\[3pt]
{\small $^{5}$Beijing Institute of Mathematical Sciences and Applications (BIMSA),}\\
{\small Beijing, China}}
\date{}
\begin{document}
\maketitle
\begin{abstract}
We compute at all loop orders the kernels of the holomorphic Feynman graphs governing loop corrections to supersymmetry transformations and regularized operator products in the holomorphic twist of four-dimensional $\mathcal N=1$ supersymmetric gauge theories. Interpreting their current equations as those of an electrical network---with Schwinger parameters as resistances and holomorphic momenta as injected currents---we iterate series, parallel and star--triangle reductions on Schwinger space to obtain the fan, spider and cubic product-forest kernels at all loop orders. Assuming that twisting commutes with quantization for composite operators, we obtain every perturbative correction to the supersymmetry transformation in $\mathcal N=4$ Yang--Mills theory, acting on operators with arbitrary holomorphic derivatives. Using the marked higher operations of Gaiotto, Kulp and Wu, we construct the quantum product and prove that on cohomology of the quantum supercharge it is associative, graded commutative and unital, furnishing the multiplication of the quantum BPS ring; in $\mathcal N=4$ Yang--Mills theory every multi-graviton operator thus admits an explicit finite all-order completion. For the classical gauge groups, every perturbative differential image in the multi-trace ring vanishes upon Cartan restriction. In $\mathcal N=4$ Yang--Mills theory, we find new fortuitous classes and one-loop lifts across several gauge groups, no two-loop lift in the sectors tested, and rationally certified three-loop lifts in $SO(11)$, $SO(12)$ and $SO(13)$. The lowest example we find lies in the level-$34$ BMN sector of $SO(13)$. The glueball lifts in $SO(11)$ and $SO(13)$ remove the excess classical powers relative to their symplectic duals. Beyond $\mathcal N=4$, we find two-loop lifts in the Leigh--Strassler and $\beta$-deformations, and two- to four-loop lifts in abelian and $\mathcal N=2$ gauge theories.
\end{abstract}
{\small\setcounter{tocdepth}{2}\tableofcontents}
\bigskip

\section{Introduction}\label{sec-introduction}

In a unitary superconformal theory, the cohomology of a nilpotent supercharge $Q$ is isomorphic to the corresponding BPS sector. Each class has a unique harmonic representative, annihilated by $Q$ and its radial adjoint $Q^\dagger=S$. For $Q=Q^4_-$ in $\mathcal N=4$ Yang--Mills theory, this is the $1/16$-BPS sector \cite[Section 2]{ChangYin}. The scaling dimensions of these states are fixed by their angular momenta and R-charges through the BPS bound. Quantum corrections to the supercharge, the anomalous supersymmetry transformation, determine which classical classes persist in the quantum BPS spectrum and constrain the microscopic states of dual AdS black holes \cite{KonishiBH}.

In the physical theory, a complete treatment requires a careful definition of the renormalized products of local operators and of their counterterms, which is technically demanding. The holomorphic twist of the theory is a much more efficient framework. The holomorphic twist organizes supersymmetry cohomology in the Batalin--Vilkovisky (BV) formalism \cite{semichiral,BK}. Homotopy transfer takes the physical BV complex to the complex of holomorphic local operators, where the action and the BV differential\footnote{Here the BV differential is the sum of the Koszul--Tate differential, which imposes the equations of motion, the gauge BRST transformation, and the nilpotent supercharge under study.} simplify considerably. The graph integrals of the twisted theory are ultraviolet finite \cite[Theorem 5 and Corollary 4]{Wang}. In the twisted theory, the higher brackets give the quantum corrections to the BV differential on composite operators \cite{higher,BK}. Assuming that twisting commutes with quantization for local operators, we identify this cohomology with the perturbative quantum BPS sector. Computing it requires the higher brackets at every loop order.\footnote{The exact spectrum may also receive nonperturbative corrections: instanton effects can lift perturbative cohomology classes \cite[Section 1]{semichiral} or, equivalently, correct the supercharges \cite[Section 2.2.1]{BK}.} Quantum corrections to operator relations already appear in the Konishi anomaly and the four-dimensional chiral ring \cite{Konishi,CDSW}; the complete one-loop correction $Q_1$ was computed in \cite{BK}.

We further develop the holomorphic graph-integral and higher-operation constructions of \cite{Wang,operatope,higher,BK} to evaluate these corrections at higher loops. The graph integrals that evaluate the higher brackets are integrals over a region of Schwinger space called the operatope \cite{operatope}.\footnote{Williams proved one-loop finiteness for holomorphic theories on $\mathbb C^d$ \cite[Theorem 3.4]{Williams}. Wang established ultraviolet finiteness at all loop orders for the heat-kernel regularized graph integrals, the holomorphic locality of their boundary integrals and the Kirchhoff structure of their Schwinger integrands \cite[Sections 2--4]{Wang}; the operatope supplied the combined Schwinger-space propagator, the edge-current variables and the Laman constraints \cite[Section 2]{operatope}; the higher operations are those of Gaiotto, Kulp and Wu \cite{higher}, and the holomorphic supercharge construction is that of \cite{semichiral,BK}.} Intuitively, these are cut Feynman diagrams, a familiar structure in the computation of anomalies \cite{Wang,BK}.

We observe that the current equations and the loop constraints of the operatope \cite[Equations (2.26), (2.31), (2.33)]{operatope} are Ohm's and Kirchhoff's laws of an electrical network: the Schwinger time of each edge becomes a resistance, the holomorphic momenta of the output letters become injected currents, and Wang's weighted Laplacian \cite[Definition 12, Remark 6]{Wang} becomes the conductance matrix.\footnote{The reading of Schwinger parameters as resistances of a network is classical for Feynman integrals \cite{Mathews,PSS}.} Building on this analogy and on the locality of Ohm's law, we use the circuit reductions iteratively, transferring the sources at each step, as changes of variables on Schwinger space. For fans, spiders and cubic product forests, these moves cancel the spanning-tree denominators and reduce the graph to one loop or a tree.\footnote{A single star--triangle move is a classical change of Schwinger variables \cite[Lemma 2.4.23]{Panzer}; here, series, parallel, and star--triangle moves are iterated.} Our circuit iteration interprets and extends the cube-integral recursion of Gui, Wang and Williams \cite[Theorem 4.13]{GWW}: it gives the recursion an explicit physical meaning as series and parallel reductions of resistors, and adds the star--triangle transformations.

The operatope is defined in arbitrary dimension $d$, for all graphs that satisfy the Laman condition together with the corresponding bound on every subgraph. Two families of graphs are of particular interest in the holomorphic twist of four-dimensional $\mathcal N=1$ gauge theories, the fans and the spiders: they give all perturbative corrections to the supersymmetry transformation of composite operators, at arbitrary loop order, in every such theory with a Lagrangian description whose superpotential is treated perturbatively, as a polynomial or a formal power series. They are drawn in Figure \ref{fig-sum-families} and evaluated in Sections \ref{sec-fans} and \ref{sec-forests}. In a theory whose superpotential is at most cubic, with an arbitrary K\"ahler potential, such as $\mathcal N=4$ Yang--Mills theory, every loop correction to the supersymmetry transformation is a fan. We derive the fan and spider kernels in closed form at every loop order. For these graphs, the successive series and parallel reductions, read in reverse, construct the positive edge resistances from dimensionless ratios, the fractions in which a current divides between parallel resistors, and the network reduction gives the kernels in these simplex coordinates.\footnote{In these coordinates, the currents are polynomial and the spanning-tree denominator of the Kirchhoff solution cancels, so every coefficient in the kernel's displacement expansion is a polynomial integral over explicit products of intervals and simplices, hence a finite factorial sum.}

The anomalous supersymmetry transformation\footnote{We use the terms anomalous supersymmetry transformation, loop correction to the supercharge and quantum BV differential interchangeably; strictly they are different notions, but here they refer to the same operator.} violates the Leibniz rule: at one loop it acts on two letters at once, like a second-order differential operator, and at $n$ loops on $n+1$ letters. The holomorphic twist also provides a perturbative quantum product, built from the higher brackets with a pair of auxiliary fields \cite{higher} and analogous to normal ordering in an interacting conformal field theory. When the quantum master equation holds, that is, when the gauge anomaly cancels, the perturbative BV differential obeys the Leibniz rule for this product. We prove that the analogy is exact: the quantum product is cohomologous to the integral over an arbitrary three-sphere against the Bochner--Martinelli form, the propagator. It is associative and graded commutative on $\mathbf Q$-cohomology, but neither on the full BV complex. The quantum product is not needed to compute the cohomology as a vector space; it supplies the ring structure. For example, the test of quantum lifting \cite[Section 1]{semichiral} then needs to be carried out only on the generators of the BPS ring. The explicit form of the product also shows that the multi-gravitons of the $1/16$-BPS sector of $SU(N)$ $\mathcal N=4$ Yang--Mills theory remain quantum cocycles at arbitrary loop order. The quantum product is evaluated by the same kind of graph integrals; it involves more types of kernels than the supersymmetry transformation. From three loops on, the quantum product also contains interaction forests that are neither fans nor spiders; for renormalizable theories, series, parallel, and zero-source star--triangle moves keep their currents polynomial.\footnote{In the original Schwinger times the same integrands are rational, with powers of the spanning-tree polynomial in the denominator \cite[Corollaries 1--2 and proof of Theorem 6]{Wang}, while the operatope coordinates simplify the integrand at the price of an intricate integration region \cite[Sections 2.3 and 2.6]{operatope}. Fans and spiders are Laman graphs of type~I$'$; Gui, Wang and Williams wrote their weights recursively from the $L_\infty$ relations as integrals over unit cubes with polynomial integrands at each order in the displacements \cite[Theorem 4.13]{GWW}, leaving the relation of the cube parameters to Schwinger parameters open \cite[Remark 4.14]{GWW}, and evaluated the fans explicitly through three loops, the last with computer algebra \cite[Sections 4.4--4.6]{GWW}. We supply this relation and its physical interpretation.}

As an explicit application, we give the all-order supercharge of $\mathcal N=4$ Yang--Mills theory in \eqref{eq-sum-N4}. We also improve the enumeration algorithm for fortuitous classes and study perturbative lifting in $\mathcal N=4$ Yang--Mills theory with gauge groups $SU(N)$, $SO(N)$, $Sp(N)$, $G_2$ and $F_4$. The results and a description of the algorithm are collected in the repository \cite{Repo}. With the explicit $Q_2,Q_3,\ldots$ derived here, we test the lifting of classical cohomology classes beyond one loop. While we find no two-loop lift in the $\mathcal N=4$ sectors tested, polarized Pfaffian and Gram-determinant relations produce rationally certified three-loop lifts ($d_3\ne0$) in $SO(11)$, $SO(12)$ and $SO(13)$ $\mathcal N=4$ Yang--Mills theory---both in the zero-derivative BMN sector of $SO(13)$ at level $34$ (and its level-$40$ mixed descendant) and on powers of the defining-representation gaugino bilinear where $d_1=d_2=0$. At one loop the spectra of the orthogonal and symplectic duals agree in the low-level sectors tested. In $SO(11)$ and $SO(13)$, the one-loop Pfaffian family and the three-loop lifts together remove all classical glueball powers absent in $Sp(5)$ and $Sp(6)$, respectively. At higher rank we obtain lower bounds on the order of any perturbative lift. We also find two-loop lifts in the Leigh--Strassler deformation \cite{LeighStrassler} of $SU(3)$ $\mathcal N=4$ Yang--Mills theory, with antichiral superpotential
\begin{equation*}
\bar W=\operatorname{tr}(\gamma^1\gamma^2\gamma^3)-q\operatorname{tr}(\gamma^1\gamma^3\gamma^2)+\frac{h}{3}\left[\operatorname{tr}((\gamma^1)^3)+\operatorname{tr}((\gamma^2)^3)+\operatorname{tr}((\gamma^3)^3)\right].
\end{equation*}
We find two-, three- and four-loop lifts in abelian theories of SQED type, and two- and three-loop lifts in $\mathcal N=2$ $SU(2)$ and $SU(3)$ gauge theories and their singlet deformations.

Section~\ref{sec-summary} summarizes our results: the all-order supercharge of $\mathcal N=4$ Yang--Mills theory, the theorems on quantum lifting, on multi-gravitons and on the quantum BPS ring, the census of fortuitous classes and of their lifts at one loop and beyond, and the circuit picture behind the kernels. As a by-product of the census, we find the threshold fortuitous class of $SU(4)$, at $2(E+J_L)=28$. Section~\ref{sec-BV} reviews the holomorphic twist of the physical theory, the higher brackets and their reduction to a current integral, the starting point of our computations, for readers who have not met them. Section~\ref{sec-restrictions} proves constraints on the contributing graphs in four dimensions and classifies the graphs of $Q_n$. Section~\ref{sec-fans} computes the fan kernel, the kernel most used in practice, together with the circuit reductions on which it rests. Section~\ref{sec-product} treats the quantum product and its Leibniz rule, with explicit products through two loops. Section~\ref{sec-forests} gives the further kernels, those of spiders and cubic forests, which enter the quantum product and theories with superpotentials of higher degree. Section~\ref{sec-supercharge} writes the $n$-loop supercharge $Q_n$ of $\mathcal N=4$ Yang--Mills theory as a differential operator of order $n+1$. Section~\ref{sec-lifting} proves that the multi-gravitons remain closed under the full supercharge and sets up the successive homotopy transfer on which the practical computation of the cohomology ring rests. Appendix~\ref{sec-notation} collects the $\mathcal N=4$ superfield and the charge sectors and gradings used in the lifting tests. Appendix~\ref{app-twist-cohomology} proves that the cohomology of the twisted complex is the $Q$-cohomology of the physical operators. Appendix~\ref{sec-B-spanning} recalls the global spanning-tree solution of the circuit for comparison with our local reductions. Appendix~\ref{app-covariant-derivatives} shows that superspace covariant derivatives in the superpotential decouple into a $Q_-$-exact full-superspace term and leave the twisted operations unchanged, and Appendix~\ref{app-sourced-star} records the sourced star--triangle reduction for Laman graphs.

\section{Summary of results}\label{sec-summary}

This section states our principal results together with the definitions needed to read them independently; Sections~\ref{sec-BV}--\ref{sec-lifting} supply the full constructions and proofs.

We work in the holomorphic twist of a four-dimensional $\mathcal N=1$ supersymmetric gauge theory on $\mathbb R^4\cong\mathbb C^2$, reviewed in Section~\ref{sec-BV}. In Euclidean spinor coordinates $x^{a\dot a}$ ($a=+,-$ and $\dot a=\dot+,\dot-$), the coordinates $x^{+\dot a}$ are holomorphic and $x^{-\dot a}$ are antiholomorphic, with
\begin{equation}\label{eq-sum-spinor-conventions}
\frac{\partial}{\partial x^{a\dot a}}=-\frac12\partial_{a\dot a},\qquad
\epsilon^{\dot+\dot-}=1=-\epsilon_{\dot+\dot-},\qquad
\bar\partial=dx^{-\dot a}\frac{\partial}{\partial x^{-\dot a}},\qquad
\Omega=2\,dx^{+\dot+}\wedge dx^{+\dot-},
\end{equation}
and dotted indices raised and lowered by $\partial_+{}^{\dot a}=\epsilon^{\dot a\dot b}\partial_{+\dot b}$ and $z^+{}_{\dot b}=\epsilon_{\dot b\dot c}z^{+\dot c}$. Let $\mathfrak g$ be a compact Lie algebra with generators $[t_B,t_C]=iC^A{}_{BC}t_A$, invariant bilinear form $\delta_{AB}$ setting $C_{ABC}=\delta_{AD}C^D{}_{BC}=C_{[ABC]}$, gauge coupling $g$, and matter representation matrices $(t_A)^j{}_i$.

Selecting one nilpotent supercharge $Q_-$ in the physical Batalin--Vilkovisky (BV) complex and eliminating contractible field pairs leaves four Dolbeault forms in $dx^{-\dot a}$ on $\mathbb C^2$: the odd ghost form $c^A$ (ghost number $1$), the even antifield form $b_A$ (ghost number $-2$), the even matter form $\gamma_i$ (ghost number $0$), and the odd antifield form $\beta^i$ (ghost number $-1$). The BV pairing pairs $c^A$ with $b_A$ and $\gamma_i$ with $\beta^i$. Writing a superscript $(0)$ for the $dx^{-\dot a}=0$ zero-form part, the zero-form fields and their holomorphic derivatives $\partial_{+\dot a}$ represent the physical $Q_-$-cohomology letters (the BPS letters of \cite[Section 2]{ChangYin}): $\gamma_i^{(0)}=\bar\phi_i$ is the antichiral scalar, $\beta^{(0)i}=-\tfrac i{\sqrt2}\psi_+^i+\bar{\mathcal F}_{\mathrm{old}}^{\ddagger i}$ and $b_A^{(0)}=\tfrac14\delta_{AB}f_{++}^B-\tfrac i2\lambda_{A,\mathrm{old}}^{\ddagger-}$ are the BV-completed chiral fermion and self-dual field strength, $\tfrac12\partial_{+\dot a}c^{A(0)}$ represents the antichiral gaugino $\bar\lambda_{\dot a}^A$, and $\partial_{+\dot a}$ represents the symmetrized covariant derivative $\nabla_{+\dot a}$ (Proposition~\ref{b-00159} and \eqref{eq-B-14}). Physical local operators at an insertion point $x$ are $\mathfrak g$-invariant polynomials in these zero-form letters and their holomorphic derivatives, independent of the undifferentiated ghost $c^{A(0)}(x)=c_{\mathrm{gh}}^A(x)$. These are the standard steps of the holomorphic twist, which we review in detail in Section~\ref{sec-BV}.

Each theory is specified by its gauge algebra $\mathfrak g$, its matter representation and its gauge-invariant antichiral potential $\bar W(\gamma,w)$, which enter the twisted action
\begin{equation}\label{local-ch9-common-action}
\begin{aligned}
S_{\mathrm{tw}}=\int_{\mathbb C^2}\Omega\wedge\Big[
&b_A\left(\bar\partial c^A-\frac g2C^A{}_{BC}c^Bc^C\right)\\
&+\left(\bar\partial\gamma_i-igc^A(t_A)^j{}_i\gamma_j\right)\beta^i
+\bar W(\gamma,w)\Big]_{(0,2)},
\end{aligned}
\end{equation}
where $[\cdot]_{(0,2)}$ selects antiholomorphic form degree two and $w_{\dot a}^A=\tfrac12\partial_{+\dot a}c^A$ is the image of the antichiral field-strength superfield under homotopy transfer \eqref{footnote-16--eq3.4}. Because $\partial_{+\dot a}\partial_+{}^{\dot a}=0$, a quadratic gauge kinetic term with constant coupling, $-\tfrac14\delta_{AB}w_{\dot a}^Aw^{B\dot a}=-\tfrac1{16}\delta_{AB}\partial_{+\dot a}(c^A\partial_+{}^{\dot a}c^B)$, is a holomorphic total derivative \eqref{b-00931} and drops out of $S_{\mathrm{tw}}$; gauge kinetic terms contribute to $\bar W(\gamma,w)$ only when $w$ multiplies matter fields $\gamma$. Superspace covariant derivatives in the superpotential likewise reduce to \eqref{local-ch9-common-action} (Appendix~\ref{app-covariant-derivatives}).

Splitting $S_{\mathrm{tw}}$ into its free $\bar\partial$ kinetic term and integrated interaction $I=\int_{\mathbb C^2}\Omega\wedge[\mathcal I]_{(0,2)}$, the interaction density is
\begin{equation}\label{eq-sum-interaction-density}
\mathcal I=-\frac g2C^A{}_{BC}b_Ac^Bc^C
-igc^A(t_A)^j{}_i\gamma_j\beta^i
+\bar W(\gamma,w).
\end{equation}
In heat-kernel regularization on $\mathbb C^2$ \cite{Costello,Wang}, all Feynman graph integrals of $S_{\mathrm{tw}}$ are ultraviolet finite, and the BV Ward identity at colliding vertices defines the $k$-ary higher brackets $\{\mathcal O_1,\ldots,\mathcal O_k\}$ \cite{semichiral,operatope,higher,BK}: signed sums over connected Feynman graphs with $k-1$ vertices integrated against $\Omega$, one edge carrying the ultraviolet heat kernel $K_\varepsilon$ as $\varepsilon\downarrow 0$, and all remaining edges carrying the Bochner--Martinelli propagator $P_{\varepsilon,L}$---which in total reduces to an integral over the ultraviolet boundary of Schwinger-parameter space, the worldline moduli space of the graph (Definition~\ref{b-00151} and Section~\ref{positive-schwinger-times-and-the-operatope}). With $\hbar$ counting loops, the full quantum BV differential on holomorphic local operators is
\begin{equation}\label{eq-sum-Q-expansion}
\mathbf Q=\sum_{\substack{n\ge0\\\text{loop orders}}}\hbar^nQ_n,\qquad
Q_n\mathcal O=\frac{1}{(n+1)!}\bigl\{\underbrace{\mathcal I,\ldots,\mathcal I}_{n+1},\mathcal O\bigr\}.
\end{equation}
At tree level ($n=0$), $Q_0\mathcal O=\{\mathcal I,\mathcal O\}$ acts as a graded derivation on the zero-form fields (written without the $(0)$ superscript below) by $[Q_0,\partial_{+\dot a}]=0$ and \eqref{eq-B-15}:
\begin{equation}\label{eq-sum-Q0-explicit}
\begin{aligned}
Q_0c^A&=-\frac g2 C^A{}_{BC}c^Bc^C,&
Q_0\gamma_i&=-igc^A(t_A)^j{}_i\gamma_j,\\
Q_0\beta^i&=igc^A(t_A)^i{}_j\beta^j-\frac{\partial\bar W(\gamma,0)}{\partial\gamma_i},&
Q_0b_A&=-gC^B{}_{AC}c^Cb_B-ig(t_A)^j{}_i\gamma_j\beta^i.
\end{aligned}
\end{equation}
Classically, $\mathrm{BPS}_{\hbar^0}=H(Q_0)$ on physical local operators is isomorphic to the $Q_-$-cohomology of gauge-invariant operators modulo the equations of motion (Proposition~\ref{prop-twist-cohomology}).

We present, in turn, the explicit all-order supercharge $Q_n$ of $\mathcal N=4$ Yang--Mills theory (Section~\ref{the-all-order-supercharge-of-mathcal-n4-yangmills-theory}), the general theorems on quantum lifting, multi-graviton protection and the quantum BPS ring (Section~\ref{the-quantum-product-and-the-quantum-bps-ring}), the computer census of fortuitous classes and of one-, two-, three- and four-loop lifts across $\mathcal N=4$, $\mathcal N=2$ and $\mathcal N=1$ gauge theories (Section~\ref{sec-applications}), and the electrical-network reduction that evaluates the fan, spider and cubic product-forest kernels in polynomial form at all loop orders (Section~\ref{sec-sum-circuit}).

\subsection{The all-order supercharge of \texorpdfstring{$\mathcal N=4$}{N=4} Yang--Mills theory}\label{the-all-order-supercharge-of-mathcal-n4-yangmills-theory}

In $\mathcal N=4$ Yang--Mills theory with gauge algebra $\mathfrak g$, the three adjoint matter multiplets $(\gamma^I,\beta_I)$ ($I=1,2,3$) combine with $(c,b)$ into one odd superfield in three Grassmann coordinates $\theta_I$ \cite{ChangYin,BK}:
\begin{equation*}
C^A(\theta)=ig\,c^A+\theta_I\delta^{AB}(\gamma^I)_B-\frac12\varepsilon^{IJK}\theta_I\theta_J(\beta_K)^A+\frac{\theta_1\theta_2\theta_3}{ig}\delta^{AB}b_B.
\end{equation*}
The odd coordinates obey $\theta_I\bar\partial=-\bar\partial\theta_I$, $\int d^3\theta\,\theta_1\theta_2\theta_3=1$ and $\varepsilon^{123}=\varepsilon_{123}=1$. The twisted action \eqref{local-ch9-common-action} becomes cubic, with the structure constants $C_{ABC}$ as couplings,
\begin{equation}\label{local-ch9-su-action}
S_{\mathrm{tw}}=\int_{\mathbb C^2}\Omega\wedge
\left[\int d^3\theta\left(\frac12\delta_{AB}C^A\bar\partial C^B
+\frac{i}{3!}C_{ABC}C^AC^BC^C\right)\right]_{(0,2)},
\qquad Q_0C^A=\frac i2C^A{}_{BC}C^BC^C.
\end{equation}
Functional derivatives with respect to the zero-form superfield at displaced points are normalized by \eqref{eq-2-10}:
\begin{equation}\label{eq-sum-deltaC-norm}
\frac{\delta C^B(y,\theta')}{\delta C^A(y',\theta)}
=\delta_A{}^B\,\delta^{(4)}(y-y')\prod_{I=1}^3(\theta_I-\theta'_I),
\end{equation}
where $\int d^4z_j\,\delta^{(4)}(z_j)\,f(z_j^{+\dot a})=f(0)$ extracts holomorphic Taylor coefficients at the insertion point $x$ via \eqref{eq-sum-N4-letters}.

Because every interaction vertex in \eqref{local-ch9-su-action} is cubic, every connected $n$-loop graph contributing to $Q_n\mathcal O$ ($n\ge 1$) is a \emph{fan} (Corollary~\ref{b-00222}): the $n+1$ interaction vertices form a path $1\to 2\to\cdots\to n+1$, each joined by one spoke to the insertion $\mathcal O$ at vertex $n+2$, leaving one surviving output superfield at each endpoint $1$ and $n+1$. Evaluating the $n$-loop fan kernel in current-division coordinates (Theorem~\ref{thm-sum-all-kernels} below and Section~\ref{sec-fans}) yields $Q_n$ in closed form: $Q_n$ replaces $n+1$ input letters at $x+z_j$ by two differentiated output letters.\footnote{At one loop, $Q_1$ reproduces the operator of Budzik and Kulp \cite[Equations (3.64)--(3.70)]{BK} up to normalization; its component \eqref{eq-2-26} is the Konishi anomaly.}

\rthm{Theorem}{b-00088}{All-loop supercharge of $\mathcal N=4$ Yang--Mills theory} \emph{For the cubic superfield action \eqref{local-ch9-su-action} and a holomorphic local operator inserted at $x$, the $n$-loop supercharge $Q_n$ ($n\ge1$) is}
\begin{equation}\label{eq-sum-N4}
\begin{aligned}
Q_n={}&\frac{(-1)^ni^{n+1}}{2(32\pi^2)^n}
\int dr_1\cdots dr_{2n}\,
\left[\prod_{j=1}^{n-1}r_{2j-1}^{n-j}r_{2j}(1-r_{2j})^{j-1}\right]r_{2n}^{n-1}
\int d^3\theta\int\prod_{j=1}^{n+1}d^4z_j\\
&\times C_{A_1DR_1}
\left(\prod_{j=2}^{n}C_{A_jR_{j-1}R_j}\right)C_{A_{n+1}R_nE}\\
&\times\big(\partial_{+\dot a_1}\cdots\partial_{+\dot a_n}C^D\big)(x+x_1(r,z),\theta)
\big(\partial_+{}^{\dot a_1}\cdots\partial_+{}^{\dot a_n}C^E\big)(x+x_{n+1}(r,z),\theta)\\
&\times\frac{\delta}{\delta C^{A_1}(x+z_1,\theta)}\cdots
\frac{\delta}{\delta C^{A_{n+1}}(x+z_{n+1},\theta)},
\end{aligned}
\end{equation}
\emph{integrated over the product of $2n-2$ unit intervals and a two-simplex,}
\begin{equation}\label{eq-sum-integration-domain}
0<r_j<1\quad(1\le j\le2n-2),\qquad r_{2n-1},r_{2n}>0,\qquad r_{2n-1}+r_{2n}<1,
\end{equation}
\emph{where the displayed derivatives act on the output fields before evaluation at the translated positions}
{\footnotesize
\begin{equation}\label{eq-sum-N4-points}
\begin{aligned}
x_1^{+\dot a}(r,z)
={}&\sum_{\substack{j=1\\\text{elimination steps}}}^{n}(1-r_{2j-1})
\left(\prod_{k=1}^{j-1}r_{2k-1}\right)\\
&\times\left[
\left(\prod_{k=1}^{j-1}(1-r_{2k})\right)z_1^{+\dot a}
+\sum_{\substack{i=2\\\text{path vertices}}}^{j}r_{2i-2}
\left(\prod_{k=i}^{j-1}(1-r_{2k})\right)z_i^{+\dot a}
\right]
+\left(\prod_{k=1}^{n}r_{2k-1}\right)z_{n+1}^{+\dot a},\\
x_{n+1}^{+\dot a}(r,z)
={}&r_{2n}\left[
\left(\prod_{k=1}^{n-1}(1-r_{2k})\right)z_1^{+\dot a}
+\sum_{\substack{i=2\\\text{path vertices}}}^{n}r_{2i-2}
\left(\prod_{k=i}^{n-1}(1-r_{2k})\right)z_i^{+\dot a}
\right]
+(1-r_{2n})z_{n+1}^{+\dot a}.
\end{aligned}
\end{equation}
}

Each coordinate $r_j$ is a parallel current-division ratio in the circuit reduction of Section~\ref{sec-sum-circuit}: for $1\le i<n$, $r_{2i-1}$ is the fraction of the source current at vertex $i$ passed along the path to $i+1$, $r_{2i}$ is the fraction of the merged spoke current carried by the original spoke $(i+1,n+2)$, and $r_{2n-1},r_{2n}$ are the fractions of the sources at $n$ and $n+1$ flowing through the final edge $(n,n+1)$.

\paragraph{Taylor expansion.}

Expanding \eqref{eq-sum-N4} in the displacements $z_1,\ldots,z_{n+1}$ produces a polynomial in $r_1,\ldots,r_{2n}$ at every total derivative order $m$, and term-by-term integration yields rational coefficients. Using $\tfrac{\partial}{\partial x^{+\dot b}}=-\tfrac12\partial_{+\dot b}$ so that $z^{+\dot b}\tfrac{\partial}{\partial x^{+\dot b}}=\tfrac12z^{+}{}_{\dot b}\partial_+{}^{\dot b}$, and writing $\partial_{1,+}{}^{\dot a},\partial_{n+1,+}{}^{\dot a}$ for derivatives acting on the first and second output factors at $(x,\theta)$, we obtain
\begin{equation}\label{eq-sum-N4-taylor}
\begin{aligned}
Q_n={}&\frac{(-1)^ni^{n+1}}{2(32\pi^2)^n}
\sum_{\substack{m=0\\\text{input derivative order}}}^{\infty}
\sum_{\substack{r_0,s_0,r_{ik},s_i\ge 0\\
r_0+s_0+\sum_{i=1}^n s_i+\sum_{1\le i\le k\le n}r_{ik}=m\\\text{moment partitions}}}
c^{(n)}_{r_0,s_0,\{r_{ik}\},\{s_i\}}
\int d^3\theta\int\prod_{j=1}^{n+1}d^4z_j\\
&\times C_{A_1DR_1}\left(\prod_{j=2}^nC_{A_jR_{j-1}R_j}\right)C_{A_{n+1}R_nE}\\
&\times\frac{\bigl(\tfrac12 z_1{}^{+}{}_{\dot b}\,\partial_{1,+}{}^{\dot b}\bigr)^{r_0}
\bigl(\tfrac12 z_{n+1}{}^{+}{}_{\dot c}\,\partial_{n+1,+}{}^{\dot c}\bigr)^{s_0}}{r_0!\,s_0!}
\prod_{1\le i\le k\le n}
\frac{\bigl[\tfrac12(z_{i+1}-z_i)^{+}{}_{\dot d}\,\partial_{1,+}{}^{\dot d}\bigr]^{r_{ik}}}{r_{ik}!}\\
&\times\prod_{i=1}^n
\frac{\bigl[\tfrac12(z_i-z_{i+1})^{+}{}_{\dot e}\,\partial_{n+1,+}{}^{\dot e}\bigr]^{s_i}}{s_i!}
\,\big(\partial_{+\dot a_1}\cdots\partial_{+\dot a_n}C^D\big)(x,\theta)
\big(\partial_+{}^{\dot a_1}\cdots\partial_+{}^{\dot a_n}C^E\big)(x,\theta)\\
&\times\frac{\delta}{\delta C^{A_1}(x+z_1,\theta)}\cdots
\frac{\delta}{\delta C^{A_{n+1}}(x+z_{n+1},\theta)},
\end{aligned}
\end{equation}
with the explicit rational factorial ratio
\begin{equation}\label{eq-sum-N4-coeff}
\begin{aligned}
c^{(n)}_{r_0,s_0,\{r_{ik}\},\{s_i\}}
={}&\frac{\bigl(\sum_{i=1}^n r_{in}\bigr)!\,\bigl(n-1+\sum_{i=1}^n s_i\bigr)!}
{\bigl(n+1+\sum_{i=1}^n r_{in}+\sum_{i=1}^n s_i\bigr)!}
\prod_{j=1}^{n-1}\frac{1}{n-j+1+\sum_{k=j}^n\sum_{i=1}^k r_{ik}}\\
&\times\prod_{j=1}^{n-1}
\frac{\bigl(1+\sum_{i=1}^j r_{ij}\bigr)!\,
\bigl(j-1+\sum_{i=1}^j\bigl[s_i+\sum_{k=j+1}^n r_{ik}\bigr]\bigr)!}
{\bigl(j+1+\sum_{i=1}^j\bigl[s_i+\sum_{k=j}^n r_{ik}\bigr]\bigr)!}.
\end{aligned}
\end{equation}

On a polynomial local operator, each displacement monomial differentiates with respect to a holomorphic-derivative letter,
\begin{equation}\label{eq-sum-N4-letters}
\int d^4z_j\,
\frac{\bigl(-\tfrac12 z_j^{+\dot+}\bigr)^{p_j}\bigl(-\tfrac12 z_j^{+\dot-}\bigr)^{q_j}}{p_j!\,q_j!}
\frac{\delta}{\delta C^{A_j}(x+z_j,\theta)}
=\frac{\delta}{\delta\bigl(\partial_{+\dot+}^{p_j}\partial_{+\dot-}^{q_j}C^{A_j}\bigr)(x,\theta)},
\end{equation}
so only finitely many orders $m$ act on any fixed operator. On letters without holomorphic derivatives ($m=0$), $c^{(n)}_{0,0,\{0\},\{0\}}=1/[(n!)^3(n+1)]$, giving
\begin{equation}\label{eq-sum-Qnzero}
\begin{aligned}
Q_n\big|_{\text{no input derivatives}}
={}&\frac{(-1)^ni^{n+1}}{2(32\pi^2)^n(n!)^3(n+1)}
\int d^3\theta\,
C_{A_1DR_1}\left(\prod_{j=2}^nC_{A_jR_{j-1}R_j}\right)
C_{A_{n+1}R_nE}\\
&\times(\partial_{+\dot a_1}\cdots\partial_{+\dot a_n}C^D)
(\partial_+{}^{\dot a_1}\cdots\partial_+{}^{\dot a_n}C^E)
\prod_{j=1}^{n+1}\frac{\delta}{\delta C^{A_j}(\theta)}.
\end{aligned}
\end{equation}

The proof of Theorem~\ref{b-00088} and explicit one-, two- and three-loop expressions and component actions are given in Section~\ref{sec-supercharge}. The same fan kernel evaluates $Q_n$ in every four-dimensional $\mathcal N=1$ gauge theory whose interaction \eqref{eq-sum-interaction-density} has degree at most three; for higher-degree superpotentials $\bar W$, the remaining graphs are spiders with a single $\bar W$ vertex, evaluated in closed form in Theorem~\ref{thm-sum-all-kernels} and Section~\ref{sec-forests}.

\subsection{Quantum lifting, multi-gravitons and the quantum BPS ring}\label{the-quantum-product-and-the-quantum-bps-ring}

A formal series $\mathcal O=\sum_{\substack{s\ge0\\\text{loop orders}}}\hbar^s\mathcal O_s$ is a perturbative quantum cocycle ($\mathbf Q\mathcal O=0$) when its coefficients solve $\sum_{\substack{j=0\\\text{loop orders}}}^{s}Q_j\mathcal O_{s-j}=0$ for every $s\ge0$; we will show that whenever the leading term $\mathcal O_0$ is a polynomial local operator in the original fields and their derivatives, such a completion can always be chosen to be a finite sum in $\hbar$ (Proposition~\ref{ch9-finiteness} and \eqref{local-ch8-finite-correction-degree}). Cochains are graded by $k=\mathrm{gh}-2J_L$, where $\mathrm{gh}$ is the BV ghost number and $J_L$ is the undotted spin; every $Q_n$ raises $k$ by one (Section~\ref{sec-homotopy-transfer}). Starting from the classical BPS cohomology $\mathrm{BPS}_{\hbar^0}=H(Q_0)$, successive homotopy transfer (Proposition~\ref{prop-ch8-transfer-validity}) defines the spaces $\mathrm{BPS}_{\hbar^r}=H(\mathrm{BPS}_{\hbar^{r-1}},d_r)$ ($r\ge1$), where $[\mathcal O_0]_{\hbar^{r-1}}$ denotes the class in $\mathrm{BPS}_{\hbar^{r-1}}$, $d_1[\mathcal O_0]_{\hbar^0}=[Q_1\mathcal O_0]_{\hbar^0}$, $d_2[\mathcal O_0]_{\hbar}=[Q_2\mathcal O_0+Q_1\mathcal O_1]_{\hbar}$ for any $\mathcal O_1$ solving $Q_0\mathcal O_1=-Q_1\mathcal O_0$, and in general $d_r[\mathcal O_0]_{\hbar^{r-1}}=[R_r]_{\hbar^{r-1}}$ in terms of the order-$r$ residual $R_r=\sum_{\substack{j=1\\\text{loop orders}}}^{r}Q_j\mathcal O_{r-j}$ once the lower equations $Q_0\mathcal O_s=-R_s$ ($1\le s<r$) are solved. Thus $d_r[\mathcal O_0]_{\hbar^{r-1}}=0$ if and only if $R_r\in\operatorname{im}Q_0$ for a suitable choice of $\mathcal O_1,\ldots,\mathcal O_{r-1}$. A classical class is \emph{lifted} at $r$ loops when either $d_r$ is nonzero on it or it is $d_r$-exact; the source and target classes of a nonzero $d_r$ leave the perturbative BPS spectrum together.

\rthm{Lemma}{b-00094}{Derivative grading and $d_r$-exactness bound} \emph{For an interaction without holomorphic derivatives, $Q_n$ raises both holomorphic derivative totals---the numbers $n_+$ and $n_-$ of $\partial_{+\dot+}$ and $\partial_{+\dot-}$ derivatives---by $n$ \cite[Theorem 6]{Wang}, \cite[Section 2.8]{operatope}.} \emph{Choosing the transfer homotopies homogeneous in $(n_+,n_-)$, the obstruction differential $d_r$ raises both totals by $r$ (Proposition~\ref{appE2-0024}); consequently, a nonzero class in $\mathrm{BPS}_{\hbar^{r-1}}$ can be $d_r$-exact only if $\min(n_+,n_-)\ge r$ on its leading representative.} \emph{In particular, a class whose leading representative has $\min(n_+,n_-)=0$ is $d_r$-exact for no $r\ge1$.}

The same derivative shift holds after eliminating a massive neutral multiplet, as in the massive SQED lift of Section~\ref{sec-results-SQED}.

\rthm{Theorem}{b-00096}{Finite loop order; Corollary~\ref{b-00235}} \emph{For the twisted interaction \eqref{eq-sum-interaction-density} with $\bar W$ a polynomial or formal power series of arbitrary degree, and every local-operator monomial $\mathcal O$,}
\begin{equation}\label{eq-sum-finite}
Q_n\mathcal O=0\qquad\text{for }n>\#(b,\beta\text{ in }\mathcal O).
\end{equation}

By the quadratic $L_\infty$ identities of the higher brackets \cite[Section 3.2]{operatope}, \cite[Theorem 7]{Wang}, the quantum differential satisfies $\mathbf Q^2=0$ whenever the interaction obeys the quantum master equation $\sum_{\substack{n\ge0\\\text{loop orders}}}\tfrac{\hbar^n}{n+2}Q_n\mathcal I=0$ (Lemma~\ref{b-00452}). For the interaction \eqref{eq-sum-interaction-density} with $\mathfrak g$-invariant $\bar W$, Theorem~\ref{b-00465} evaluates $Q_n\mathcal I$ at all loop orders:
\begin{equation}\label{eq-sum-QI}
Q_0\mathcal I=0,\qquad
Q_1\mathcal I=\frac{ig^3}{128\pi^2}
\operatorname{Tr}_R\!\left(t_A\{t_B,t_C\}\right)
c^A(\partial_{+\dot a}c^B)
(\partial_+{}^{\dot a}c^C),\qquad
Q_{n\ge2}\mathcal I=0.
\end{equation}
Cancellation of the one-loop cubic gauge anomaly, $\operatorname{Tr}_R(t_A\{t_B,t_C\})=0$, is therefore necessary and sufficient for $\mathcal I$ to satisfy the quantum master equation at all loop orders, ensuring $\mathbf Q^2=0$.

\paragraph{The quantum product.}

Because $Q_n$ ($n\ge1$) differentiates $n+1$ input letters simultaneously, it violates the Leibniz rule on ordinary products $AB$. Compatibility with $\mathbf Q$ is restored by the perturbative quantum product $A\star B=AB+O(\hbar)$, constructed from the marked higher operations of Gaiotto, Kulp and Wu \cite[Sections 4.3 and 4.5]{higher} by dressing $A,B$ with a decoupled free auxiliary pair $(a,b_{\mathrm{aux}})$ and retaining only graphs in which $(a,b_{\mathrm{aux}})$ contracts along a single marked edge $\gamma$ between the last two slots, evaluated at zero relative displacement (Section~\ref{sec-dressed-chain}):
\begin{equation}\label{eq-sum-star}
A\star B=\sum_{\substack{n\ge0\\\text{loop orders}}}\frac{\hbar^n}{n!}
\bigl\{\underbrace{\mathcal I,\ldots,\mathcal I}_{n},A,B\bigr\}^{\gamma,0}
=\left.\sum_{\substack{n\ge0\\\text{loop orders}}}\frac{\hbar^n}{n!}
\bigl\{\underbrace{\mathcal I,\ldots,\mathcal I}_{n},Aa,b_{\mathrm{aux}}B\bigr\}\right|_{a=b_{\mathrm{aux}}=0}.
\end{equation}
In slot notation for $\{\mathcal O_1,\ldots,\mathcal O_{n+2}\}$, the $n$ copies of $\mathcal I$ occupy slots $1,\ldots,n$ and $A,B$ occupy slots $n+1,n+2$, so the marked edge is $\gamma=\{(n+1,n+2)\}$. On $\hbar$-independent polynomial inputs, the order-$n$ term vanishes once $n>\#(b,\beta\text{ in }A)+\#(b,\beta\text{ in }B)$ (Corollary~\ref{local-ch8-finite-product-corollary}).

\rthm{Theorem}{b-00106}{Quantum BPS ring; Theorems~\ref{thm-leibniz-rule}, \ref{b-00513} and \ref{b-00598}} \emph{Suppose that the interaction $\mathcal I$ satisfies the quantum master equation.}

(i) \emph{For homogeneous operators $A,B$, with $|A|$ the fermion number of $A$, $\mathbf Q$ is a graded derivation of $\star$:}
\[
\mathbf Q(A\star B)=(\mathbf QA)\star B+(-1)^{|A|}A\star(\mathbf QB).
\]

(ii) \emph{On $H(\mathbf Q)$, the product $\star$ induces an associative, graded-commutative multiplication with strict cochain-level unit $1\star A=A\star 1=A$.\footnote{On cochains, associativity and graded commutativity hold up to explicit $\mathbf Q$-exact homotopies \eqref{eq-ring-common-root-homotopy} and \eqref{eq-ring-scalar-associator}.}}

(iii) \emph{On the spaces $\mathrm{BPS}_{\hbar^0}=H(Q_0)$ and $\mathrm{BPS}_{\hbar^r}=H(\mathrm{BPS}_{\hbar^{r-1}},d_r)$ ($r\ge1$) of successive homotopy transfer, $\star$ induces the ordinary product of $\hbar^0$ leading representatives, and every differential $d_r$ is a graded derivation of that ring.}

(iv) \emph{For homogeneous quantum cocycles $A,B$ and every infrared heat-kernel cutoff $L>0$, $(A\star B)[L]$ is cohomologous in the scale-$L$ effective-observable complex to the quantum normal-ordering product of \cite[Section 2.3]{semichiral}---the Bochner--Martinelli $S^3$ contour integral \eqref{eq-contour-product-definition} of the scale-$L$ two-insertion expansion of $A(x)$ around $B(0)$.}\footnote{The marked operation is defined in \cite[Sections 4.3 and 4.5]{higher}. Part~(i) follows from the one-mark quadratic identities \cite[Section 3.2]{operatope}, \cite[Theorem 7]{Wang}, \cite[Equations (4.19), (4.42)]{higher}; a ring structure for contour normal ordering was anticipated in \cite[Section 2.2 and footnote 28]{semichiral}, and the two-mark associator topology appears in \cite[Figure 7]{higher}. The all-order common-root homotopy, strict unit, transfer-derivation property, and contour comparison (iv) are proved in Section~\ref{sec-product}.}

Each space $\mathrm{BPS}_{\hbar^r}=H(\mathrm{BPS}_{\hbar^{r-1}},d_r)$ is therefore a subquotient ring of the classical BPS ring $\mathrm{BPS}_{\hbar^0}=H(Q_0)$, while $H(\mathbf Q)$ retains the full $\hbar$-completed multiplication. Beginning with the classical product,
\begin{equation}\label{eq-sum-product-expansion}
A\star B=AB+\hbar\,\{\mathcal I,A,B\}^{\{(2,3)\},0}+\frac{\hbar^2}{2}\{\mathcal I,\mathcal I,A,B\}^{\{(3,4)\},0}+O(\hbar^3),
\end{equation}
the one-loop correction in $\mathcal N=4$ Yang--Mills theory for $B$ at the origin (and general position $x$ in \eqref{eq-2-34}) is
\begin{equation}\label{eq-sum-product1}
\begin{aligned}
\{\mathcal I,A,B\}^{\{(2,3)\},0}
={}&\frac{i}{32\pi^2}\int_0^1dr_1\int_0^{1-r_1}dr_2
\int d^3\theta\,C_{XYZ}\int d^4z_1\,d^4z_2\\
&\times(\partial_{+\dot a}C^X)\big(r_1z_1+(1-r_1)z_2,\theta\big)\\
&\times\left.
\partial_{2,+}{}^{\dot a}
\left(\frac{\delta A(x_2)}{\delta C^Y(x_2+z_1,\theta)}\right)
\right|_{x_2=r_2(z_2-z_1)}
\frac{\delta B(0)}{\delta C^Z(z_2,\theta)},
\end{aligned}
\end{equation}
where $\partial_{2,+}{}^{\dot a}=\epsilon^{\dot a\dot b}\partial_{2,+\dot b}$ differentiates $A(x_2)$ with respect to its insertion coordinate $x_2^{+\dot b}$ before setting $x_2=r_2(z_2-z_1)$. The complete two-loop product in $\mathcal N=4$ Yang--Mills theory is given in \eqref{eq-product2A}. At $n$ loops, the contributing graphs of $\{\mathcal I,\ldots,\mathcal I,A,B\}^{\gamma,0}$ are marked product forests (Theorem~\ref{thm-sum-all-kernels}); their kernels and Taylor tables through seven loops are archived in \cite{KernelRepo}.

\paragraph{Monotone classes, fortuitous classes and uses of the quantum product.}\label{uses-of-the-quantum-product}

Following the distinction of Chang and Lin \cite[Section 2.3]{ChangLinCover} (see also \cite{microstates,ChangChevalley}), a classical class in $\mathrm{BPS}_{\hbar^0}=H(Q_0)$ for a matrix gauge group family ($SU(N)$, $SO(N)$, $Sp(N)$) is \emph{monotone} if it lies in the image of the large-rank stable cohomology, where no finite-rank trace identity holds and $Q_0$-closure follows from universal Lie-algebra identities alone; in $\mathcal N=4$ Yang--Mills theory the multi-graviton classes are monotone. A class is \emph{fortuitous} otherwise: every representative $\mathcal O_0$ satisfies $Q_0\mathcal O_0=0$ by virtue of finite-rank trace relations (such as Cayley--Hamilton identities) that fail above a threshold rank.

Writing $\mathcal O[\mathcal O_0]=\mathcal O_0+\hbar\mathcal O_1+\cdots$ for a closed completion ($\mathbf Q\mathcal O[\mathcal O_0]=0$), Theorem~\ref{b-00106} yields five structural consequences:

(a) \emph{Multi-gravitons remain quantum cocycles at all loop orders (Theorem~\ref{b-00630}).} In $\mathcal N=4$ Yang--Mills theory, the single-graviton operators $G$ are the $\theta$-components of $\operatorname{Tr}(dC)^p$ ($p\ge2$) \cite[Section 4]{ChangYin}, where $dC=dx^{+\dot a}\tfrac{\partial C}{\partial x^{+\dot a}}+d\theta_I\tfrac{\partial_L C}{\partial\theta_I}$. Odd-coordinate counting gives $Q_nG=0$ for every $n\ge0$ \eqref{eq-2-60}. Hence any classical product $G_1\cdots G_m$ of single gravitons admits an explicit closed completion given by their iterated quantum product, under any bracketing:
\begin{equation}\label{eq-sum-multigraviton}
\mathcal O=G_1\star\cdots\star G_m,\qquad
\mathbf Q\mathcal O=0,\qquad
\left.\mathcal O\right|_{\hbar=0}=G_1\cdots G_m.
\end{equation}
With $\deg_\theta$ equal to $0,1,2,3$ on $c,\gamma,\beta,b$, each loop order lowers $\deg_\theta$ by $3$, so $\mathcal O$ terminates once $3n>\deg_\theta(G_1\cdots G_m)$.

(b) \emph{Graviton-dressed classes complete by a single product.} Any closed completion $\mathcal O[\mathcal O_0]$ of a fortuitous core $\mathcal O_0$ immediately furnishes the closed completion $G\star\mathcal O[\mathcal O_0]$ of $G\mathcal O_0$ \cite[supplement, Section S8.3]{Repo}.

(c) \emph{Fortuitous bound on the total rank of higher differentials.} Every completed cocycle is annihilated by $d_r$ in every $\mathrm{BPS}_{\hbar^{r-1}}$ in which its leading class survives, though it can still be paired and lifted if its leading class becomes $d_r$-exact (as occurs at one and three loops in $SO(N)$, Section~\ref{sec-results-SOSp}).\footnote{If $\mathbf QX=\hbar^r\mathcal O$, then $[\mathcal O]_{\mathbf Q}$ is $\hbar^r$-torsion over $\mathbb C[[\hbar]]$ and vanishes after inverting $\hbar$. Conversely, the quantum product of two non-closed classes can become closed, producing a quantum analogue of fortuity.} Consequently, every multi-graviton class lies in $\ker d_r$, and in any finite-dimensional charge sector of $\mathcal N=4$ Yang--Mills theory at cochain degree $k$ and through any loop order $R$ (Section~\ref{sec-lifting}),
\begin{equation}\label{eq-sum-rank}
\begin{aligned}
\sum_{\substack{r=1\\\text{loop orders}}}^{R}\operatorname{rank}d_r^k
&\le\dim\mathrm{BPS}_{\hbar^0}^k-
\dim\operatorname{span}\left\{[\mathcal O_0]_{\hbar^0}\in\mathrm{BPS}_{\hbar^0}^k:\mathcal O_0\text{ is a multi-graviton}\right\}\\
&=\dim\{\text{fortuitous classes in }\mathrm{BPS}_{\hbar^0}^k\}.
\end{aligned}
\end{equation}

(d) \emph{Explicit ring multiplication.} Theorem~\ref{b-00106} and \eqref{eq-sum-star} evaluate the multiplication on $H(\mathbf Q)$ graph by graph without solving auxiliary renormalization conditions.

(e) \emph{Reduction of closure tests to ring generators.} Because $d_r$ is a derivation on $\mathrm{BPS}_{\hbar^{r-1}}$ and vanishes on all multi-graviton generators, testing whether $d_r=0$ across an entire sector reduces to evaluating $d_r$ on its fortuitous generators.

(f) \emph{All-order Cartan invisibility of perturbative differential images (Proposition~\ref{prop-cartan-image-vanishing}).} Combining multi-graviton closure (a), the derivation rule for $d_r$ on $\mathrm{BPS}_{\hbar^{r-1}}$, and the classical scalar-discriminant localization of \cite[Section 6.2]{KvLM}, every image class in $\operatorname{im}d_r$ ($r\ge1$) within the multi-trace ring (which equals the entire gauge-invariant complex for $SU(N)$, $SO(2n+1)$, and $Sp(n)$, and the $O(2n)$-even direct summand for $SO(2n)$) has vanishing Cartan restriction; hence no Cartan-visible multi-trace class is $d_r$-exact at any loop order, and $d_r=0$ into every target sector with $\HQ=\HQC$.

\subsection{Fortuitous classes and quantum lifting}\label{sec-applications}

Using the explicit loop differentials $Q_1,\ldots,Q_4$ constructed here, we enumerate the classical cohomology $\mathrm{BPS}_{\hbar^0}=H(Q_0)$ sector by sector, isolate its fortuitous classes, and compute the successive obstructions $d_r$ across $\mathcal N=4$ Yang--Mills theory with gauge groups $SU(N)$, $SO(N)$, $Sp(N)$, $G_2$ and $F_4$, its Leigh--Strassler and $\beta$-deformations, and a family of $\mathcal N=2$ and $\mathcal N=1$ gauge theories. Table~\ref{tab-census-overview} summarizes the entire census; Sections~\ref{sec-results-SU}--\ref{sec-results-higher} record each theory's sectors, classes and certificates, with full data and algorithms archived in \cite{Repo}.

\paragraph{Principal physical findings of the census.}

Four structural results emerge before entering the sector-by-sector data:

(I) \emph{Symmetry pattern of higher-loop lifting.} Throughout all scanned sectors of $SU(N)$, $Sp(N)$, $G_2$ and $F_4$ $\mathcal N=4$ Yang--Mills theory, of $SO(7)$ and $SO(9)$ through $2(E+J_L)=23$, and of the $\mathcal N=2$ superconformal theories tested ($SU(N)\times SU(N)$ orbifold quivers and $SU(N)$ SQCD with $N_f=2N$), every computed higher differential $d_r$ with $r\ge2$ vanishes, and \emph{no two-loop lift ($d_2\ne0$) occurs in any tested sector of undeformed $\mathcal N=4$ Yang--Mills theory} (moreover, deforming the reverse Taylor shear by its one-loop partner proves over $\mathbb Q$ that $d_2=0$ on every source sector with $J_L\le1$ and $\Theta+n_++n_-<12$ for any simple gauge algebra; Section~\ref{sec-results-G2}). In higher-rank orthogonal $\mathcal N=4$ Yang--Mills theory, however, \emph{three-loop lifting ($d_3\ne0$) turns on where $d_2=0$}: exact rational certificates establish $d_3\ne0$ in the zero-derivative BMN sector $(7,5,5;0,0)_{16}$ of $SO(13)$ at level $34$ (and its level-$40$ descendant into $(2,0,0;6,6)_{14}$), as well as into the glueball powers $[S^6]$ in $SO(11)$ (level $36$), $[S^7]$ in $SO(12)$ and $SO(13)$ (level $42$), and $[S^8]$ in $SO(13)$ (level $48$), as well as scalar-dressed products in $SO(13)$ at levels $46,47$. A further $SO(14)$ lift at level $48$ is reported in \cite[discussion D014]{Repo} (Section~\ref{sec-results-SOSp}). On the explicit pure-glueball chains, $\deg_\theta$ counting kills the direct higher-loop fans ($Q_2\mathcal O_0=Q_2\mathcal O_1=Q_3\mathcal O_0=0$), so $d_3[\mathcal O_0]_{\hbar^2}=[Q_1\mathcal O_2]_{\hbar^2}\ne0$ arises from threefold iterated one-loop transfer. Breaking to $\mathcal N=1$ superconformal symmetry activates two-loop lifting ($d_2\ne0$): in $12$ sector pairs of the $SU(3)$ Leigh--Strassler theory through level $22$ (Section~\ref{sec-results-LS}) and in one sector pair of the $SU(3)$ $\beta$-deformation through level $21$ (Section~\ref{sec-results-beta}), lifting classes on which $d_1=0$ (again via $Q_2\mathcal O_0=0$ and $d_2[\mathcal O_0]_{\hbar}=[Q_1\mathcal O_1]_{\hbar}\ne0$ on the lowest lifted classes). Breaking conformal symmetry in abelian theories with a neutral chiral multiplet and in nonconformal $\mathcal N=2$ $SU(2),SU(3)$ gauge theories produces lifts at two, three and four loops ($d_2,d_3,d_4\ne0$; Sections~\ref{sec-results-SQED}--\ref{sec-results-higher}).

(II) \emph{One-loop and higher-loop restoration of Montonen--Olive S-duality.} For $n\ge3$, the classical $Q_0$-cohomologies $\mathrm{BPS}_{\hbar^0}$ of $\mathcal N=4$ Yang--Mills theory with S-dual gauge groups $SO(2n+1)$ and $Sp(n)$ disagree \cite{ChangLin,GLRT}. Through level $2(E+J_L)=22$, our full-letter census finds $d_1\ne0$ in $16$ sectors of $SO(7)$ and $1$ sector of $SO(9)$ (and on the level-$26$ threshold of $SO(11)$), while $d_1=0$ throughout $Sp(3)$ and $Sp(4)$, bringing $\mathrm{BPS}_{\hbar}(SO(7))\cong\mathrm{BPS}_{\hbar}(Sp(3))$ (certified over $\mathbb Q$ across all sectors at even levels $18,20,22$, and on the level-$24$ fortuitous threshold block $(11;5,2,2;1,0)$ where S-duality mixes fortuitous and Cartan-null multi-graviton slices) and $\mathrm{BPS}_{\hbar}(SO(9))\cong\mathrm{BPS}_{\hbar}(Sp(4))$ into exact charge-by-charge agreement (Table~\ref{tab-sosp-sduality}). At higher levels, $SO(2n+1)$ carries $n-2$ classical powers $[S^{n+1}],\ldots,[S^{2n-2}]$ of the defining-trace gaugino bilinear $S=\operatorname{tr}_{2n+1}\bigl((\partial_{+\dot+}c)(\partial_{+\dot-}c)\bigr)$ that vanish classically in $Sp(n)$: a closed-form one-loop Pfaffian family lifts $[S^4]$ in $SO(7)$ and the top two powers $[S^{2n-2}],[S^{2n-3}]$ for every $n\ge4$, whereas for $n=5,6$ the remaining excess powers $[S^6]$ in $SO(11)$ and $[S^7],[S^8]$ in $SO(13)$ satisfy $d_1=d_2=0$ and are lifted at three loops ($d_3\ne0$, Table~\ref{tab-orthogonal-d3-lifts}), while \cite[discussion D014]{Repo} reports that $[S^{10}]$ in $SO(19)$ and $[S^{13}]$ in $SO(25)$ survive through $\mathrm{BPS}_{\hbar^4}$ and $\mathrm{BPS}_{\hbar^5}$ respectively. Any perturbative lift of these two powers must therefore occur at order five or higher in $SO(19)$ and six or higher in $SO(25)$.

(III) \emph{New non-BMN fortuitous thresholds in $\mathcal N=4$ Yang--Mills theory.} Complete full-letter enumeration lowers the fortuitous threshold in $SU(3)$ from level $30$ (in the zero-derivative BMN subsector \cite{CCKLL,KLLS,BraneFused}) to level $24$ ($X_7^{\mathrm{a}}$, with $n_+=n_-=2$), and discovers the $SU(4)$ threshold at level $28$ ($(4,2,2;2,2)_9^{\mathrm{a}}$, eight levels below the level-$36$ BMN construction of \cite{KLLS}); both classes survive at every perturbative order in the finite-field computations. We also identify the index-predicted $SU(2)$ generators $\mathrm{P39}^{\mathrm{b}}$ and $\mathrm{X40}^{\mathrm{b}}$, establish the full-letter thresholds of $SO(5)=Sp(2)$, $Sp(3)$ and $SO(9)$, reproduce the level-$26$ BMN class of $SO(11)$ \cite[Sections 6.2--6.3]{FortuityAI}, and prove that no fortuitous class exists for $SU(5)$--$SU(15)$ within the complete scan ranges $2(E+J_L)\le 28$--$36$ (Table~\ref{tab-su-frontier}).

(IV) \emph{Bidirectional counterexamples to Cartan restriction.} Restricting operators to the Cartan subalgebra neither detects only monotone classes nor annihilates all fortuitous classes: the fortuitous classes $\mathrm{X40}^{\mathrm{b}}$ ($SU(2)$) and $X_7^{\mathrm{a}}$ ($SU(3)$) have nonzero Cartan restriction, whereas the $G_2$ scalar Gram determinant \eqref{local-ch9-g2-gram-class} and the first Cartan-invisible glueball powers $[S^3]$ in $G_2$ and $[S^5]$ in $F_4$ are monotone classes lying entirely in the kernel of Cartan restriction and surviving every perturbative order.

\begin{samepage}
Superscripts on fortuitous classes mark their provenance: \textsuperscript{a}~new class discovered in this work; \textsuperscript{b}~class previously indicated by an index or proposed construction and explicitly identified or verified here; \textsuperscript{c}~previously constructed class reproduced here. Multiplet members and descendants inherit their parent's mark, and a mark on a mixed sector refers to its newly identified classes.\par
\end{samepage}

\paragraph{Sectors, conserved charges and the BMN subsector.}
For the $\mathcal N=4$ theories, cochain sectors are labelled $(\lambda)_Y=(\theta_1,\theta_2,\theta_3;n_+,n_-)_Y$ (Appendix~\ref{sec-E-sectors}): here the integers $\theta_1,\theta_2,\theta_3\in\mathbb Z_{\ge0}$ record the total $\theta_I$-monomial weights of the letters ($\theta(\gamma^I)=e_I$, $\theta(\beta_I)=(1,1,1)-e_I$, $\theta(b)=(1,1,1)$, $\theta(c)=0$, distinct from the Grassmann coordinates $\theta_I$ themselves), $n_+,n_-$ count $\partial_{+\dot+},\partial_{+\dot-}$ derivatives, and $Y$ is the number of letters. The \emph{BMN sector} \cite{ChangYin,microstates} is the zero-derivative subsector $n_+=n_-=0$ built from $(\gamma^I,\beta_I,b)$ without gauginos $\partial_{+\dot a}c$. With $\Theta=\theta_1+\theta_2+\theta_3$, the scaling dimension $E$, $SU(4)_R$ charges $(R_1,R_2,R_3)$ and spins $(J_L,J_R)$ are
\begin{equation*}
\begin{gathered}
E=\tfrac12(\Theta+Y)+n_++n_-,\qquad R_I=\theta_I-\tfrac12(\Theta-Y),\\
J_L=\tfrac12(\Theta-Y+n_++n_-),\qquad J_R=\tfrac12(n_+-n_-),\qquad
2(E+J_L)=2\Theta+3(n_++n_-).
\end{gathered}
\end{equation*}
Every $Q_n$ maps $(\lambda)_Y\to(\lambda-n\mathbf d)_{Y+1-n}$ with $\mathbf d=(1,1,1;-1,-1)$ and preserves the \emph{level} $2(E+J_L)$; a census is complete through a given level when every sector up to that bound has been computed. The Leigh--Strassler deformation conserves $\Theta$ and $c=\theta_2+2\theta_3\pmod3$ in place of the individual $\theta_I$; the adjoint-mass deformations use the infrared weight $w=\theta_1+\theta_2+2\theta_3+2(n_++n_-)$; and the abelian and nonconformal $\mathcal N=2$ scans assign superpotential weights $w_g$ to $\gamma_g$, $D-w_g$ to $\beta^g$, $D$ to $b$ and $0$ to $c$, where $D$ is the weight of $\bar W$.

\paragraph{Cartan restriction, lifting criterion and certification.}
Write $\HQ=\dim\mathrm{BPS}_{\hbar^0}$ and $\HQC$ for the rank of restriction to the Cartan subalgebra, so that $\HQ-\HQC$ is the Cartan-kernel dimension, and write $f$ for the fortuitous dimension ($\dim\mathrm{BPS}_{\hbar^0}$ modulo the stable-cohomology image \cite[supplement, Section S1.5]{Repo}). To certify all-order survival of a class in $\mathrm{BPS}_{\hbar^0}$, we check both that $d_r=0$ on the class and that the class is $d_r$-exact for no $r\ge1$ allowed by the grading. Rank computations carry the field labels $[p]$ (computed modulo $p=2^{31}-1$, the default when unmarked), $[p,p']$ (independently confirmed at a second prime), and $[\mathbb Q]$ (certified over $\mathbb Q$ either by exact rational elimination or by exhibiting a rational chain $(\mathcal O_0,\ldots,\mathcal O_{r-1})$ together with a dual rational linear functional vanishing on $\operatorname{im}Q_0+\sum_{j=1}^{r-1}\text{lower images}$ and nonvanishing on $R_r$, \cite[supplement, Section S1.7]{Repo}). For an integral matrix a modular rank is a lower bound on its rational rank. Conclusions labelled $[p]$ or $[p,p']$ refer to the indicated fields unless a separate characteristic-zero argument is given. Because $d_r$ acts on the subquotient $\mathrm{BPS}_{\hbar^{r-1}}$, its rank is independent of representative and homotopy choices (Section~\ref{sec-homotopy-transfer}).

\needspace{8\baselineskip}
{\footnotesize\renewcommand{\arraystretch}{1.28}\setlength{\tabcolsep}{3.5pt}
\begin{xltabular}{\linewidth}{|>{\raggedright\arraybackslash}p{2.35cm}|>{\raggedright\arraybackslash}p{2.15cm}|>{\raggedright\arraybackslash}X|>{\raggedright\arraybackslash}X|>{\raggedright\arraybackslash}p{1.75cm}|}
\caption{Overview of the classical census and quantum lifting tests. Second column: complete scan cutoff in $2(E+J_L)$ unless another conserved weight is stated. Third column: threshold fortuitous class or benchmark class, with sector and charges $(E;R_1,R_2,R_3;J_L,J_R)$. Last column: ground field of the rank verification.}\label{tab-census-overview}\\
\hline
Theory & Scan cutoff & Threshold / benchmark class & Quantum lifting results & Field\\
\hline\endfirsthead
\hline
Theory & Scan cutoff & Threshold / benchmark class & Quantum lifting results & Field\\
\hline\endhead
$\mathcal N=4$, $SU(2)$ & $33$ & $\mathcal O_0$\textsuperscript{c}: $(4,4,4;0,0)_7$, $(\tfrac{19}2;\tfrac32,\tfrac32,\tfrac32;\tfrac52,0)$ \cite{KKLP,ChangLinWords}; generators $\mathcal O_1$\textsuperscript{c}, $\mathrm{P39}$\textsuperscript{b}, $\mathrm{X40}$\textsuperscript{b} & $d_1\ne0$ on $\mathcal O_1$\textsuperscript{c} \cite{KKLP}; $\mathcal O_0$\textsuperscript{c}, $\mathrm{P39}$\textsuperscript{b}, $\mathrm{X40}$\textsuperscript{b} survive at every order & $\mathcal O_0$\textsuperscript{c}: $\mathbb Q$; others: $p,p'$\\
\hline
$\mathcal N=4$, $SU(3)$ & $30$ & $X_7$\textsuperscript{a}: $(2,2,2;2,2)_7$, $(\tfrac{21}2;\tfrac52,\tfrac52,\tfrac52;\tfrac32,0)$; towers $F_1$\textsuperscript{c}, $B_1$\textsuperscript{c} \cite{KLLS,BraneFused}; four $10,11$-letter non-BMN sectors & no lift found; infinite towers $F_1$\textsuperscript{c}, $B_1$\textsuperscript{c} ($m\ge3$), $X_7$\textsuperscript{a}, and all $4$ non-BMN sectors survive at all orders (Table~\ref{local-ch9-su-classes}) & $F_1,B_1$: $\mathbb Q$; others: $p$\\
\hline
$\mathcal N=4$, $SU(4)$ & $28$ & $(4,2,2;2,2)_9$\textsuperscript{a}, $(\tfrac{25}2;\tfrac92,\tfrac52,\tfrac52;\tfrac32,0)$; unique fortuitous class through $28$ & no lift; survives at every order & $f=1$: $\mathbb Q$; lifting: $p$\\
\hline
$\mathcal N=4$, $SU(5)$--$SU(15)$ & $28$ to $36$ (Table~\ref{tab-su-frontier}) & no fortuitous class in scan range & all classes monotone ($d_r=0$) & $p$\\
\hline
$\mathcal N=4$, $SO(5)=Sp(2)$ & $23$ & $(3,2,2;1,1)_7$\textsuperscript{a}, $(9;3,2,2;1,0)$ & no lift; survives at every order & $p$\\
\hline
$\mathcal N=4$, $SO(7)$ & $23$ (targeted to $32$) & $O_f$\textsuperscript{c}: $(3,3,3;0,0)_8$, $(\tfrac{17}2;\tfrac52,\tfrac52,\tfrac52;\tfrac12,0)$ \cite{GLRT} & $d_1\ne0$ in $16$ sectors through $22$ (first on $O_f$\textsuperscript{c} \cite{KonishiBH}), restoring S-duality with $Sp(3)$ (and at $(11;5,2,2;1,0)$, level $24$); $d_{r\ge2}=0$ & levels $18,20,22,24$: $\mathbb Q$; odd levels: $p$\\
\hline
$\mathcal N=4$, $Sp(3)$ & $24$ (targeted at $28$) & $(5,2,2;1,1)_9$\textsuperscript{a}, $(11;5,2,2;1,0)$ & no lift; threshold $(5,2,2;1,1)_9$\textsuperscript{a} survives at every order ($[\mathbb Q]$, matching $SO(7)$ block dim.\ $150$) & threshold \& $18,20,22$: $\mathbb Q$; others: $p$\\
\hline
$\mathcal N=4$, $SO(9)$ & $23$ & $(5,3,3;0,0)_{10}$\textsuperscript{c}, $(\tfrac{21}2;\tfrac92,\tfrac52,\tfrac52;\tfrac12,0)$ & $\operatorname{rank}d_1=1$ on threshold, restoring S-duality with $Sp(4)$ through $22$ & $p$\\
\hline
$\mathcal N=4$, $Sp(4)$, $Sp(5)$ & $24$ & no fortuitous class in scan range & no lift found & $p$\\
\hline
$\mathcal N=4$, $SO(11)$ & $24$ (partial at $25,26$) & $(7,3,3;0,0)_{12}$\textsuperscript{c}, $(\tfrac{25}2;\tfrac{13}2,\tfrac52,\tfrac52;\tfrac12,0)$ & $\operatorname{rank}d_1=1$ on this level-$26$ class & $p$\\
\hline
$\mathcal N=4$, $SO(10)$--$SO(26)$ higher-loop sectors & targeted to $84$ (Table~\ref{tab-orthogonal-d3-lifts}) & $SO(13)$ BMN $(7,5,5;0,0)_{16}$ (level $34$); glueball powers $S^p\in(0,0,0;p,p)_{2p}$ & $S^{N-3},S^{N-4}$ hit by $d_1$ in the stated range; $d_3\ne0$ in $SO(13)$ at levels $34,40,46,47$ and on $S^6$ ($SO(11)$), $S^7$ ($SO(12),SO(13)$), $S^8$ ($SO(13)$); further results reported in D014: $S^8$ ($SO(14)$) hit by $d_3$, $S^{10}$ ($SO(19)$) in $\mathrm{BPS}_{\hbar^4}$; $S^{13},S^{14}$ ($SO(25),SO(26)$) in $\mathrm{BPS}_{\hbar^5}$ & $\mathbb Q$\\
\hline
$\mathcal N=4$, $G_2$ & $23$ (targeted at $24$, BMN to $28$) & Cartan-kernel monotone \eqref{local-ch9-g2-gram-class} and $[S^3]$; level-$24$ fortuitous $(5,2,2;1,1)_9$\textsuperscript{a}, $(11;5,2,2;1,0)$ & no lift: \eqref{local-ch9-g2-gram-class}, $[S^3]$ and $(5,2,2;1,1)_9$\textsuperscript{a} survive at all orders ($[\mathbb Q]$); $d_{1,2,3}=0$ through $23$ & named classes: $\mathbb Q$; scan to $23$: $p$\\
\hline
$\mathcal N=4$, $F_4$ & targeted at $30,36$ & Cartan-kernel glueball powers $[S^5]\in(0,0,0;5,5)_{10}$, $[S^6]\in(0,0,0;6,6)_{12}$ & $[S^5]$ survives at all perturbative orders ($d_r$-exact for no $r$); $[S^6]$ survives $d_1$ ($\mathrm{BPS}_{\hbar}\ne0$; $d_2$ restricted to $\le138$ indecomposable classes in $(4,4,3,3)$) & $\mathbb Q$\\
\hline
$SU(3)$ Leigh--Strassler & $22$ & $P_{14}$\textsuperscript{a}: $(\Theta,c;n_+,n_-)_Y=(7,0;0,0)_5$, $(E,J_L,J_R)=(6,1,0)$, $f=5$ & $d_1\ne0$ in $53$ pairs; $d_2\ne0$ in $12$ pairs (first on $P_{14}$); no $d_3$ pair & first $3$ $d_2$: $\mathbb Q$; others: $p$\\
\hline
$SU(3)$ $\beta$-deformation & $21$ & lift source: $(5,2,2;1,0)_8$, $(\tfrac{19}2;\tfrac92,\tfrac32,\tfrac32;1,\tfrac12)$ & $\operatorname{rank}d_2=1$ at level $21$; all other $d_2,d_3$ candidates have rank $0$ & $p,p'$\\
\hline
$\mathcal N=2$ quiver, $SU(2)^2$, $SU(3)^2$ & $21$, $18$ & full sector census (stable split pending) & $d_1\ne0$ in $61$ and $43$ blocks; all $d_2,d_3$ candidates have rank $0$ & $p$\\
\hline
$\mathcal N=2$ SQCD, $SU(2)$, $SU(3)$ & $24$, $18$ & flavour-singlet full census & $SU(2)$: $d_1=0$; $SU(3)$: $d_1\ne0$ in $8$ sectors; all $d_2$ ranks $0$ & $p$\\
\hline
Adjoint-mass deformations & $w\le 13$ to $16$ (\S\ref{sec-results-masses}) & $\mathcal O_7$\textsuperscript{c}, $\mathcal O_{16}$\textsuperscript{c} of \cite{CK}; both survive at every order & $d_1\ne0$ in $4$ sectors; all tested $d_2,d_3$ candidates have rank $0$ & $p$\\
\hline
Abelian theories with neutral $\gamma_0$ & $\mathrm{wt}\le 24$ (\S\ref{sec-results-higher}) & explicit SQED source \eqref{local-ch9-sqed-source} and $7$ models & $d_2\ne0$, $d_3\ne0$, and $d_4\ne0$ (Table~\ref{local-ch9-higher-lifts}) & $\mathbb Q$; one quiver: $p$\\
\hline
$\mathcal N=2$ $SU(2)$, $SU(3)$ \& singlet flip & $\mathrm{wt}\le 14$ ($SU(2)$); $12$, partial ($SU(3)$) & gauge-invariant component census & $d_2\ne0$ in all three; $d_3\ne0$ with singlet (Table~\ref{local-ch9-higher-lifts}) & first $SU(2)$ lifts: $\mathbb Q$; $SU(3)$: $p$\\
\hline
\end{xltabular}}

The classical cohomology and its fortuitous classes are organized following \cite{ChangYin,microstates,ChangLinWords,ChangChevalley}, extending the enumerations of \cite{ChangLin,GLRT,KLLS,BraneFused} and reproducing the published $SU(2)$ census \cite{BPSCounting}, the one-loop lifts of \cite{KonishiBH,KKLP} and the mass-deformed classes of \cite{CK}. Classes in the kernel of Cartan restriction are analysed via Lie-algebra commuting schemes using higher-dimensional Chevalley restriction \cite{ChenNgo,ChenNgoSp,SXX,Vaccarino,Domokos,Hunziker,Richardson,LNY,SemplinerYang,Goyal}, the Cachazo--Douglas--Seiberg--Witten chiral-ring theorems \cite{CDSW,EtingofKac,CederwallFerretti,Kumar,WittenSpSO,EtingofII}, the super-commuting schemes of \cite{KvLM} and classical invariant theory \cite{Sergeev,BereleRegev,GoodmanWallach,LehrerZhang,Bourbaki}. The Leigh--Strassler family follows \cite{LeighStrassler,BundzikMansson}; our one- and two-loop operators were cross-checked against \cite{BK,KKLP}, \cite[Section 4.5]{GWW} and \cite{FortuityAI}, with exact integer and rational linear algebra performed in FLINT \cite{FLINT}.

\subsubsection{$\mathcal N=4$ Yang--Mills: $SU(N)$}\label{sec-results-SU}

For $\mathfrak g=\mathfrak{su}(N)$ \cite[results report, \S1.1]{Repo}, Table~\ref{tab-su-frontier} records the complete full-letter scan cutoffs alongside the trace-relation bound $2(E+J_L)\le 2N+1$ (below which no trace identity enters and every class is a stable multi-graviton \cite{ChangYin,ChangLinWords}) and compares the new non-BMN thresholds with earlier BMN constructions.\footnote{In $SU(4)$, the earlier BMN construction first produces a fortuitous class at level $36$ \cite[Sections 3.3 and 4.3]{KLLS}, and multi-core BMN classes appear at still higher charges \cite{MultiCore}; our level-$28$ threshold $(4,2,2;2,2)_9^{\mathrm{a}}$ lies outside the BMN sector ($n_+=n_-=2$).} No fortuitous class occurs for $5\le N\le 15$ within these complete ranges \cite[supplement, Section S2.1]{Repo}.

\needspace{8\baselineskip}
{\footnotesize\renewcommand{\arraystretch}{1.26}\setlength{\tabcolsep}{4pt}
\begin{xltabular}{\linewidth}{|l|>{\centering\arraybackslash}p{2.75cm}|>{\centering\arraybackslash}p{1.85cm}|>{\raggedright\arraybackslash}X|>{\raggedright\arraybackslash}X|}
\caption{Complete scan frontier and fortuitous thresholds in $SU(N)$ $\mathcal N=4$ Yang--Mills theory, measured by the level $2(E+J_L)$.}\label{tab-su-frontier}\\
\hline
$G$ & Trivial ($=$ stable multi-gravitons) for $2(E+J_L)\le$ & Complete through & Threshold in full-letter census & Earliest class in prior constructions\\
\hline\endfirsthead
\hline
$G$ & Trivial ($=$ stable multi-gravitons) for $2(E+J_L)\le$ & Complete through & Threshold in full-letter census & Earliest class in prior constructions\\
\hline\endhead
$SU(2)$ & $5$ & $33$ & $24$: $\mathcal O_0$\textsuperscript{c} & $24$ \cite{ChangLinWords,microstates}\\
\hline
$SU(3)$ & $7$ & $30$ & $24$: $X_7$\textsuperscript{a} (non-BMN) & $30$: BMN tower $F_1$\textsuperscript{c} \cite{CCKLL,KLLS,BraneFused}\\
\hline
$SU(4)$ & $9$ & $28$ & $28$: $(4,2,2;2,2)_9$\textsuperscript{a} (non-BMN) & $36$: double brane-fused BMN \cite[Section 4.3]{KLLS}\\
\hline
$SU(5)$ & $11$ & $28$ & none in scan & asymptotic BMN existence \cite{EveryN}\\
\hline
$SU(6)$--$SU(7)$ & $13$, $15$ & $30$ & none in scan & asymptotic BMN existence \cite{EveryN}\\
\hline
$SU(8)$, $SU(9)$, $SU(10)$ & $17$, $19$, $21$ & $31$, $33$, $34$ & none in scan & asymptotic BMN existence \cite{EveryN}\\
\hline
$SU(11)$--$SU(15)$ & $2N+1$ & $36$ & none in scan & asymptotic BMN existence \cite{EveryN}\\
\hline
\end{xltabular}}

Table~\ref{local-ch9-su-classes} lists the fortuitous generators and selected multiplet sectors for $SU(2),SU(3),SU(4)$ \cite[supplement, Sections S2.2--S2.3 and S8.2--S8.3]{Repo}; by consequence~(e) of Theorem~\ref{b-00106}, vanishing of $d_r$ on these generators implies vanishing on their quantum products. For $SU(2)$, the superconformal index of \cite[Section 3.2]{microstates} requires additional core states by level $39$ and at the level-$40$ charges of $\mathrm{X40}^{\mathrm{b}}$; we identify $\mathrm{P39}^{\mathrm{b}}$ beyond the $12$ products and descendants in $(5,5,5;3,0)_9$, and $\mathrm{X40}^{\mathrm{b}}$ as a Cartan-visible fortuitous class ($\HQ=\HQC=8$ vs.\ graviton dimension $7$). Both $\mathrm{X40}^{\mathrm{b}}$ and the $SU(3)$ threshold $X_7^{\mathrm{a}}$ ($\HQ=\HQC=26$ vs.\ graviton dimension $25$) provide explicit counterexamples to the expectation that fortuitous classes vanish on the Cartan subalgebra. Every tested sector has $d_r=0$ for all $r\ge2$. Moreover, in $SU(2)$, Cartan restriction is injective ($\HQ=\HQC$) at every Taylor degree both at $J_L=0$ (zero second jets, which holds for all $SU(N)$ and $Sp(n)$ by \cite{Vaccarino,ChenNgoSp}) and at $J_L=\tfrac12$ (one second jet, proved over $\mathbb Q$ from quadratic minor syzygies and exact rank saturation in degrees $3,4,5,7$ \cite{Repo}): by Proposition~\ref{prop-cartan-image-vanishing}, every classical class in $SU(2)$ with $J_L\in\{0,\tfrac12\}$ survives at all perturbative orders, so any nonzero $d_r$ ($r\ge1$) in $SU(2)$ requires source $J_L\ge\tfrac32$ and target $J_L\ge1$ (and at $J_L=1$, all $53$ dominant dotted-singlet BMN $d_2$ targets through level $30$ likewise satisfy $\HQ=\HQC$).

\needspace{8\baselineskip}
{\footnotesize\renewcommand{\arraystretch}{1.28}\setlength{\tabcolsep}{3.5pt}
\begin{xltabular}{\linewidth}{|l|>{\raggedright\arraybackslash}p{3.55cm}|>{\raggedright\arraybackslash}X|>{\raggedright\arraybackslash}X|}
\caption{Fortuitous generators and selected sectors of $SU(N)$ $\mathcal N=4$ Yang--Mills theory and their quantum lifting tests. Second column: sector $(\theta_1,\theta_2,\theta_3;n_+,n_-)_Y$ and charges $(E;R_1,R_2,R_3;J_L,J_R)$. For target sectors with $\HQ=\HQC$, Proposition~\ref{prop-cartan-image-vanishing} proves $d_r=0$ directly from $\operatorname{im}d_r\subseteq\ker\rho_0$.}\label{local-ch9-su-classes}\\
\hline
$G$ & Class, sector, charges & Classical characterization & Quantum lifting tests \& certification\\
\hline\endfirsthead
\hline
$G$ & Class, sector, charges & Classical characterization & Quantum lifting tests \& certification\\
\hline\endhead
$SU(2)$ & $\mathcal O_0$\textsuperscript{c}: $(4,4,4;0,0)_7$\newline $(\tfrac{19}2;\tfrac32,\tfrac32,\tfrac32;\tfrac52,0)$ & Threshold class \cite{ChangLinWords}; explicit representative in \cite[Section 3.1]{microstates} & Closed completion through $\hbar^4$; survives at every order $[\mathbb Q]$\\
\cline{2-4}
 & $\mathcal O_n$\textsuperscript{c} ($1\le n\le4$): $(6,6,6;0,0)_9$\newline $(\tfrac{27}2;\tfrac32,\tfrac32,\tfrac32;\tfrac92,0)$ at $n=1$ & Heavy-core BMN tower $(2n+4,2n+4,2n+4;0,0)_{2n+7}$ ($2(E+J_L)=24+12n$) \cite[Section 3.1]{microstates} & $n=1$: $d_1\ne0$ \cite{KKLP} $[p,p']$; for $n=2,3,4$ (levels $48,60,72$), $\dim_{\mathbb Q}\mathrm{BPS}_{\hbar^0}=2$ and $\operatorname{rank}_{\mathbb Q}d_1=2$ (and in $(9,8,8;0,0)_{11}$ at level $50$) $[\mathbb Q]$ \cite{Repo}\\
\cline{2-4}
 & $\mathrm{P39}$\textsuperscript{b}: $(5,5,5;3,0)_9$\newline $(15;2,2,2;\tfrac92,\tfrac32)$ & Index-predicted core \cite[Section 3.2]{microstates}; $1$ class beyond $12$ products/descendants & $d_1=0$; $R_2,R_3,R_4\in\operatorname{im}Q_0$; $5$th target empty; $d_r$-exact for no $r$. Survives at all orders $[p,p']$\\
\cline{2-4}
 & $\mathrm{X40}$\textsuperscript{b}: $(7,5,5;1,1)_{10}$\newline $(\tfrac{31}2;\tfrac72,\tfrac32,\tfrac32;\tfrac92,0)$ & Cartan-visible fortuitous class ($\HQ=\HQC=8$, graviton dim.\ $7$) \cite[Section 3.2]{microstates} & $d_r=0$ for all $r$; source sectors have $\HQ=0$ ($d_r$-exact for no $r$). Survives at all orders ($[p,p']$ for $r\le2$, $[p]$ above)\\
\hline
$SU(3)$ & $X_7$\textsuperscript{a}: $(2,2,2;2,2)_7$\newline $(\tfrac{21}2;\tfrac52,\tfrac52,\tfrac52;\tfrac32,0)$ & New non-BMN threshold; Cartan-visible ($\HQ=\HQC=26$, graviton dim.\ $25$); $C$-odd & $d_r=0$ for all $r$ and $d_r$-exact for no $r$. Survives at every order $[p]$\\
\cline{2-4}
 & $X_7$\textsuperscript{a} partner: $(0,0,0;5,5)_7$\newline $(\tfrac{27}2;\tfrac72,\tfrac72,\tfrac72;\tfrac32,0)$ & Member of the $X_7$\textsuperscript{a} superconformal multiplet; relative Lie-algebra class $[\mathbb Q]$ & Survives at every order with the $X_7$\textsuperscript{a} multiplet $[p]$\\
\cline{2-4}
 & $F_1$\textsuperscript{c} ($m\ge3$): $(7,4,4;0,0)_{10}$\newline $(\tfrac{25}2;\tfrac92,\tfrac32,\tfrac32;\tfrac52,0)$ at $m=3$ & Fermionic BMN tower ($2(E+J_L)=24+2m$) \cite[Section 2.2]{BraneFused} & Local graviton-syzygy repair + Prop.~\ref{prop-cartan-image-vanishing} + $SU(2)$ block dual: entire tower ($m\ge3$) survives at all orders $[\mathbb Q]$ \cite{Repo}\\
\cline{2-4}
 & $(5,4,4;1,1)_{10}$\textsuperscript{a}\newline $(\tfrac{27}2;\tfrac72,\tfrac52,\tfrac52;\tfrac52,0)$ & Non-BMN sector; $\HQ=2$, $\HQC=0$ & $d_1=d_2=d_3=0$ on both classes; $d_4=0$ (Prop.~\ref{prop-cartan-image-vanishing}); $d_r$-exact for no $r$. Survives at all orders $[p]$\\
\cline{2-4}
 & $(7,4,4;1,0)_{11}$\textsuperscript{a}\newline $(14;5,2,2;\tfrac52,\tfrac12)$ & $\HQ=2$, $\HQC=0$; $1$ class beyond the $F_1$\textsuperscript{c} descendant & $d_1=d_3=0$; $d_2=d_4=0$ (Prop.~\ref{prop-cartan-image-vanishing}); $d_r$-exact for no $r$. Survives at all orders $[p]$\\
\cline{2-4}
 & $(6,4,4;1,1)_{11}$\textsuperscript{a}\newline $(\tfrac{29}2;\tfrac92,\tfrac52,\tfrac52;\tfrac52,0)$ & Non-BMN sector; $\HQ=3$, $\HQC=0$ & $d_3=0$; $d_4=0$ (Prop.~\ref{prop-cartan-image-vanishing}); $d_r$-exact for no $r$. Survives at all orders $[p]$\\
\cline{2-4}
 & $(7,5,4;1,0)_{11}$\textsuperscript{a}\newline $(\tfrac{29}2;\tfrac92,\tfrac52,\tfrac32;3,\tfrac12)$ & Non-BMN sector; $\HQ=4$, $\HQC=0$ & $d_4=0$ (Prop.~\ref{prop-cartan-image-vanishing}); $d_r$-exact for no $r$. Survives at all orders $[p]$\\
\cline{2-4}
 & $B_1$\textsuperscript{c} ($m\ge3$): $(8,5,4;0,0)_{11}$\newline $(14;5,2,1;3,0)$ at $m=3$ & Bosonic BMN tower ($2(E+J_L)=28+2m$) \cite[Section 4.1]{KLLS} & Seeds at levels $34,36$ + marked scalar products + Prop.~\ref{prop-cartan-image-vanishing}: entire tower ($m\ge3$) survives at all orders $[\mathbb Q]$ \cite{Repo}\\
\hline
$SU(4)$ & $(4,2,2;2,2)_9$\textsuperscript{a}\newline $(\tfrac{25}2;\tfrac92,\tfrac52,\tfrac52;\tfrac32,0)$ & New non-BMN threshold; $f=1$ $[\mathbb Q]$; $C$-odd; unique fortuitous class through level $28$ & $R_1,R_2\in\operatorname{im}Q_0$; higher targets empty; $d_1$ from all $453$ classes of $(5,3,3;1,1)_9$ has rank $0$; $d_2$ source has $\HQ=0$. Survives at all orders $[p]$\\
\hline
\end{xltabular}}

\subsubsection{$\mathcal N=4$ Yang--Mills: $SO(N)$ and $Sp(n)$}\label{sec-results-SOSp}

The twisted action is \eqref{local-ch9-su-action} with $\mathfrak g=\mathfrak{so}_N$ ($N$-dimensional defining representation) or $\mathfrak{sp}_{2n}$ ($2n$-dimensional defining representation). Although $SO(2n+1)$ and $Sp(n)$ are Montonen--Olive S-dual, their Lie algebras are non-isomorphic for $n\ge3$ and their classical $Q_0$-cohomologies $\mathrm{BPS}_{\hbar^0}$ differ \cite{ChangLin,GLRT}. Table~\ref{local-ch9-sosp-classes} lists the threshold fortuitous classes and their lifting tests \cite[supplement, Sections S3.1--S3.2]{Repo}.

\needspace{8\baselineskip}
{\footnotesize\renewcommand{\arraystretch}{1.28}\setlength{\tabcolsep}{3.5pt}
\begin{xltabular}{\linewidth}{|l|>{\raggedright\arraybackslash}p{3.55cm}|>{\raggedright\arraybackslash}X|>{\raggedright\arraybackslash}X|}
\caption{Threshold and related fortuitous sectors of $SO(2n+1)$ and $Sp(n)$ $\mathcal N=4$ Yang--Mills theory.}\label{local-ch9-sosp-classes}\\
\hline
$G$ & Source sector, charges & Classical characterization & Quantum lifting tests \& certification\\
\hline\endfirsthead
\hline
$G$ & Source sector, charges & Classical characterization & Quantum lifting tests \& certification\\
\hline\endhead
$SO(5)=Sp(2)$ & $(3,2,2;1,1)_7$\textsuperscript{a}\newline $(9;3,2,2;1,0)$ & New full-letter threshold; $f=1$, $\HQC=0$ & $d_r=0$ for all $r$ and $d_r$-exact for no $r$. Survives at every order $[p]$\\
\hline
$SO(7)$ & $O_f$\textsuperscript{c}: $(3,3,3;0,0)_8$\newline $(\tfrac{17}2;\tfrac52,\tfrac52,\tfrac52;\tfrac12,0)$ & BMN threshold \cite[Section 4.3]{GLRT} & $\operatorname{rank}d_1=1$ into $(2,2,2;1,1)_8$ \cite{KonishiBH}: lifted at one loop\\
\hline
$Sp(3)$ & $(5,2,2;1,1)_9$\textsuperscript{a}\newline $(11;5,2,2;1,0)$ & New full-letter threshold; $\HQ=f=1$ & $d_r=0$ for all $r$ and $d_r$-exact for no $r$. Survives at every order $[\mathbb Q]$ \cite{Repo}\\
\cline{2-4}
 & $(5,0,0;3,3)_9$\textsuperscript{a}\newline $(13;7,2,2;1,0)$ & Related non-BMN sector (level $28$); $\HQ=f=1$, $\HQC=0$ ($\HQ=0$ for $Sp(n\ne3)$) $[\mathbb Q]$ & Certified outgoing $d_r=0$ for all $r$ and incoming $d_1=0$ $[p]$ \cite{Repo}\\
\hline
$SO(9)$ & $(5,3,3;0,0)_{10}$\textsuperscript{c}\newline $(\tfrac{21}2;\tfrac92,\tfrac52,\tfrac52;\tfrac12,0)$ & Higher-rank threshold ($E=\tfrac{21}2,J_L=\tfrac12$); $\HQ=6$, one-class excess over $Sp(4)$ & $\operatorname{rank}d_1=1$ into $(4,2,2;1,1)_{10}$ $[p]$: lifted at one loop\\
\hline
$SO(11)$ & $(7,3,3;0,0)_{12}$\textsuperscript{c}\newline $(\tfrac{25}2;\tfrac{13}2,\tfrac52,\tfrac52;\tfrac12,0)$ & Lowest class found (level $26$); $\HQ=23$, $f=1$ & $\operatorname{rank}d_1=1$ into $(6,2,2;1,1)_{12}$ $[p]$: lifted at one loop\\
\hline
\end{xltabular}}

For $n=3,4,5$, the threshold fortuitous class of $SO(2n+1)$ sits at $(2n-3,3,3;0,0)_{2n+2}$ ($E=\tfrac{4n+5}2$, $J_L=\tfrac12$) and is lifted at one loop by $d_1:(2n-3,3,3;0,0)_{2n+2}\to(2n-4,2,2;1,1)_{2n+2}$, in agreement with the BMN computations of \cite[Sections 6.2 and 6.6]{FortuityAI} and the higher-rank mechanism anticipated in \cite[Section 5.3]{KvLM}. Each rank-one $d_1$ map removes one source--target pair of conserved charges $(E;R_1,R_2,R_3;J_L,J_R)$ from $\mathrm{BPS}_{\hbar^0}$, accounting for two discrepant charge tuples. As Table~\ref{tab-sosp-sduality} shows, these one-loop lifts remove every classical discrepancy between $SO(7)$ and $Sp(3)$ and between $SO(9)$ and $Sp(4)$ through $2(E+J_L)=22$, with no $d_2$ candidate pairs and vanishing rank on all $d_3$ candidates \cite[supplement, Section S3.3]{Repo}: the classical $SO(7)$ and $Sp(3)$ counts at levels $18,20,22$ agree with \cite[Section 6.7]{FortuityAI}. At these levels, superconformal return matrices from $[S^4]$ and Cartan-product spanning minors across all $38$ nonempty balanced weights prove over $\mathbb Q$ that $d_1$ pairs $(1,5,17)$ independent $SU(3)_R$ multiplets in $SO(7)$ and that the remaining $(983,2578,6713)$ classes in both $SO(7)$ and $Sp(3)$ are spanned by multi-graviton products ($d_{r\ge2}=0$). At level $24$, evaluating the complete conserved-charge block $(11;5,2,2;1,0)$ across its four cochain slices $(6-k,3-k,3-k;k,k)_{10-k}$ ($0\le k\le3$) over $\mathbb Q$ gives $\dim\mathrm{BPS}_{\hbar^0}\to\dim\mathrm{BPS}_{\hbar^\infty}$ equal to $(21,0,130,0)\to(21,0,129,0)$ in $SO(7)$ versus $(20,1,129,0)\to(20,1,129,0)$ in $Sp(3)$, yielding total dimension $150$ in both theories and matching $Sp(3)$'s fortuitous threshold $(5,2,2;1,1)_9^{\mathrm a}$ with $SO(7)$'s extra Cartan-null multi-graviton class in $(6,3,3;0,0)_{10}$ across the letter/derivative filtration \cite{Repo}. In $SO(7)$ at higher BMN levels, exact lowering returns to the level-$26$ $d_1$ triplet certify five further $d_1\ne0$ flavour multiplets across levels $28,30,32$ over $\mathbb Q$ and exhaust all $145$ nonempty level-$28$ BMN blocks (forcing outgoing BMN $d_2=0$ at level $28$), while in $SO(5)=Sp(2)$ primary-index deficits through level $48$ require at least four odd BMN fortuitous generator multiplets \cite{Repo}.

\begin{table}[ht!]
\centering
\footnotesize\renewcommand{\arraystretch}{1.25}\setlength{\tabcolsep}{4.2pt}
\caption{Restoration of S-duality by one-loop lifting $d_1$ through $2(E+J_L)=22$. Each nonzero rank-one $d_1$ sector removes two charge tuples (its source and target) from $\mathrm{BPS}_{\hbar^0}$, leaving $\mathrm{BPS}_{\hbar}(SO(2n+1))\cong\mathrm{BPS}_{\hbar}(Sp(n))$ charge by charge.}\label{tab-sosp-sduality}
\begin{tabular}{|l|c|c|c|c|c||l|}
\hline
Dual pair \textbackslash\ Level $2(E+J_L)$ & $18$ & $19$ & $20$ & $21$ & $22$ & $\mathrm{BPS}_{\hbar}$ discrepancy through $22$\\
\hline
$SO(7)$ vs.\ $Sp(3)$: discrepant $\mathrm{BPS}_{\hbar^0}$ charges & $2$ & $2$ & $6$ & $7^\ast$ & $12$ & $0$ (exact match; $d_2,d_3=0$)\\
$SO(7)$ sectors with $\operatorname{rank}d_1=1$ ($Sp(3)$: $0$) & $1$ & $1$ & $3$ & $4^\ast$ & $7$ & $16$ $d_1$ sectors (levels $18,20,22$: $[\mathbb Q]$)\\
\hline
$SO(9)$ vs.\ $Sp(4)$: discrepant $\mathrm{BPS}_{\hbar^0}$ charges & $0$ & $0$ & $0$ & $0$ & $2$ & $0$ (exact match; $d_2,d_3=0$)\\
$SO(9)$ sectors with $\operatorname{rank}d_1=1$ ($Sp(4)$: $0$) & $0$ & $0$ & $0$ & $0$ & $1$ & threshold lifted $[p]$\\
\hline
\end{tabular}

\smallskip
{\footnotesize$^\ast$Pairs crossing flavour-permuted or coincident charge tuples account for the total of $16$ source sectors across levels $18$--$22$; see \cite[supplement, Section S3.3]{Repo}.}
\end{table}

\paragraph{Higher-rank glueball powers and three-loop lifts in $SO(N)$.}
Beyond the complete scan frontier, higher-loop lifting in $SO(N)$ is probed analytically by powers of the defining-representation gaugino bilinear (the glueball operator)
\begin{equation}\label{eq-orthogonal-glueball-def}
S=\operatorname{tr}_N\bigl((\partial_{+\dot+}c)(\partial_{+\dot-}c)\bigr)\in C^2(0,0,0;1,1),
\end{equation}
which carries conserved charges $(E;R_1,R_2,R_3;J_L,J_R)=(3;1,1,1;0,0)$, level $2(E+J_L)=6$, fermion number $0$, and $\deg_\theta S=0$. The classical differential acts on the first ghost jets by a gauge transformation, so $Q_0S^p=0$ by trace invariance. Since $\deg_\theta(S^p)=0$, equation~\eqref{eq-graviton-differential-degree} gives $Q_m S^p=0$ for every $m\ge1$. Thus $S^p$ is an all-order $\mathbf Q$-cocycle ($\mathbf Q(S^p)=0$) representing a monotone multi-graviton class in $(0,0,0;p,p)_{2p}$ at level $6p$. At $\Theta=0$, classical predecessors in $C^{2p-1}(0,0,0;p,p)$ contain one second jet $\partial_{+\dot a}\partial_{+\dot b}c$ and $2p-2$ gaugino letters $\partial_{+\dot\pm}c$, imposing the classical gaugino anticommutator relations of the $\mathcal N=1$ chiral ring \cite{CDSW,WittenSpSO,EtingofII}: in $Sp(n)$ ($h^\vee=n+1$), the symplectic-trace bilinear satisfies $[S^p]_{\hbar^0}\ne0$ for $1\le p\le n$ and $[S^{n+1}]_{\hbar^0}=0$, whereas in $SO(N)$ ($N\ge5$, $h^\vee=N-2$, rank $\lfloor N/2\rfloor$), $[S^p]_{\hbar^0}\ne0$ for $1\le p\le N-3$ and $[S^{N-2}]_{\hbar^0}=0$, with $[S^p]_{\hbar^0}$ lying in the kernel of Cartan restriction ($\HQC=0$) for $\lfloor N/2\rfloor<p\le N-3$ \cite{KvLM}. For every S-dual pair $SO(2n+1)$ and $Sp(n)$ ($n\ge3$), $SO(2n+1)$ therefore possesses $n-2$ excess classical glueball powers $[S^{n+1}],\ldots,[S^{2n-2}]$ at levels $6(n+1),\ldots,6(2n-2)$ that vanish classically in $Sp(n)$.

By the orthogonal second fundamental theorem \cite{GoodmanWallach,LehrerZhang}, finite-colour trace identities in $SO(N)$ are generated by polarized skew-matrix Pfaffians and Gram determinants of degree exceeding $N$. For $p\ge4$ and $N\le p+4$, taking the odd-trace reversal $Q_0$-primitive of a polarized Pfaffian of degree $p+5$ produces a classical cocycle $\mathcal O_0^{(1)}\in C^{2p}(1,1,1;p-1,p-1)$ whose one-loop differential in the full relative complex evaluates over $\mathbb Q$ to \cite{Repo}
\begin{equation}\label{eq-orthogonal-d1-family}
d_1[\mathcal O_0^{(1)}]_{\hbar^0}
=c_1\,\frac{3\,2^{p-5}}{(p-4)!}(2N-3p-4)[S^p]_{\hbar^0}
\qquad(p\ge4,\;N\le p+4),
\end{equation}
where $c_1\in\mathbb Q^\times$ is an $(N,p)$-independent normalization constant (intertwined by odd flavour-to-dotted substitutions with two companion infinite $d_1$ families into $S^p\prod_{I=1}^m\operatorname{tr}_N((\partial_{+\dot+}c)\gamma^I)$ for $m=2,3$ \cite{Repo}; at $(N,p)=(8,4)$, $2N-3p-4=0$ vanishes in agreement with Proposition~\ref{prop-cartan-image-vanishing} since $[S^4]$ is Cartan-visible in $SO(8)$). Because $2N-3p-4$ equals $5-N\ne0$ at $p=N-3$ ($N\ge7$) and $8-N\ne0$ at $p=N-4$ ($N\ge9$), equation~\eqref{eq-orthogonal-d1-family} lifts $[S^4]$ in $SO(7)$ \cite{KvLM}, the $O(10)$-reflection-even classes $[S^6],[S^7]$ in $SO(10)$, and both top excess powers $[S^{N-3}]$ and $[S^{N-4}]$ for every $N\ge9$---removing all excess glueball powers in $SO(7)$ and $SO(9)$, while leaving $[S^6]$ in $SO(11)$ and $[S^7],[S^8]$ in $SO(13)$.

Once $p\le N-5$, the degree-$(p+5)$ Pfaffian no longer vanishes in $N$ colours, and $[S^p]$ survives both $d_1$ and $d_2$ in the full relative complex for $(SO(11),S^6)$, $(SO(12),S^7)$, $(SO(13),S^7)$, $(SO(13),S^8)$, and $(SO(14),S^8)$. At orders $0$ and $1$---in $(0,0,0;p,p)_{2p}$ and $(1,1,1;p-1,p-1)_{2p}$---the finite-colour trace-relation spaces have dimension $0$, so the stable transposed functional $L_1$ solving $\lambda_{S^p}Q_1=L_1Q_0$ (for the coefficient functional $\lambda_{S^p}$) descends directly to $N$ colours and annihilates $\operatorname{im}d_1$. At order $2$---in $(2,2,2;p-2,p-2)_{2p+1}$---the flavour-singlet finite-colour relation spaces have character dimensions $29,114,3,301,4$ in the five cases above (Table~\ref{tab-orthogonal-d3-lifts}) and are spanned over $\mathbb Q$ by explicit polarized Pfaffians of degree $>N$. On the entire relation space, the stable second functional $L_2$ solving $-L_1Q_1=L_2Q_0$ agrees over $\mathbb Q$ with an explicit linear combination of commuting Cartan-plane evaluations; subtracting this Cartan functional (which vanishes on $\operatorname{im}Q_0$) yields a finite-colour dual functional annihilating all incoming $d_1$ and $d_2$ images and certifying $[S^p]_{\hbar^2}\ne0$ over $\mathbb Q$ (Proposition~\ref{prop-ch9-subspace-certificate}). The same transposed-Pfaffian construction certifies $\mathrm{BPS}_{\hbar^2}$-survival of the one-scalar classes $[S^5\operatorname{tr}_{11}((\partial_{+\dot+}c)\gamma^1)]$ in $SO(11)$ (level $35$, $123$ relations) and $[S^6\operatorname{tr}_{13}((\partial_{+\dot+}c)\gamma^1)]$ in $SO(13)$ (level $41$, $17$ relations), as well as the quadratic and cubic scalar-dressed sectors in $SO(13)$--$SO(16)$ \cite{Repo}.

Along a three-loop chain into $(0,0,0;p,p)_{2p}$, the cochains $\mathcal O_k\in C^{Y_k}(3-k,3-k,3-k;p-3+k,p-3+k)$ ($k=0,1,2$, $Y_k=2p+2-k$) obey $\Theta_k+n_{+,k}+n_{-,k}=2p+3-k=Y_k+1$. Since every relative letter in Table~\ref{appE2-0020} has $\deg_\theta+n_++n_-\ge1$, each monomial of $\mathcal O_k$ consists of $Y_k-1$ weight-$1$ letters ($\gamma^I$ or $\partial_{+\dot\pm}c$, $\deg_\theta\le1$) and a single weight-$2$ letter ($\beta_I$, $\partial_{+\dot\pm}\gamma^I$, or $\partial_{+\dot a}\partial_{+\dot b}c$, $\deg_\theta\le2$). Any $r+1$ letters therefore carry combined degree $\deg_\theta\le r+2<3r$ for every $r\ge2$, so~\eqref{eq-graviton-differential-degree} kills all direct higher-loop fans on the chain, reducing $d_3$ to threefold iterated one-loop transfer:
\begin{equation}\label{eq-orthogonal-d3-chain}
\begin{gathered}
Q_r\mathcal O_k=0\qquad(r\ge2,\;k\in\{0,1,2\}),\\
Q_0\mathcal O_0=0,\qquad Q_0\mathcal O_1=-Q_1\mathcal O_0,\qquad Q_0\mathcal O_2=-Q_1\mathcal O_1,\\
d_3[\mathcal O_0]_{\hbar^2}=[Q_1\mathcal O_2]_{\hbar^2}=c\,[S^p]_{\hbar^2}\ne0,
\qquad
\mathbf Q\bigl(\mathcal O_0+\hbar\mathcal O_1+\hbar^2\mathcal O_2+\hbar^3\mathcal O_3\bigr)=c\,\hbar^3 S^p,
\end{gathered}
\end{equation}
where $\mathcal O_3=-h_0(Q_1\mathcal O_2-c\,S^p)\in C^{2p-1}(0,0,0;p,p)$ satisfies $Q_{r\ge1}\mathcal O_3=0$ by $\deg_\theta\mathcal O_3=0$. For $SO(11)$, $SO(12)$ and $SO(13)$, the rational chains and nonzero pairings are archived in \cite{Repo}. Table~\ref{tab-orthogonal-d3-lifts} also records the $SO(14)$ lift and the higher-rank survival bounds reported in discussion D014 of \cite{Repo}.
\begin{enumerate}
\item $SO(11)$ at level $36$ ($p=6$), where a degree-$15$ polarized Pfaffian primitive and a signed $12\times12$ Gram-determinant primitive in $(3,3,3;3,3)_{14}$ span the same line modulo the image of stable classical cohomology \cite{Repo}. Both hit $[S^6]_{\hbar^2}$, removing the remaining excess glueball power of $SO(11)$ over $Sp(5)$;
\item $SO(12)$ at level $42$ ($p=7$), where an $O(12)$-even source in $(3,3,3;4,4)_{16}$ hits $[S^7]_{\hbar^2}$. For $SO(14)$ at level $48$, \cite[discussion D014]{Repo} reports an $O(14)$-even lift of $[S^8]_{\hbar^2}$ that includes the signed odd-cycle contribution to the ordered tail pairing;
\item $SO(13)$ at level $42$ ($p=7$, promoted Gram-$14$ source $x_0\in(3,3,3;4,4)_{16}$), lifting $[S^7]_{\hbar^2}$, together with $SO(13)$ at level $48$ ($p=8$): multiplying $(x_0,x_1,x_2,x_3)$ by the closed even glueball $S$ gives $Q_0(Sx_k)=S\,Q_0x_k$, $Q_1(Sx_k)=S\,Q_1x_k$ (since $\deg_\theta S=0$ and every letter of $x_k$ has $\deg_\theta\le2$, so no pair involving $S$ reaches $\deg_\theta=3$), and $Q_{r\ge2}(Sx_k)=0$ ($\deg_\theta\le r+2<3r$), whence $\mathbf Q\bigl(S(x_0+\hbar x_1+\hbar^2x_2+\hbar^3x_3)\bigr)=c\,\hbar^3S^8$. Combined with the $301$-Pfaffian dual certificate $[S^8]_{\hbar^2}\ne0$, this realizes the derivation rule~\eqref{eq-2-78} of the quantum BPS ring (Theorem~\ref{b-00106}),
\begin{equation}\label{eq-so13-s8-product-lift}
d_3\bigl([S]_{\hbar^2}[x_0]_{\hbar^2}\bigr)
=[S]_{\hbar^2}\,d_3[x_0]_{\hbar^2}
=c\,[S^8]_{\hbar^2}\ne0\qquad(c\in\mathbb Q^\times),
\end{equation}
lifting both remaining excess glueball powers $[S^7]$ and $[S^8]$ of $SO(13)$ over $Sp(6)$ at three loops (on $\mathrm{BPS}_{\hbar^2}$, the derivation rule~\eqref{eq-2-78} applied to the product of $[x_0]_{\hbar^2}$ with the closed graviton class of $\operatorname{tr}_{13}((\gamma^1)^2)$ or $\operatorname{tr}_{13}((\partial_{+\dot+}c)\gamma^1)$ likewise gives $d_3\ne0$ into $[S^7\operatorname{tr}_{13}((\gamma^1)^2)]$ at level $46$ and $[S^7\operatorname{tr}_{13}((\partial_{+\dot+}c)\gamma^1)]$ at level $47$); and
\item $SO(13)$ in the zero-derivative BMN sector $(7,5,5;0,0)_{16}$ at level $34$, with charges $(\tfrac{33}2;\tfrac{13}2,\tfrac92,\tfrac92;\tfrac12,0)$, and its six-substitution mixed descendant in $(5,3,3;3,3)_{16}$ at level $40$: two products of a scalar bilinear with a degree-$15$ Pfaffian primitive span a two-dimensional fortuitous subspace on which $\operatorname{rank}d_1=1$, while the unique kernel class $[O_0]_{\hbar}$ has $\mathrm{BPS}_{\hbar^0}=0$ in its odd-length $d_2$ target $(5,3,3;2,2)_{15}$ and satisfies $d_3[O_0]_{\hbar^2}\ne0$ into $(4,2,2;3,3)_{14}$---certified over $\mathbb Q$ via $d_3[DO_0]_{\hbar^2}\ne0$ into the two-scalar Cartan-kernel sector $(2,0,0;6,6)_{14}$ at level $40$ \cite{Repo}. At higher ranks, \cite[discussion D014]{Repo} reports that the first Cartan-invisible glueball power $[S^{\lfloor N/2\rfloor+1}]$ survives past three loops, reaching $\mathrm{BPS}_{\hbar^4}$ for $[S^{10}]$ in $SO(19)$ (level $60$) and $\mathrm{BPS}_{\hbar^5}$ for $[S^{13}]$ in $SO(25)$ (level $78$) and $[S^{14}]$ in $SO(26)$ (level $84$).
\end{enumerate}

\needspace{8\baselineskip}
{\footnotesize\renewcommand{\arraystretch}{1.25}\setlength{\tabcolsep}{3.5pt}
\begin{xltabular}{\linewidth}{|l|>{\raggedright\arraybackslash}p{4.05cm}|c|>{\raggedright\arraybackslash}X|}
\caption{One-loop and three-loop lifts and higher-page survival in $SO(N)$ $\mathcal N=4$ Yang--Mills theory \cite{Repo}. Here $S=\operatorname{tr}_N\bigl((\partial_{+\dot+}c)(\partial_{+\dot-}c)\bigr)$, and the third column gives finite-colour trace-relation dimensions at dual orders $(0,1,2)$ unless marked $\dagger$. A dash denotes an unreported entry.}\label{tab-orthogonal-d3-lifts}\\
\hline
$G$ & Level, target sector \& class & Rel.\ dims. & Perturbative fate and source\\
\hline\endfirsthead
\hline
$G$ & Level, target sector \& class & Rel.\ dims. & Perturbative fate and source\\
\hline\endhead
$SO(7)$ & $24$: $(0,0,0;4,4)_8$, $[S^4]$ & $(0,\text{--},\text{--})$ & Lifted at $1$ loop ($d_1\ne0$) from $(1,1,1;3,3)_8$ via~\eqref{eq-orthogonal-d1-family} \cite{KvLM}; restores glueball S-duality with $Sp(3)$\\
\hline
$SO(9)$ & $30,36$: $(0,0,0;p,p)_{2p}$, $[S^5],[S^6]$ & $(0,\text{--},\text{--})$ & Both lifted at $1$ loop ($d_1\ne0$) from $(1,1,1;p-1,p-1)_{2p}$ via~\eqref{eq-orthogonal-d1-family}; restores glueball S-duality with $Sp(4)$\\
\hline
$SO(10)$ & $36,42$: $(0,0,0;p,p)_{2p}$, $[S^6],[S^7]$ & $(0,\text{--},\text{--})$ & Both lifted at $1$ loop ($d_1\ne0$, $O(10)$-even) via~\eqref{eq-orthogonal-d1-family}\\
\hline
$SO(11)$ & $36$: $(0,0,0;6,6)_{12}$, $[S^6]$ & $(0,0,29)$ & $d_1=d_2=0$; \textbf{lifted at $3$ loops ($d_3\ne0$)} from $(3,3,3;3,3)_{14}$ (Pfaffian and Gram sources)\\
\cline{2-4}
 & $42,48$: $(0,0,0;p,p)_{2p}$, $[S^7],[S^8]$ & $(0,\text{--},\text{--})$ & Both lifted at $1$ loop ($d_1\ne0$) via~\eqref{eq-orthogonal-d1-family}; with $[S^6]$ at $3$ loops, lifts all $SO(11)$ excess over $Sp(5)$\\
\cline{2-4}
 & $35$: $(1,0,0;6,5)_{12}$, $[S^5\operatorname{tr}_{11}((\partial_{+\dot+}c)\gamma^1)]$ & $(0,0,123)$ & $d_1=d_2=0$: certified nonzero in $\mathrm{BPS}_{\hbar^2}$\\
\hline
$SO(12)$ & $42$: $(0,0,0;7,7)_{14}$, $[S^7]$ & $(0,0,114)$ & $d_1=d_2=0$; \textbf{lifted at $3$ loops ($d_3\ne0$)} from $O(12)$-even promoted Gram-$13$ source in $(3,3,3;4,4)_{16}$\\
\hline
$SO(13)$ & $34$: $(4,2,2;3,3)_{14}$ (from BMN $(7,5,5;0,0)_{16}$) & $\text{--}$ & $2$D source has $\operatorname{rank}d_1=1$; kernel class has $d_2=0$ and is \textbf{lifted at $3$ loops ($d_3\ne0$)} (via level-$40$ descendant)\\
\cline{2-4}
 & $40$: $(2,0,0;6,6)_{14}$ (from $(5,3,3;3,3)_{16}$) & $(0,0,76^\dagger)$ & $d_1=d_2=0$ on descendant $DO_0$; \textbf{lifted at $3$ loops ($d_3\ne0$)} into Cartan-kernel two-scalar sector\\
\cline{2-4}
 & $42$: $(0,0,0;7,7)_{14}$, $[S^7]$ & $(0,0,3)$ & $d_1=d_2=0$; \textbf{lifted at $3$ loops ($d_3\ne0$)} from promoted Gram-$14$ source $x_0\in(3,3,3;4,4)_{16}$\\
\cline{2-4}
 & $46$: $[S^7\operatorname{tr}_{13}((\gamma^1)^2)]$ & $(0,\text{--},208^\dagger)$ & Incoming $d_1$ has rank $1$ and $d_2$ rank $0$, leaving dimension $2$ in $\mathrm{BPS}_{\hbar^2}$. The displayed product is hit by $d_3$.\\
\cline{2-4}
 & $47$: $[S^7\operatorname{tr}_{13}((\partial_{+\dot+}c)\gamma^1)]$ & $\text{--}$ & The displayed product survives through $\mathrm{BPS}_{\hbar^2}$ and is hit by $d_3$.\\
\cline{2-4}
 & $48$: $(0,0,0;8,8)_{16}$, $[S^8]$ & $(0,0,301)$ & $d_1=d_2=0$; \textbf{lifted at $3$ loops ($d_3\ne0$)} by quantum-ring product $[S]_{\hbar^2}[x_0]_{\hbar^2}\in(3,3,3;5,5)_{18}$ via~\eqref{eq-so13-s8-product-lift}\\
\cline{2-4}
 & $54,60$: $(0,0,0;p,p)_{2p}$, $[S^9],[S^{10}]$ & $(0,\text{--},\text{--})$ & Both lifted at $1$ loop ($d_1\ne0$) via~\eqref{eq-orthogonal-d1-family}; with $[S^7],[S^8]$ at $3$ loops, lifts all $SO(13)$ excess over $Sp(6)$\\
\cline{2-4}
 & $41$: $(1,0,0;7,6)_{14}$, $[S^6\operatorname{tr}_{13}((\partial_{+\dot+}c)\gamma^1)]$ & $(0,0,17)$ & $d_1=d_2=0$: certified nonzero in $\mathrm{BPS}_{\hbar^2}$\\
\hline
$SO(14)$ & $48$: $(0,0,0;8,8)_{16}$, $[S^8]$ & $(0,0,4)$ & $d_1=d_2=0$; $d_3\ne0$ reported in \cite[discussion D014]{Repo} ($O(14)$-even source, signed odd-cycle pairing)\\
\hline
$SO(19)$ & $60$: $(0,0,0;10,10)_{20}$, $[S^{10}]$ & $\text{--}$ & First Cartan-invisible glueball power: survival reported in D014 through $\mathrm{BPS}_{\hbar^4}$ ($d_1=\cdots=d_4=0$)\\
\hline
$SO(25),SO(26)$ & $78,84$: $[S^{13}]$, $[S^{14}]$ & $\text{--}$ & First Cartan-invisible glueball powers: survival reported in D014 through $\mathrm{BPS}_{\hbar^5}$ ($d_1=\cdots=d_5=0$)\\
\hline
\end{xltabular}}
{\footnotesize$^\dagger$Reduced auxiliary-weight determinant representatives spanning the $(4,2,2)$ finite-colour relations modulo $Q_0$ \cite{Repo}.\par}

\subsubsection{$\mathcal N=4$ Yang--Mills: exceptional groups $G_2$ and $F_4$}\label{sec-results-G2}

For $\mathfrak g=\mathfrak g_2$ in the action \eqref{local-ch9-su-action}, the six-scalar Gram determinant with traces in the $7$-dimensional fundamental representation,
\begin{equation}\label{local-ch9-g2-gram-class}
\left[\det_{1\le I,J\le 3}\operatorname{tr}_7(\gamma^I\gamma^J)\right]
=\left[\frac16\varepsilon_{IJK}\varepsilon_{LMN}
\operatorname{tr}_7(\gamma^I\gamma^L)
\operatorname{tr}_7(\gamma^J\gamma^M)
\operatorname{tr}_7(\gamma^K\gamma^N)\right]\ne0,
\end{equation}
defines a nonzero \emph{monotone} (multi-graviton) class in the $SL(3)$ flavour singlet of sector $(2,2,2;0,0)_6$ with charges $(6;2,2,2;0,0)$ (within the singlet, $\dim C^6=4$, the incoming $Q_0$ has rank $3$, and $\HQ=1$) that vanishes identically upon Cartan restriction ($\HQC=0$, certified over $\mathbb Q$ \cite[supplement, Section S5.2]{Repo}). By contrast, the same determinant is $Q_0$-exact for $SU(3)$ and $SO(5)=Sp(2)$, and Cartan-visible for $SU(4)$, $SO(7)$ and $Sp(3)$. Having a closed multi-graviton completion and $n_+=n_-=0$, \eqref{local-ch9-g2-gram-class} survives at every perturbative order. Likewise, the glueball cube $[S^3]\in(0,0,0;3,3)_6$ at level $18$, with charges $(9;3,3,3;0,0)$, is classically nonzero ($\HQ=1$ \cite{EtingofKac,CederwallFerretti}), vanishes on the rank-$2$ Cartan subalgebra ($\HQC=0$), and survives every perturbative order over $\mathbb Q$ ($d_1=0$, empty $d_2,d_3$ source cohomologies, and $n_+=n_-=3$ excludes $d_{r\ge4}$ by Lemma~\ref{b-00094}), showing that non-Cartan glueball powers need not lift perturbatively outside $SO(N)$ \cite{Repo}. Across the complete $G_2$ scan through $2(E+J_L)=23$ (reaching $\sum_{\text{sectors}}\HQ=2075$ and $\sum_{\text{sectors}}(\HQ-\HQC)=170$ at level $23$, \cite[supplement, Section S5.3]{Repo}), $d_1=0$ in every sector, no $d_2$ candidate pair exists, and every $d_3$ candidate has rank $0$ $[p]$; in the zero-derivative BMN subsector of $G_2$, graviton products span $\mathrm{BPS}_{\hbar^0}$ over $\mathbb Q$ across all $617$ nonvacuum highest-weight blocks through level $28$ (with $\HQ=0$ at odd $Y$). At level $24$, sector $(5,2,2;1,1)_9$---with charges $(11;5,2,2;1,0)$---contains a one-dimensional fortuitous class ($\HQ=f=1$ and $d_1=0$ over $\mathbb Q$). Deforming the reverse Taylor shear $\bar P_0$ by its one-loop partner $\bar P_1$ ($\{Q_0,\bar P_0\}=0$, $\{Q_1,\bar P_0\}=-\{Q_0,\bar P_1\}$, and $\{Q_1,\bar P_1\}=0$ for $2J_L=\Theta+n_++n_--Y\le2$) proves that $d_2$ intertwines the linear Taylor $\mathfrak{gl}(2|3)$ action on $J_L\le1$ sources and shifts weights by twice the Berezinian $(2,2\mid-2,-2,-2)$; super Schur--Weyl duality over $\mathbb Q$ \cite{Sergeev,BereleRegev} requires the degree-$12$ partition $(3,3,3,3)$, whence $d_2=0$ over $\mathbb Q$ whenever $J_L\le1$ and $\Theta+n_++n_-<12$ for any simple $\mathfrak g$ \cite{Repo}. Since $(5,2,2;1,1)_9$ has $\Theta+n_++n_-=11<12$ and $J_L=1$ (with empty higher source and target sectors), this fortuitous class survives every perturbative order over $\mathbb Q$.

For $\mathfrak g=\mathfrak f_4$ ($\operatorname{rk}=4$, $h^\vee=9$), writing $S=\operatorname{tr}_{26}\bigl((\partial_{+\dot+}c)(\partial_{+\dot-}c)\bigr)$ in the $26$-dimensional minimal representation, the classical pure-glueball ring is $\mathbb C[S]/(S^9)$ \cite{EtingofKac,CederwallFerretti,Kumar}, and $[S^p]_{\hbar^0}$ lies in the Cartan kernel ($\HQC=0$) for $5\le p\le8$. At level $36$ ($p=6$, charges $(18;6,6,6;0,0)$), solving $\lambda_{S^6}Q_1=L_1Q_0$ across $(1,1,1;5,5)_{12}$ certifies $[S^6]_{\hbar}\ne0$ over $\mathbb Q$, while reverse-Taylor equivariance confines any incoming $d_2$ from $(2,2,2;4,4)_{13}$ to the Taylor partition $(4,4,3,3)$ ($\dim=3{,}720$, where rational minors establish $\operatorname{rank}_{\mathbb Q}Q_0|_{13}=703$ and $\operatorname{rank}_{\mathbb Q}Q_0|_{12}\ge2{,}879$, leaving $\HQ\le138$, and the first-page factor tests of \cite{Repo}, together with the derivation rule and survival of $[S^5]$ below, give $d_2=0$ on all positive-degree classical products, restricting any two-loop lift of $[S^6]$ to indecomposable $13$-letter classes) \cite{Repo}. At level $30$ ($p=5$, charges $(15;5,5,5;0,0)$), the derivation property of $d_1$ on $[S][S^5]=[S^6]$ immediately gives $[S^5]\notin\operatorname{im}d_1$; the only Taylor component of the source $(2,2,2;3,3)_{11}$ that can map to $[S^5]$ under $d_2$, of shape $(3,3,3,3)$, has zero classical multiplicity over $\mathbb Q$, and reverse-Taylor equivariance excludes $d_{r\ge3}$, proving over $\mathbb Q$ that the first Cartan-invisible glueball power $[S^5]$ in $F_4$ survives every perturbative order \cite{Repo}.

\subsubsection{$SU(3)$ Leigh--Strassler deformation}\label{sec-results-LS}

This $\mathcal N=1$ superconformal deformation of $SU(3)$ $\mathcal N=4$ Yang--Mills theory \cite{LeighStrassler} gave the first two-loop lifts found in a non-abelian gauge theory. With three adjoint chiral multiplets and $\operatorname{tr}(t_At_B)=\tfrac12\delta_{AB}$ \cite[results report, \S6.1]{Repo}, the antichiral superpotential in \eqref{local-ch9-common-action} is\footnote{As noted after \eqref{local-ch9-common-action}, the constant-coupling gauge kinetic density $-\tfrac14\delta_{AB}w_{\dot a}^Aw^{B\dot a}$ is a holomorphic total derivative and contributes no interaction vertex here or in Sections~\ref{sec-results-beta}--\ref{sec-results-SQED}.}
\begin{equation}\label{local-ch9-ls-action}
\bar W(\gamma)=
\operatorname{tr}(\gamma^1\gamma^2\gamma^3)
-q\operatorname{tr}(\gamma^1\gamma^3\gamma^2)
+\frac h3\left[\operatorname{tr}((\gamma^1)^3)
+\operatorname{tr}((\gamma^2)^3)+\operatorname{tr}((\gamma^3)^3)\right].
\end{equation}

The deformation preserves $\Theta=\theta_1+\theta_2+\theta_3$ and $c=\theta_2+2\theta_3\pmod3$, so sectors are $(\Theta,c;n_+,n_-)_Y$ with conserved charges $(E,J_L,J_R,\sum_{I=1}^3R_I,c)$, and $d_r$ maps $(\Theta,c;n_+,n_-)_Y\to(\Theta-3r,c;n_++r,n_-+r)_{Y+1-r}$. Evaluating at generic rational couplings $\rho=(1-q)/(1+q)=12345/678$ and $h_e=2h/(1+q)=4321/97$ (with independent parameter and prime checks in \cite[supplement, Section S6.1]{Repo}), the complete census through $2(E+J_L)=22$ finds $d_1\ne0$ in $53$ sector pairs, $d_2\ne0$ in the $12$ sector pairs of Table~\ref{local-ch9-ls-lifts}, and no candidate pair for $d_3$. Writing $P_m$ for a primary source sector at level $m$, Table~\ref{local-ch9-ls-fortuity} compares the finite-rank dimension $h=\dim\mathrm{BPS}_{\hbar^0}$ against the trace-relation-free stable cohomology of $SU(N)$ ($N\ge Y+1$), giving $f\ge h_{SU(3)}-h_{SU(N)}$ with equality whenever $h_{SU(N)}=0$ \cite[supplement, Section S6.6]{Repo}.

\needspace{8\baselineskip}
{\footnotesize\renewcommand{\arraystretch}{1.26}\setlength{\tabcolsep}{3.5pt}
\begin{xltabular}{\linewidth}{|>{\raggedright\arraybackslash}X|l|l|l|c|}
\caption{Source sectors of the two-loop lifts in the $SU(3)$ Leigh--Strassler theory: classical dimension $h=\dim\mathrm{BPS}_{\hbar^0}$, large-$N$ stable comparison, fortuitous dimension $f$, and $\operatorname{rank}d_2$.}\label{local-ch9-ls-fortuity}\\
\hline
Source sector & $SU(3)$ $\dim\mathrm{BPS}_{\hbar^0}$ & Stable $\mathrm{BPS}_{\hbar^0}$ & Fortuitous dim.\ $f$ & $\operatorname{rank}d_2$\\
\hline\endfirsthead
\hline
Source sector & $SU(3)$ $\dim\mathrm{BPS}_{\hbar^0}$ & Stable $\mathrm{BPS}_{\hbar^0}$ & Fortuitous dim.\ $f$ & $\operatorname{rank}d_2$\\
\hline\endhead
$P_{14}:(7,0;0,0)_5$ & $h=5$ & $h_{SU(6)}=0$ & $f=5$\textsuperscript{a} (purely fortuitous) & $1$\\
\hline
$P_{16}:(8,0;0,0)_6$ & $h=4$ & $h_{SU(7)}=2$ & $f\ge2$\textsuperscript{a} & $1$\\
\hline
$\partial_{+\dot+}P_{14}:(7,0;1,0)_5$ & $h=5$ & $h_{SU(6)}=0$ & $f=5$\textsuperscript{a} (purely fortuitous) & $1$\\
\hline
$\partial_{+\dot+}^2P_{14}:(7,0;2,0)_5$ & $h=5$ & $h_{SU(6)}=0$ & $f=5$\textsuperscript{a} (purely fortuitous) & $1$\\
\hline
$\partial_{+\dot+}P_{16}:(8,0;1,0)_6$ & $h=6$ & $h_{SU(7)}=4$ & $f\ge2$\textsuperscript{a} & $1$\\
\hline
$P_{19}:(8,0;1,0)_7$ & $h=19$ & $h_{SU(8)}=6$ & $f\ge13$\textsuperscript{a} & $1$\\
\hline
$P_{21}:(9,0;1,0)_7$ & $h=40$ & $h_{SU(8)}=5$ & $f\ge35$\textsuperscript{a} & $6$\\
\hline
$P_{21}:(9,1;1,0)_7$ & $h=36$ & $h_{SU(8)}=0$ & $f=36$\textsuperscript{a} (purely fortuitous) & $6$\\
\hline
\end{xltabular}}

\needspace{8\baselineskip}
{\footnotesize\renewcommand{\arraystretch}{1.26}\setlength{\tabcolsep}{3.5pt}
\begin{xltabular}{\linewidth}{|l|l|l|l|c|>{\raggedright\arraybackslash}X|l|}
\caption{All $12$ two-loop lifts ($d_2\ne0$) in the $SU(3)$ Leigh--Strassler theory through $2(E+J_L)=22$. Source charges are $(E,J_L,J_R,\sum_{I=1}^3R_I)$; for $\Theta\not\equiv0\pmod3$, $c=0$ represents three $\mathbb Z_3$-cyclic images.}\label{local-ch9-ls-lifts}\\
\hline
$2(E+J_L)$ & Source & Source charges & Target & $\operatorname{rank}d_2$ & Multiplet origin & Field\\
\hline\endfirsthead
\hline
$2(E+J_L)$ & Source & Source charges & Target & $\operatorname{rank}d_2$ & Multiplet origin & Field\\
\hline\endhead
$14$ & $(7,0;0,0)_5$ & $(6,1,0,4)$ & $(1,0;2,2)_4$ & $1$ & $P_{14}$ & $[\mathbb Q]$\\
\hline
$16$ & $(8,0;0,0)_6$ & $(7,1,0,5)$ & $(2,0;2,2)_5$ & $1$ & $P_{16}$ & $[\mathbb Q]$\\
\hline
$17$ & $(7,0;1,0)_5$ & $(7,\tfrac32,\tfrac12,4)$ & $(1,0;3,2)_4$ & $1$ & $\partial_{+\dot+}P_{14}$ & $[\mathbb Q]$\\
\hline
$19$ & $(8,0;1,0)_6$ & $(8,\tfrac32,\tfrac12,5)$ & $(2,0;3,2)_5$ & $1$ & $\partial_{+\dot+}P_{16}$ & $[p]$\\
\hline
$19$ & $(8,0;1,0)_7$ & $(\tfrac{17}2,1,\tfrac12,\tfrac{13}2)$ & $(2,0;3,2)_6$ & $1$ & $P_{19}$ & $[p]$\\
\hline
$20$ & $(7,0;2,0)_5$ & $(8,2,1,4)$ & $(1,0;4,2)_4$ & $1$ & $\partial_{+\dot+}^2P_{14}$ & $[p]$\\
\hline
$21$ & $(9,c;1,0)_7$, $c=0,1$ & $(9,\tfrac32,\tfrac12,6)$ & $(3,c;3,2)_6$ & $6$ each & $P_{21}$ & $[p]$\\
\hline
$22$ & $(8,0;1,1)_7$ & $(\tfrac{19}2,\tfrac32,0,\tfrac{13}2)$ & $(2,0;3,3)_6$ & $5$ & level-$22$ mixed block & $[p]$\\
\hline
$22$ & $(8,0;2,0)_6$ & $(9,2,1,5)$ & $(2,0;4,2)_5$ & $1$ & $\partial_{+\dot+}^2P_{16}$ & $[p]$\\
\hline
$22$ & $(8,0;2,0)_7$ & $(\tfrac{19}2,\tfrac32,1,\tfrac{13}2)$ & $(2,0;4,2)_6$ & $5$ & includes $\partial_{+\dot+}P_{19}$ & $[p]$\\
\hline
$22$ & $(8,0;2,0)_8$ & $(10,1,1,8)$ & $(2,0;4,2)_7$ & $1$ & level-$22$ primary & $[p]$\\
\hline
\end{xltabular}}

The three lowest two-loop lifts ($P_{14}$, $P_{16}$ and $\partial_{+\dot+}P_{14}$) are certified over $\mathbb Q$ by explicit pairs $(\mathcal O_0,\mathcal O_1)$ and rational target functionals vanishing on $\operatorname{im}Q_0+Q_1\ker Q_0$ and nonvanishing on $Q_1\mathcal O_1$. On the lowest purely fortuitous sector $P_{14}$, the cohomology dimensions evolve as
\begin{equation}\label{local-ch9-ls-first-pages}
\begin{gathered}
P_{14}:\quad
\dim\mathrm{BPS}_{\hbar^0}=5\ \xrightarrow{\ \operatorname{rank}d_1=3\ }\
\dim\mathrm{BPS}_{\hbar}=2\ \xrightarrow{\ \operatorname{rank}d_2=1\ }\
\dim\mathrm{BPS}_{\hbar^2}=1,\\
Q_2\mathcal O_0=0,\qquad
Q_0\mathcal O_1=-Q_1\mathcal O_0,\qquad
d_2[\mathcal O_0]_{\hbar}=[Q_1\mathcal O_1]_{\hbar}\ne0.
\end{gathered}
\end{equation}
Here the direct two-loop fan $Q_2$ annihilates $\mathcal O_0$, yet $Q_1\mathcal O_0$ is a nonzero $Q_0$-coboundary requiring a compensating one-loop completion $\mathcal O_1$, and a second application of $Q_1$ to $\mathcal O_1$ produces the non-exact two-loop obstruction $[Q_1\mathcal O_1]_{\hbar}\ne0$. Of the five fortuitous classes in $P_{14}$, three lift at one loop, one lifts at two loops, and one survives to all perturbative orders because no $d_{r\ge3}$ target exists. Scheme independence and special parameter loci where $d_2$ vanishes are analysed in \cite[supplement, Sections S6.4--S6.5]{Repo}.

\subsubsection{$SU(3)$ $\beta$-deformation}\label{sec-results-beta}

Setting $h=0$ in \eqref{local-ch9-ls-action} gives the $\beta$-deformed superpotential
\begin{equation}\label{local-ch9-beta-action}
\bar W(\gamma)=\operatorname{tr}(\gamma^1\gamma^2\gamma^3)-q\operatorname{tr}(\gamma^1\gamma^3\gamma^2),
\end{equation}
which conserves all three weights $(\theta_1,\theta_2,\theta_3)$ separately. Scanning at $\rho=(1-q)/(1+q)=12345/678$ through $2(E+J_L)=21$ \cite[supplement, Section S6.7]{Repo}, every $d_2$ candidate through level $20$ and every $d_3$ candidate through level $21$ has rank $0$, while at $2(E+J_L)=21$ the single classical class in $(5,2,2;1,0)_8$---charges $(\tfrac{19}2;\tfrac92,\tfrac32,\tfrac32;1,\tfrac12)$---is lifted into $(3,0,0;3,2)_7$ ($\dim\mathrm{BPS}_{\hbar}=4$) with $\operatorname{rank}d_2=1$ $[p,p']$, again purely through iterated one-loop transfer:
\begin{equation}\label{local-ch9-beta-first-lift}
Q_2\mathcal O_0=0,\qquad
d_2[\mathcal O_0]_{\hbar}=[Q_1\mathcal O_1]_{\hbar}\ne0.
\end{equation}

\subsubsection{$\mathcal N=2$ orbifold quiver: $SU(N)\times SU(N)$}\label{sec-results-quiver}

At equal gauge couplings $g_1=g_2=g$, the twisted theory is the $\mathfrak{su}(2N)$ superfield action \eqref{local-ch9-su-action} restricted to superfields obeying the $\mathbb Z_2$ orbifold projection \cite[results report, \S3.1]{Repo}:
\begin{equation}\label{local-ch9-quiver-action}
\operatorname{diag}(\mathbf1_N,-\mathbf1_N)\,
C(\theta_1,\theta_2,\theta_3)\,
\operatorname{diag}(\mathbf1_N,-\mathbf1_N)
=C(-\theta_1,-\theta_2,\theta_3).
\end{equation}
Thus $c,b,\gamma^3,\beta_3$ are block diagonal (with each $N\times N$ block traceless) and $\gamma^{1,2},\beta_{1,2}$ are off diagonal, furnishing two $\mathfrak{su}(N)$ vector multiplets and two bifundamental hypermultiplets. Rescaling fields shows that the cohomology spaces are independent of the nonzero coupling ratio:
\begin{equation}\label{local-ch9-quiver-couplings}
\kappa=\frac{g_2}{g_1}\ne0
\quad\Longrightarrow\quad \mathrm{BPS}_{\hbar^{r-1}}(\kappa)\cong\mathrm{BPS}_{\hbar^{r-1}}(1)
\qquad(r\ge1).
\end{equation}

\needspace{8\baselineskip}
{\footnotesize\renewcommand{\arraystretch}{1.26}\setlength{\tabcolsep}{3.5pt}
\begin{xltabular}{\linewidth}{|l|>{\raggedright\arraybackslash}p{2.6cm}|>{\raggedright\arraybackslash}X|>{\raggedright\arraybackslash}X|}
\caption{Quantum lifting tests in the $\mathcal N=2$ $SU(N)\times SU(N)$ orbifold quiver \cite[supplement, Section S7.1]{Repo}.}\label{local-ch9-quiver-results}\\
\hline
Gauge group & Complete scan through & One-loop $d_1$ & Higher-loop tests ($r\ge2$)\\
\hline\endfirsthead
\hline
Gauge group & Complete scan through & One-loop $d_1$ & Higher-loop tests ($r\ge2$)\\
\hline\endhead
$SU(2)\times SU(2)$ & $2(E+J_L)=21$ & Nonzero in $61$ source blocks & All $37$ $d_2$ candidates have rank $0$ (checked at $\kappa^2=1,9$) $[p]$\\
\hline
$SU(3)\times SU(3)$ & $2(E+J_L)=18$ & Nonzero in $43$ source blocks & All $26$ $d_2$ candidates and the level-$18,20$ $d_3$ candidates have rank $0$ $[p]$\\
\hline
\end{xltabular}}

\subsubsection{$\mathcal N=2$ SQCD: $SU(N)$ with $N_f=2N$}\label{sec-results-SQCD}

Freezing the second gauge node of \eqref{local-ch9-quiver-action} keeps $c,b,\gamma^3,\beta_3$ in the first traceless $\mathfrak{su}(N)$ diagonal block while retaining $\gamma^{1,2},\beta_{1,2}$ in both off-diagonal blocks as $N_f=2N$ fundamental hypermultiplets \cite[results report, \S4.1]{Repo}. Inheriting $C_{ABC}$ and $\delta^{AB}$ from $\mathfrak{su}(2N)$, with $D,E$ running over the off-diagonal matter blocks and $F$ over the dynamical $\mathfrak{su}(N)$ adjoint, the superpotential is
\begin{equation}\label{local-ch9-sqcd-action}
\bar W(\gamma)=iC_{ABC}\delta^{AD}\delta^{BE}\delta^{CF}
(\gamma^1)_D(\gamma^2)_E(\gamma^3)_F,
\qquad G=SU(N),\quad N_f=2N.
\end{equation}

\needspace{8\baselineskip}
{\footnotesize\renewcommand{\arraystretch}{1.26}\setlength{\tabcolsep}{3.5pt}
\begin{xltabular}{\linewidth}{|l|>{\raggedright\arraybackslash}p{2.6cm}|>{\raggedright\arraybackslash}X|>{\raggedright\arraybackslash}X|}
\caption{Flavour-singlet lifting tests in superconformal $\mathcal N=2$ SQCD \cite[supplement, Section S7.1]{Repo}.}\label{local-ch9-sqcd-results}\\
\hline
Gauge group & Complete scan through & One-loop $d_1$ & Higher-loop tests ($r\ge2$)\\
\hline\endfirsthead
\hline
Gauge group & Complete scan through & One-loop $d_1$ & Higher-loop tests ($r\ge2$)\\
\hline\endhead
$SU(2)$, $N_f=4$ & $2(E+J_L)=24$ & $d_1=0$ throughout the census & All $42$ $d_2$ candidates have rank $0$ $[p]$\\
\hline
$SU(3)$, $N_f=6$ & $2(E+J_L)=18$ & Nonzero in $8$ sectors & All $13$ $d_2$ candidates have rank $0$ $[p]$\\
\hline
\end{xltabular}}

\subsubsection{Adjoint-mass deformations}\label{sec-results-masses}

Adding a nonzero mass $M\ne0$ for the third adjoint chiral multiplet modifies the one-node twisted actions by \cite[results report, \S5.1]{Repo}
\begin{equation}\label{local-ch9-mass-action}
\Delta S_{\mathrm{tw}}
=\int_{\mathbb C^2}\Omega\wedge
\left[\frac M2\delta^{AB}(\gamma^3)_A(\gamma^3)_B\right]_{(0,2)},
\end{equation}
and with opposite masses $(M,-M)$ on the two blocks of the $SU(2)\times SU(2)$ quiver. Integrating out $(\gamma^3)_F$ produces the infrared quartic $\mathcal N=1$ superpotential \cite[supplement, Section S7.2]{Repo} (where multi-loop $Q_n$ also receives spider contributions; Section~\ref{sec-forests}):
\begin{equation}\label{local-ch9-mass-ir-potential}
\bar W_{\mathrm{IR}}(\gamma)=\frac1{2M}
C_{ABC}C_{DEF}\delta^{CF}\delta^{AG}\delta^{BH}\delta^{DI}\delta^{EJ}
(\gamma^1)_G(\gamma^2)_H(\gamma^1)_I(\gamma^2)_J.
\end{equation}

Breaking the third flavour symmetry leaves the sector tuple $(q_A,q_B;n_+,n_-;\mathrm{gh})$---written $(A,B;n_+,n_-;\mathrm{gh})$ in \cite{Repo}, with integer flavour combinations $q_A=\theta_1+\theta_3$, $q_B=\theta_2+\theta_3$ and $\mathrm{gh}=Y-\Theta$---together with the conserved infrared weight $w_{\mathrm{IR}}=q_A+q_B+2(n_++n_-)$. Table~\ref{local-ch9-mass-classes} tracks three benchmark $SU(2)$ classes from \cite{CK} (including the square $w_{\mathrm{grav}}^2$ of the weight-$6$ single-graviton generator $w_{\mathrm{grav}}$ in $(3,3;0,0;-2)$), and Table~\ref{local-ch9-mass-census} summarizes the full mass-deformed scans \cite[supplement, Sections S7.2 and S8.4]{Repo}.

\needspace{8\baselineskip}
{\footnotesize\renewcommand{\arraystretch}{1.26}\setlength{\tabcolsep}{3.5pt}
\begin{xltabular}{\linewidth}{|l|l|>{\raggedright\arraybackslash}X|>{\raggedright\arraybackslash}X|}
\caption{All-order survival of benchmark classes in the mass-deformed $SU(2)$ $\mathcal N=4$ theory.}\label{local-ch9-mass-classes}\\
\hline
Class & Sector $(q_A,q_B;n_+,n_-;\mathrm{gh})$ & Classical origin & Quantum lifting tests \& certification\\
\hline\endfirsthead
\hline
Class & Sector $(q_A,q_B;n_+,n_-;\mathrm{gh})$ & Classical origin & Quantum lifting tests \& certification\\
\hline\endhead
$w_{\mathrm{grav}}^2$ ($w^2$ in \cite{CK}) & $(6,6;0,0;-4)$ & Product of the $(3,3;0,0;-2)$ graviton & Explicit closed completion through $\hbar^3$; survives at all orders $[\mathbb Q]$\\
\hline
$\mathcal O_7$\textsuperscript{c} & $(5,2;0,0;-2)$ & Cartan-visible fortuitous class \cite[Section 4.3.1]{CK} & $d_1=0$; higher target sectors empty; $d_r$-exact for no $r$. Survives at all orders $[p]$\\
\hline
$\mathcal O_{16}$\textsuperscript{c} & $(8,8;0,0;-5)$ & Cartan-invisible fortuitous class \cite[Section 4.4]{CK} & $d_1=d_2=d_3=d_4=0$; higher targets empty; $d_r$-exact for no $r$. Survives at all orders $[p]$\\
\hline
\end{xltabular}}

\needspace{8\baselineskip}
{\footnotesize\renewcommand{\arraystretch}{1.26}\setlength{\tabcolsep}{3.5pt}
\begin{xltabular}{\linewidth}{|>{\raggedright\arraybackslash}X|l|>{\raggedright\arraybackslash}p{3.2cm}|>{\raggedright\arraybackslash}X|}
\caption{Census of adjoint-mass-deformed theories ordered by $w_{\mathrm{IR}}$.}\label{local-ch9-mass-census}\\
\hline
Ultraviolet theory and mass & Complete through & Sectors with $d_1\ne0$ & Higher-loop results ($r\ge2$)\\
\hline\endfirsthead
\hline
Ultraviolet theory and mass & Complete through & Sectors with $d_1\ne0$ & Higher-loop results ($r\ge2$)\\
\hline\endhead
$\mathcal N=4$ $SU(2)$, mass $M$ & $w_{\mathrm{IR}}\le16$ & $(6,6;1,1;-3)$ & All tested $d_2,d_3$ candidate ranks $0$ $[p]$\\
\hline
$\mathcal N=4$ $SU(3)$, mass $M$ & $w_{\mathrm{IR}}\le13$ & None in scan & All tested $d_2,d_3$ candidate ranks $0$ $[p]$\\
\hline
$\mathcal N=2$ SQCD $SU(2)$, $N_f=4$, mass $M$ & $w_{\mathrm{IR}}\le15$ & None in scan & All tested $d_2,d_3$ candidate ranks $0$ $[p]$\\
\hline
$\mathcal N=2$ SQCD $SU(3)$, $N_f=6$, mass $M$ & $w_{\mathrm{IR}}\le13$ & None in scan & All tested $d_2$ candidate ranks $0$ $[p]$\\
\hline
$SU(2)\times SU(2)$ quiver, $(M,-M)$ & $w_{\mathrm{IR}}\le13$ & $(2,2;k,1;1)$, $k=1,2,3$ & All tested $d_2,d_3$ candidate ranks $0$ $[p]$\\
\hline
\end{xltabular}}

\subsubsection{Massive SQED with a cubic superpotential}\label{sec-results-SQED}

In this toy model, the gauge group is $G=U(1)$ coupled to chiral multiplets $(\gamma_+,\gamma_-,\gamma_0)$ of charges $(+1,-1,0)$ (representation matrix $t=\operatorname{diag}(-1,1,0)$) with nonvanishing couplings $gmy\ne0$ \cite[results report, \S5.4]{Repo}:
\begin{equation}\label{local-ch9-sqed-action}
\begin{aligned}
S_{\mathrm{tw}}=\int_{\mathbb C^2}\Omega\wedge\Big[
&b\bar\partial c
+(\bar\partial\gamma_+)\beta^+
+(\bar\partial\gamma_-)\beta^-
+(\bar\partial\gamma_0)\beta^0\\
&+igc(\gamma_+\beta^+-\gamma_-\beta^-)
+\frac m2\gamma_0^2+y\gamma_0\gamma_+\gamma_-\Big]_{(0,2)}.
\end{aligned}
\end{equation}

Its lift is the first two-loop lift that we found. In the weight-$14$ sector with derivative counts $(n_+,n_-)=(1,0)$, the classical cohomology is one-dimensional ($\dim\ker Q_0=18$ modulo $\dim\operatorname{im}Q_0=17$, so $\dim\mathrm{BPS}_{\hbar^0}=1$), spanned by the gauge-invariant cocycle with a free dotted index $\dot a$,
\begin{equation}\label{local-ch9-sqed-source}
\mathcal O_0=ig\Big[\beta^+\beta^-b^2
+\frac ym(\gamma_-\beta^--\gamma_+\beta^+)\beta^0b^2
+\frac{y^2}{3m\,ig}\gamma_+\gamma_-b^3\Big]
\partial_{+\dot a}c,
\qquad Q_0\mathcal O_0=0.
\end{equation}
Because $n_-=0$, Lemma~\ref{b-00094} excludes $d_r$-exactness of $[\mathcal O_0]_{\hbar^0}$ for every $r\ge1$. At one loop, the obstruction vanishes,
\begin{equation}\label{local-ch9-sqed-d1}
d_1[\mathcal O_0]_{\hbar^0}=[Q_1\mathcal O_0]_{\hbar^0}=0,\qquad Q_1\mathcal O_0=-Q_0\mathcal O_1,
\end{equation}
with $\mathcal O_1$ constructed explicitly in \cite[supplement, Section S8.1]{Repo}. At two loops, however, appending $Q_2\mathcal O_0+Q_1\mathcal O_1$ increases the exact rational rank of $\operatorname{im}Q_0+Q_1\ker Q_0$ in the target block from $394$ to $395$ $[\mathbb Q]$, certifying the explicit nonvanishing two-loop obstruction
\begin{equation}\label{local-ch9-sqed-lift}
d_2[\mathcal O_0]_{\hbar}
=-\frac{2(ig)^2y^4}{m^2(64\pi^2)^2}\Big[
\partial_{+\dot a}(\gamma_+\gamma_-)
\,\partial_{+\dot b}\partial_{+\dot c}(\gamma_+\gamma_-)
\,\partial_+{}^{\dot b}\partial_+{}^{\dot c}(\gamma_+\gamma_-)
\Big]_{\hbar}\ne0.
\end{equation}

\subsubsection{Lifts at two, three and four loops in abelian and $\mathcal N=2$ theories}\label{sec-results-higher}

Higher-loop lifts up to four loops occur across abelian gauge theories coupled to a neutral chiral multiplet $\gamma_0$ and in nonconformal $\mathcal N=2$ $SU(2)$ and $SU(3)$ gauge theories evaluated in components on the gauge-invariant subcomplex \cite[results report, \S8--\S9]{Repo}. Sectors are labelled $(\mathrm{wt};n_+,n_-;\mathrm{gh})_f$, where $\mathrm{wt}$ assigns $w_g$ to $\gamma_g$, $D-w_g$ to $\beta^g$, $D$ to $b$ and $0$ to $c$ ($D$ being the weight of $\bar W$), $\mathrm{gh}$ equals $0,-1,-2,1$ on $\gamma,\beta,b,c$, and $f$ is the flavour charge; then $R=2\,\mathrm{wt}/D+\mathrm{gh}$, $J_L=\tfrac12(n_++n_--\mathrm{gh})$, $J_R=\tfrac12(n_+-n_-)$, and $d_r$ maps $(\mathrm{wt};n_+,n_-;\mathrm{gh})\to(\mathrm{wt}-rD;n_++r,n_-+r;\mathrm{gh}+2r+1)$. Table~\ref{local-ch9-higher-lifts} records the lowest block with $d_r\ne0$ at each loop order $r$ found in each theory (alongside the weight-$14$ SQED block of \eqref{local-ch9-sqed-lift}).

\needspace{8\baselineskip}
{\footnotesize\renewcommand{\arraystretch}{1.26}\setlength{\tabcolsep}{3.5pt}
\begin{xltabular}{\linewidth}{|>{\raggedright\arraybackslash}X|l|l|c|c|l|}
\caption{Lowest blocks with nonvanishing higher-loop differentials $d_r$ ($r=2,3,4$) in abelian and nonconformal $\mathcal N=2$ gauge theories. For pure $\mathcal N=2$ Yang--Mills ($\bar W=0$), no superpotential fixes $D$, so the third column lists $(J_L,J_R)$ alone.}\label{local-ch9-higher-lifts}\\
\hline
Theory and potential $\bar W(\gamma)$ & Block $(\mathrm{wt};n_+,n_-;\mathrm{gh})_f$ & $(R;J_L,J_R)$ & $r$ & $\operatorname{rank}d_r$ & Field\\
\hline\endfirsthead
\hline
Theory and potential $\bar W(\gamma)$ & Block $(\mathrm{wt};n_+,n_-;\mathrm{gh})_f$ & $(R;J_L,J_R)$ & $r$ & $\operatorname{rank}d_r$ & Field\\
\hline\endhead
$U(1)$, $\tfrac m2\gamma_0^2+y\gamma_0\gamma_+\gamma_-$ (massive SQED) & $(12;1,1;-4)$ & $(2;3,0)$ & $2$ & $1$ & $[\mathbb Q]$\\
\cline{2-6}
 & $(14;1,0;-5)$ & $(2;3,\tfrac12)$ & $2$ & $1$ & $[\mathbb Q]$\\
\hline
$U(1)$, $\lambda_4\gamma_+^2\gamma_-^2$ (quartic SQED, coupling $\lambda_4\ne0$) & $(12;1,1;-4)$ & $(2;3,0)$ & $2$ & $1$ & $[\mathbb Q]$\\
\hline
$U(1)$, $y\gamma_0\gamma_+\gamma_-$ ($\mathcal N=2$ SQED, $N_f=1$) & $(11;1,1;-5)_{(0,-1)}$ & $(\tfrac73;\tfrac72,0)$ & $3$ & $1$ & $[\mathbb Q]$\\
\hline
$U(1)$, $y\gamma_0\gamma_+\gamma_-+\tfrac h3\gamma_0^3$ & $(9;1,1;-4)$ & $(2;3,0)$ & $2$ & $1$ & $[\mathbb Q]$\\
\cline{2-6}
 & $(11;1,0;-5)$ & $(\tfrac73;3,\tfrac12)$ & $3$ & $1$ & $[\mathbb Q]$\\
\cline{2-6}
 & $(16;1,1;-8)$ & $(\tfrac83;5,0)$ & $4$ & $1$ & $[\mathbb Q]$\\
\hline
$U(1)$, $y\gamma_0^2\gamma_+\gamma_-$ & $(11;1,1;-3)_{(0,-1)}$ & $(\tfrac52;\tfrac52,0)$ & $2$ & $1$ & $[\mathbb Q]$\\
\hline
$U(1)^2$, two cubic terms (triangle) & $(10;1,1;-4)_{(-1,0,-1,0)}$ & $(\tfrac83;3,0)$ & $2$ & $1$ & $[\mathbb Q]$\\
\hline
$U(1)^2$ triangle with massive neutral $\gamma_0$ & $(12;0,0;-4)$ & $(2;2,0)$ & $2$ & $1$ & $[\mathbb Q]$\\
\hline
$\mathcal N=2$ $U(1)^2$ linear quiver & $(11;1,1;-4)_{(-1,1,1,1)}$ & $(\tfrac{10}3;3,0)$ & $2$ & $1$ & $[p]$\\
\hline
Pure $\mathcal N=2$ $SU(2)$ Yang--Mills ($\bar W=0$) & $(10;0,0;-5)_{-2}$ & $(\tfrac52,0)$ & $2$ & $1$ & $[\mathbb Q]$\\
\hline
Pure $\mathcal N=2$ $SU(3)$ Yang--Mills ($\bar W=0$) & $(11;0,0;-5)_{-1}$ & $(\tfrac52,0)$ & $2$ & $1$ & $[p]$\\
\hline
$SU(2)$ with adjoint $\gamma_\Phi$ and singlet $\gamma_S$, $\tfrac12\gamma_S\operatorname{tr}\gamma_\Phi^2$ & $(9;1,1;-3)_0$ & $(3;\tfrac52,0)$ & $2$ & $1$ & $[\mathbb Q]$\\
\cline{2-6}
 & $(13;1,1;-5)_1$ & $(\tfrac{11}3;\tfrac72,0)$ & $3$ & $1$ & $[\mathbb Q]$\\
\cline{2-6}
 & $(14;1,0;-6)_2$ & $(\tfrac{10}3;\tfrac72,\tfrac12)$ & $3$ & $1$ & $[\mathbb Q]$\\
\hline
\end{xltabular}}

Three features deserve emphasis. First, in $U(1)$ with $\bar W=y\gamma_0\gamma_+\gamma_-+\tfrac h3\gamma_0^3$, the weight-$16$ block carries an explicit \emph{four-loop lift} $d_4\ne0$: its source is a two-term class on which $d_1,d_2,d_3$ all vanish, certified over $\mathbb Q$ across all $106{,}587$ monomials of the participating blocks. Second, $\mathcal N=2$ SQED with $N_f=1$---the Seiberg--Witten infrared theory at the monopole point of pure $\mathcal N=2$ $SU(2)$ Yang--Mills theory---has $d_2=0$ on every block where $\ker d_1\ne0$, so its first quantum lift occurs at \emph{three loops} ($d_3\ne0$). Third, pure $\mathcal N=2$ $SU(2)$ Yang--Mills exhibits $d_2\ne0$ in $12$ blocks through weight $14$ ($5$ certified over $\mathbb Q$), and coupling the flipping singlet $\gamma_S$ yields $d_2\ne0$ in $7$ blocks through weight $12$ ($5$ over $\mathbb Q$) together with three-loop lifts $d_3\ne0$ at weights $13$ and $14$.

\subsection{The circuit picture, local reductions and all-loop graph kernels}\label{sec-sum-circuit}

All the operator formulas and lifting tests above rest on evaluating the spacetime kernels $\mathcal I_\Gamma(\lambda;z)$ of $n$-loop holomorphic Feynman graphs $\Gamma=(V,I)$ ($|V|=n+2$, $|I|=2n+1$). In the edge-current representation of \cite{Wang,operatope} (Proposition~\ref{b-00169} and \eqref{eq-operatope}), each oriented edge $e=(e(0),e(1))\in I$ carries a Schwinger time $t_e>0$, a holomorphic displacement $z_e^{+\dot a}$ (generating input-letter derivatives), and a two-component current $y_{e,+\dot a}$ ($\dot a=\dot+,\dot-$); each vertex $v\in V$ carries the holomorphic momentum $\lambda_{v,+\dot a}$ of its surviving output letters ($\sum_{v\in V}\lambda_v=0$). Choosing any spanning tree $T\subset I$, the kernel is the current-space integral
\begin{equation}\label{eq-sum-operatope-def}
\mathcal I_\Gamma(\lambda;z)
=\int_{\Delta_\Gamma(\lambda)}
e^{-\sum_{\substack{e\in I\\\text{edges}}}y_{e,+\dot a}z_e^{+\dot a}}
\bigwedge_{e\in I\setminus T}(-dy_{e,+\dot+}\wedge dy_{e,+\dot-}),
\end{equation}
where $\Delta_\Gamma(\lambda)$ is the $2n$-dimensional manifold swept out by the solutions of Kirchhoff's current and cycle-voltage equations as the projective resistance ratios $(t_e)$ range over $\mathbb{RP}^{|I|-1}_{>0}$:
\begin{equation}\label{eq-sum-kirchhoff-laws}
\sum_{\substack{e:e(0)=v\\\text{outgoing}}}y_{e,+\dot a}
-\sum_{\substack{e:e(1)=v\\\text{incoming}}}y_{e,+\dot a}
=\lambda_{v,+\dot a},\qquad
\sum_{\substack{e\in C\\\text{along }C}}t_ey_{e,+\dot a}
-\sum_{\substack{e\in C\\\text{against }C}}t_ey_{e,+\dot a}=0
\quad(\text{every cycle }C).
\end{equation}
These are precisely Ohm's and Kirchhoff's laws of a linear electrical network on $\Gamma$ with resistances $t_e$ (Table~\ref{tab-sum-dictionary}).

\needspace{8\baselineskip}
{\small\renewcommand{\arraystretch}{1.25}\setlength{\tabcolsep}{4pt}
\begin{xltabular}{\linewidth}{@{}>{\raggedright\arraybackslash}X>{\raggedright\arraybackslash}Xl@{}}
\caption{Dictionary between holomorphic Feynman graph kernels \eqref{eq-sum-operatope-def}--\eqref{eq-sum-kirchhoff-laws} and electrical networks. $V_S$ and $I_S$ denote the vertex and edge sets of an induced subgraph $S\subseteq\Gamma$.}\label{tab-sum-dictionary}\\
\toprule
Graph kernel & Electrical network & \\
\midrule\endfirsthead
\toprule
Graph kernel & Electrical network & \\
\midrule\endhead
Schwinger time $t_e>0$ & resistance of resistor $e$ & \pictResistor\\
edge current $y_{e,+\dot a}$, $\dot a=\dot+,\dot-$ & two-component current through $e$ & \pictCurrent\\
antiholomorphic coordinate $\frac12x^-_{v,\dot a}$ & node potential at $v$ ($t_ey_e=\Delta(\tfrac12x^-)$: Ohm's law) & \\
output momentum $\lambda_{v,+\dot a}$ & external current injected at node $v$ & \pictSource\\
vertex delta functions & Kirchhoff's current conservation law & \pictKirchhoff\\
cycle constraint $\sum_{e\in C}\pm t_ey_e=0$ & Kirchhoff's loop voltage law & \pictLoop\\
weighted Laplacian \cite[Remark 6]{Wang} & network conductance matrix & \pictConductance\\
$-\partial/\partial z_e^{+\dot a}$ acting on $\mathcal I_\Gamma$ & moment insertion of the current $y_{e,+\dot a}$ & \pictInsert\\
Laman bound $|I_S|\le2|V_S|-3$ ($|V_S|\ge2$) \cite{operatope} & local Ohm/Kirchhoff bound on independent loop currents & \pictLaman\\
kernel $\mathcal I_\Gamma(\lambda;0)$ at zero displacement & oriented volume swept by $n$ independent chord currents & \pictVolume\\
\bottomrule
\end{xltabular}}

\begin{figure}[ht!]
\centering
\makebox[2em][l]{(a)}\raisebox{-.5\height}{\includegraphics[width=0.75\linewidth]{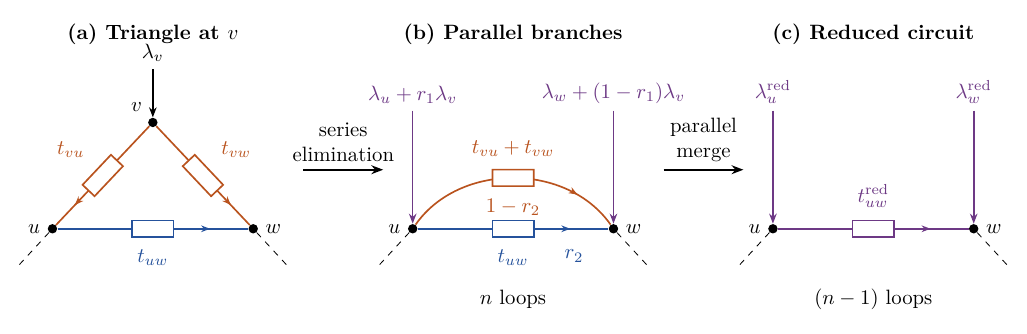}}\\[5pt]
\makebox[2em][l]{(b)}\raisebox{-.5\height}{\includegraphics[width=0.75\linewidth]{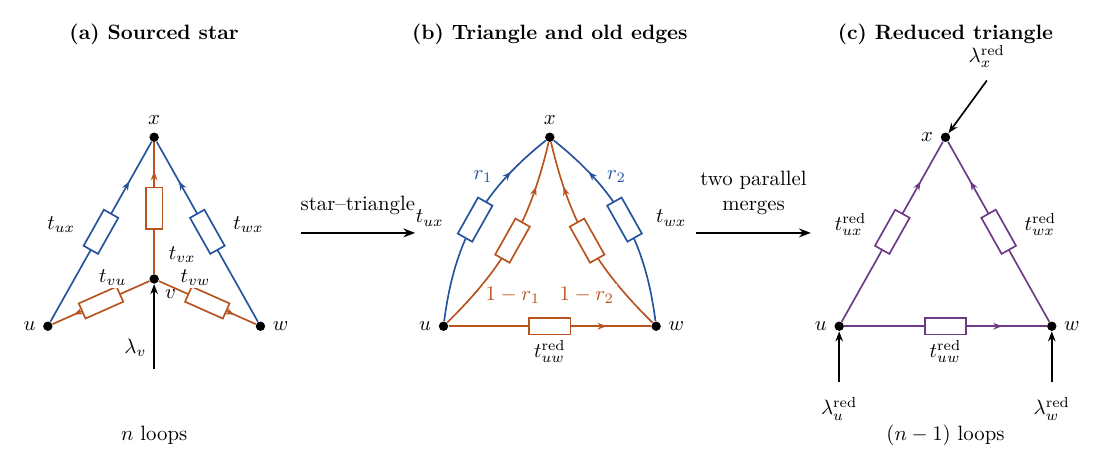}}\\[5pt]
\makebox[2em][l]{(c)}\raisebox{-.5\height}{\includegraphics[width=0.75\linewidth]{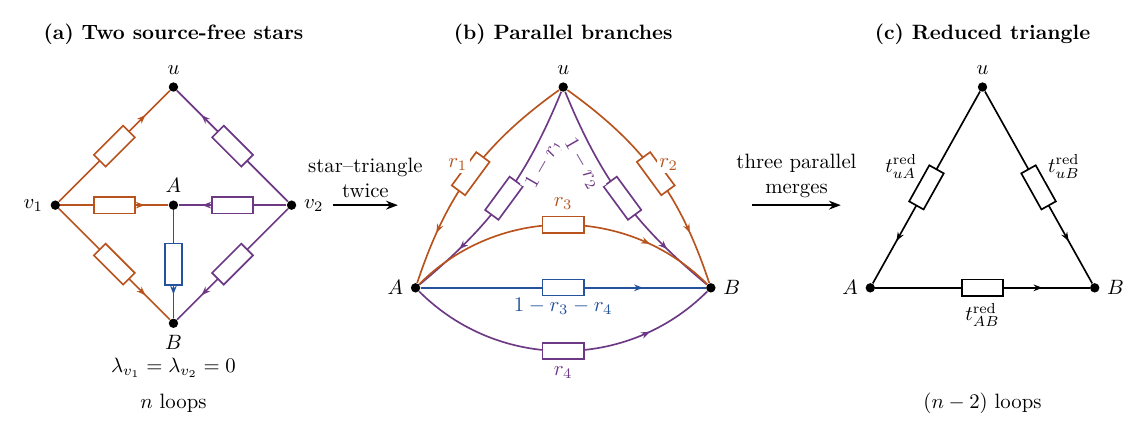}}
\caption{The three local network-reduction moves. (a) Series--parallel move: $\lambda_v$ divides between $u$ and $w$, and the series branch (orange) merges in parallel with the existing $uw$ resistor (blue), removing one loop. (b) Star--triangle ($Y$--$\Delta$) move: a degree-three vertex $v$ becomes a triangle on $u,w,x$ whose edges merge in parallel with $ux,wx$. (c) Paired source-free star--triangle move: two degree-three vertices sharing $(u,A,B)$ reduce simultaneously, removing two loops.}
\label{fig-sum-moves-a}
\label{fig-sum-moves-b}
\label{fig-sum-moves}
\end{figure}

Two structural properties make \eqref{eq-sum-operatope-def} computable in closed polynomial form at arbitrary loop order:

First, \emph{Ohm's law is local}. Any subnetwork of resistors is a linear black box determined for the rest of the graph solely by the relation between currents and potentials at its boundary terminals. Replacing that subnetwork by an electrically equivalent one---via series elimination, parallel merging, or the star--triangle ($Y$--$\Delta$) transformation, accompanied by current-source transfer when an eliminated node carries $\lambda_v\ne0$ (Figure~\ref{fig-sum-moves})---leaves every current outside the subnetwork unchanged.

Second, \emph{the integration domain $\mathbb{RP}^{|I|-1}_{>0}$ sweeps independently over all positive resistances}. Under each local replacement, the equivalent resistances of the reduced network again range over all of $\mathbb R_{>0}$ independently of the eliminated internal ratios, while the eliminated degrees of freedom become dimensionless current-division fractions $r\in(0,1)$ or standard simplex coordinates.

Boxes~(a) and~(b) record the resistance, source and reverse-current maps of the basic moves.

\begin{tcolorbox}[enhanced,colback=white,colframe=black!45,boxrule=0.5pt,arc=1pt,left=4pt,right=4pt,top=2pt,bottom=2pt,fonttitle=\small\bfseries,title={(a) Series--parallel move (Figure \ref{fig-sum-moves}(a)): a degree-two vertex $v$ between $u$ and $w$}]
\begin{equation}\label{eq-sum-moves}
\begin{gathered}
r_1=\frac{t_{vw}}{t_{vu}+t_{vw}},\qquad t^{(1)}_{uw}=t_{vu}+t_{vw},\qquad
\lambda^{\mathrm{red}}_u=\lambda_u+r_1\lambda_v,\qquad
\lambda^{\mathrm{red}}_w=\lambda_w+(1-r_1)\lambda_v,\\
\frac1{t^{\mathrm{red}}_{uw}}=\frac1{t^{(1)}_{uw}}+\frac1{t_{uw}},\qquad
r_2=\frac{t^{\mathrm{red}}_{uw}}{t_{uw}}=\frac{t_{vu}+t_{vw}}{t_{vu}+t_{vw}+t_{uw}},\\
y_{uw}=r_2y^{\mathrm{red}}_{uw},\qquad
y_{vu}=r_1\lambda_v-(1-r_2)y^{\mathrm{red}}_{uw},\qquad
y_{vw}=(1-r_1)\lambda_v+(1-r_2)y^{\mathrm{red}}_{uw}.
\end{gathered}
\end{equation}
The series step replaces $u$--$v$--$w$ by $t^{(1)}_{uw}$ and divides $\lambda_v$ between $u$ and $w$ in the ratio $r_1:(1-r_1)$; the parallel merge combines $t^{(1)}_{uw}$ with the existing resistor $t_{uw}$ between $u$ and $w$, so that the original edge carries $r_2y^{\mathrm{red}}_{uw}$ and the series branch carries $(1-r_2)y^{\mathrm{red}}_{uw}$.
\end{tcolorbox}

\begin{tcolorbox}[enhanced,colback=white,colframe=black!45,boxrule=0.5pt,arc=1pt,left=4pt,right=4pt,top=2pt,bottom=2pt,fonttitle=\small\bfseries,title={(b) Star--triangle move (Figure \ref{fig-sum-moves}(b)): a degree-three vertex $v$ with neighbours $u,w,x$}]
For a source-free star ($\lambda_v=0$), with $ux,wx$ already present and $uw$ absent:
\begin{equation}\label{eq-sum-star-triangle}
\begin{gathered}
t_{uw}^{\mathrm{red}}=t_{vu}+t_{vw}+\frac{t_{vu}t_{vw}}{t_{vx}},\qquad
\frac1{t_{ux}^{\mathrm{red}}}=\frac1{t_{ux}}+\frac{t_{vw}}{\Sigma},\qquad
\frac1{t_{wx}^{\mathrm{red}}}=\frac1{t_{wx}}+\frac{t_{vu}}{\Sigma},\\
\Sigma=t_{vu}t_{vw}+t_{vw}t_{vx}+t_{vx}t_{vu},\qquad
r_1=\frac{t_{ux}^{\mathrm{red}}}{t_{ux}},\qquad
r_2=\frac{t_{wx}^{\mathrm{red}}}{t_{wx}},\\
y_{vu}=-y_{uw}^{\mathrm{red}}-(1-r_1)y_{ux}^{\mathrm{red}},\qquad
y_{vw}=y_{uw}^{\mathrm{red}}-(1-r_2)y_{wx}^{\mathrm{red}},\\
y_{vx}=(1-r_1)y_{ux}^{\mathrm{red}}+(1-r_2)y_{wx}^{\mathrm{red}},\qquad
y_{ux}=r_1y_{ux}^{\mathrm{red}},\qquad y_{wx}=r_2y_{wx}^{\mathrm{red}}.
\end{gathered}
\end{equation}
Here $r_1,r_2\in(0,1)$ are the current fractions carried by the original $ux,wx$ resistors after parallel merging; the paired move of Figure~\ref{fig-sum-moves}(c) introduces four fractions $(r_1,r_2,r_3,r_4)\in(0,1)^2\times\Delta^2$ (Theorem~\ref{b-01173}). When $\lambda_v\ne0$, $\lambda_v$ divides rationally among $u,w,x$ as $\lambda_u^{\mathrm{red}}=\lambda_u+\frac{t_{vw}t_{vx}}{\Sigma}\lambda_v$ (and cyclically; Appendix~\ref{app-sourced-star}).
\end{tcolorbox}

Move~(a) supplies the concrete Schwinger-parameter meaning of the algebraic cube recursion found for type-I$'$ Laman graphs by Gui, Wang and Williams \cite[Theorem 4.13 and Remark 4.14]{GWW}, while moves~(b) and~(c) extend polynomial solvability beyond type-I$'$ graphs to all cubic product forests.

\rthm{Theorem}{thm-sum-all-kernels}{Classification and all-loop kernels of four-dimensional holomorphic graphs; Corollary~\ref{b-00247}, Theorems~\ref{thm-fan-kernel}, \ref{thm-spider-kernel}, \ref{b-00249} and \ref{b-00405}} \emph{For the twisted four-dimensional $\mathcal N=1$ interaction \eqref{eq-sum-interaction-density}, iterative circuit reduction classifies and evaluates the contributing $n$-loop graphs ($n\ge1$) as follows:}

(i) \emph{\textbf{Fans ($Q_n$ for $\deg\bar W\le3$).} When $\bar W$ has degree at most three (including all renormalizable $\mathcal N=1$ gauge theories, massless or massive), every nonzero connected graph of $Q_n\mathcal O$ is a fan. In the coordinates \eqref{eq-sum-integration-domain}, its kernel at $\lambda_2=\cdots=\lambda_n=0$, $z_{i,i+1}=0$ and $z_{i,n+2}=-z_i$ is}
\begin{equation}\label{eq-sum-fan-kernel-explicit}
\begin{aligned}
\mathcal I_{\mathrm{fan}}(\lambda;z)
={}&(\lambda_{1,+\dot a}\lambda_{n+1,+}{}^{\dot a})^n
\int_{\eqref{eq-sum-integration-domain}}d^{2n}r\,
\left[\prod_{j=1}^{n-1}r_{2j-1}^{n-j}r_{2j}(1-r_{2j})^{j-1}\right]r_{2n}^{n-1}\\
&\hspace{38mm}\times
e^{\lambda_{1,+\dot b}x_1^{+\dot b}(r,z)+\lambda_{n+1,+\dot b}x_{n+1}^{+\dot b}(r,z)},\\
\mathcal I_{\mathrm{fan}}(\lambda;0)
={}&\frac{(\lambda_{1,+\dot a}\lambda_{n+1,+}{}^{\dot a})^n}{(n!)^3(n+1)},
\end{aligned}
\end{equation}
\emph{with $x_1(r,z),x_{n+1}(r,z)$ in \eqref{eq-sum-N4-points} and general Taylor coefficients \eqref{eq-sum-N4-coeff}.}

(ii) \emph{\textbf{Spiders ($Q_n$ for higher-degree $\bar W$).} For $\bar W$ of arbitrary degree, every nonzero connected graph of $Q_n\mathcal O$ contains at most one $\bar W$ vertex and is either a fan or a spider of $a\ge3$ cubic arms of lengths $\ell_1,\ldots,\ell_a$ ($\sum_{i=1}^a\ell_i=n$) meeting at $A=\bar W$ and wired to $B=\mathcal O$. On $0<r_{i,j}<1$ ($j<2\ell_i$), $r_{i,2\ell_i}>0$, $\sum_{i=1}^a r_{i,2\ell_i}<1$, its $Q_n$ kernel ($\lambda_{v_{i,j}}=0$ for $j>1$) is polynomial in $r$:}
\begin{equation}\label{eq-sum-spider-kernel-explicit}
\begin{aligned}
\mathcal I_{\mathrm{spider}}(\lambda;z)
={}&\int\prod_{i=1}^a\prod_{j=1}^{2\ell_i}dr_{i,j}\,
\left(1-\sum_{\substack{i=1\\\text{arms}}}^{a}r_{i,2\ell_i}\right)
\prod_{i=1}^a\Bigg[r_{i,2\ell_i}^{\ell_i-1}
\prod_{j=1}^{\ell_i-1}r_{i,2j-1}^{\ell_i-j}r_{i,2j}^{j-1}(1-r_{i,2j})\\
&\hspace{46mm}\times\bigl(\lambda_{v_{i,1},+\dot a}y_{AB,+}^{\mathrm{red}}{}^{\dot a}\bigr)^{\ell_i}\Bigg]
e^{-\sum_{\substack{e\in I\\\text{edges}}}y_{e,+\dot c}(r)z_e^{+\dot c}},
\end{aligned}
\end{equation}
\emph{where $y_{AB}^{\mathrm{red}}=\lambda_A+\sum_{\substack{i=1\\\mathrm{arms}}}^a\bigl(\prod_{k=1}^{\ell_i}r_{i,2k-1}\bigr)\lambda_{v_{i,1}}$ and all edge currents $y_e(r)$ are reconstructed polynomially by \eqref{eq-1-51}.}

(iii) \emph{\textbf{Cubic product forests ($A\star B$ for $\deg\bar W\le3$).} Removing the insertions $\{A,B\}$ from any nonzero $n$-loop marked graph of $A\star B$ leaves a forest $\coprod_j T_j$ of trees meeting both $A$ and $B$, with at most one $\bar W$ vertex per tree. When $\deg\bar W\le3$, each tree $T_j$ carries a single source at its unique degree-two vertex, and iterating source-free single and paired star--triangle moves from the leaves to the root followed by one series--parallel move reduces $\Gamma$ to the marked edge $AB$. Every edge current and the Jacobian density are polynomials on an explicit product of unit intervals and two-simplices, so every displacement Taylor coefficient is a finite sum of Dirichlet factorial ratios \eqref{eq-Dirichlet}.}

\needspace{8\baselineskip}
{\small\renewcommand{\arraystretch}{1.25}\setlength{\tabcolsep}{4.5pt}
\begin{xltabular}{\linewidth}{|>{\raggedright\arraybackslash}p{3.0cm}|>{\raggedright\arraybackslash}X|>{\raggedright\arraybackslash}X|>{\raggedright\arraybackslash}p{3.1cm}|}
\caption{Comparison between the global Kirchhoff spanning-tree representation in Schwinger times \cite{Wang} (Appendix~\ref{sec-B-spanning}) and iterated network reduction in current-division coordinates $r$.}\label{tab-sum-complexity}\\
\hline
Graph family ($n$ loops) & Schwinger-time integrand (original method) & Iterated circuit reduction (this work) & Exact evaluation of each Taylor term\\
\hline\endfirsthead
\hline
Graph family ($n$ loops) & Schwinger-time integrand (original method) & Iterated circuit reduction (this work) & Exact evaluation of each Taylor term\\
\hline\endhead
Fan ($Q_n$, $\deg\bar W\le3$) & Rational function with spanning-tree denominator ($F_{2n+2}$ trees, growing exponentially with $n$) & Factored polynomial density on $(0,1)^{2n-2}\times\Delta^2$ (series--parallel moves) & Closed factorial ratio \eqref{eq-sum-N4-coeff}; $O(n)$ arithmetic operations per term\\
\hline
Spider ($Q_n$, general $\bar W$, $a$ arms) & Rational function over $\prod_{i=1}^a F_{2\ell_i+1}$--scale spanning-tree polynomials & Polynomial density \eqref{eq-sum-spider-kernel-explicit} on $(0,1)^{2n-a}\times\Delta^a$ (arm-wise series--parallel) & Finite sum of Beta and Dirichlet factorial ratios\\
\hline
Cubic product forest ($A\star B$, $n\ge3$ non-type-I$'$) & Rational function over $2n$ resistances; outside the type-I$'$ cube recursion of \cite{GWW} & Polynomial currents \& density on intervals $\times$ two-simplices (adds single/paired $Y$--$\Delta$) & Exact rational moments \eqref{eq-Dirichlet} (archived through $7$ loops \cite{KernelRepo})\\
\hline
Higher-$\bar W$ forest with multi-source star & Rational spanning-tree quotient & Sourced $Y$--$\Delta$ reduction (Appendix~\ref{app-sourced-star}): rational transfer in remaining $t^{\mathrm{red}}$ & Reduces dimension by $2$ per eliminated star\\
\hline
\end{xltabular}}

\section{Review of holomorphic twist}\label{sec-BV}

This section is a self-contained review designed solely for readers unfamiliar with the holomorphic twist~\cite{semichiral,BK} and its perturbative graph operations~\cite{Costello,Wang,operatope,higher}. Its final goal is to derive the $d^2y_e$ operatope current integral~\eqref{eq-C-current-delta-integral} (and its independent-current form~\eqref{eq-operatope}), which serves as the starting point for all loop computations in the remainder of the paper. Readers already familiar with the holomorphic twist and the operatope integral~\eqref{eq-C-current-delta-integral}--\eqref{eq-operatope} may skip directly to Section~\ref{sec-restrictions}.

The starting point is the BV formulation of the physical theory, in which the fields, ghosts, and their antifields carry the differential $(S_{\mathrm{BV}},\,\cdot\,)$. Without the supercharge, its cohomology on local operators consists of gauge-invariant operators modulo the equations of motion; including one nilpotent supercharge $Q_-$ in $S_{\mathrm{BV}}$ yields a complex whose cohomology consists of gauge-invariant operators annihilated by $Q_-$ modulo $Q_-$-exact ones, namely the physical BPS sector~\eqref{eq-twist-cohomology}.

The holomorphic twist simplifies this complex in three steps. First, canonical changes of BV variables remove the $Q_-$-exact part of the action and pair most fields into contractible doublets $(u,v)$ with $(S_{\mathrm{BV}},u)\propto v$---explicitly $u=\{V_{+\dot a}^A,\lambda_+^A,\phi^i,\psi_+^i\}$ and $v=\{\bar\lambda_{\dot a}^{\prime A},D^{\prime A},\psi_-^{\prime i},\mathcal F^{\prime i}\}$ (Table~\ref{b-00904})---leaving the twisted action $S_{\mathrm{tw}}$ on a small retained subcomplex $\mathcal S$ (Section~\ref{canonical-changes-of-variables}). Second, the retained fields and antifields assemble into four Dolbeault forms $c,b,\gamma,\beta$ on $\mathbb C^2$, in terms of which $S_{\mathrm{tw}}$ is first order with a polynomial interaction (Section~\ref{the-retained-fields}), while their zero-form components reproduce the BV-completed BPS letters of~\cite{ChangYin} (Section~\ref{bv-completion-of-the-changyin-letters}). Third, passing to the cohomology of the free differential $\bar\partial$ leaves polynomials in the zero-form letters and their holomorphic derivatives (Section~\ref{dolbeault-contraction-and-wick-derivatives}), on which the interaction acts quantum-mechanically through higher brackets built from heat-kernel and Bochner--Martinelli graph integrals (Sections~\ref{sec-brackets} and~\ref{sec-brackets-origin}), culminating in the $d^2y_e$ operatope integral of Section~\ref{positive-schwinger-times-and-the-operatope}.

\subsection{Canonical changes of variables and the twisted action}\label{canonical-changes-of-variables}

We use the Euclidean spinor and Lie-algebra conventions~\eqref{eq-sum-spinor-conventions} of Section~\ref{sec-summary}, obtained under $x^0=ix_E^0$ via $x^{a\dot a}=x_E^\mu\bar\sigma_\mu^{\dot a a}$ ($\partial_\mu=\partial/\partial x_E^\mu$, $\partial_{a\dot a}=\sigma^\mu_{a\dot a}\partial_\mu=-2\,\partial/\partial x^{a\dot a}$, $\epsilon^{+-}=1=-\epsilon_{+-}$, and $dx^-_{\dot a}=\epsilon_{\dot a\dot b}dx^{-\dot b}$). In the untwisted off-shell $\mathcal N=1$ gauge theory~\cite[Chapters 26--27]{Weinberg}, the vector multiplet and BRST sector comprise the 0-form gauge ghost $c_{\mathrm{gh}}^A$, gauge field $V_\mu^A$, gauginos $\lambda_a^A,\bar\lambda_{\dot a}^A$, and auxiliary field $D^A$, with field strength $f_{\mu\nu}^A=\partial_\mu V_\nu^A-\partial_\nu V_\mu^A+gC^A{}_{BC}V_\mu^BV_\nu^C$ and self-dual component $f_{++}^A=\partial_{+\dot a}V_+^{A\dot a}+\tfrac g2C^A{}_{BC}V_{+\dot a}^BV_+^{C\dot a}$; a chiral matter multiplet has components $(\phi^i,\psi_a^i,\mathcal F^i)$ with $\mathcal F^i$ auxiliary, and $\bar W_{\dot a}^A(x,\theta,\bar\theta)$ denotes the antichiral field-strength superfield~\cite[Eqs.~(27.2.12), (27.3.16)]{Weinberg}. Starting from this physical action, we perform two BV canonical transformations (``adding $Q_-$-exact terms to the action'') to obtain the twisted action $S_{\mathrm{tw}}$ in~\eqref{eq-B-5} and the cohomology isomorphism~\eqref{eq-twist-cohomology}.

Write $\Phi^{\mathcal A}$ for all component fields and gauge ghosts, $\Phi^\ddagger_{\mathcal A}$ for their antifields, and $s$ for the gauge BRST differential. The BV antibracket is
\begin{equation}\label{eq-A-8}
(F,G)=\int d^4x\left[
\frac{\delta_RF}{\delta\Phi^\ddagger_{\mathcal A}}
\frac{\delta_LG}{\delta\Phi^{\mathcal A}}
-\frac{\delta_RF}{\delta\Phi^{\mathcal A}}
\frac{\delta_LG}{\delta\Phi^\ddagger_{\mathcal A}}\right].
\end{equation}
Let $K(\bar\Phi,\Phi)$ be the K\"ahler potential and $W(\Phi,W_a)$ the chiral superpotential including the gauge kinetic term, where $\nabla_a,\bar\nabla_{\dot a}$ are gauge-covariant superspace derivatives and a vertical bar evaluates at $\theta=\bar\theta=0$. Define the odd gauge-fixing functional
\begin{equation}\label{eq-Psi}
\Psi=\int d^4x\left[\frac18\nabla_+\bar\nabla^2K(\bar\Phi,\Phi)
+\frac12\nabla_+W(\Phi,W_a)\right]\Big|_{\theta=\bar\theta=0}.
\end{equation}
For the chosen nilpotent supercharge $Q_-$, the physical action equals $Q_-\Psi$ up to the antichiral superpotential:
\begin{equation*}
\begin{aligned}
S_{\mathrm{phys}}
&=\int d^4x\,\Bigl[
\frac18\nabla_-\nabla_+\bar\nabla^2K(\bar\Phi,\Phi)
+\frac12\nabla_-\nabla_+W(\Phi,W_a)
+\frac14\bar\nabla^2\bar W(\bar\Phi,\bar W_{\dot a})
\Bigr]\Big|_{\theta=\bar\theta=0},\\[4pt]
Q_-\Psi
&=\int d^4x\,\nabla_-\!\left[
\frac18\nabla_+\bar\nabla^2K(\bar\Phi,\Phi)
+\frac12\nabla_+W(\Phi,W_a)
\right]\Big|_{\theta=\bar\theta=0}\\
&=S_{\mathrm{phys}}-\frac14\int d^4x\,
\bar\nabla^2\bar W(\bar\Phi,\bar W_{\dot a})\Big|_{\theta=\bar\theta=0}.
\end{aligned}
\end{equation*}
Writing $S_{\mathrm{free}}$ for the free quadratic kinetic action and $[\bar W]_{\bar{\mathcal F}}=\tfrac14\bar\nabla^2\bar W|_{\theta=\bar\theta=0}$ for the antichiral top component, the physical and BV actions decompose as
\begin{equation}\label{eq-B-1}
\begin{aligned}
S_{\mathrm{phys}}
&=Q_-\Psi+\int d^4x\,[\bar W(\bar\Phi,\bar W_{\dot a})]_{\bar{\mathcal F}},\\
S_{\mathrm{BV}}
&=S_{\mathrm{phys}}+\int d^4x\,((s+Q_-)\Phi^{\mathcal A})\Phi_{\mathcal A}^{\ddagger},\\
Q_{\mathrm{BV}}&=(S_{\mathrm{free}},\,\cdot),\qquad
s\Psi=0,\qquad(s+Q_-)^2=0.
\end{aligned}
\end{equation}

Performing the BV canonical transformation generated by $\Psi$,
\begin{equation}\label{eq-B-2}
\begin{aligned}
\Phi_{\mathcal A,\mathrm{new}}^{\ddagger}
&=\Phi_{\mathcal A,\mathrm{old}}^{\ddagger}
+\frac{\delta_L\Psi}{\delta\Phi^{\mathcal A}},\qquad \Phi^{\mathcal A}_{\mathrm{new}}=\Phi^{\mathcal A}_{\mathrm{old}},\\
S_{\mathrm{BV,new}}
&=S_{\mathrm{BV,old}}\left(\Phi,
\Phi_{\mathrm{new}}^{\ddagger}-\frac{\delta_L\Psi}{\delta\Phi}\right)
=\int d^4x\left([
\bar W(\bar\Phi,\bar W_{\dot a})]_{\bar{\mathcal F}}
+((s+Q_-)\Phi^{\mathcal A})\Phi_{\mathcal A,\mathrm{new}}^{\ddagger}\right),
\end{aligned}
\end{equation}
cancels $Q_-\Psi$ from the action, leaving only the antichiral superpotential term $[\bar W]_{\bar{\mathcal F}}$ alongside the $(s+Q_-)$ variations.

Most component fields now arrange into the four upper-triangular pairs $(u,v)$ of Table~\ref{b-00904}, satisfying $(S_{\mathrm{BV}},u)=av$ for a nonzero constant $a$ and $(S_{\mathrm{BV}},v)=0$ by nilpotence. Changing coordinates via $(\bar\lambda,D,\psi_-,\mathcal F)\mapsto(\bar\lambda',D',\psi_-',\mathcal F')$ has unit Jacobian. Let $\mathcal T$ collect these pairs $(u,v)$ together with all their spacetime derivatives and antifields, and let $\mathcal S=\{c_{\mathrm{gh}},V_-,\lambda_-;\bar\phi,\bar\psi,\bar{\mathcal F}\}$ denote the remaining fields, which transform solely among themselves.

\needspace{8\baselineskip}
{\small\renewcommand{\arraystretch}{1.25}\setlength{\tabcolsep}{4pt}
\begin{xltabular}{\linewidth}{@{}l>{\raggedright\arraybackslash}Xl@{}}
\caption{Contractible pairs of the BV complex.}\label{b-00904}\\
\toprule
$u$ & $v$ & $(S_{\mathrm{BV}},u)=av$\\
\midrule\endfirsthead
\toprule
$u$ & $v$ & $(S_{\mathrm{BV}},u)=av$\\
\midrule\endhead
$V_{+\dot a}^A$ & $\bar\lambda_{\dot a}^{\prime A} =\bar\lambda_{\dot a}^A-\tfrac12\nabla_{+\dot a}c_{\mathrm{gh}}^A$ & $a=-2$\\
$\lambda_+^A$ & $D^{\prime A} =D^A+gC^A{}_{BC}c_{\mathrm{gh}}^B\lambda_+^C-if_{+-}^A$ & $a=-1$\\
$\phi^i$ & $\psi_-^{\prime i} =\psi_-^i-\tfrac g{\sqrt2}c_{\mathrm{gh}}^A(t_A)^i{}_j\phi^j$ & $a=-i\sqrt2$\\
$\psi_+^i$ & $\mathcal F^{\prime i} =\mathcal F^i+\tfrac g{\sqrt2}c_{\mathrm{gh}}^A(t_A)^i{}_j\psi_+^j$ & $a=i\sqrt2$\\
\bottomrule
\end{xltabular}}

\rthm{Remark}{rem-ch3-S-vs-CY}{Retained fields versus physical BPS letters}
Eliminating the contractible subcomplex $\mathcal T$ retains $\mathcal S$ and its antifields. Note that the bare fields in $\mathcal S$ are not directly the physical BPS letters $\{\bar\phi_i,\psi_+^i,f_{++}^A,\bar\lambda_{\dot a}^A\}$ and $\nabla_{+\dot a}$ of~\cite{ChangYin}: instead, $\psi_+^i$ and $f_{++}^A$ survive inside the shifted zero-form antifields $\beta^{(0)i}$ and $b_A^{(0)}$, as made explicit in~\eqref{eq-B-13} and~\eqref{eq-B-14}.

On field functionals, the contracting homotopy $h_{\mathcal T}$ of the pairs $(u,v)$ (summed over all spacetime derivatives) defines a second antifield translation $\Psi'$ that eliminates the $\mathcal T$-dependence of $[\bar W]_{\bar{\mathcal F}}$:
\begin{equation}\label{eq-B-4}
\begin{aligned}
h_{\mathcal T}\mathcal O
&=\int_0^1\frac{dt}{t}
\left.\left(\sum_{(u,v)}\frac ua\frac{\partial_L\mathcal O}{\partial v}\right)
\right|_{(u,v)\mapsto(tu,tv)},\\
\Psi'&=h_{\mathcal T}\int d^4x\,[\bar W(\bar\Phi,\bar W_{\dot a})]_{\bar{\mathcal F}},\\
(s+Q_-)\Psi'
&=\int d^4x\,[\bar W]_{\bar{\mathcal F}}
-\left.\int d^4x\,[\bar W]_{\bar{\mathcal F}}\right|_{\mathcal T=0}.
\end{aligned}
\end{equation}
Applying the canonical transformation generated by $\Psi'$ as in~\eqref{eq-B-2} splits the BV action into two decoupled pieces:
\begin{equation}\label{eq-B-5}
\begin{gathered}
S_{\mathrm{BV,new}}=S_{\mathrm{tw}}+S_{\mathcal T},\qquad
S_{\mathcal T}=\int d^4x\sum_{(u,v)}av\,u^\ddagger,\\
(S_{\mathrm{tw}},S_{\mathcal T})=0,\qquad
(S_{\mathrm{tw}},S_{\mathrm{tw}})=(S_{\mathcal T},S_{\mathcal T})=0,
\end{gathered}
\end{equation}
where $S_{\mathrm{tw}}$ depends only on the retained fields $\mathcal S$ and their antifields (written out in~\eqref{eq-B-8}), and $S_{\mathcal T}$ is the trivial quadratic action of the contractible pairs.

\begin{proofbox}
\textbf{Derivation of the $\mathcal T$-contraction~\eqref{eq-T-representative}.} Extending $h_{\mathcal T}$ to the antifields $(u^\ddagger,v^\ddagger)$ and all spacetime derivatives gives
\begin{equation}\label{eq-B-6}
\begin{aligned}
(S_{\mathrm{BV}},v^\ddagger)&=(-1)^{|u|}a u^\ddagger,
&(S_{\mathrm{BV}},u^\ddagger)&=0,\\
h_{\mathcal T}\mathcal O
&=\int_0^1\frac{dt}{t}\left.
\sum_{(u,v)}\left(\frac ua\frac{\partial_L}{\partial v}
+\frac{(-1)^{|u|}v^\ddagger}{a}
\frac{\partial_L}{\partial u^\ddagger}\right)\mathcal O
\right|_{\mathcal T\mapsto t\mathcal T},\\
(S_{\mathrm{BV}},h_{\mathcal T}\mathcal O)
+h_{\mathcal T}(S_{\mathrm{BV}},\mathcal O)
&=\mathcal O-\mathcal O|_{\mathcal T=0},\qquad [h_{\mathcal T},\partial_\mu]=0.
\end{aligned}
\end{equation}
Consequently, every $(S_{\mathrm{BV}},\,\cdot\,)$-cocycle $\mathcal O$ is cohomologous to its restriction to $\mathcal T=0$:
\begin{equation}\label{eq-T-representative}
\mathcal O=\mathcal O|_{\mathcal T=0}
+(S_{\mathrm{BV}},h_{\mathcal T}\mathcal O),
\end{equation}
whose representative $\mathcal O|_{\mathcal T=0}$ depends only on the retained complex $\mathcal S$.~\hfill\qedsymbol
\end{proofbox}

In particular, setting $V_{+\dot a}^A=0$ and $\bar\lambda^{\prime A}_{\dot a}=0$ in Table~\ref{b-00904} projects
\begin{equation}\label{eq-T-projection}
\nabla_{+\dot a}\longmapsto\partial_{+\dot a},\qquad
\bar\lambda_{\dot a}^A\longmapsto
\frac12\partial_{+\dot a}c_{\mathrm{gh}}^A
=-\frac{\partial c_{\mathrm{gh}}^A}{\partial x^{+\dot a}}.
\end{equation}
Classically, the complex of $S_{\mathrm{tw}}$ computes the physical $Q_-$-cohomology: on local operators invariant under constant gauge transformations and independent of the undifferentiated ghost $c_{\mathrm{gh}}^A$, in each finite-dimensional $2(E+J_L)$ eigenspace we have
\begin{equation}\label{eq-twist-cohomology}
H\bigl((S_{\mathrm{tw}},\,\cdot\,)\bigr)
\;\cong\;H\bigl((S_{\mathrm{BV}},\,\cdot\,)\bigr)
\;\cong\;H\Bigl(Q_-\,;\ \frac{\text{gauge-invariant local operators}}{\text{equations of motion}}\Bigr).
\end{equation}
The first isomorphism follows from~\eqref{eq-T-representative}; the second identifies the cohomology of the combined BRST--Koszul--Tate--$Q_-$ differential with $Q_-$-cohomology on physical gauge-invariant operators modulo equations of motion~\cite{CostelloTwist,SaberiWilliams,ESW,semichiral,BK} (proved in Proposition~\ref{prop-twist-cohomology} and its perturbative counterpart Proposition~\ref{prop-twist-cohomology-quantum} in Appendix~\ref{app-twist-cohomology}).

\subsection{The retained fields as Dolbeault forms}\label{the-retained-fields}

In the retained complex $\mathcal S$, the fields and antifields assemble into four superfields in $\bar\theta^{\dot a}$, equivalently Dolbeault forms in $dx^{-\dot a}=4\bar\theta^{\dot a}$ on $\mathbb C^2$:
\begin{equation}\label{eq-B-7}
\begin{aligned}
c^A&=c_{\mathrm{gh}}^A-2\bar\theta_{\dot a}V_-^{A\dot a}
+2i\bar\theta_{\dot a}\bar\theta^{\dot a}\lambda_-^A,\\
\gamma_i&=\bar\phi_i-\sqrt2\bar\theta_{\dot a}\bar\psi_i^{\dot a}
+\bar\theta_{\dot a}\bar\theta^{\dot a}\bar{\mathcal F}_i,\\
\beta^i&=\bar{\mathcal F}^{\ddagger i}
+\sqrt2\bar\theta_{\dot a}\bar\psi^{\ddagger\dot ai}
+\bar\theta_{\dot a}\bar\theta^{\dot a}\bar\phi^{\ddagger i},\\
b_A&=-\frac i2\lambda_A^{\ddagger-}
+\bar\theta_{\dot a}V_A^{\ddagger-\dot a}
+\bar\theta_{\dot a}\bar\theta^{\dot a}c_{\mathrm{gh},A}^{\ddagger},\\
\bar\partial&=2\bar\theta_{\dot a}\partial_-{}^{\dot a},
\qquad dx^{-\dot a}=4\bar\theta^{\dot a}.
\end{aligned}
\end{equation}
Here $\bar\psi^\ddagger_{\dot a}{}^i$ is BV-dual to $\bar\psi_i^{\dot a}$, with dotted index raised as in~\eqref{eq-sum-spinor-conventions} so that $\bar\psi^\ddagger_{\dot a}{}^i((s+Q_-)\bar\psi_i^{\dot a})=-((s+Q_-)\bar\psi_{i\dot a})\bar\psi^{\ddagger\dot a i}$ and $\sqrt2\bar\theta_{\dot a}\bar\psi^{\ddagger\dot a i}=-\sqrt2\bar\psi^\ddagger_{\dot a}{}^i\bar\theta^{\dot a}$. With $dx^{-\dot a}=4\bar\theta^{\dot a}$, we have $\bar\partial=-\frac12dx^{-\dot a}\partial_{-\dot a}=2\bar\theta_{\dot a}\partial_-{}^{\dot a}$, and the relative coefficients in~\eqref{eq-B-7} are fixed so that the free BV differential acts on all four forms simply as $\bar\partial$ (for instance, $sV_-^{A\dot a}=\partial_-{}^{\dot a}c_{\mathrm{gh}}^A+\cdots$ gives $(S_{\mathrm{BV}},-2\bar\theta_{\dot a}V_-^{A\dot a})=2\bar\theta_{\dot a}\partial_-{}^{\dot a}c_{\mathrm{gh}}^A+\cdots$, matching the $(0,1)$-form part of $\bar\partial c^A$).

Using $dx^{-\dot a}=4\bar\theta^{\dot a}$ together with the Euclidean volume relation
\begin{equation}\label{b-00919}
\Omega\wedge dx^{-\dot+}\wedge dx^{-\dot-}=-8\,d^4x_E,
\qquad d^4x_E=dx_E^0\wedge dx_E^1\wedge dx_E^2\wedge dx_E^3,
\end{equation}
rewrites $S_{\mathrm{tw}}$ of~\eqref{eq-B-5} as a holomorphic--Dolbeault action on $\mathbb C^2$:

\rthm{Proposition}{b-00164}{Twisted action and interaction density~\cite{semichiral,BK}} With $(t_A)^j{}_i$ the matter-representation generators, the action of the holomorphic twist is
\begin{equation}\label{eq-B-8}
\begin{aligned}
S_{\mathrm{tw}}=\int_{\mathbb C^2}\Omega\wedge\Big[
&b_A\left(\bar\partial c^A
-\frac g2C^A{}_{BC}c^Bc^C\right)\\
&+\left(\bar\partial\gamma_i-igc^A(t_A)^j{}_i\gamma_j\right)\beta^i
+\bar W(\gamma,w)\Big]_{(0,2)},
\end{aligned}
\end{equation}
where $[\cdot]_{(0,2)}$ projects onto antiholomorphic form degree two, $\Omega=2\,dx^{+\dot+}\wedge dx^{+\dot-}$, and $\bar W(\gamma,w)$ collects the antichiral superpotential and gauge kinetic terms restricted to $\mathcal T=0$.

In the main text, $\bar W$ is a polynomial (or formal power series read coefficientwise); superpotentials with covariant derivatives are treated in Appendix~\ref{app-covariant-derivatives}. Separating $S_{\mathrm{tw}}=S_{\mathrm{free}}+I$, the free action and the interaction density $\mathcal I$ entering the higher brackets are
\begin{equation}\label{eq-BV-free-action}
\begin{aligned}
S_{\mathrm{free}}&=\int_{\mathbb C^2}\Omega\wedge
[b_A\bar\partial c^A+(\bar\partial\gamma_i)\beta^i]_{(0,2)},\\
Q_{\mathrm{BV}}&=(S_{\mathrm{free}},\cdot)=\bar\partial,
\qquad (S_{\mathrm{tw}},\cdot)=Q_{\mathrm{BV}}+(I,\cdot),
\end{aligned}
\end{equation}
\begin{equation}\label{eq-B-10}
\mathcal I=-\frac g2C^A{}_{BC}b_Ac^Bc^C
-igc^A(t_A)^j{}_i\gamma_j\beta^i
+\bar W(\gamma,w),\qquad
I=\int_{\mathbb C^2}\Omega\wedge[\mathcal I]_{(0,2)}.
\end{equation}
Note that $\mathcal I$ in~\eqref{eq-B-10} is viewed as a polynomial in the full inhomogeneous Dolbeault fields $(c,b,\gamma,\beta)$: in the higher brackets~\eqref{eq-A-28}, Wick contractions act before the final projection onto $H(\bar\partial)$, so the propagator and heat-kernel forms supply the $(0,2)$ antiholomorphic top degree at each integrated vertex while uncontracted fields reduce to their zero-form holomorphic Taylor jets.

Writing a superscript $(0)$ for the $dx^{-\dot a}=0$ zero-form part, the two antifield shifts~\eqref{eq-B-2} generated by $\Psi$ and $\Psi'$ relate $\beta^{(0)i}$ and $b_A^{(0)}$ to the original physical fields and antifields by
\begin{equation}\label{eq-B-11}
\begin{aligned}
\beta^{(0)i}
&=\bar{\mathcal F}_{\mathrm{old}}^{\ddagger i}
+\frac{\delta_L(\Psi+\Psi')}{\delta\bar{\mathcal F}_i},\\
b_A^{(0)}
&=-\frac i2\left(
\lambda_{A,\mathrm{old}}^{\ddagger-}
+\frac{\delta_L(\Psi+\Psi')}{\delta\lambda_-^A}\right).
\end{aligned}
\end{equation}
For canonical matter and quadratic gauge kinetic terms, the combined generator is
\begin{equation}\label{eq-B-12}
\begin{aligned}
\Psi+\Psi'=\int d^4x\,\Big[
&-\frac1{\sqrt2}\phi^i\nabla_{+\dot a}\bar\psi_i^{\dot a}
+g\bar\phi_j(t_A)^j{}_i\phi^i\lambda_+^A\\
&-\frac i{\sqrt2}\bar{\mathcal F}_i\psi_+^i
-\frac i{\sqrt2}\psi_+^i\frac{\partial W(\phi,0)}{\partial\phi^i}\\
&+\frac i2\delta_{AB}f_{++}^A\lambda_-^B
-\frac12\delta_{AB}(D^A+if_{+-}^A)\lambda_+^B\Big],
\end{aligned}
\end{equation}
which yields
\begin{equation}\label{eq-B-13}
\beta^{(0)i}
=\bar{\mathcal F}_{\mathrm{old}}^{\ddagger i}
-\frac i{\sqrt2}\psi_+^i,\qquad
 b_A^{(0)}
=-\frac i2\lambda_{A,\mathrm{old}}^{\ddagger-}
+\frac14\delta_{AB}f_{++}^B.
\end{equation}
Combining~\eqref{eq-B-7} and~\eqref{eq-B-13} gives the full component dictionary.

\rthm{Proposition}{b-00159}{Component dictionary} \emph{For canonical matter and quadratic gauge kinetic terms, the Dolbeault fields and the untwisted antichiral field-strength form $\bar W_{\dot a}^A|_{\theta=0,\bar\theta=dx^-/4}$ expand in physical components as}
\begin{equation}\label{eq-field-form-dictionary}
\begin{aligned}
c^A={}&c_{\mathrm{gh}}^A-\frac12dx^-_{\dot a}V_-^{A\dot a}
 +\frac i4dx^{-\dot+}\wedge dx^{-\dot-}\lambda_-^A,\\
\gamma_i={}&\bar\phi_i-\frac1{2\sqrt2}dx^-_{\dot a}\bar\psi_i^{\dot a}
 +\frac18dx^{-\dot+}\wedge dx^{-\dot-}\bar{\mathcal F}_i,\\
\beta^i={}&\bar{\mathcal F}_{\mathrm{old}}^{\ddagger i}
 -\frac i{\sqrt2}\psi_+^i
 +\frac1{2\sqrt2}dx^-_{\dot a}\bar\psi^{\ddagger\dot a i}
 +\frac18dx^{-\dot+}\wedge dx^{-\dot-}\bar\phi^{\ddagger i},\\
b_A={}&\frac14\delta_{AB}f_{++}^B
 -\frac i2\lambda_{A,\mathrm{old}}^{\ddagger-}
 +\frac14dx^-_{\dot a}V_A^{\ddagger-\dot a}
 +\frac18dx^{-\dot+}\wedge dx^{-\dot-}c_{\mathrm{gh},A}^\ddagger,\\
\left.\bar W_{\dot a}^A\right|_{\substack{\theta=0\\
 \bar\theta^{\dot b}=dx^{-\dot b}/4}}={}&\bar\lambda_{\dot a}^A+\frac i4dx^-_{\dot a}D^A
 +\frac18dx^{-\dot b}\epsilon^{de}
 \sigma^\mu_{d\dot a}\sigma^\nu_{e\dot b}f_{\mu\nu}^A
 -\frac i8dx^{-\dot+}\wedge dx^{-\dot-}
 \nabla^d{}_{\dot a}\lambda_d^A.
\end{aligned}
\end{equation}

Upon restricting $\bar W_{\dot a}^A|_{\theta=0,\bar\theta=dx^-/4}$ to $\mathcal T=0$ via Table~\ref{b-00904} (setting $V_{+\dot a}^A=0$, $\bar\lambda_{\dot a}^A=\tfrac12\partial_{+\dot a}c_{\mathrm{gh}}^A$, $\lambda_+^A=0$, and $D^A=if_{+-}^A=\tfrac i2\partial_{+\dot b}V_-^{A\dot b}$), all surviving components collapse into the holomorphic derivative of $c^A$:
\begin{equation}\label{footnote-16--eq3.4}
\begin{aligned}
w_{\dot a}^A
:=\left.\bar W_{\dot a}^A\right|_{\substack{\theta=0,\;\bar\theta^{\dot b}=dx^{-\dot b}/4\\\mathcal T=0}}
&=\frac12\partial_{+\dot a}c_{\mathrm{gh}}^A
-\frac14dx^-_{\dot b}\partial_{+\dot a}V_-^{A\dot b}
+\frac i8dx^{-\dot+}\wedge dx^{-\dot-}
 \partial_{+\dot a}\lambda_-^A\\
&=\frac12\partial_{+\dot a}c^A
=-\frac{\partial c^A}{\partial x^{+\dot a}}.
\end{aligned}
\end{equation}
In particular, on zero-forms $[\bar\lambda_{\dot a}^A]=[w_{\dot a}^{A(0)}]=\tfrac12[\partial_{+\dot a}c^{A(0)}]$, and for a constant quadratic gauge kinetic term in $\bar W(\gamma,w)$ (since $c^A$ is odd and $\epsilon^{\dot a\dot b}$ is antisymmetric),
\begin{equation}\label{b-00931}
-\frac14\delta_{AB}w_{\dot a}^Aw^{B\dot a}
=-\frac1{16}\delta_{AB}(\partial_{+\dot a}c^A)(\partial_+{}^{\dot a}c^B)
=-\frac1{16}\delta_{AB}\partial_{+\dot a}
  (c^A\partial_+{}^{\dot a}c^B),
\end{equation}
which is a holomorphic total derivative and therefore drops out of $S_{\mathrm{tw}}$.

\subsection{BV completion of the Chang--Yin letters}\label{bv-completion-of-the-changyin-letters}

For the canonical kinetic terms of~\eqref{eq-B-12}, the physical BPS letters of Chang and Yin~\cite[Section 2]{ChangYin} are $\bar\phi_i$, $\psi_+^i$, $f_{++}^A$, and $\bar\lambda_{\dot a}^A$, modulo the auxiliary $\bar{\mathcal F}$ and gaugino $\lambda^-$ equations of motion:
\begin{equation}\label{eq-CY-component-equations}
\mathcal F^i+\frac{\partial\bar W(\bar\phi,0)}{\partial\bar\phi_i}=0,
\qquad
\nabla_{+\dot a}\bar\lambda^{A\dot a}
-\sqrt2g\delta^{AB}\bar\phi_j(t_B)^j{}_i\psi_+^i=0.
\end{equation}
Within the BV complex, the original antifields implement these equations via
\begin{equation}\label{eq-CY-antifield-equations}
\begin{aligned}
(S_{\mathrm{BV}},\bar{\mathcal F}_{\mathrm{old}}^{\ddagger i})
&=igc_{\mathrm{gh}}^A(t_A)^i{}_j\bar{\mathcal F}_{\mathrm{old}}^{\ddagger j}
-\mathcal F^i-\frac{\partial\bar W(\bar\phi,0)}{\partial\bar\phi_i},\\
(S_{\mathrm{BV}},\lambda_{A,\mathrm{old}}^{\ddagger-})
&=-gC^B{}_{AC}c_{\mathrm{gh}}^C\lambda_{B,\mathrm{old}}^{\ddagger-}
+i\delta_{AB}\nabla_{+\dot a}\bar\lambda^{B\dot a}
-i\sqrt2g\bar\phi_j(t_A)^j{}_i
 \left(\psi_+^i+i\sqrt2\bar{\mathcal F}_{\mathrm{old}}^{\ddagger i}\right).
\end{aligned}
\end{equation}
The BV-completed fermion and self-dual curvature are precisely the zero-form fields~\eqref{eq-B-13}:
\begin{equation}\label{eq-CY-completed-letters}
\psi_+^i+i\sqrt2\bar{\mathcal F}_{\mathrm{old}}^{\ddagger i}
=i\sqrt2\beta^{(0)i},\qquad
f_{++}^A-2i\delta^{AB}\lambda_{B,\mathrm{old}}^{\ddagger-}
=4\delta^{AB}b_B^{(0)}.
\end{equation}
Combining $Q_-\psi_+^i=i\sqrt2\mathcal F^i$ and $Q_-f_{++}^A=-2\nabla_{+\dot a}\bar\lambda^{A\dot a}$ with~\eqref{eq-CY-antifield-equations} cancels both $\mathcal F^i$ and $\nabla_{+\dot a}\bar\lambda^{A\dot a}$ identically, keeping the retained complex $\mathcal S$ closed under $(S_{\mathrm{BV}},\,\cdot\,)$:
\begin{equation}\label{eq-CY-completion-images}
\begin{aligned}
(S_{\mathrm{BV}},
 \psi_+^i+i\sqrt2\bar{\mathcal F}_{\mathrm{old}}^{\ddagger i})
&=igc_{\mathrm{gh}}^A(t_A)^i{}_j\left(\psi_+^j+i\sqrt2\bar{\mathcal F}_{\mathrm{old}}^{\ddagger j}\right)
-i\sqrt2\frac{\partial\bar W(\bar\phi,0)}{\partial\bar\phi_i},\\
(S_{\mathrm{BV}},f_{++}^A-2i\delta^{AB}\lambda_{B,\mathrm{old}}^{\ddagger-})
&=-4g\delta^{AB}C^C{}_{BD}c_{\mathrm{gh}}^Db_C^{(0)}
-4ig\delta^{AB}\bar\phi_j(t_B)^j{}_i\beta^{(0)i}.
\end{aligned}
\end{equation}

At an insertion point $x=0$, physical operators are gauge-invariant polynomials independent of the undifferentiated ghost $c^{A(0)}(0)=c_{\mathrm{gh}}^A(0)$. Writing $\longleftrightarrow$ for the image under the $\mathcal T=0$ contraction~\eqref{eq-T-representative} followed by projection onto zero-forms and their holomorphic derivatives (with parentheses denoting weight-one symmetrization of dotted indices, $n\ge0$), the complete letter dictionary is
\begin{equation}\label{eq-B-14}
\begin{aligned}
\nabla_{+(\dot a_1}\cdots\nabla_{+\dot a_n)}\bar\phi_i
&\longleftrightarrow
\partial_{+\dot a_1}\cdots\partial_{+\dot a_n}\gamma_i^{(0)},\\
\nabla_{+(\dot a_1}\cdots\nabla_{+\dot a_n)}
\left(\psi_+^i+i\sqrt2\bar{\mathcal F}_{\mathrm{old}}^{\ddagger i}\right)
&\longleftrightarrow
 i\sqrt2\partial_{+\dot a_1}\cdots\partial_{+\dot a_n}\beta^{(0)i},\\
\nabla_{+(\dot a_1}\cdots\nabla_{+\dot a_n)}
\left(f_{++}^A-2i\delta^{AB}\lambda_{B,\mathrm{old}}^{\ddagger-}\right)
&\longleftrightarrow
 4\delta^{AB}\partial_{+\dot a_1}\cdots\partial_{+\dot a_n}b_B^{(0)},\\
2\nabla_{+(\dot a_1}\cdots\nabla_{+\dot a_n}\bar\lambda_{\dot a_0)}^A
&\longleftrightarrow
\partial_{+\dot a_0}\cdots\partial_{+\dot a_n}c^{A(0)}.
\end{aligned}
\end{equation}

\begin{proofbox}
\textbf{Derivation of unsymmetrized covariant-derivative transfers.} Ordered covariant derivatives decompose into symmetrized derivatives plus $f_{++}^A$ curvature commutators, which require BV completion by $\lambda_{B,\mathrm{old}}^{\ddagger-}$ to lie in $\mathcal S$. Using $\nabla_{+\dot a}\bar\phi_i=\partial_{+\dot a}\bar\phi_i+igV_{+\dot a}^A(t_A)^j{}_i\bar\phi_j$, we compute
\begin{equation}\label{eq-ordered-scalar-variation}
\begin{aligned}
(S_{\mathrm{BV}},\nabla_{+\dot a}\nabla_+{}^{\dot b}\bar\phi_i)
={}&-igc_{\mathrm{gh}}^A(t_A)^j{}_i\nabla_{+\dot a}\nabla_+{}^{\dot b}\bar\phi_j\\
&-2ig\Big[
 \frac12\left(\nabla_{+\dot a}\bar\lambda^{A\dot b}
 +\nabla_+{}^{\dot b}\bar\lambda_{\dot a}^A\right)\bar\phi_j
 +\bar\lambda_{\dot a}^A\nabla_+{}^{\dot b}\bar\phi_j
 +\bar\lambda^{A\dot b}\nabla_{+\dot a}\bar\phi_j
 \Big](t_A)^j{}_i\\
&-ig\delta_{\dot a}{}^{\dot b}
 (\nabla_{+\dot c}\bar\lambda^{A\dot c})
 (t_A)^j{}_i\bar\phi_j,
\end{aligned}
\end{equation}
whose gaugino-divergence term cancels against the variation of $g\delta_{\dot a}{}^{\dot b}\delta^{AB}\lambda_{B,\mathrm{old}}^{\ddagger-}(t_A)^j{}_i\bar\phi_j$:
\begin{equation}\label{eq-ordered-scalar-complement}
\begin{aligned}
&\left(S_{\mathrm{BV}},
g\delta_{\dot a}{}^{\dot b}\delta^{AB}\lambda_{B,\mathrm{old}}^{\ddagger-}
 (t_A)^j{}_i\bar\phi_j\right)
=-ig^2\delta_{\dot a}{}^{\dot b}\delta^{AB}c_{\mathrm{gh}}^C
\lambda_{B,\mathrm{old}}^{\ddagger-}(t_C)^j{}_i(t_A)^k{}_j\bar\phi_k\\
&\qquad+ig\delta_{\dot a}{}^{\dot b}
 (\nabla_{+\dot c}\bar\lambda^{A\dot c})(t_A)^j{}_i\bar\phi_j
-i\sqrt2g^2\delta_{\dot a}{}^{\dot b}\delta^{AB}
 \bar\phi_k(t_B)^k{}_\ell
 \left(\psi_+^\ell+i\sqrt2\bar{\mathcal F}_{\mathrm{old}}^{\ddagger\ell}\right)
 (t_A)^j{}_i\bar\phi_j.
\end{aligned}
\end{equation}
Adding the two gives the transferred representative
\begin{equation}\label{eq-B-16}
\begin{aligned}
&\nabla_{+\dot a}\nabla_+{}^{\dot b}\bar\phi_i
+g\delta_{\dot a}{}^{\dot b}\delta^{AB}\lambda_{B,\mathrm{old}}^{\ddagger-}
 (t_A)^j{}_i\bar\phi_j\\
&\quad=\frac12\left(\nabla_{+\dot a}\nabla_+{}^{\dot b}
+\nabla_+{}^{\dot b}\nabla_{+\dot a}\right)\bar\phi_i
+\frac{ig}{2}\delta_{\dot a}{}^{\dot b}
 \left(f_{++}^A-2i\delta^{AB}\lambda_{B,\mathrm{old}}^{\ddagger-}\right)
 (t_A)^j{}_i\bar\phi_j\\
&\quad\longleftrightarrow
\partial_{+\dot a}\partial_+{}^{\dot b}\gamma_i^{(0)}
+2ig\delta_{\dot a}{}^{\dot b}\delta^{AB}b_B^{(0)}
 (t_A)^j{}_i\gamma_j^{(0)}.
\end{aligned}
\end{equation}
Likewise, for the antisymmetric derivative of the gaugino~\cite[Eq.~(2.4)]{ChangYin}, equation~\eqref{eq-CY-antifield-equations} supplies the explicit BV coboundary
\begin{equation}\label{eq-B-17}
\begin{aligned}
&\nabla_{+\dot a}\bar\lambda^{A\dot b}
+\frac{ig}{2}\delta_{\dot a}{}^{\dot b}\delta^{AB}
 C^C{}_{BD}c_{\mathrm{gh}}^D\lambda_{C,\mathrm{old}}^{\ddagger-}\\
&\quad-\left[\frac12\left(\nabla_{+\dot a}\bar\lambda^{A\dot b}
+\nabla_+{}^{\dot b}\bar\lambda_{\dot a}^A\right)
+\frac g{\sqrt2}\delta_{\dot a}{}^{\dot b}\delta^{AB}
 \bar\phi_j(t_B)^j{}_i
 \left(\psi_+^i+i\sqrt2\bar{\mathcal F}_{\mathrm{old}}^{\ddagger i}\right)\right]\\
&=\left(S_{\mathrm{BV}},
 -\frac i2\delta_{\dot a}{}^{\dot b}\delta^{AB}\lambda_{B,\mathrm{old}}^{\ddagger-}\right),
\end{aligned}
\end{equation}
so the bracketed representative transfers to
\begin{equation}\label{eq-CY-gaugino-transferred}
\begin{aligned}
&\frac12\left(\nabla_{+\dot a}\bar\lambda^{A\dot b}
+\nabla_+{}^{\dot b}\bar\lambda_{\dot a}^A\right)
+\frac g{\sqrt2}\delta_{\dot a}{}^{\dot b}\delta^{AB}
\bar\phi_j(t_B)^j{}_i
\left(\psi_+^i+i\sqrt2\bar{\mathcal F}_{\mathrm{old}}^{\ddagger i}\right)\\
&\qquad\longleftrightarrow
\frac12\partial_{+\dot a}\partial_+{}^{\dot b}c^{A(0)}
+ig\delta_{\dot a}{}^{\dot b}\delta^{AB}(t_B)^j{}_i
\gamma_j^{(0)}\beta^{(0)i}.\quad\qedsymbol
\end{aligned}
\end{equation}
\end{proofbox}

On the four zero-form fields (suppressing the superscript $(0)$ from here on), the induced classical differential $Q_0$ acts by
\begin{equation}\label{eq-B-15}
\begin{aligned}
Q_0c^A&=-\frac g2 C^A{}_{BC}c^Bc^C,
&Q_0\gamma_i&=-igc^A(t_A)^j{}_i\gamma_j,\\
Q_0\beta^i&=igc^A(t_A)^i{}_j\beta^j
-\frac{\partial\bar W(\gamma,0)}{\partial\gamma_i},
&Q_0b_A&=-gC^B{}_{AC}c^Cb_B
-ig(t_A)^j{}_i\gamma_j\beta^i,
\end{aligned}
\end{equation}
extended to arbitrary holomorphic derivatives via $[Q_0,\partial_{+\dot a}]=0$.

\subsection{Holomorphic local operators}\label{dolbeault-contraction-and-wick-derivatives}

The free differential of $S_{\mathrm{tw}}$ is $Q_{\mathrm{BV}}=\bar\partial$~\eqref{eq-BV-free-action}. Passing to $\bar\partial$-cohomology on local operators---one step prior to $\mathrm{BPS}_{\hbar^0}=H(Q_0)$ in the homotopy-transfer sequence of Section~\ref{sec-homotopy-transfer}---eliminates all antiholomorphic derivatives $\partial_{-\dot a}$ and form differentials $dx^{-\dot a}$ from the jet algebra at the insertion point.

Viewing local jets at the origin as polynomial differential forms in $(x^{+\dot a},x^{-\dot a},dx^{-\dot a})$, let $\iota$ include the purely holomorphic zero-form polynomials, $\pi$ project onto $x^{-\dot a}=dx^{-\dot a}=0$, and $\iota_{-\dot a}$ contract with $\partial/\partial x^{-\dot a}$. By Cartan's magic formula, the operator $N$ counting total degree in $(x^{-\dot a},dx^{-\dot a})$ is $\bar\partial$-exact:
\begin{equation}\label{eq-B-20}
\begin{gathered}
\{\bar\partial,\iota_{-\dot a}\}=\frac{\partial}{\partial x^{-\dot a}},
\qquad
N=\{\bar\partial,\iota_{x^{-\dot a}\partial/\partial x^{-\dot a}}\}
=x^{-\dot a}\frac{\partial}{\partial x^{-\dot a}}
+dx^{-\dot a}\iota_{-\dot a},\\
\pi X=X\big|_{x^{-\dot a}=dx^{-\dot a}=0},
\qquad
h_{\bar\partial}=\begin{cases}
\iota_{x^{-\dot a}\partial/\partial x^{-\dot a}}N^{-1},&N>0,\\
0,&N=0,
\end{cases}\\
\bar\partial h_{\bar\partial}+h_{\bar\partial}\bar\partial=1-\iota\pi,
\qquad h_{\bar\partial}^2=h_{\bar\partial}\iota=\pi h_{\bar\partial}=0.
\end{gathered}
\end{equation}
Thus $H(\bar\partial)$ consists precisely of polynomials in the zero-form letters and their holomorphic derivatives $\partial_{+\dot a}$, and for any $\bar\partial$-closed surviving output field $X$, spacetime translation projects onto purely holomorphic Taylor shifts:
\begin{equation}\label{eq-B-21}
\pi X(x_v+x)
=\pi\exp\!\left(x_v^{+\dot a}\frac{\partial}{\partial x^{+\dot a}}
+x_v^{-\dot a}\frac{\partial}{\partial x^{-\dot a}}\right)X(x)
=\exp\!\left(x_v^{+\dot a}\frac{\partial}{\partial x^{+\dot a}}\right)\pi X(x).
\end{equation}

\subsection{Perturbative quantization and scale transport}\label{sec-brackets}\label{sec-setup}

Following~\cite{Costello,CostelloBV,Wang,semichiral,operatope,higher,BK}, we now quantize the twisted action $S_{\mathrm{tw}}=S_{\mathrm{free}}+I$ on $\mathbb C^2$ using heat-kernel regularization (see Appendix~\ref{app-twist-cohomology} for the perturbative comparison with the untwisted complex).

\paragraph{BV pairing, $\bar\partial^*$ gauge fixing, and propagator.}
Let $e=(c,b,\gamma,\beta)$ denote the collective Dolbeault BV fields on $E$. The odd symplectic pairing $\omega_E$ pairs $b_A$ with $c^A$ and $\beta^i$ with $\gamma_i$ under $\int_{\mathbb C^2}\Omega\wedge[\cdot]_{(0,2)}$, so that
\begin{equation}\label{eq-C-quadratic-action}
S_{\mathrm{free}}(e)
=\frac12\omega_E(e,Q_{\mathrm{BV}}e)
=\int_{\mathbb C^2}\Omega\wedge
\bigl[b_A\bar\partial c^A
+(\bar\partial\gamma_i)\beta^i\bigr]_{(0,2)},
\qquad (S_{\mathrm{free}},\,\cdot\,)=Q_{\mathrm{BV}}=\bar\partial,
\end{equation}
in the $e^{S/\hbar}$ convention of~\cite{CostelloBV}. The inverse pairing $\omega_E^{-1}=[K_0(x_v,x_w)]_\Omega\,\Delta_{vw}$ factors into the $(0,2)$-form Dirac distribution $\int_{\mathbb C^2}\Omega_{x_v}\wedge[K_0(x_v,x_w)]_\Omega\,f(x_v)=f(x_w)$ and the field-space contraction
\begin{equation}\label{eq-A-20}
\Delta_{vw}=
\frac{\partial_L}{\partial b_{A,v}}
\frac{\partial_L}{\partial c_w^A}
+\frac{\partial_L}{\partial c_v^A}
\frac{\partial_L}{\partial b_{A,w}}
-\frac{\partial_L}{\partial\beta_v^i}
\frac{\partial_L}{\partial\gamma_{i,w}}
-\frac{\partial_L}{\partial\gamma_{i,v}}
\frac{\partial_L}{\partial\beta_w^i}.
\end{equation}
Here the subscript $\Omega$ on any kernel strips off the pairing volume $\Omega$ and $\Delta_{vw}$; for an odd smoothed kernel $K_t$ we write $\Delta_t=\partial_{K_t}$ for the associated signed BV Laplacian, and for an even propagator $P$ we write $\partial_P$ for unordered-pair Wick contraction.\footnote{In the raw symmetric-tensor convention of~\cite[Sections 6.4 and 10]{CostelloBV}, $\Delta_t=-\partial_{K_t}$; our signed convention writes $\Delta_t=\partial_{K_t}$, leaving $\partial_P$ unchanged.} On a local functional, contracting two fields at the same point via $\Delta=\partial_{K_0}$ hits $K_0(x,x)$ on the diagonal, which heat-kernel regularization resolves.

On flat $\mathbb C^2\cong\mathbb R^4$ with Euclidean metric $\delta_{\mu\nu}$, we choose the degree-$-1$ gauge-fixing operator $Q^{\mathrm{GF}}=\bar\partial^*$ so that $(Q^{\mathrm{GF}})^2=0$, $H=\{\bar\partial,\bar\partial^*\}$ is elliptic, and on nonzero modes $e=Q_{\mathrm{BV}}Q^{\mathrm{GF}}H^{-1}e+Q^{\mathrm{GF}}Q_{\mathrm{BV}}H^{-1}e$ splits $E$ into isotropic subspaces with Lagrangian gauge-fixing subspace $\mathcal L=\operatorname{im}\bar\partial^*$:
\begin{equation}\label{eq-dbar-star}
\begin{aligned}
Q^{\mathrm{GF}}=\bar\partial^*
&=-2\,\epsilon^{\dot a\dot b}\,\iota_{-\dot a}
\frac{\partial}{\partial x^{+\dot b}},\\
H=\{\bar\partial,\bar\partial^*\}
&=-2\,\epsilon^{\dot a\dot b}
\frac{\partial}{\partial x^{-\dot a}}
\frac{\partial}{\partial x^{+\dot b}}
=-\frac12\delta^{\mu\nu}
\frac{\partial}{\partial x_E^\mu}
\frac{\partial}{\partial x_E^\nu}.
\end{aligned}
\end{equation}
With Schwinger proper-time cutoffs $0<\varepsilon<L$ ($\varepsilon=\Lambda_{\mathrm{UV}}^{-2}$, where $\sqrt\varepsilon$ and $\sqrt L$ are the ultraviolet and effective resolution lengths), $H$ has regulated inverse
\begin{equation}\label{eq-C-proper-time-inverse}
\int_\varepsilon^L e^{-tH}\,dt
=H^{-1}(e^{-\varepsilon H}-e^{-LH}),\qquad
\int_\varepsilon^L e^{-tp_E^2/2}\,dt
=\frac{2}{p_E^2}\bigl(e^{-\varepsilon p_E^2/2}-e^{-Lp_E^2/2}\bigr)\quad(p_E^2>0).
\end{equation}
Smoothing $\omega_E^{-1}$ by $e^{-tH}$ yields the heat kernel $K_t=(e^{-tH}\otimes1)\omega_E^{-1}$. Using the Gaussian Fourier integral $\int_{\mathbb R^4}\frac{d^4p}{(2\pi)^4}e^{ip\cdot x_E-tp_E^2/2}=\frac{e^{-|x_E|^2/(2t)}}{(2\pi t)^2}$ and~\eqref{b-00919}, its $(0,2)$-form kernel and the regulated propagator $P_{\varepsilon,L}$ read~\cite{Costello,Wang}
\begin{equation}\label{eq-C-heat-spatial}
[K_t(x_1,x_2)]_\Omega
=-\frac{e^{-|x_1-x_2|_E^2/(2t)}}{32\pi^2t^2}
 d(x_1-x_2)^{-\dot+}\wedge d(x_1-x_2)^{-\dot-},
\qquad\int_{\mathbb C^2}\Omega\wedge[K_t]_\Omega=1,
\end{equation}
\begin{equation}\label{eq-C-heat-homotopy}
\begin{aligned}
P_{\varepsilon,L}
&=\int_\varepsilon^L(\bar\partial^*\otimes1)K_t\,dt
=\bigl(\bar\partial^*H^{-1}(e^{-\varepsilon H}-e^{-LH})\otimes1\bigr)\omega_E^{-1},\\
(\bar\partial\otimes1+1\otimes\bar\partial)P_{\varepsilon,L}
&=-\int_\varepsilon^L\partial_tK_t\,dt
=K_\varepsilon-K_L.
\end{aligned}
\end{equation}
Concretely, in terms of free gauge-fixed two-point contractions~\eqref{eq-A-20} on zero-form fields, $[P_{\varepsilon,L}]_\Omega$ is the regulated propagator and~\eqref{eq-C-heat-homotopy} is its $\bar\partial$ Green identity:
\begin{equation}\label{eq-C-free-two-point}
\begin{aligned}
\bigl\langle c^A(x_1)\,b_B(x_2)\bigr\rangle_{\varepsilon,L}
&=\hbar\,\delta_B{}^A\,[P_{\varepsilon,L}(x_1,x_2)]_\Omega,\\
\bigl\langle\gamma_i(x_1)\,\beta^j(x_2)\bigr\rangle_{\varepsilon,L}
&=-\hbar\,\delta_i{}^j\,[P_{\varepsilon,L}(x_1,x_2)]_\Omega,\\
(\bar\partial_{x_1}+\bar\partial_{x_2})\bigl\langle c^A(x_1)\,b_B(x_2)\bigr\rangle_{\varepsilon,L}
&=\hbar\,\delta_B{}^A\bigl([K_\varepsilon(x_1,x_2)]_\Omega-[K_L(x_1,x_2)]_\Omega\bigr)\\
&\xrightarrow{\;\varepsilon\downarrow0,\,L\to\infty\;}
\hbar\,\delta_B{}^A\,[K_0(x_1,x_2)]_\Omega.
\end{aligned}
\end{equation}
Here $[P_{\varepsilon,L}]_\Omega$ has antiholomorphic form degree $1$ and approaches the Bochner--Martinelli kernel $[P_{0,\infty}]_\Omega$ as $\varepsilon\downarrow0,L\to\infty$, which satisfies $\bar\partial[P_{0,\infty}(x_1,x_2)]_\Omega=0$ away from the diagonal ($x_1\ne x_2$) and obeys the distributional contact identity~\cite{Wang}
\begin{equation}\label{eq-C-contact-boundary}
(\bar\partial\otimes1+1\otimes\bar\partial)P_{0,\infty}=K_0,
\end{equation}
whereas $[K_t]_\Omega$ has antiholomorphic form degree $2$.

\paragraph{Comparison with the formal unregularized path integral and Ward identity.}
Before imposing cutoffs, it is instructive to recall the formal path-integral identities that our regulated formulas refine. Generating unordered Wick contractions by $e^{\hbar\partial_P}=\sum_{m\ge0}\frac{\hbar^m}{m!}\partial_P^m$ (where each edge contributes $\hbar P$ and each interaction vertex contributes $I/\hbar$), the finite-dimensional Gaussian integral on nonzero modes~\cite[Section 6.6]{CostelloBV} gives the formal unregularized correspondence
\begin{equation}\label{eq-C-Gaussian-Wick}
\frac{\displaystyle\int_{\mathcal L}\mathcal O(e)\,
 e^{(S_{\mathrm{free}}(e)+I(e))/\hbar}\,\mathcal D e}
 {\displaystyle\int_{\mathcal L}e^{S_{\mathrm{free}}(e)/\hbar}\,\mathcal D e}
=\left.e^{\hbar\partial_{P_{0,\infty}}}
 \bigl(\mathcal O\,e^{I/\hbar}\bigr)\right|_{e=0},
\qquad
P_{0,\infty}=\bigl(Q^{\mathrm{GF}}H^{-1}\otimes1\bigr)\omega_E^{-1},
\end{equation}
where leaving $e$ unevaluated retains external legs, and dividing by the $\mathcal O=1$ partition function removes vacuum bubbles. Writing $S=S_{\mathrm{free}}+I$, the formal BV Laplacian $\Delta=\partial_{K_0}$ satisfies
\begin{equation}\label{eq-C-formal-QME}
\hbar^2\Delta e^{S/\hbar}
=\left[\frac12(S,S)+\hbar\Delta S\right]e^{S/\hbar},\qquad
\int_{\mathcal L}\Delta\bigl(\mathcal O e^{S/\hbar}\bigr)\,\mathcal D e=0,
\end{equation}
where the second identity is the formal BV Stokes theorem on the Lagrangian subspace $\mathcal L$. Expanding $\hbar\Delta(\mathcal O e^{S/\hbar})$ yields the formal Ward identity with a possible anomaly insertion:
\begin{equation}\label{eq-C-naive}
0=\hbar\int_{\mathcal L}\Delta\bigl(\mathcal O e^{S/\hbar}\bigr)\,\mathcal D e
=\int_{\mathcal L}\left[
 (S,\mathcal O)+\hbar\Delta\mathcal O
 +\underbrace{\frac1\hbar\left(\frac12(S,S)+\hbar\Delta S\right)\mathcal O}_{\text{QME-defect insertion}}
\right]e^{S/\hbar}\,\mathcal D e.
\end{equation}
When $S$ formally satisfies the quantum master equation $\frac12(S,S)+\hbar\Delta S=0$, the QME-defect term drops out and~\eqref{eq-C-naive} becomes the standard Slavnov--Taylor / BRST Ward identity, stating that quantum BV-exact insertions decouple:
\begin{equation}\label{eq-C-formal-Ward}
\frac{\displaystyle\int_{\mathcal L}
 \bigl[(S,\mathcal O)+\hbar\Delta\mathcal O\bigr]e^{S/\hbar}\,\mathcal D e}
 {\displaystyle\int_{\mathcal L}e^{S/\hbar}\,\mathcal D e}
=\frac{\displaystyle\hbar\int_{\mathcal L}
 \Delta\bigl(\mathcal O e^{S/\hbar}\bigr)\,\mathcal D e}
 {\displaystyle\int_{\mathcal L}e^{S/\hbar}\,\mathcal D e}=0.
\end{equation}
In continuum field theory, however,~\eqref{eq-C-Gaussian-Wick}--\eqref{eq-C-formal-Ward} are singular as written: $\Delta=\partial_{K_0}$ evaluates $K_0(x,x)$ at coincident points, and even in the holomorphic twist---where loop integrals of $[P_{0,\infty}]_\Omega$ are UV finite---acting with $Q_{\mathrm{BV}}=\bar\partial$ on products of propagators hits the contact distribution $K_0$ along collision diagonals~\eqref{eq-C-free-two-point}. Keeping $\varepsilon>0$ replaces the ill-defined diagonal operator $\Delta$ and the singular collision boundaries by smooth heat-kernel insertions at $t_e=\varepsilon$, revealing finite anomalous operator corrections $(Q_1,Q_2,\ldots)$ that are invisible to naive formal manipulation.

\paragraph{Regularized effective action, quantum BV differential, and scale transport.}
Replacing $P_{0,\infty}$ in~\eqref{eq-C-Gaussian-Wick} by $P_{\varepsilon,L}$ separates the UV limit $\varepsilon\downarrow0$ at fixed resolution $\sqrt L$ from the IR integration $L\to\infty$ over the remaining covariance $P_{L,\infty}$: for local cutoff functionals $I(\varepsilon),\mathcal O(\varepsilon)$ (including any counterterms), the regularized vacuum-normalized correlator is
\begin{equation}\label{eq-C-cutoff-correlator}
\langle\mathcal O\rangle
=\lim_{L\to\infty}\lim_{\varepsilon\downarrow0}
\frac{\left.e^{\hbar\partial_{P_{\varepsilon,L}}}
 \bigl(\mathcal O(\varepsilon)e^{I(\varepsilon)/\hbar}\bigr)\right|_{e=0}}
 {\left.e^{\hbar\partial_{P_{\varepsilon,L}}}
 e^{I(\varepsilon)/\hbar}\right|_{e=0}}.
\end{equation}
Stopping at a finite scale $L>0$ and retaining uncontracted fields as arguments defines Costello's effective interaction $I[L]$ and effective operator insertion $\mathcal O[L]$~\cite{CostelloBV}:
\begin{equation}\label{eq-C-local-to-effective}
e^{I[L]/\hbar}
=\lim_{\varepsilon\downarrow0}
 e^{\hbar\partial_{P_{\varepsilon,L}}}e^{I(\varepsilon)/\hbar},\qquad
\mathcal O[L]e^{I[L]/\hbar}
=\lim_{\varepsilon\downarrow0}
 e^{\hbar\partial_{P_{\varepsilon,L}}}
 \bigl(\mathcal O(\varepsilon)e^{I(\varepsilon)/\hbar}\bigr).
\end{equation}
From~\eqref{eq-C-heat-homotopy}, the contraction operators intertwine $Q_{\mathrm{BV}}$ with the smoothed BV Laplacians $\Delta_t=\partial_{K_t}$ (whose associated smeared antibracket is $(\,\cdot\,,\,\cdot\,)_t$):
\begin{equation}\label{eq-C-finite-scale-intertwine}
[Q_{\mathrm{BV}},\partial_{P_{\varepsilon,L}}]
=\Delta_\varepsilon-\Delta_L,\qquad
(Q_{\mathrm{BV}}+\hbar\Delta_L)e^{\hbar\partial_{P_{\varepsilon,L}}}
=e^{\hbar\partial_{P_{\varepsilon,L}}}
 (Q_{\mathrm{BV}}+\hbar\Delta_\varepsilon).
\end{equation}
The regularized QME demands $(Q_{\mathrm{BV}}+\hbar\Delta_L)e^{I[L]/\hbar}=0$ after the $\varepsilon\downarrow0$ limit (a bare $I(\varepsilon)$ need not solve the cutoff QME at finite $\varepsilon$):
\begin{equation}\label{eq-C-regularized-QME}
\begin{aligned}
&Q_{\mathrm{BV}}I[L]+\frac12(I[L],I[L])_L+\hbar\Delta_LI[L]\\
&\qquad=\lim_{\varepsilon\downarrow0}e^{-I[L]/\hbar}e^{\hbar\partial_{P_{\varepsilon,L}}}
\left[\left(Q_{\mathrm{BV}}I(\varepsilon)+\frac12(I(\varepsilon),I(\varepsilon))_\varepsilon+\hbar\Delta_\varepsilon I(\varepsilon)\right)e^{I(\varepsilon)/\hbar}\right]=0.
\end{aligned}
\end{equation}
When~\eqref{eq-C-regularized-QME} holds, the scale-$L$ quantum BV differential $\mathbf Q_L$ acting on effective insertions,
\begin{equation}\label{eq-C-effective-charge}
\mathbf Q_L\mathcal O[L]
=e^{-I[L]/\hbar}(Q_{\mathrm{BV}}+\hbar\Delta_L)\bigl(\mathcal O[L]e^{I[L]/\hbar}\bigr)
=Q_{\mathrm{BV}}\mathcal O[L]+(I[L],\mathcal O[L])_L+\hbar\Delta_L\mathcal O[L],
\end{equation}
satisfies $\mathbf Q_L^2=0$, $\mathbf Q_L1=0$, and $\mathbf Q_L(FG)=(\mathbf Q_LF)G+(-1)^{|F|}F\mathbf Q_LG+\hbar(-1)^{|F|}(F,G)_L$, and is related to the UV cutoff by~\eqref{eq-C-finite-scale-intertwine}:
\begin{equation}\label{eq-C-regularized-insertion}
(\mathbf Q_L\mathcal O[L])e^{I[L]/\hbar}
=\lim_{\varepsilon\downarrow0}
 e^{\hbar\partial_{P_{\varepsilon,L}}}
 \left[\left\{\begin{aligned}
 &Q_{\mathrm{BV}}\mathcal O(\varepsilon)
 +(I(\varepsilon),\mathcal O(\varepsilon))_\varepsilon
 +\hbar\Delta_\varepsilon\mathcal O(\varepsilon)\\
 &+\frac1\hbar\left(Q_{\mathrm{BV}}I(\varepsilon)
 +\frac12(I(\varepsilon),I(\varepsilon))_\varepsilon
 +\hbar\Delta_\varepsilon I(\varepsilon)\right)\mathcal O(\varepsilon)
 \end{aligned}\right\}e^{I(\varepsilon)/\hbar}\right].
\end{equation}
Comparing~\eqref{eq-C-regularized-insertion} directly with the formal identity~\eqref{eq-C-naive}, the first line in braces is the regularized BV variation of $\mathcal O(\varepsilon)$, while the second is the scale-$\varepsilon$ QME-defect insertion---which must remain inside the Wick contractions $e^{\hbar\partial_{P_{\varepsilon,L}}}$ with $\mathcal O(\varepsilon)$ even when its standalone expectation vanishes by~\eqref{eq-C-regularized-QME}.

For local insertions $F(\varepsilon),G(\varepsilon)$ supported at distinct points, the Gaussian kernel~\eqref{eq-C-heat-spatial} vanishes exponentially between their supports as $\varepsilon\downarrow0$, giving
\begin{equation}\label{eq-C-separated-insertions}
\lim_{\varepsilon\downarrow0}(F(\varepsilon),G(\varepsilon))_\varepsilon=0,
\end{equation}
so the local quantum differential acts as a graded derivation on separated operators, while coincident composite insertions retain local contact contractions. Finally, since $P_{L,L''}=P_{L,L'}+P_{L',L''}$ and $(Q_{\mathrm{BV}}\otimes1+1\otimes Q_{\mathrm{BV}})P_{L,L'}=K_L-K_{L'}$ for $0<L<L'<L''$, RG flow $e^{\hbar\partial_{P_{L,L'}}}$ is an invertible chain map identifying the effective complexes across scales:
\begin{equation}\label{eq-C-RG-transport}
\begin{aligned}
e^{I[L']/\hbar}&=e^{\hbar\partial_{P_{L,L'}}}e^{I[L]/\hbar},\\
\mathcal O[L']e^{I[L']/\hbar}&=e^{\hbar\partial_{P_{L,L'}}}\bigl(\mathcal O[L]e^{I[L]/\hbar}\bigr),\\
(\mathbf Q_{L'}\mathcal O[L'])e^{I[L']/\hbar}&=e^{\hbar\partial_{P_{L,L'}}}\bigl((\mathbf Q_L\mathcal O[L])e^{I[L]/\hbar}\bigr).
\end{aligned}
\end{equation}
Integrating the remaining long-time modes $P_{L,L'}$ reproduces~\eqref{eq-C-cutoff-correlator} independently of $L$, $\langle\mathcal O\rangle=\lim_{L'\to\infty}\bigl[e^{\hbar\partial_{P_{L,L'}}}(\mathcal O[L]e^{I[L]/\hbar})\bigr|_{e=0}\big/e^{\hbar\partial_{P_{L,L'}}}e^{I[L]/\hbar}\bigr|_{e=0}\bigr]$. Under an infrared prescription where the $K_{L'}$ boundary vanishes after vacuum normalization as $L'\to\infty$ (and $Q_{\mathrm{BV}}|_{e=0}=0$), intertwining $e^{\hbar\partial_{P_{\varepsilon,L'}}}(Q_{\mathrm{BV}}+\hbar\Delta_\varepsilon)|_{e=0}=\hbar\Delta_{L'}e^{\hbar\partial_{P_{\varepsilon,L'}}}|_{e=0}$ gives the regularized counterpart of the formal Ward identity~\eqref{eq-C-formal-Ward}: $\langle\mathbf Q_L\mathcal O[L]\rangle=0$.

\subsection{Heat-edge origin and definition of the higher brackets}\label{sec-brackets-origin}

We now specialize~\eqref{eq-C-regularized-insertion} to the holomorphic twist~\eqref{eq-B-10} and a local operator $\mathcal O(0)$ polynomial in the zero-form fields $(c,b,\gamma,\beta)$ and their holomorphic derivatives. By Wang's finiteness theorem~\cite[Theorem 5 and Corollary 4]{Wang}, all holomorphic graph integrals admit finite $\varepsilon\downarrow0$ limits without divergent counterterms, so we set $I(\varepsilon)=I$ and $\mathcal O(\varepsilon)=\mathcal O$. Even without UV divergences, however, applying $\bar\partial$ to $P_{\varepsilon,L}$ produces finite UV collision boundary terms~\cite[Theorem 6]{Wang}.

\begin{figure}[H]
\centering
\begin{tikzpicture}[x=1cm,y=1cm]
\draw[rounded corners=6pt,draw=black!18,fill=black!3] (-2.75,-1.35) rectangle (12.3,2.75);
\draw[cutp] (-2.4,2.35)--(-1.5,2.35); \node[anchor=west,font=\scriptsize] at (-1.45,2.35) {propagator $P_\varepsilon$};
\draw[cutI] (1.3,2.35)--(2.2,2.35); \node[anchor=west,font=\scriptsize] at (2.25,2.35) {heat edge from $\tfrac12(I,I)_\varepsilon$};
\draw[cutO] (6.3,2.35)--(7.2,2.35); \node[anchor=west,font=\scriptsize] at (7.25,2.35) {heat edge from $(I,\mathcal O)_\varepsilon$};
\draw[black!18] (-2.5,2.0)--(12.05,2.0);
\node[anchor=west,font=\small\bfseries] at (-2.55,.5) {one loop};
\cutone{0.6cm}{cutI}{cutp}{cutp}\node at (2.0,.5) {$-$};
\cutone{3.4cm}{cutp}{cutO}{cutp}\node at (4.8,.5) {$+$};
\cutone{6.2cm}{cutp}{cutp}{cutO}
\draw[decorate,decoration={brace,mirror,amplitude=4pt}] (-.15,-.45)--(1.35,-.45) node[midway,below=5pt] {\scriptsize$\tfrac1{2}(I,I)_\varepsilon\,\mathcal O$};
\draw[decorate,decoration={brace,mirror,amplitude=4pt}] (2.65,-.45)--(6.95,-.45) node[midway,below=5pt] {\scriptsize$(I,\mathcal O)_\varepsilon\,I$};
\begin{scope}[yshift=-4.35cm]
\draw[rounded corners=6pt,draw=black!18,fill=black!3] (-2.75,-1.35) rectangle (12.3,1.9);
\node[anchor=west,font=\small\bfseries] at (-2.55,.5) {two loops};
\cuttwo{0.6cm}{cutI}{cutp}{cutp}{cutp}{cutp}\node at (1.9,.5) {$-$};
\cuttwo{3.2cm}{cutp}{cutI}{cutp}{cutp}{cutp}\node at (4.5,.5) {$+$};
\cuttwo{5.8cm}{cutp}{cutp}{cutO}{cutp}{cutp}\node at (7.1,.5) {$-$};
\cuttwo{8.4cm}{cutp}{cutp}{cutp}{cutO}{cutp}\node at (9.7,.5) {$+$};
\cuttwo{11.0cm}{cutp}{cutp}{cutp}{cutp}{cutO}
\draw[decorate,decoration={brace,mirror,amplitude=4pt}] (-.45,-.45)--(4.25,-.45) node[midway,below=5pt] {\scriptsize$\tfrac12(I,I)_\varepsilon\,I\,\mathcal O$};
\draw[decorate,decoration={brace,mirror,amplitude=4pt}] (4.75,-.45)--(12.05,-.45) node[midway,below=5pt] {\scriptsize$\tfrac12(I,\mathcal O)_\varepsilon\,I^2$};
\end{scope}
\end{tikzpicture}
\caption{Origin of the heat edge in $Q_1\mathcal O$ and $Q_2\mathcal O$. Dots are interaction vertices and the square is $\mathcal O$. The heat edge, marked by a cut, carries $K_\varepsilon$: between two interaction vertices (orange) it comes from the defect $\frac12(I,I)_\varepsilon$ of the master equation at scale $\varepsilon$, and at $\mathcal O$ (violet) from $(I,\mathcal O)_\varepsilon$. The signs are $(-1)^{j-1}$ for the edge orders $(12,1\mathcal O,2\mathcal O)$ and $(12,23,1\mathcal O,2\mathcal O,3\mathcal O)$.}
\label{fig-cut-origin}
\end{figure}
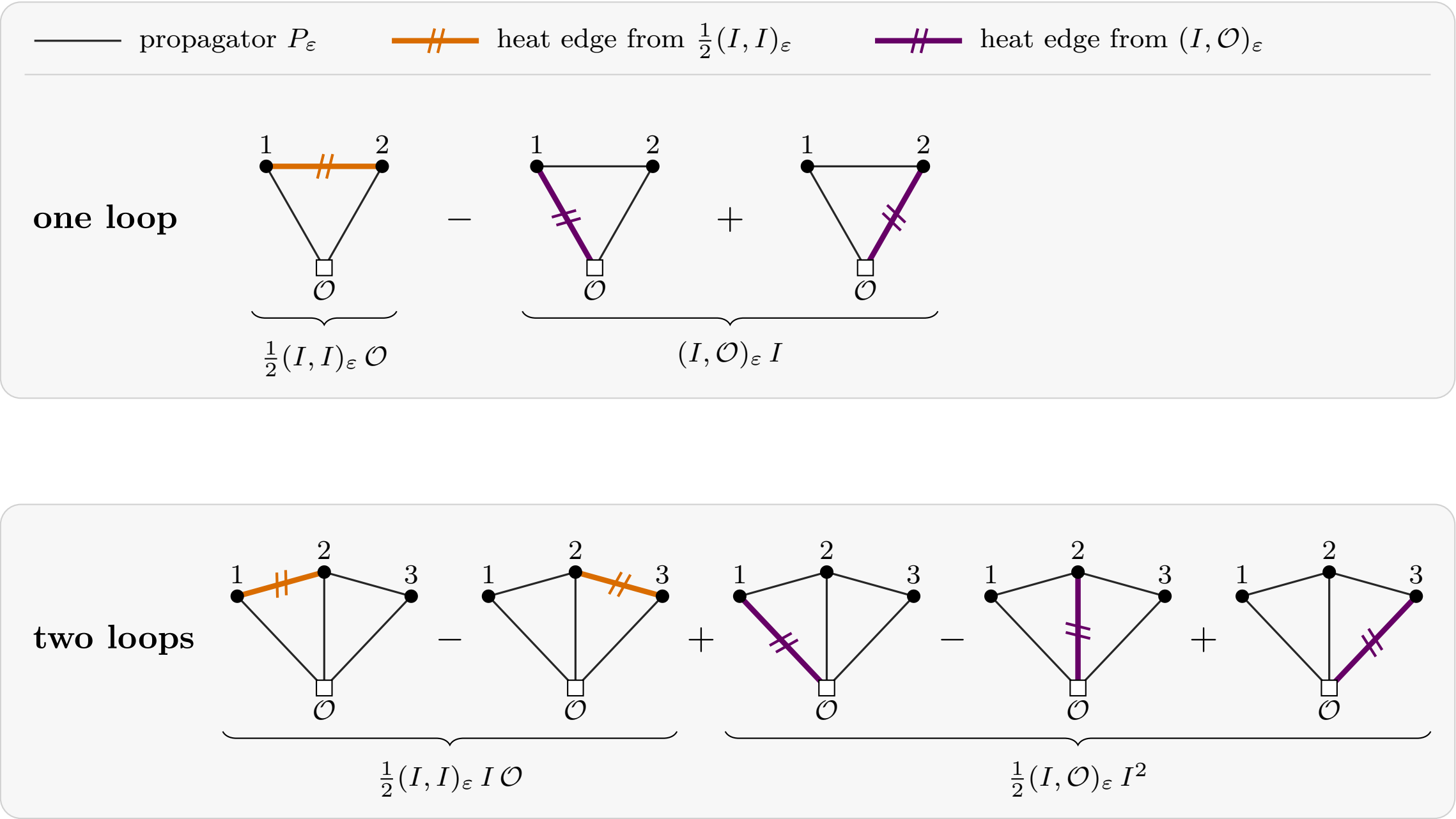

Writing $P_\varepsilon=P_{\varepsilon,\infty}$ for the local UV analysis (at finite $L$ the smooth $K_L$ face contributes to the nonlocal effective tail in~\eqref{eq-C-effective-charge}, not to the local operator at the origin), we have $Q_{\mathrm{BV}}e^{\hbar\partial_{P_\varepsilon}}=e^{\hbar\partial_{P_\varepsilon}}(Q_{\mathrm{BV}}+\hbar\Delta_\varepsilon)$. Crucially, in the braces of~\eqref{eq-C-regularized-insertion}, four terms vanish identically: $Q_{\mathrm{BV}}I=Q_{\mathrm{BV}}\mathcal O=0$ because $I$ and the zero-form jet $\mathcal O$ are $\bar\partial$-closed, and $\Delta_\varepsilon I=\Delta_\varepsilon\mathcal O=0$ because $d(x_1-x_2)^{-\dot+}\wedge d(x_1-x_2)^{-\dot-}$ in~\eqref{eq-C-heat-spatial} vanishes at $x_1=x_2$. Thus~\eqref{eq-C-regularized-insertion} collapses to just two heat-edge contributions~\cite{Costello,semichiral,higher,BK}:
\begin{equation}\label{eq-C-two-cuts}
e^{-I/\hbar}\bigl(Q_{\mathrm{BV}}+\hbar\Delta_\varepsilon\bigr)\bigl(\mathcal O\,e^{I/\hbar}\bigr)
=\underbrace{(I,\mathcal O)_\varepsilon}_{\substack{\text{action bracketed with }\mathcal O:\\ \text{heat edge between }\mathcal I\text{ and }\mathcal O}}
+\underbrace{\frac1{2\hbar}(I,I)_\varepsilon\,\mathcal O}_{\substack{\text{QME defect at scale }\varepsilon:\\ \text{heat edge between }\mathcal I\text{ and }\mathcal I}}.
\end{equation}
On the unnormalized Wick functional $\langle\mathcal O\rangle_\varepsilon=e^{\hbar\partial_{P_\varepsilon}}(\mathcal O e^{I/\hbar})$, this gives the cutoff Ward identity
\begin{equation}\label{eq-C-Ward}
Q_{\mathrm{BV}}\,\langle\mathcal O\rangle_\varepsilon
=e^{\hbar\partial_{P_\varepsilon}}
\Bigl(\Bigl[(I,\mathcal O)_\varepsilon+\frac1{2\hbar}(I,I)_\varepsilon\,\mathcal O\Bigr]e^{I/\hbar}\Bigr),
\end{equation}
whose Feynman graphs carry $K_\varepsilon$ on one edge---either an $(\mathcal I,\mathcal O)$ spoke from $(I,\mathcal O)_\varepsilon$ or an $(\mathcal I,\mathcal I)$ internal edge from $\frac12(I,I)_\varepsilon$---and $P_\varepsilon$ on all remaining edges (Figure~\ref{fig-cut-origin}). Equivalently, on the ordered product of propagator $(0,1)$-forms along $e_1,\ldots,e_{|I|}$, the graded Leibniz rule for $\bar\partial$ places $[K_{\varepsilon;e_j}]_\Omega$ on each edge in turn~\cite[\S2.3]{BK}:
\begin{equation}\label{eq-C-heat-edge-cut}
\bar\partial\bigwedge_{i=1}^{|I|}[P_{\varepsilon;e_i}]_\Omega
=\sum_{\substack{j=1,\ldots,|I|\\\mathrm{heat\ edges}}}(-1)^{j-1}
\left(\bigwedge_{i<j}[P_{\varepsilon;e_i}]_\Omega\right)
\wedge[K_{\varepsilon;e_j}]_\Omega
\wedge\left(\bigwedge_{i>j}[P_{\varepsilon;e_i}]_\Omega\right).
\end{equation}

\begin{figure}[H]
\centering
\begin{tikzpicture}[x=1cm,y=1cm]
\draw[rounded corners=6pt,draw=black!18,fill=black!3] (-2.55,-1.0) rectangle (12.55,3.75);
\draw[hfprop] (-2.2,3.3)--(-1.1,3.3); \node[anchor=west,font=\scriptsize] at (-1.05,3.3) {propagator $P_{\varepsilon,L}$, from $c$ to $b$};
\draw[hfheat] (3.35,3.3)--(4.45,3.3); \node[anchor=west,font=\scriptsize] at (4.5,3.3) {heat edge $K_\varepsilon$ (cut)};
\node[hfv] at (7.75,3.3) {}; \node[anchor=west,font=\scriptsize] at (7.85,3.3) {$bcc$};
\node[hfo] at (8.95,3.3) {}; \node[anchor=west,font=\scriptsize] at (9.1,3.3) {$bb$};
\draw[hfout] (9.95,3.2)--(10.5,3.4); \node[anchor=west,font=\scriptsize] at (10.5,3.3) {output letter};
\draw[black!18] (-2.3,2.95)--(12.3,2.95);
\bbtri{0cm}{hfheat}{hfprop}{hfprop}\node at (2.43,.85) {\large$-$};
\bbtri{4.85cm}{hfprop}{hfheat}{hfprop}\node at (7.28,.85) {\large$+$};
\bbtri{9.7cm}{hfprop}{hfprop}{hfheat}
\end{tikzpicture}
\caption{The three heat-edge terms of one contraction graph in $\{bcc,bcc,bb\}$, with signs $(-1)^{j-1}$ for edge order $(12,1\mathcal O,2\mathcal O)$. The interaction vertices sit at $x_1,x_2$ and the insertion $bb$ at the origin; surviving output letters are $b$ at $x_1$ and $c$ at $x_2$.}
\label{b-00153}
\end{figure}
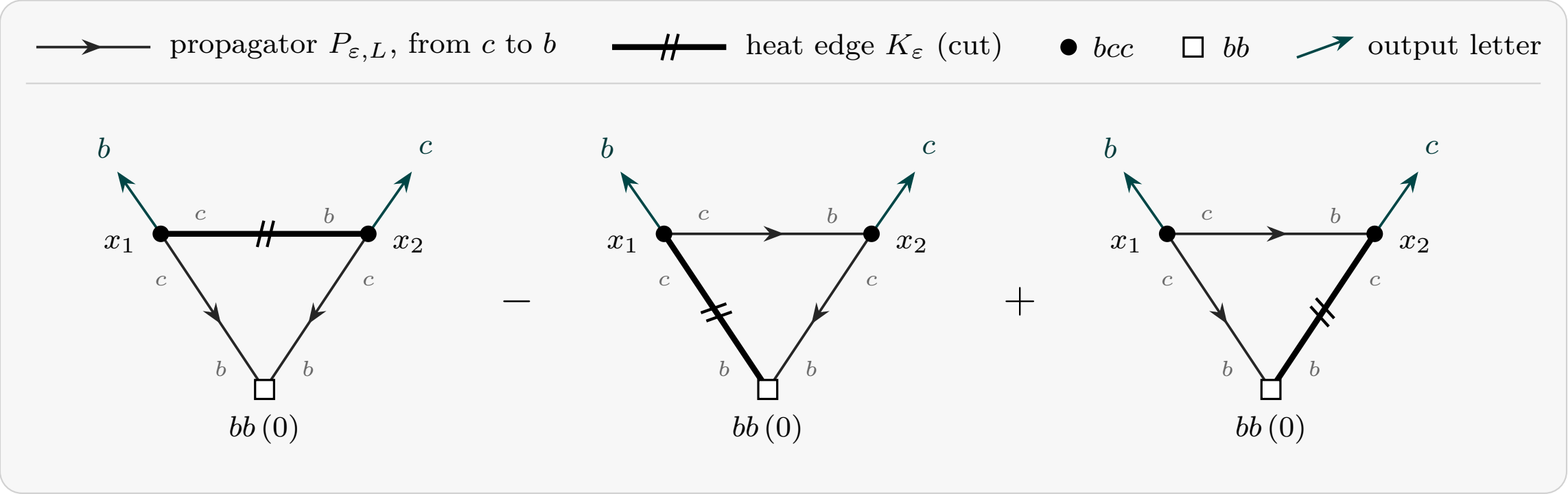

As $\varepsilon\downarrow0$, $K_\varepsilon$ concentrates on the diagonal and forces the connected subgraph containing the heat edge to collapse to a local differential operator~\cite[Theorem 6]{Wang} (much as BRST Ward identities in string theory localize to boundary integrals on worldsheet moduli space~\cite[arXiv version, Section 7.1]{WittenSuperstring}). Pure-interaction clusters give $\{\mathcal I,\ldots,\mathcal I\}$ and cancel when the QME $\lim_{\varepsilon\downarrow0}e^{\hbar\partial_{P_\varepsilon}}((I,I)_\varepsilon e^{I/\hbar})=0$ holds (which reduces at one loop to the triangle gauge anomaly $\operatorname{Tr}_R(t_A\{t_B,t_C\})=0$, Theorem~\ref{b-00465}), while clusters collapsing $n+1$ interaction vertices onto $\mathcal O(0)$ define the higher brackets and the local quantum differential $\mathbf Q\mathcal O$:

\Needspace{8\baselineskip}
\rthm{Definition}{b-00151}{Brackets and the full quantum differential} Let $\mathcal O_1,\ldots,\mathcal O_k$ be local operators of fermion numbers $|\mathcal O_i|$, with $\mathcal O_k$ placed at the origin. Their higher bracket is
\begin{equation}\label{eq-A-28}
\begin{aligned}
\{\mathcal O_1,\ldots,\mathcal O_k\}
={}&(-1)^{\sum_{i=1}^{k}(k-i)|\mathcal O_i|}
\sum_\Gamma\lim_{\varepsilon\downarrow0}
\sum_{j=1}^{|I|}(-1)^{j-1}
\int_{(\mathbb C^2)^{k-1}}
\left(\bigwedge_{v<k}\Omega_v\right)\\
&\times W_\Gamma\!\left[
\left(\bigwedge_{a<j}[P_{\varepsilon,L;e_a}]_\Omega\right)
\wedge[K_{\varepsilon;e_j}]_\Omega
\wedge\left(\bigwedge_{a>j}[P_{\varepsilon,L;e_a}]_\Omega\right);\right.\\
&\hspace{7em}\left.
\mathcal O_1(x_1),\ldots,\mathcal O_k(0)
\right]\Big|_{H(\bar\partial)}.
\end{aligned}
\end{equation}
Here $\Gamma$ runs over connected graphs on $\{1,\ldots,k\}$ with ordered edges $I=(e_1,\ldots,e_{|I|})$, $\Omega_v=2\,dx_v^{+\dot+}\wedge dx_v^{+\dot-}$, $e_j$ is the \emph{heat (cut) edge} carrying $[K_{\varepsilon;e_j}]_\Omega$ (Figure~\ref{b-00153}), all remaining edges carry $[P_{\varepsilon,L;e_a}]_\Omega$, and the final projection onto $H(\bar\partial)$ uses~\eqref{eq-B-20}.

For operators without input holomorphic derivatives and spacetime form $F$, the ordered Wick operator acts from right to left by
\begin{equation}\label{eq-W-fibre-explicit}
W_\Gamma[F;\mathcal O_1,\ldots,\mathcal O_k]
=F\,\prod_{a=1}^{|I|}\!{}^{\mathrm{edge\ order}}
\Delta_{v_aw_a}\,
(\mathcal O_1\cdots\mathcal O_k),
\end{equation}
with $\Delta_{vw}$ given in~\eqref{eq-A-20}. When contracted letters carry holomorphic derivatives, each edge contraction is promoted to displaced functional derivatives ($\delta X(y)/\delta X(y')=\delta^{(4)}(y-y')$):
\begin{equation}\label{eq-W-jet-edge}
\frac{\partial_L}{\partial X_v}\frac{\partial_L}{\partial Y_w}
\;\longmapsto\;
\int d^4z_v\,d^4z_w\,
\frac{\delta_L}{\delta X(x_v+z_v)}\,
\frac{\delta_L}{\delta Y(x_w+z_w)},
\end{equation}
with the edge kernel evaluated at $x_v-x_w+z_e$ ($z_e=z_v-z_w$), so derivatives on input letters integrate by parts onto $z_e$-derivatives of $F$---for example, $W_{(v,w)}[F;(\partial_{+\dot a}b_A)(x_v),c^B(x_w)]=-2\delta_A{}^B\left.\partial F/\partial z_e^{+\dot a}\right|_{z_e=0}$---while surviving uncontracted letters are translated by~\eqref{eq-B-21}.

\begin{figure}[H]
\centering
\includegraphics[width=0.98\linewidth]{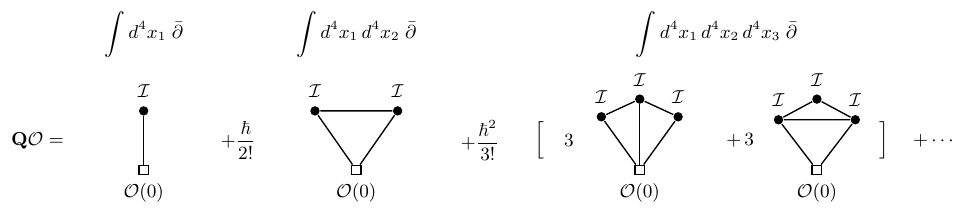}
\caption{The higher-bracket expansion of $\mathbf Q\mathcal O$ through two loops, before imposing the vertex restrictions of Section~\ref{sec-restrictions}. The square is the fixed insertion; dots are interactions integrated over $\mathbb C^2$. Each two-loop topology has three interaction labellings.}
\label{fig-higher-bracket-series}
\end{figure}

For an action with interaction density $\mathcal I$ satisfying the QME, $\lim_{\varepsilon\downarrow0}Q_{\mathrm{BV}}\langle\mathcal O\rangle_\varepsilon=\lim_{\varepsilon\downarrow0}\langle\mathbf Q\mathcal O\rangle_\varepsilon$, where the full local quantum BV differential is (Figure~\ref{fig-higher-bracket-series})
\begin{equation}\label{eq-1-1}
\mathbf Q=\sum_{n\geq0}\hbar^nQ_n,\qquad
Q_n\mathcal O=\frac1{(n+1)!}\bigl\{\underbrace{\mathcal I,\ldots,\mathcal I}_{n+1},\mathcal O\bigr\}.
\end{equation}

\subsection{Positive Schwinger times and the operatope}\label{positive-schwinger-times-and-the-operatope}

Evaluating a higher bracket~\eqref{eq-A-28} reduces to computing the spacetime integral kernel $\mathcal I_\Gamma(\lambda;z)$ of each contributing graph $\Gamma$. By~\eqref{eq-B-21}, translating the surviving output fields at vertex $v<k$ inserts $\exp(x_v^{+\dot a}\partial/\partial x_v^{+\dot a})$: representing these holomorphic derivatives by plane-wave momenta $\lambda_{v,+\dot a}$, momentum conservation fixes the root momentum $\lambda_k$, and
\begin{equation}\label{eq-D-lambda}
\lambda_{k,+\dot a}=-\sum_{\substack{v=1,\ldots,k-1\\\mathrm{integrated\ vertices}}}\lambda_{v,+\dot a},\qquad
\lambda_{v,+\dot a}\longmapsto\frac{\partial}{\partial x_v^{+\dot a}}=-\frac12\partial_{v,+\dot a}\qquad(v<k),
\end{equation}
acting by the Leibniz rule if several output letters survive at $v$, and setting $\lambda_v=0$ if $v$ is fully contracted.

\rthm{Proposition}{b-00169}{Current representation; external input~\cite[Section 2.7, Equations (2.32)--(2.33)]{operatope}, \cite[Lemma 2, Proposition 9]{Wang}} \emph{Let $\Gamma$ be an $n$-loop graph on $V=\{1,\ldots,k\}$ ($k=n+2$, fixed insertion $x_k=0$) with $2n+1$ ordered oriented edges $I=(e_1,\ldots,e_{2n+1})$ and Schwinger times $t_e>0$. With vertex sources $\lambda_v$~\eqref{eq-D-lambda}, edge displacements $z_e$~\eqref{eq-W-jet-edge}, and edge currents $y_{e,+\dot a}$~\eqref{eq-1-8}, the graph kernel~\eqref{eq-1-7} equals}
\begin{equation}\label{eq-C-current-delta-integral}
\begin{aligned}
\mathcal I_\Gamma(\lambda;z)
={}&(-1)^n\int_{\Delta_\Gamma}
 e^{-\sum_ez_e^{+\dot a}y_{e,+\dot a}}
 \bigwedge_{e\in I}d^2y_e\\
&\times\prod_{v<k}\prod_{\dot a=\dot+,\dot-}
\delta\!\left(\lambda_{v,+\dot a}
-\sum_{e:e(0)=v}y_{e,+\dot a}
+\sum_{e:e(1)=v}y_{e,+\dot a}\right),
\end{aligned}
\end{equation}
\emph{where the currents satisfy the cycle voltage law around every oriented loop $C$,}
\begin{equation}\label{eq-C-oriented-cycle}
\sum_{\substack{e\in C\\\mathrm{along}\ C}}t_ey_{e,+\dot a}
-\sum_{\substack{e\in C\\\mathrm{against}\ C}}t_ey_{e,+\dot a}=0,
\end{equation}
\emph{for a common projective tuple of positive Schwinger times $(t_e)\in\mathbb{RP}^{|I|-1}_{>0}$ shared by both dotted components, sweeping out the operatope domain $\Delta_\Gamma$ oriented by the UV boundary~\eqref{eq-UV-boundary}:}
\begin{equation}\label{eq-C-projective-times}
\Delta_\Gamma=\bigcup_{(t_e)\in\mathbb{RP}^{|I|-1}_{>0}}
\Bigl\{(y_{e,+\dot a})_{e\in I}:\ %
\sum_{\substack{e\in C\\\mathrm{along}\ C}}t_ey_{e,+\dot a}
-\sum_{\substack{e\in C\\\mathrm{against}\ C}}t_ey_{e,+\dot a}=0
\ \text{ for every cycle }C\Bigr\}.
\end{equation}

\subsubsection*{Derivation via the UV Schwinger boundary and Fourier integration}\label{sec-B-kernel}

Let $V=\{1,\ldots,n+2\}$ and $I=(e_1,\ldots,e_{2n+1})$ (the unique edge count for which the integral can be nonzero, Lemma~\ref{b-00180}), with $x_{n+2}=0$ and $\lambda_{n+2}=-\sum_{v<n+2}\lambda_v$. Replacing surviving output letters in~\eqref{eq-A-28} by plane waves $e^{\lambda_{v,+\dot a}x_v^{+\dot a}}$ and writing each propagator as a Schwinger integral~\eqref{eq-C-heat-homotopy} converts the heat-edge sum into an integral over the ultraviolet boundary of the Schwinger-time cube~\cite{Wang,operatope}:
\begin{equation}\label{eq-1-7}
\begin{aligned}
\mathcal I_\Gamma(\lambda;z)
={}&(-8\pi^2)^n\lim_{\varepsilon\downarrow0}
\sum_{j=1}^{2n+1}(-1)^{j-1}
\int_{(\mathbb C^2)^{n+1}}
\left(\bigwedge_{v<n+2}\Omega_v\right)
e^{\sum_{v<n+2}\lambda_{v,+\dot a}x_v^{+\dot a}}\\
&\times\left(\bigwedge_{i<j}[P_{\varepsilon,L;e_i}]_\Omega\right)
\wedge[K_{\varepsilon;e_j}]_\Omega
\wedge\left(\bigwedge_{i>j}[P_{\varepsilon,L;e_i}]_\Omega\right)\\
={}&(-1)^n(-8\pi^2)^n\lim_{\varepsilon\downarrow0}
\int_{(\mathbb C^2)^{n+1}}
\left(\bigwedge_{v<n+2}\Omega_v\right)
e^{\sum_{v<n+2}\lambda_{v,+\dot a}x_v^{+\dot a}}\\
&\times\int_{\partial_{\mathrm{UV}}[\varepsilon,L]^{2n+1}}
\bigwedge_{i=1}^{2n+1}\Bigl([K_{t_{e_i};e_i}]_\Omega
-dt_{e_i}\wedge[\bar\partial^*K_{t_{e_i};e_i}]_\Omega\Bigr),
\end{aligned}
\end{equation}
\begin{equation}\label{eq-UV-boundary}
\partial_{\mathrm{UV}}[\varepsilon,L]^{2n+1}
=\bigcup_{j=1}^{2n+1}\bigl\{t_{e_j}=\varepsilon,\ \ \varepsilon\le t_{e_i}\le L\ (i\ne j)\bigr\},
\end{equation}
where facet $j$ carries orientation $(-1)^{j-1}\bigwedge_{i\ne j}dt_{e_i}$, and reordering the $2n$ differentials $dt_{e_i}$ past the propagator one-forms produces $(-1)^{n(2n-1)}=(-1)^n$ (while the $t_{e_j}=L$ facets vanish for local operators and are absent at $L=\infty$).

On each oriented edge $e=(v,w)$, introducing the two-component current $y_{e,+\dot a}$ combines the two terms in the last line of~\eqref{eq-1-7} into a single Gaussian two-form:
\begin{equation}\label{eq-C-explicit-propagator}
[P_{\varepsilon,L;e}]_\Omega
=-\frac{(x_v-x_w)^-_{\dot a}\,d(x_v-x_w)^{-\dot a}}{32\pi^2}
 \int_\varepsilon^L\frac{dt_e}{t_e^3}\,
 \exp\!\left[-\frac{(x_v-x_w)^-_{\dot a}(x_v-x_w+z_e)^{+\dot a}}{2t_e}\right],
\end{equation}
\begin{equation}\label{eq-1-8}
y_{e,+\dot a}
=\frac{(x_{e(0)}-x_{e(1)})^-_{\dot a}}{2t_e},\qquad
 d^2y_e=dy_{e,+\dot+}\wedge dy_{e,+\dot-},
\end{equation}
\begin{equation}\label{eq-C-current-Jacobian}
[K_{t_e;e}]_\Omega
-dt_e\wedge[\bar\partial^*K_{t_e;e}]_\Omega
=-\frac{e^{-y_{e,+\dot a}(x_{e(0)}-x_{e(1)}+z_e)^{+\dot a}}}{8\pi^2}\,d^2y_e.
\end{equation}
Substituting~\eqref{eq-C-current-Jacobian} into~\eqref{eq-1-7} (using $(-1)^n(-8\pi^2)^n(-1/(8\pi^2))^{2n+1}=-1/(8\pi^2)^{n+1}$) gives
\begin{equation}\label{eq-kernel-current-form}
\begin{aligned}
\mathcal I_\Gamma(\lambda;z)
={}&-\frac{1}{(8\pi^2)^{n+1}}\lim_{\varepsilon\downarrow0}
\int_{(\mathbb C^2)^{n+1}}
\left(\bigwedge_{v<n+2}\Omega_v\right)
e^{\sum_{v<n+2}\lambda_{v,+\dot a}x_v^{+\dot a}}\\
&\times\int_{\partial_{\mathrm{UV}}[\varepsilon,L]^{2n+1}}
e^{-\sum_{e\in I}y_{e,+\dot a}(x_{e(0)}-x_{e(1)}+z_e)^{+\dot a}}
\bigwedge_{e\in I}d^2y_e.
\end{aligned}
\end{equation}
Deforming the integration contour to purely imaginary $x_v^{+\dot a}\in i\mathbb R$ and real $x_v^{-\dot a}$ (for real $\lambda_v$ and imaginary $z_e$) regroups the exponent by vertex,
\begin{equation}\label{eq-C-fourier-exponent}
\begin{aligned}
&\sum_{v<n+2}\lambda_{v,+\dot a}x_v^{+\dot a}
-\sum_ey_{e,+\dot a}(x_{e(0)}-x_{e(1)}+z_e)^{+\dot a}\\
&\qquad=-\sum_ey_{e,+\dot a}z_e^{+\dot a}
+\sum_{v<n+2}\left(\lambda_{v,+\dot a}
-\sum_{e:e(0)=v}y_{e,+\dot a}
+\sum_{e:e(1)=v}y_{e,+\dot a}\right)x_v^{+\dot a},
\end{aligned}
\end{equation}
so integrating $\Omega_v=2\,dx_v^{+\dot+}\wedge dx_v^{+\dot-}$ over $(i\mathbb R)^2$ yields $2(2\pi i)^2=-8\pi^2$ times the current-conservation delta function at each vertex $v<n+2$:
\begin{equation}\label{eq-1-9}
\sum_{e:e(0)=v}y_{e,+\dot a}
-\sum_{e:e(1)=v}y_{e,+\dot a}
=\lambda_{v,+\dot a}.
\end{equation}
Since $-\frac{1}{(8\pi^2)^{n+1}}(-8\pi^2)^{n+1}=(-1)^n$ and $t_ey_{e,+\dot a}=\tfrac12(x_{e(0)}-x_{e(1)})^-_{\dot a}$ sums to zero around every cycle $C$, extending coefficientwise in $(\lambda,z)$ proves~\eqref{eq-C-current-delta-integral}.

\subsubsection*{Independent currents and the electrical-network dictionary}\label{sec-B-network}

Choosing any spanning tree $T\subset I$ of $\Gamma$, its grounded incidence matrix has determinant $\pm1$, so integrating out the tree currents $(y_e)_{e\in T}$ against the vertex delta functions in~\eqref{eq-C-current-delta-integral} has unit Jacobian for both dotted components (cancelling $(-1)^n$ upon writing $-d^2y_e$ on the $n$ complementary chords $e\notin T$). This yields the independent-current operatope integral used throughout Sections~\ref{sec-restrictions}--\ref{sec-lifting}:
\begin{equation}\label{eq-operatope}
\mathcal I_\Gamma(\lambda;z)
=\int_{\substack{y\in\Delta_\Gamma\\
\sum_{e:e(0)=v}y_e-\sum_{e:e(1)=v}y_e=\lambda_v}}
e^{-\sum_ey_{e,+\dot a}z_e^{+\dot a}}
\bigwedge_{e\notin T}(-dy_{e,+\dot+}\wedge dy_{e,+\dot-}).
\end{equation}

\rthm{Remark}{b-00176}{Circuit dictionary}
Equations~\eqref{eq-C-current-delta-integral}--\eqref{eq-operatope} describe a linear electrical network on $\Gamma$ (summarized in Table~\ref{tab-sum-dictionary} of Section~\ref{sec-sum-circuit}): $y_e$ is the current on edge $e$ and $\lambda_v$ is the external current injected at vertex $v$; equation~\eqref{eq-1-8} is Ohm's law with vertex potential $\tfrac12x^-_{v,\dot a}$ and edge resistance $t_e$; equation~\eqref{eq-1-9} is Kirchhoff's current law at every node and~\eqref{eq-C-oriented-cycle} is Kirchhoff's voltage law around every loop; and $z_e$ is the generating source whose derivatives $-\partial/\partial z_e^{+\dot a}$ insert moments of $y_{e,+\dot a}$ (while at $z=0$, $\mathcal I_\Gamma(\lambda;0)$ is the oriented volume swept out by the $n$ independent loop currents as $(t_e)$ ranges over $\mathbb{RP}^{|I|-1}_{>0}$). This dictionary brings classical series, parallel, and star--triangle circuit reductions directly to holomorphic Feynman integrals, where iterating them with forward source transfer provides polynomial changes of variables on Schwinger space.

\section{Constraints and classification of \texorpdfstring{$Q_n$}{Q\_n} graphs}\label{sec-restrictions}

\begin{figure}[ht!]
\centering
\includegraphics[width=0.82\linewidth]{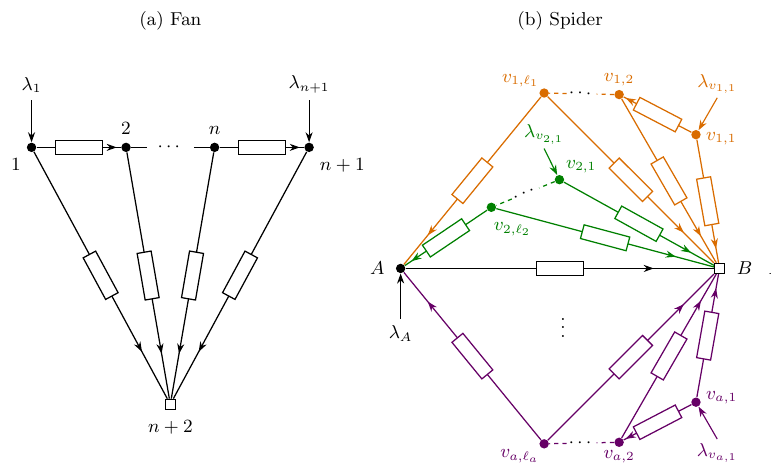}
\caption{(a) A \emph{fan}: the interaction vertices $1,\ldots,n+1$ form a path, and each of them is joined to the insertion $n+2$ by a spoke. (b) A \emph{spider}: several such paths, the arms, meet at a vertex $A$, and every vertex, $A$ included, is joined to a second vertex $B$; in $Q_n$, $A$ is the $\bar W$ vertex and $B=\mathcal O$ (Definition~\ref{b-00380}).}
\label{fig-sum-families}
\end{figure}

Only a few graphs can have a nonzero contribution to \eqref{eq-C-current-delta-integral}; they must satisfy the so-called Laman condition (Section \ref{sec-laman}). In addition, the diagrams in the four-dimensional holomorphic twist of an $\mathcal N=1$ theory separate further into a few classes. This comes from the fact that each propagator lowers the number of $b$ and $\beta$ factors by one, while each vertex raises it by at most one. The result of this section is the classification of the graphs of $Q_n$ (Corollary~\ref{b-00247}): for the interaction \eqref{eq-B-10} with $\bar W$ of arbitrary degree, every nonzero connected graph of $Q_n$, $n\ge1$, is a fan or a spider whose arms meet at its single $\bar W$ vertex (Figure \ref{fig-sum-families}); for interactions of degree at most three, including all renormalizable $\mathcal N=1$ gauge theories, only fans contribute (Section \ref{sec-only-fans}). The same counting shows that $Q_n\mathcal O$ vanishes beyond a finite loop order (Corollary~\ref{b-00235}).

\subsection{Laman constraints}\label{sec-laman}

For an induced subgraph $S\subseteq\Gamma$, write $V_S\subseteq V$ for its vertex set and $I_S\subseteq I$ for its edge set. Because several constraints have a direct electrical-network reading (Table~\ref{tab-sum-dictionary}), we present both the form-degree proof and the circuit proof where helpful.

\rthm{Lemma}{b-00180}{Laman count \cite[Equation (2.22)]{operatope}, \cite[Theorem 6]{Wang}} \emph{Every nonzero graph satisfies}
\begin{equation}\label{eq-1-12}
|I|=2|V|-3 .
\end{equation}

\begin{proofbox}
\textbf{Proof (graph integral).} The heat edge has antiholomorphic form degree two and every other propagator degree one; together they saturate the relative antiholomorphic integration directions:
\begin{equation*}
\underbrace{2}_{\text{heat kernel}}+\underbrace{(|I|-1)}_{P}
=\underbrace{2(|V|-1)}_{|V|-1\text{ points}}.
\end{equation*}
\par\vspace{-\medskipamount}\noindent\hspace*{\fill}\qedsymbol

\smallskip\noindent\textbf{Proof (circuit).} The resistance ratios sweep out the real components of the independent loop currents in \eqref{eq-C-current-delta-integral}:
\begin{equation*}
\underbrace{|I|-1}_{\text{independent resistance ratios}}
=\underbrace{2}_{\text{holomorphic dimension}}\,
\underbrace{(|I|-|V|+1)}_{\text{loops}}.
\end{equation*}
\par\vspace{-\medskipamount}\noindent\hspace*{\fill}\qedsymbol
\end{proofbox}

\rthm{Lemma}{b-00194}{Laman bound on subgraphs \cite[Equation (2.23)]{operatope}, \cite[Proposition 5]{Wang}} \emph{Every $S\subseteq\Gamma$ with $I_S\ne\emptyset$ satisfies}
\begin{equation*}
|I_S|\le2|V_S|-3 .
\end{equation*}

\begin{proofbox}
\textbf{Proof (graph integral).} Fix $v_0\in V_S$; the coordinates $(x_v-x_{v_0})^{-\dot a}$, $v\in V_S\setminus\{v_0\}$, give $2(|V_S|-1)$ one-form directions in the numerators of the propagators. The total form degree cannot exceed the number of one-form directions. If the heat kernel edge $K$ lies in $S$, then $|I_S|+1\le2(|V_S|-1)$. If the heat edge lies outside $S$, the radial vector annihilates every internal propagator one-form \cite[Proposition 5, Lemma 2]{Wang}, \cite[Equation (2.15)]{operatope}:
\begin{equation*}
\begin{gathered}
R=\sum_{u\in V_S\setminus\{v_0\}}(x_u-x_{v_0})^{-\dot c}\frac{\partial}{\partial x_u^{-\dot c}},\\
\iota_R\bigl[-(x_v-x_w)^-_{\dot a}d(x_v-x_w)^{-\dot a}\bigr]
=-(x_v-x_w)^-_{\dot a}(x_v-x_w)^{-\dot a}=0,
\end{gathered}
\end{equation*}
so the number of antiholomorphic directions drops by one and $|I_S|\le2(|V_S|-1)-1$.~\hfill\qedsymbol

\smallskip\noindent\textbf{Proof (circuit).} Ohm's law is local. For connected $S$, fix the currents entering its vertices from outside. Kirchhoff's law leaves $2(|I_S|-|V_S|+1)$ independent loop-current components, while the internal resistance ratios range over $\mathbb{RP}^{|I_S|-1}_{>0}$. If their variation cannot span these loop-current directions, the pullback of the loop-current volume form vanishes identically, and so does the graph kernel. Thus its form degree cannot exceed the dimension of the resistance-ratio space:
\begin{equation*}
\underbrace{2(|I_S|-|V_S|+1)}_{\text{loop-current form degree}}
\le\underbrace{|I_S|-1}_{\dim\mathbb{RP}^{|I_S|-1}_{>0}}
\quad\Longleftrightarrow\quad |I_S|\le2|V_S|-3.
\end{equation*}
For disconnected $S$, apply the bound to each connected component containing an edge and sum the bounds.~\hfill\qedsymbol
\end{proofbox}

\rthm{Corollary}{b-00188}{No self-loops or parallel edges \cite[Proposition 5]{Wang}, \cite[Section 2.1]{operatope}} \emph{A nonzero graph has no self-loop and at most one edge between any two vertices.}

\begin{proofbox}
\textbf{Proof (propagators).} A self-loop carries $d(x_v-x_v)^{-\dot a}=0$ \cite[Section 2.3]{Wang}. Two edges between $v$ and $w$, with holomorphic shifts $z,z'$, share the one-form direction of $x_v-x_w$ \cite[Section 2.1]{operatope}:
\begin{equation}\label{eq-1-13}
\begin{aligned}
[P_{\varepsilon,L;vw}(x_v-x_w+z)]_\Omega
\wedge[P_{\varepsilon,L;vw}(x_v-x_w+z')]_\Omega&=0,\\
[K_{\varepsilon;vw}(x_v-x_w+z)]_\Omega
\wedge[P_{\varepsilon,L;vw}(x_v-x_w+z')]_\Omega&=0.
\end{aligned}
\end{equation}
\par\vspace{-\medskipamount}\noindent\hspace*{\fill}\qedsymbol

\smallskip\noindent\textbf{Proof (circuit).} A self-loop is shorted: $t_ey_{e,+\dot a}=0$ with $t_e>0$ gives $y_e=0$. Two resistors between the same nodes are in parallel: $t_ey_e=t_{e'}y_{e'}$, so their four current components vary in only three directions.~\hfill\qedsymbol
\end{proofbox}

Corollary~\ref{b-00188} is the $|V_S|\le2$ specialization of Lemma~\ref{b-00194}; together, Lemmas~\ref{b-00180} and~\ref{b-00194} are the Laman constraints in two holomorphic dimensions \cite[Equations (2.22)--(2.23)]{operatope}, \cite[Section 3.3]{higher}, \cite[Definition 16, Proposition 5, Theorem 6]{Wang}.

\rthm{Lemma}{b-00207}{Counts at $n$ loops} \emph{A nonzero connected $n$-loop graph satisfies}
\begin{equation}\label{eq-1-11}
\begin{gathered}
|V|=n+2,\qquad |I|=2n+1,\qquad
|I|-|V|+1=n,\\
|I_S|\le2|V_S|-3\qquad(S\subseteq\Gamma,\ |V_S|\ge2).
\end{gathered}
\end{equation}

\begin{proofbox}
\textbf{Proof.} Lemma~\ref{b-00180} and $n=|I|-|V|+1$ give the first line; Lemma~\ref{b-00194} gives the second.~\hfill\qedsymbol
\end{proofbox}

\subsection{Cubic vertices: only fans}\label{sec-only-fans}

Turning to the four-dimensional field content, \eqref{eq-1-1} shows that the Feynman graphs of the loop-corrected supercharge $Q_n$ have $n+1$ interaction vertices and one insertion $\mathcal O$. We call an edge that joins an interaction vertex to $\mathcal O$ a \emph{spoke}, and write $\deg_\Gamma\mathcal O$ for the number of spokes of a graph $\Gamma$, the degree of $\mathcal O$; by Corollary~\ref{b-00188} it is at most $n+1$.

\begin{figure}[ht!]
\centering
\includegraphics[width=0.91\linewidth]{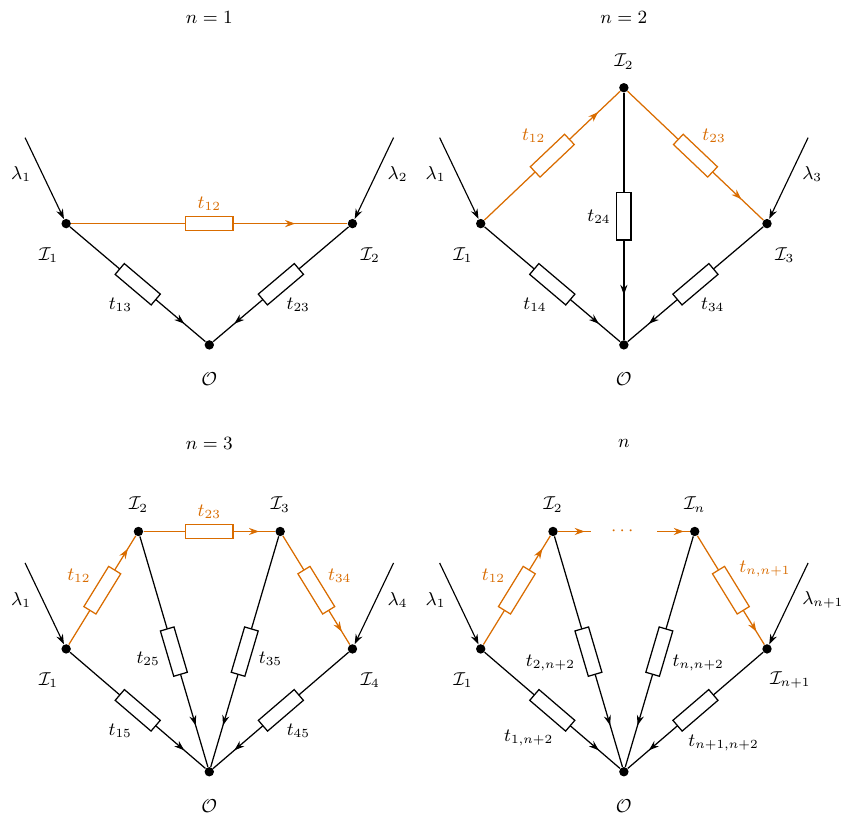}
\caption{Fans at one, two, three and $n$ loops as circuits, with the interaction path in orange, resistances $t_{vw}$ and injected currents $\lambda_1,\lambda_{n+1}$.}
\label{fig-fan-family}
\end{figure}

\rthm{Proposition}{b-00210}{Graphs of $Q_n$ with cubic vertices are fans} \emph{For a translation-invariant interaction $\mathcal I$ whose vertices have degree at most three, every nonzero connected $n$-loop kernel with one local-operator insertion is a fan of cubic vertices (Figure~\ref{fig-fan-family}), for $n\ge1$ (at $n=0$, the single-edge tree represents the classical supersymmetry transformation $Q_0$). The interaction vertices form a path; each is joined to the insertion, and one output field remains at each endpoint of the path.}

By Lemma~\ref{b-00207}, there are $n+1$ interaction vertices besides the insertion $\mathcal O$; Proposition~\ref{b-00210} follows from two lemmas.

\rthm{Lemma}{b-00213}{$\deg_\Gamma\mathcal O$ is maximal} \emph{In a graph of Proposition~\ref{b-00210}, $\deg_\Gamma\mathcal O=n+1$: every interaction vertex is joined to $\mathcal O$ by one edge and exactly two interaction fields remain uncontracted.}

\begin{proofbox}
\textbf{Proof (circuit).} The uncontracted interaction fields inject the external currents. By Corollary~\ref{b-00188} and Lemma~\ref{b-00207}, with equality in the first step only for cubic vertices,
\begin{equation}\label{eq-cubiccount}
\begin{aligned}
\#(\text{sources})
&\le\underbrace{3}_{\text{degree bound}}\,
\underbrace{(n+1)}_{\mathcal I\text{ vertices}}
-\underbrace{2\bigl(2n+1-\deg_\Gamma\mathcal O\bigr)}_{\substack{\text{letters contracted between}\\\mathcal I\text{ vertices}}}
-\underbrace{\deg_\Gamma\mathcal O}_{\substack{\text{edges from }\mathcal O\\\text{to }\mathcal I\text{ vertices}}}\\
&=\underbrace{\deg_\Gamma\mathcal O}_{\substack{\le n+1\\\text{by Corollary~\ref{b-00188}}}}-n+1\le2.
\end{aligned}
\end{equation}
If there is only one source, every current is a real multiple of the source current and sweeps at most $n<2n$ dimensions; a nonzero kernel therefore has at least two sources, forcing both inequalities to be equalities.~\hfill\qedsymbol
\end{proofbox}

\rthm{Lemma}{b-00217}{No interaction cycles} \emph{If every interaction vertex is joined to $\mathcal O$, no subgraph $S$ on interaction vertices carries a cycle.}

\begin{proofbox}
\textbf{Proof.} A loop of resistors on the interaction nodes $V_S$, together with their $|V_S|$ spokes, gives a subcircuit on $V_S\cup\{\mathcal O\}$ with more resistors than the local count of Ohm's law allows (Lemma~\ref{b-00194}):
\begin{equation*}
|I_S|+|V_S|\ge2|V_S|>2(|V_S|+1)-3 .
\end{equation*}
\par\vspace{-\medskipamount}\noindent\hspace*{\fill}\qedsymbol
\end{proofbox}

\begin{proofbox}
\textbf{Proof of Proposition~\ref{b-00210}.} By Lemma~\ref{b-00213}, each of the $n+1$ cubic interaction nodes is wired to $\mathcal O$ by one spoke, which leaves $2n+1-(n+1)=n$ internal resistors; they close no loop (Lemma~\ref{b-00217}) and form a tree. Each node spends one terminal on its spoke and keeps at most two for internal resistors, so the tree is a chain: its interior nodes wire all three terminals, and each end keeps one free terminal, the two sources of Lemma~\ref{b-00213}.~\hfill\qedsymbol
\end{proofbox}

\rthm{Corollary}{b-00222}{Renormalizable theories} \emph{For every renormalizable $\mathcal N=1$ gauge theory, massless or massive, every nonzero connected graph of $Q_n$, $n\ge1$, with one insertion $\mathcal O$ is a fan of cubic vertices.}

\begin{proofbox}
\textbf{Proof.} The superpotential has degree at most three, so Proposition~\ref{b-00210} applies.~\hfill\qedsymbol
\end{proofbox}

The fan kernel is the simplest family to evaluate. When the interaction contains higher-degree superpotential vertices, additional graph topologies can appear, though antifield counting still restricts them to a single family.

\subsection{Superpotential vertices of higher degree: spiders}\label{sec-spider-classification}

\rthm{Proposition}{b-00225}{Superpotential vertices and antifield count} \emph{For the interaction \eqref{eq-B-10}, allow $\bar W(\gamma,w)$ to have arbitrary degree. Every nonzero connected $n$-loop graph of $Q_n\mathcal O$ with one insertion $\mathcal O$, $n\ge1$, contains at most one $\bar W$ vertex, and each graph term lowers the number of $b,\beta$ factors by $n$ plus the number of $\bar W$ vertices.}

We prove Proposition~\ref{b-00225} with the help of two lemmas.

\rthm{Lemma}{lem-antifield-terminals}{Antifield terminals} \emph{Every contraction pairs a $b$ or $\beta$ factor with a $c$ or $\gamma$ factor. A gauge or matter vertex of \eqref{eq-B-10} supplies one $b$ or $\beta$ factor, and a $\bar W$ vertex none.}

\begin{proofbox}
\textbf{Proof.} A gauge or matter node, $bcc$ or $c\gamma\beta$ in \eqref{eq-B-10}, has one antifield terminal, whereas a node of $\bar W(\gamma,w)$ has none.~\hfill\qedsymbol
\end{proofbox}

\rthm{Lemma}{b-00228}{Internal edges} \emph{A nonzero connected $n$-loop graph of $Q_n$ has at least $n$ edges between interaction vertices.}

\begin{proofbox}
\textbf{Proof.} Two resistors from $\mathcal O$ to the same interaction node would be in parallel (Corollary~\ref{b-00188}), so at most $n+1$ of the $2n+1$ resistors (Lemma~\ref{b-00207}) end at $\mathcal O$, leaving $2n+1-\deg_\Gamma\mathcal O\ge2n+1-(n+1)=n$ internal edges between interaction nodes without using the cubic-vertex hypothesis of Lemma~\ref{b-00213}.~\hfill\qedsymbol
\end{proofbox}

\begin{proofbox}
\textbf{Proof of Proposition~\ref{b-00225}.} By Lemmas~\ref{lem-antifield-terminals} and~\ref{b-00228},
\begin{equation}\label{eq-1-15}
\begin{gathered}
\underbrace{n}_{\substack{\deg_\Gamma\mathcal O\le n+1\\
\text{at most }n+1\text{ spokes}}}
\le
\underbrace{2n+1-\deg_\Gamma\mathcal O}_{\text{internal resistors}}
\le
\underbrace{n+1-\#(\bar W\text{ vertices})}_{\substack{
\text{antifield terminals,}\\
\text{one per internal resistor}}}
\\[3pt]
\Longrightarrow\quad\#(\bar W\text{ vertices})\le1,
\end{gathered}
\end{equation}
and each of the $2n+1$ resistors takes one antifield terminal, a $b$ or $\beta$ factor:
\begin{equation}\label{eq-antifield-loss}
\begin{aligned}
\#(b,\beta\text{ in the output})
&=\#(b,\beta\text{ in }\mathcal O)
  +\bigl[n+1-\#(\bar W\text{ vertices})\bigr]-(2n+1)\\
&=\#(b,\beta\text{ in }\mathcal O)
  -n-\#(\bar W\text{ vertices}).
\end{aligned}
\end{equation}
\par\vspace{-\medskipamount}\noindent\hspace*{\fill}\qedsymbol
\end{proofbox}

\rthm{Corollary}{b-00235}{Finite loop order} \emph{For $\bar W$ of any degree, or a formal power series read coefficientwise, $Q_n\mathcal O=0$ for $n>\#(b,\beta\text{ in }\mathcal O)$, and terms with a $\bar W$ vertex vanish already for $n+1>\#(b,\beta\text{ in }\mathcal O)$.}

\begin{proofbox}
\textbf{Proof.} The left-hand side of \eqref{eq-antifield-loss} is nonnegative.~\hfill\qedsymbol
\end{proofbox}

Corollary~\ref{b-00235} and dimensional counting make the perturbative expansion truncate at a finite order on every polynomial local operator, so that the perturbative cohomology is well defined without auxiliary cutoffs.

\rthm{Proposition}{b-00240}{Graphs with one superpotential vertex} \emph{For the interaction \eqref{eq-B-10}, a nonzero connected $n$-loop graph of $Q_n$, $n\ge1$, with one insertion $\mathcal O$ and one $\bar W$ vertex joins $\mathcal O$ to all $n+1$ interaction vertices, for any degree of $\bar W$. Removing $\mathcal O$ leaves a tree whose arms meet at $\bar W$, i.e., a spider (Definition~\ref{b-00380}).}

\begin{proofbox}
\textbf{Proof.} With one $\bar W$ vertex the two bounds in \eqref{eq-1-15} coincide:
\begin{equation}\label{eq-superpotential-spokes}
n\le2n+1-\deg_\Gamma\mathcal O\le n
\quad\Longrightarrow\quad
\deg_\Gamma\mathcal O=n+1,
\end{equation}
so every interaction node is wired to $\mathcal O$, and the $n$ internal resistors close no loop (Lemma~\ref{b-00217}) and form a tree. A cubic node keeps at most two internal resistors besides its spoke, so the tree branches only at the $\bar W$ node; with $a$ arms of $\ell_i$ cubic nodes,
\begin{equation}\label{eq-1-16}
\begin{gathered}
\Gamma\setminus\{\mathcal O\}
=\text{arms of lengths }\ell_1,\ldots,\ell_a\text{ meeting at }\bar W,\\
\ell_1+\cdots+\ell_a=n,\qquad\ell_i\ge1,\qquad\deg\bar W\ge a+1,
\end{gathered}
\end{equation}
where $\deg\bar W$, the number of field factors of the contributing monomial, covers the $a$ internal resistors at the $\bar W$ node and its spoke.~\hfill\qedsymbol
\end{proofbox}

\rthm{Remark}{b-00246}{Specialization to $\deg\bar W\le3$} With at most two arms the graph of Proposition~\ref{b-00240} reduces to a fan; the bound $a+1\le\deg\bar W$ forces $a\le2$ whenever $\deg\bar W\le3$.

\rthm{Corollary}{b-00247}{Graphs of $Q_n$} \emph{For the interaction \eqref{eq-B-10}, every nonzero connected graph of $Q_n$, $n\ge1$, with one insertion $\mathcal O$ is a fan, or a spider whose arms meet at its $\bar W$ vertex (Definition~\ref{b-00380}); for $\deg\bar W\le3$ it is a fan.}

\begin{proofbox}
\textbf{Proof.} Combine Propositions~\ref{b-00210}, \ref{b-00225} and~\ref{b-00240}.~\hfill\qedsymbol
\end{proofbox}

This completes the classification of the graphs of $Q_n$; the marked graphs of the quantum product (Section~\ref{sec-product}) involve more general forests of interaction trees attached to both insertions, classified in Section~\ref{sec-dressed-definition}.

\section{Fan kernels}\label{sec-fans}

\begin{figure}[H]
\centering
\includegraphics[width=0.91\linewidth]{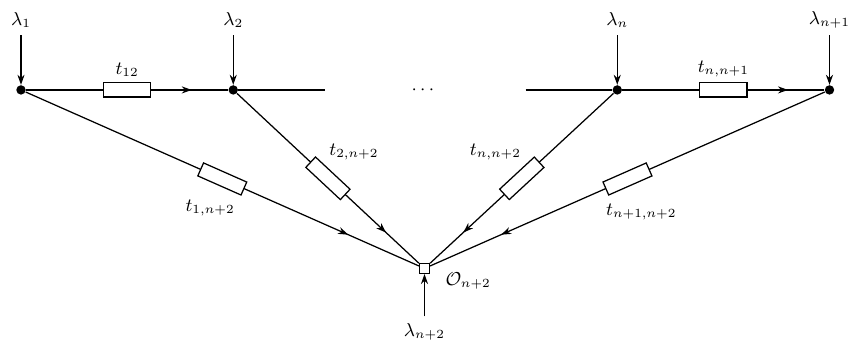}
\caption{The $n$-loop fan circuit: a path of $n+1$ cubic interactions joined to the insertion $n+2$. The resistors carry Schwinger resistances $t_{vw}$, and external arrows inject currents $\lambda_v$. Short stubs and central ellipses indicate omitted intermediate vertices.}
\label{fig-general-fan}
\end{figure}

In Section~\ref{sec-BV}, homotopy transfer eliminated all $Q_-$-exact terms from the physical $\mathcal N=1$ action, leaving the cubic gauge ($bcc$) and matter ($c\gamma\beta$) interaction vertices together with the antichiral superpotential $\bar W(\gamma,w)$ in~\eqref{eq-B-10}. In four dimensions, every renormalizable $\mathcal N=1$ supersymmetric gauge theory (massless or massive, with constant gauge coupling) has $\deg\bar W\le3$, so every interaction vertex in $S_{\mathrm{tw}}$ is at most cubic. By Corollary~\ref{b-00222}, the fan graphs of Figure~\ref{fig-general-fan} therefore determine \emph{every} perturbative loop correction $Q_n$ ($n\ge1$) to the supersymmetry transformation in renormalizable theories, as well as all one- and two-loop quantum-product graphs (Section~\ref{sec-product-loops}).

We state the complete all-loop fan theorem first and then build its circuit proof step by step: Section~\ref{the-one-loop-triangle-diagram} solves the base one-loop triangle; Section~\ref{sec-B-local-reduction} formulates local series, parallel, and series--parallel moves as variable changes on Schwinger space; Sections~\ref{sec-fan-step} and~\ref{sec-fan-positive} iterate $n-1$ series--parallel eliminations to prove~\eqref{eq-1-40}--\eqref{eq-1-39}; and Section~\ref{endpoint-sources-and-taylor-expansion} specializes to endpoint sources to prove~\eqref{eq-1-41}--\eqref{eq-1-42} and the all-order Taylor expansion~\eqref{eq-fan-full-taylor}.

\rthm{Theorem}{thm-fan-kernel}{Fan kernel at every loop order} \emph{Let $\Gamma$ be the $n$-loop fan of Figure~\ref{fig-general-fan} ($n\ge1$): a directed path $1\to2\to\cdots\to n+1$ of interaction vertices, each joined to the insertion $n+2$ by a spoke $(i,n+2)$ oriented towards $n+2$. Let the path vertices carry arbitrary external sources $\lambda_1,\ldots,\lambda_{n+1}$ (with $\lambda_{n+2}=-\sum_{v<n+2}\lambda_v$) and the edges carry arbitrary displacements $z_e$. On the $(2n-2)$-cube times two-simplex}
\begin{equation}\label{eq-1-37}
0<r_j<1\quad(1\leq j\leq2n-2),\qquad r_{2n-1}>0,\quad r_{2n}>0,\quad r_{2n-1}+r_{2n}<1,
\end{equation}
\emph{the operatope kernel~\eqref{eq-operatope} equals}
\begin{equation}\label{eq-1-40}
\begin{aligned}
\mathcal I_\Gamma(\lambda;z)
={}&\int_{\eqref{eq-1-37}}d^{2n}r\,
\prod_{i=1}^{n-1}\underbrace{r_{2i}
\left[\sum_{j=1}^{i}\left(\prod_{k=j}^{i-1}r_{2k-1}\right)
\lambda_{j,+\dot a_i}\right]
(\lambda_{i+1,+}{}^{\dot a_i}-y_{i+1,i+2,+}{}^{\dot a_i}(r))}_{\text{kernel factor from eliminating vertex }i}\\
&\times\underbrace{\left[\sum_{j=1}^{n}
\left(\prod_{k=j}^{n-1}r_{2k-1}\right)\lambda_{j,+\dot b}\right]
\lambda_{n+1,+}{}^{\dot b}}_{\text{final triangle factor}}
\,e^{-\sum_e y_{e,+\dot a}(r)z_e^{+\dot a}}.
\end{aligned}
\end{equation}
\emph{Every edge current $y_e(r)$ is a polynomial in $r$: along the interaction path,}
\begin{equation}\label{eq-1-38}
\begin{aligned}
y_{n,n+1,+\dot a}
&=r_{2n-1}\sum_{j=1}^{n}
\left(\prod_{k=j}^{n-1}r_{2k-1}\right)\lambda_{j,+\dot a}
-r_{2n}\lambda_{n+1,+\dot a},\\
y_{i,i+1,+\dot a}
&=r_{2i-1}r_{2i}\sum_{j=1}^{i}
\left(\prod_{k=j}^{i-1}r_{2k-1}\right)\lambda_{j,+\dot a}\\
&\quad-(1-r_{2i})\lambda_{i+1,+\dot a}
+(1-r_{2i})y_{i+1,i+2,+\dot a},\qquad1\le i<n,
\end{aligned}
\end{equation}
\emph{and on the spokes, by current conservation,}
\begin{equation}\label{eq-1-39}
\begin{aligned}
y_{1,n+2,+\dot a}&=\lambda_{1,+\dot a}-y_{12,+\dot a},\\
y_{i,n+2,+\dot a}&=\lambda_{i,+\dot a}+y_{i-1,i,+\dot a}-y_{i,i+1,+\dot a},
\qquad2\le i\le n,\\
y_{n+1,n+2,+\dot a}&=\lambda_{n+1,+\dot a}+y_{n,n+1,+\dot a}.
\end{aligned}
\end{equation}
\emph{For the source and displacement pattern of $Q_n$ in a renormalizable theory ($\lambda_2=\cdots=\lambda_n=0$, $z_{i,i+1}=0$, and $z_{i,n+2}=-z_i$), the measure factorizes into a single monomial:}
\begin{equation}\label{eq-1-41}
\begin{aligned}
&\mathcal I_\Gamma(\lambda;z)
\big|_{\lambda_2=\cdots=\lambda_n=0,
\ z_{i,i+1}=0,\ z_{i,n+2}=-z_i}\\
&=(\lambda_{1,+\dot a}\lambda_{n+1,+}{}^{\dot a})^n
\int_{\eqref{eq-1-37}}d^{2n}r\,
\left[\prod_{j=1}^{n-1}r_{2j-1}^{n-j}r_{2j}(1-r_{2j})^{j-1}\right]
r_{2n}^{n-1}
e^{\sum_j y_{j,n+2,+\dot a}z_j^{+\dot a}}.
\end{aligned}
\end{equation}
\emph{At zero displacement ($z=0$), this integral evaluates in closed form to}\footnote{For $n=1,2,3$, \eqref{eq-1-42} agrees with the operatope values \cite[Equations (4.20), (4.44), (4.67)]{operatope} (publicly implemented in \cite{LamanLoopstrap}) and with the fan weights of Gui, Wang, and Williams \cite[Sections 4.4--4.6]{GWW}; at $n=1$ it reproduces \cite[Equation (2.48)]{BK}.}
\begin{equation}\label{eq-1-42}
\left.\mathcal I_\Gamma(\lambda;0)\right|_{\lambda_{2,+\dot a}=\cdots=\lambda_{n,+\dot a}=0}
=\frac{(\lambda_{1,+\dot a}\lambda_{n+1,+}{}^{\dot a})^n}{(n!)^3(n+1)},
\end{equation}
\emph{and for arbitrary edge displacements at $\lambda_2=\cdots=\lambda_n=0$, every Taylor coefficient is given explicitly by the finite factorial sum~\eqref{eq-fan-full-taylor}.}

Each coordinate $r_j$ in~\eqref{eq-1-37} has a direct electrical meaning: for $1\le i\le n-1$, $r_{2i-1}\in(0,1)$ is the fraction of the accumulated source at vertex $i$ passed forward to $i+1$ during series elimination, while $r_{2i}\in(0,1)$ is the fraction of the merged spoke current carried by the original spoke $(i+1,n+2)$ after parallel merging; the final pair $(r_{2n-1},r_{2n})\in\Delta^2$ parameterizes the remaining one-loop triangle as in~\eqref{eq-1-25}. Because all currents $y_e(r)$ are polynomial in $r$---eliminating the spanning-tree denominators of the global Kirchhoff solution (Appendix~\ref{sec-B-spanning})---every Taylor coefficient in $z$ reduces to an elementary Beta and simplex integral on~\eqref{eq-1-37}. Although $\lambda_2=\cdots=\lambda_n=0$ in $Q_n$ for renormalizable theories (where interior cubic vertices have all three legs contracted), keeping $\lambda_1,\ldots,\lambda_{n+1}$ general in~\eqref{eq-1-40} simultaneously covers both two-loop quantum products (Section~\ref{sec-product-loops}) and higher-degree superpotential vertices (where at most one interior vertex carries a nonzero source by Propositions~\ref{b-00225} and~\ref{b-00240}).

\subsection{The one-loop triangle diagram}\label{the-one-loop-triangle-diagram}

\begin{figure}[ht!]
\centering
\includegraphics[width=0.92\linewidth]{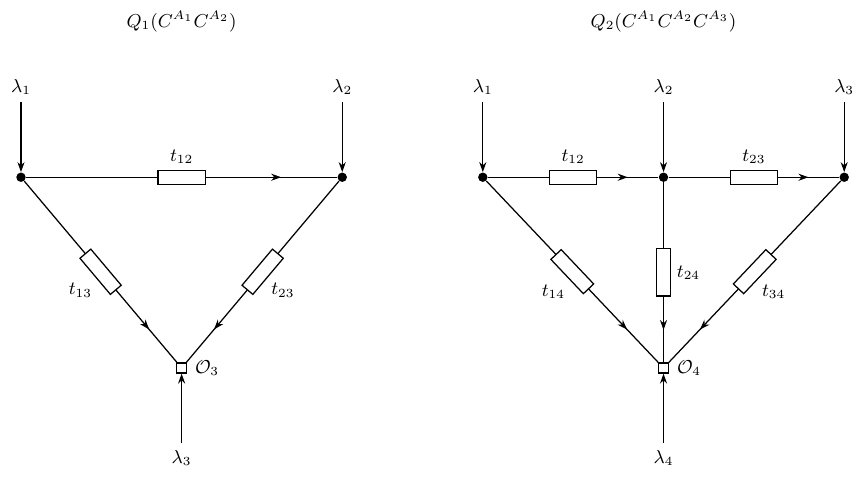}
\caption{The one- and two-loop fan circuits. Dots are interaction vertices and the square is the local insertion. Each edge carries a resistor $t_{vw}$; $\lambda_v$ is the current injected at vertex $v$.}
\label{fig-low-fans}
\end{figure}

We begin with the base $n=1$ triangle on vertices $\{1,2,3\}$ with the insertion at $3$ (Figure~\ref{fig-low-fans}).

Normalizing the projective resistances by $t_{12}+t_{13}+t_{23}=1$, current conservation at vertices $1,2$ and Kirchhoff's loop law $t_{12}y_{12}+t_{23}y_{23}-t_{13}y_{13}=0$ determine all three edge currents:\footnote{When $\lambda_2=0$, $\lambda_1$ enters at $1$ and exits at $3$ across two parallel branches of resistances $t_{13}$ and $t_{12}+t_{23}$, so branch $1\to2\to3$ carries fraction $t_{13}/(t_{12}+t_{13}+t_{23})=t_{13}$; superposing the symmetric $\lambda_2$ flow gives $y_{12}$.}
\begin{equation}\label{eq-triangle-current}
\begin{aligned}
y_{12,+\dot a}&=t_{13}\lambda_{1,+\dot a}-t_{23}\lambda_{2,+\dot a},\\
y_{13,+\dot a}&=\lambda_{1,+\dot a}-y_{12,+\dot a},\\
y_{23,+\dot a}&=\lambda_{2,+\dot a}+y_{12,+\dot a}.
\end{aligned}
\end{equation}
Passing to simplex coordinates $(t_{12},t_{13},t_{23})=(1-r_1-r_2,r_1,r_2)$ on $\Delta^2=\{r_1,r_2>0,\,r_1+r_2<1\}$, taking $T=\{13,23\}$ as the spanning tree in~\eqref{eq-operatope} gives the chord Jacobian
\begin{equation}\label{b-00267}
-dy_{12,+\dot+}\wedge dy_{12,+\dot-}
=(\lambda_{1,+\dot a}\lambda_{2,+}{}^{\dot a})\,dr_1\wedge dr_2,
\end{equation}
and hence the one-loop triangle kernel
\begin{equation}\label{eq-1-25}
\mathcal I_\triangle(\lambda;z)
=(\lambda_{1,+\dot a}\lambda_{2,+}{}^{\dot a})
\int_0^1dr_1\int_0^{1-r_1}dr_2\,
e^{-\sum_e y_{e,+\dot b}z_e^{+\dot b}}.
\end{equation}
Substituting $y_{13}=\lambda_1-y_{12}$ and $y_{23}=\lambda_2+y_{12}$ regroups the exponent into $-\lambda_{1,+\dot b}z_{13}^{+\dot b}-\lambda_{2,+\dot b}z_{23}^{+\dot b}-(r_1\lambda_1-r_2\lambda_2)_{+\dot b}(z_{12}+z_{23}-z_{13})^{+\dot b}$. Expanding in powers of $r_1,r_2$ and integrating term by term via $\int_{\Delta^2}r_1^ir_2^j\,dr_1\,dr_2=\frac{i!\,j!}{(i+j+2)!}$ cancels $i!\,j!$ and yields the complete displacement expansion~\cite[Eq.~(2.47)]{BK}:
\begin{equation}\label{eq-1-46}
\begin{aligned}
\mathcal I_\triangle(\lambda;z)
={}&(\lambda_{1,+\dot a}\lambda_{2,+}{}^{\dot a})\,
 e^{-\lambda_{1,+\dot a}z_{13}^{+\dot a}-\lambda_{2,+\dot a}z_{23}^{+\dot a}}\\
&\times\sum_{i,j\ge0}
\frac{[-\lambda_{1,+\dot b}(z_{12}+z_{23}-z_{13})^{+\dot b}]^i
[\lambda_{2,+\dot c}(z_{12}+z_{23}-z_{13})^{+\dot c}]^j}{(i+j+2)!}.
\end{aligned}
\end{equation}

\subsection{Local reductions as kernel operators}\label{sec-B-local-reduction}\label{sec-network}

We now express local circuit reductions as integral-differential operators acting on the kernel $\mathcal I_{\Gamma^{\mathrm{red}}}$ of a smaller graph $\Gamma^{\mathrm{red}}$ (indicated by superscript $\mathrm{red}$). Because Ohm's and Kirchhoff's laws are local, each move changes variables only on the edges incident to the reduced subcircuit without ever invoking the global spanning-tree denominator (Appendix~\ref{sec-B-spanning}).

\subsubsection{Series elimination}\label{local-B-series-heading}

\begin{figure}[H]
\centering
\includegraphics[width=0.95\linewidth]{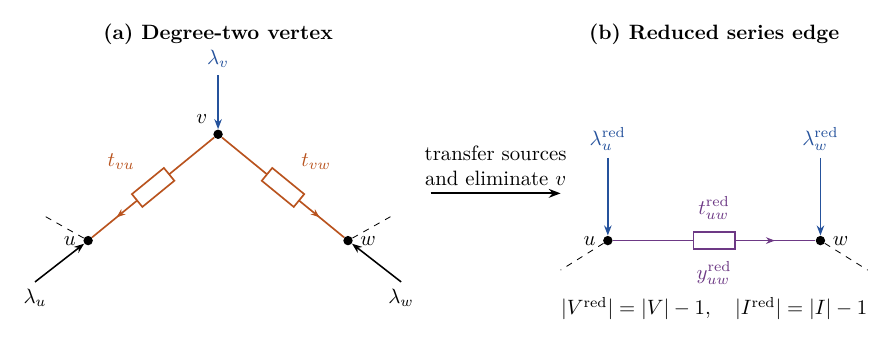}
\caption{Series elimination: the source $\lambda_v$ at a degree-two vertex $v$ is transferred to $u,w$ by~\eqref{eq-1-64}, leaving the loop number unchanged.}
\label{fig-series-resistors}
\end{figure}

Suppose a vertex $v$ has degree two in $\Gamma$, incident only to $vu$ and $vw$. We eliminate $v$ while transferring its injected source $\lambda_v$ to its neighbours $u$ and $w$ (Figure~\ref{fig-series-resistors}).

Current conservation at $v$ and the total potential drop from $u$ to $w$ give
\begin{equation}\label{eq-B-series-derivation}
\begin{aligned}
y_{vu,+\dot a}+y_{vw,+\dot a}&=\lambda_{v,+\dot a},\\
\frac{(x_u-x_w)^-_{\dot a}}2
&=-t_{vu}y_{vu,+\dot a}+t_{vw}y_{vw,+\dot a}
=t_{vw}\lambda_{v,+\dot a}-(t_{vu}+t_{vw})y_{vu,+\dot a}.
\end{aligned}
\end{equation}
Setting $r_j=t_{vw}/(t_{vu}+t_{vw})\in(0,1)$ replaces $(vu,vw)$ by a single directed series edge $u\to w$ with resistance, transferred sources, and reconstructed currents
\begin{equation}\label{eq-1-64}
\begin{gathered}
r_j=\frac{t_{vw}}{t_{vu}+t_{vw}},\qquad
t_{uw}^{\mathrm{red}}=t_{vu}+t_{vw},\\
\lambda_{u,+\dot a}^{\mathrm{red}}
=\lambda_{u,+\dot a}+r_j\lambda_{v,+\dot a},\qquad
\lambda_{w,+\dot a}^{\mathrm{red}}
=\lambda_{w,+\dot a}+(1-r_j)\lambda_{v,+\dot a},\\
y_{vu,+\dot a}=r_j\lambda_{v,+\dot a}-y_{uw,+\dot a}^{\mathrm{red}},\qquad
y_{vw,+\dot a}=(1-r_j)\lambda_{v,+\dot a}+y_{uw,+\dot a}^{\mathrm{red}}.
\end{gathered}
\end{equation}

\subsubsection{Parallel merge}\label{local-B-parallel-heading}

When series elimination (or star--triangle reduction in Section~\ref{sec-forests}) produces $q>1$ parallel edges between $u$ and $w$, orient all $q$ branches $u\to w$ ($j=1,\ldots,q$). Sharing the same potential drop, their conductances add:
\begin{equation}\label{eq-B-parallel-derivation}
t_{uw}^{(j)}y_{uw,+\dot a}^{(j)}
=\frac{(x_u-x_w)^-_{\dot a}}2,\qquad
y_{uw,+\dot a}^{\mathrm{red}}
=\sum_{j=1}^q y_{uw,+\dot a}^{(j)}
=\left(\sum_{j=1}^q\frac1{t_{uw}^{(j)}}\right)
\frac{(x_u-x_w)^-_{\dot a}}2,
\end{equation}
giving the merged resistance and current-division fractions
\begin{equation}\label{eq-1-63}
\begin{gathered}
\frac1{t_{uw}^{\mathrm{red}}}=\sum_{j=1}^q\frac1{t_{uw}^{(j)}},\qquad
r_j=\frac{t_{uw}^{\mathrm{red}}}{t_{uw}^{(j)}},\qquad\sum_{j=1}^q r_j=1,\qquad
y_{uw,+\dot a}^{(j)}=r_jy_{uw,+\dot a}^{\mathrm{red}}.
\end{gathered}
\end{equation}

\subsubsection{The series--parallel move}\label{sec-B-kernel-recursion}

\begin{figure}[H]
\centering
\includegraphics[width=0.96\linewidth]{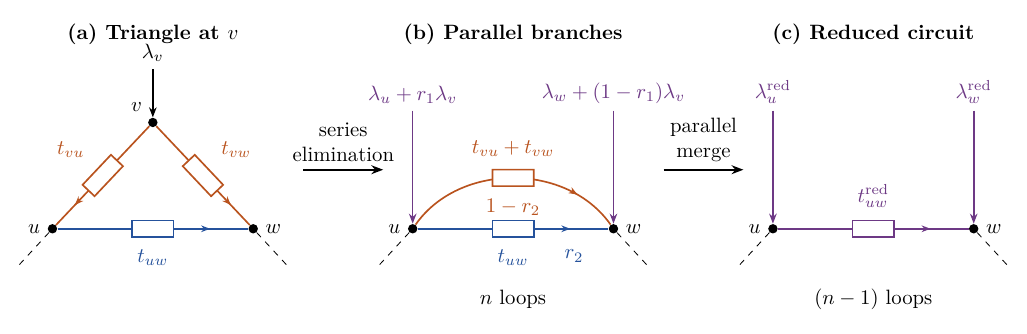}
\caption{Series--parallel reduction: eliminating a degree-two vertex $v$ merges the series branch (orange) with the existing edge $uw$ (blue), reducing the loop number by one.}
\label{fig-series-parallel-resistors}
\end{figure}

\rthm{Theorem}{b-01142}{Series--parallel reduction} \emph{Let $\Gamma$ be a Laman graph containing a degree-two vertex $v$ with oriented incident edges $vu,vw$ and base edge $uw$ (Figure~\ref{fig-series-parallel-resistors}), ordered with $(vu,vw)$ first and all remaining edges (including the merged edge $uw$) retaining their relative order in $\Gamma^{\mathrm{red}}$. Setting}
\begin{equation*}
\begin{gathered}
r_1=\frac{t_{vw}}{t_{vu}+t_{vw}}\in(0,1),\qquad
r_2=\frac{t_{vu}+t_{vw}}{t_{vu}+t_{vw}+t_{uw}}\in(0,1),\\
\lambda_u^{\mathrm{red}}=\lambda_u+r_1\lambda_v,\qquad
\lambda_w^{\mathrm{red}}=\lambda_w+(1-r_1)\lambda_v,\\
z_{uw}^{\mathrm{red}}=r_2z_{uw}+(1-r_2)(z_{vw}-z_{vu}),\qquad
z_e^{\mathrm{red}}=z_e\quad(e\ne uw),
\end{gathered}
\end{equation*}
\emph{the reduced graph $\Gamma^{\mathrm{red}}$ is again Laman, and the operatope kernel~\eqref{eq-operatope} satisfies}
\begin{equation}\label{eq-kernel-series-parallel}
\begin{aligned}
\mathcal I_\Gamma(\lambda;z)
={}&\int_0^1dr_1\int_0^1dr_2\,
e^{-\lambda_{v,+\dot a}(r_1z_{vu}+(1-r_1)z_{vw})^{+\dot a}}
\underbrace{r_2\lambda_{v,+\dot a}\left(-\epsilon^{\dot a\dot b}
\frac{\partial}{\partial z_{uw}^{\mathrm{red},+\dot b}}\right)}_{\text{local current Jacobian}}
\mathcal I_{\Gamma^{\mathrm{red}}}
(\lambda^{\mathrm{red}};z^{\mathrm{red}})\\[4pt]
={}&\int_0^1dr_1\int_0^1dr_2
\int_{\substack{y^{\mathrm{red}}\in\Delta_{\Gamma^{\mathrm{red}}}\\
\sum_{e:e(0)=v'}y_e^{\mathrm{red}}-\sum_{e:e(1)=v'}y_e^{\mathrm{red}}=\lambda_{v'}^{\mathrm{red}}}}
r_2\lambda_{v,+\dot a}y_{uw,+}^{\mathrm{red}}{}^{\dot a}\\
&\qquad\times e^{-\sum_{e\in I}y_{e,+\dot b}z_e^{+\dot b}}
\bigwedge_{e\in I^{\mathrm{red}}\setminus T^{\mathrm{red}}}(-d^2y_e^{\mathrm{red}}),
\end{aligned}
\end{equation}
\emph{where $T^{\mathrm{red}}$ is any spanning tree of $\Gamma^{\mathrm{red}}$ containing $uw$, the wedge product is taken at fixed $(r_1,r_2)$, and the original edge currents in the exponent are reconstructed polynomially by~\eqref{b-01151}.}

\begin{proofbox}
\textbf{Proof.} Combining the series elimination~\eqref{eq-1-64} ($r_j=r_1$) with the two-branch parallel merge~\eqref{eq-1-63} (where the series branch $u\to v\to w$ and original edge $uw$ carry fractions $1-r_2$ and $r_2$ of $y_{uw}^{\mathrm{red}}$) gives
\begin{equation}\label{b-01151}
y_{uw}=r_2y_{uw}^{\mathrm{red}},\qquad
y_{vu}=r_1\lambda_v-(1-r_2)y_{uw}^{\mathrm{red}},\qquad
y_{vw}=\lambda_v-y_{vu},
\end{equation}
so the displacement pairing splits into
\begin{equation*}
\sum_{e\in I}y_{e,+\dot a}z_e^{+\dot a}
=\lambda_{v,+\dot a}(r_1z_{vu}+(1-r_1)z_{vw})^{+\dot a}
+\sum_{e\in I^{\mathrm{red}}}
y_{e,+\dot a}^{\mathrm{red}}z_e^{\mathrm{red},+\dot a}.
\end{equation*}
Lift the spanning tree $T^{\mathrm{red}}\ni uw$ of $\Gamma^{\mathrm{red}}$ to $T=T^{\mathrm{red}}\cup\{vw\}$ of $\Gamma$, so $vu\notin T$ is the eliminated chord. At fixed external sources $\lambda$, Kirchhoff conservation makes $d(y_{uw,+\dot a}^{\mathrm{red}}-r_1\lambda_{v,+\dot a})$ a linear combination of the retained chord differentials $\{dy_e^{\mathrm{red}}:e\in I^{\mathrm{red}}\setminus T^{\mathrm{red}}\}$; writing $\equiv$ for equality modulo those differentials,
\begin{equation*}
\begin{aligned}
dy_{vu,+\dot a}
&=r_2\lambda_{v,+\dot a}dr_1+y_{uw,+\dot a}^{\mathrm{red}}dr_2
-(1-r_2)d\!\left(y_{uw,+\dot a}^{\mathrm{red}}-r_1\lambda_{v,+\dot a}\right)\\
&\equiv r_2\lambda_{v,+\dot a}dr_1+y_{uw,+\dot a}^{\mathrm{red}}dr_2,\\
-dy_{vu,+\dot+}\wedge dy_{vu,+\dot-}
&\equiv -r_2\left(\lambda_{v,+\dot a}y_{uw,+}^{\mathrm{red}}{}^{\dot a}\right)
 dr_1\wedge dr_2.
\end{aligned}
\end{equation*}
Since the inverse resistance map $t_{vu}=(1-r_1)t_{uw}^{\mathrm{red}}/(1-r_2)$, $t_{vw}=r_1t_{uw}^{\mathrm{red}}/(1-r_2)$, $t_{uw}=t_{uw}^{\mathrm{red}}/r_2$ reverses orientation,\footnote{Explicitly, $d(t_{vu}/t_{uw})\wedge d(t_{vw}/t_{uw})=-\frac{r_2}{(1-r_2)^3}\,dr_1\wedge dr_2$, cancelling the minus sign in $-dy_{vu,+\dot+}\wedge dy_{vu,+\dot-}$ on $(0,1)^2$.} the oriented integral over $0<r_1,r_2<1$ acquires $+r_2(\lambda_{v,+\dot a}y_{uw,+}^{\mathrm{red}}{}^{\dot a})\,dr_1\wedge dr_2$. Finally,
\begin{equation*}
\lambda_{v,+\dot a}\left(-\epsilon^{\dot a\dot b}
\frac{\partial}{\partial z_{uw}^{\mathrm{red},+\dot b}}\right)
e^{-\sum_{e\in I^{\mathrm{red}}}y_{e,+\dot c}^{\mathrm{red}}z_e^{\mathrm{red},+\dot c}}
=(\lambda_{v,+\dot a}y_{uw,+}^{\mathrm{red}}{}^{\dot a})
e^{-\sum_{e\in I^{\mathrm{red}}}y_{e,+\dot c}^{\mathrm{red}}z_e^{\mathrm{red},+\dot c}},
\end{equation*}
proving both lines of~\eqref{eq-kernel-series-parallel}.~\hfill\qedsymbol
\end{proofbox}

\subsection{One elimination step}\label{sec-fan-step}

\rthm{Proposition}{prop-fan-step}{One fan-elimination step}
\emph{After path vertices $1,\ldots,i-1$ ($1\le i<n$) of the $n$-loop fan have been eliminated, vertex $i$ has degree two with accumulated source $\lambda_i^{\mathrm{red}}$ and spoke resistance $t_{i,n+2}^{\mathrm{red}}$ (equal to $\lambda_1$ and $t_{1,n+2}$ at $i=1$). Setting}
\begin{equation}\label{eq-fan-local-ratios}
\begin{aligned}
r_{2i-1}&=\frac{t_{i,n+2}^{\mathrm{red}}}{t_{i,i+1}+t_{i,n+2}^{\mathrm{red}}},
&r_{2i}&=\frac{t_{i,i+1}+t_{i,n+2}^{\mathrm{red}}}
{t_{i,i+1}+t_{i,n+2}^{\mathrm{red}}+t_{i+1,n+2}},\\
t_{i+1,n+2}^{\mathrm{red}}
&=\left[\frac1{t_{i,i+1}+t_{i,n+2}^{\mathrm{red}}}
+\frac1{t_{i+1,n+2}}\right]^{-1},
\end{aligned}
\end{equation}
\emph{eliminating vertex $i$ via Theorem~\ref{b-01142} transfers its source to $i+1$ and $n+2$,}
\begin{equation}\label{eq-fan-source-transfer}
\begin{aligned}
\lambda_{i+1,+\dot a}^{\mathrm{red}}
&=\lambda_{i+1,+\dot a}+r_{2i-1}\lambda_{i,+\dot a}^{\mathrm{red}},\\
\left.\lambda_{n+2,+\dot a}^{\mathrm{red}}\right|_{\text{after }i}
&=\lambda_{n+2,+\dot a}
+\sum_{\substack{1\le j\le i\\\mathrm{eliminated\ vertices}}}
(1-r_{2j-1})\lambda_{j,+\dot a}^{\mathrm{red}},
\end{aligned}
\end{equation}
\emph{reconstructs the stage-$i$ currents backward from $y_{i+1,i+2}$ by}
\begin{equation}\label{eq-fan-step-currents}
\begin{aligned}
y_{i+1,n+2,+\dot a}^{\mathrm{red}}
&=\lambda_{i+1,+\dot a}^{\mathrm{red}}-y_{i+1,i+2,+\dot a},\\
y_{i,i+1,+\dot a}
&=r_{2i-1}\lambda_{i,+\dot a}^{\mathrm{red}}
-(1-r_{2i})y_{i+1,n+2,+\dot a}^{\mathrm{red}},\\
y_{i,n+2,+\dot a}^{\mathrm{red}}
&=\lambda_{i,+\dot a}^{\mathrm{red}}-y_{i,i+1,+\dot a},\qquad
y_{i+1,n+2,+\dot a}=r_{2i}y_{i+1,n+2,+\dot a}^{\mathrm{red}},
\end{aligned}
\end{equation}
\emph{and multiplies the current integrand by the local Jacobian factor}
\begin{equation}\label{b-00282}
r_{2i}\lambda_{i,+\dot a}^{\mathrm{red}}y_{i+1,n+2,+}^{\mathrm{red}}{}^{\dot a}
=r_{2i}\lambda_{i,+\dot a}^{\mathrm{red}}
\bigl(\lambda_{i+1,+}{}^{\dot a}-y_{i+1,i+2,+}{}^{\dot a}\bigr).
\end{equation}

\begin{proofbox}
\textbf{Proof.} Apply Theorem~\ref{b-01142} to $(v,u,w)=(i,i+1,n+2)$ with $(r_1,r_2)=(r_{2i-1},r_{2i})$, noting that the edge $i\to i+1$ is already oriented from $v$ to $u$ (replacing $t_{vu},t_{vw},t_{uw}$ by $t_{i,i+1},t_{i,n+2}^{\mathrm{red}},t_{i+1,n+2}$). Conservation at vertex $i+1$ in the reduced fan (where $i$ has been removed, so only $i+1\to i+2$ and the merged spoke $(i+1,n+2)$ meet $i+1$) gives the first line of~\eqref{eq-fan-step-currents}, while~\eqref{b-01151} gives the remaining lines. Substituting $y_{i+1,n+2}^{\mathrm{red}}=\lambda_{i+1}+r_{2i-1}\lambda_i^{\mathrm{red}}-y_{i+1,i+2}$ into $r_{2i}\lambda_{i,+\dot a}^{\mathrm{red}}y_{i+1,n+2,+}^{\mathrm{red}}{}^{\dot a}$ and using $\lambda_{i,+\dot a}^{\mathrm{red}}\lambda_{i,+}^{\mathrm{red}}{}^{\dot a}=0$ yields~\eqref{b-00282}.~\hfill\qedsymbol
\end{proofbox}

\subsection{Polynomial reconstruction of the fan}\label{sec-fan-positive}\label{sec-B-fan-reduction}

\begin{figure}[H]
\centering
\begin{tikzpicture}
\node[anchor=south west,inner sep=0] (img) at (0,0) {\includegraphics[width=0.96\linewidth]{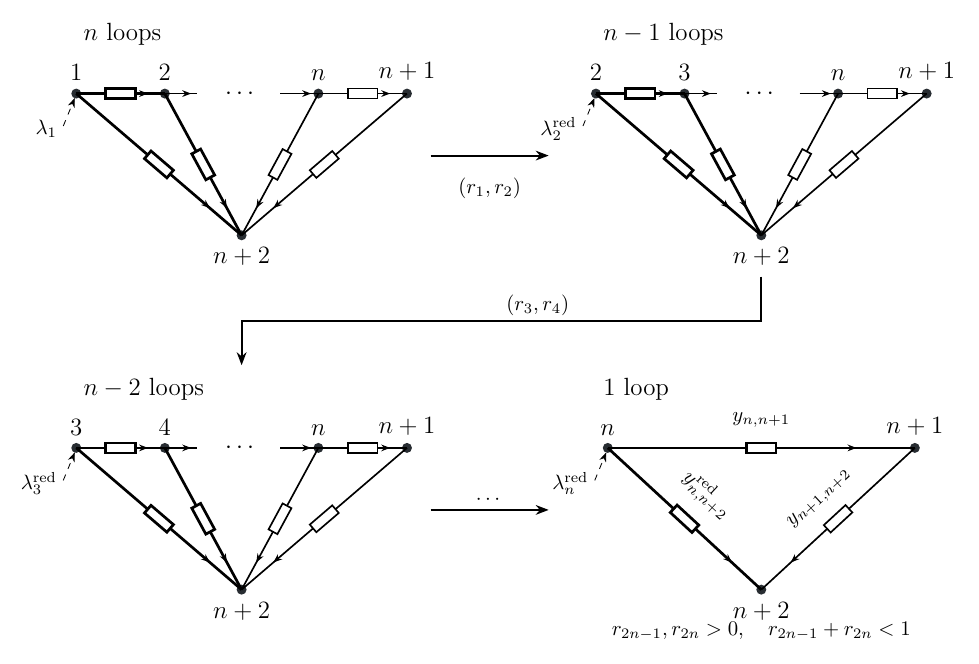}};
\begin{scope}[x={(img.south east)},y={(img.north west)}]
\node[fill=white,inner sep=1pt,minimum width=0.066\linewidth,minimum height=2.4ex,font=\footnotesize,text=black] at (0.5004,0.7889) {\eqref{eq-kernel-series-parallel}};
\node[fill=white,inner sep=1pt,minimum width=0.072\linewidth,minimum height=2.4ex,font=\footnotesize,text=black] at (0.4733,0.5397) {\eqref{eq-kernel-series-parallel},};
\node[fill=white,inner sep=1pt,minimum width=0.066\linewidth,minimum height=2.4ex,font=\footnotesize,text=black] at (0.5004,0.1824) {\eqref{eq-kernel-series-parallel}};
\end{scope}
\end{tikzpicture}
\caption{Successive series--parallel moves remove vertices $1,2,\ldots,n-1$. Sources pass towards the final triangle; its currents reconstruct the fan backwards by \eqref{eq-B-fan-general-fractions}.}
\label{local-B4-fan-chain}
\end{figure}
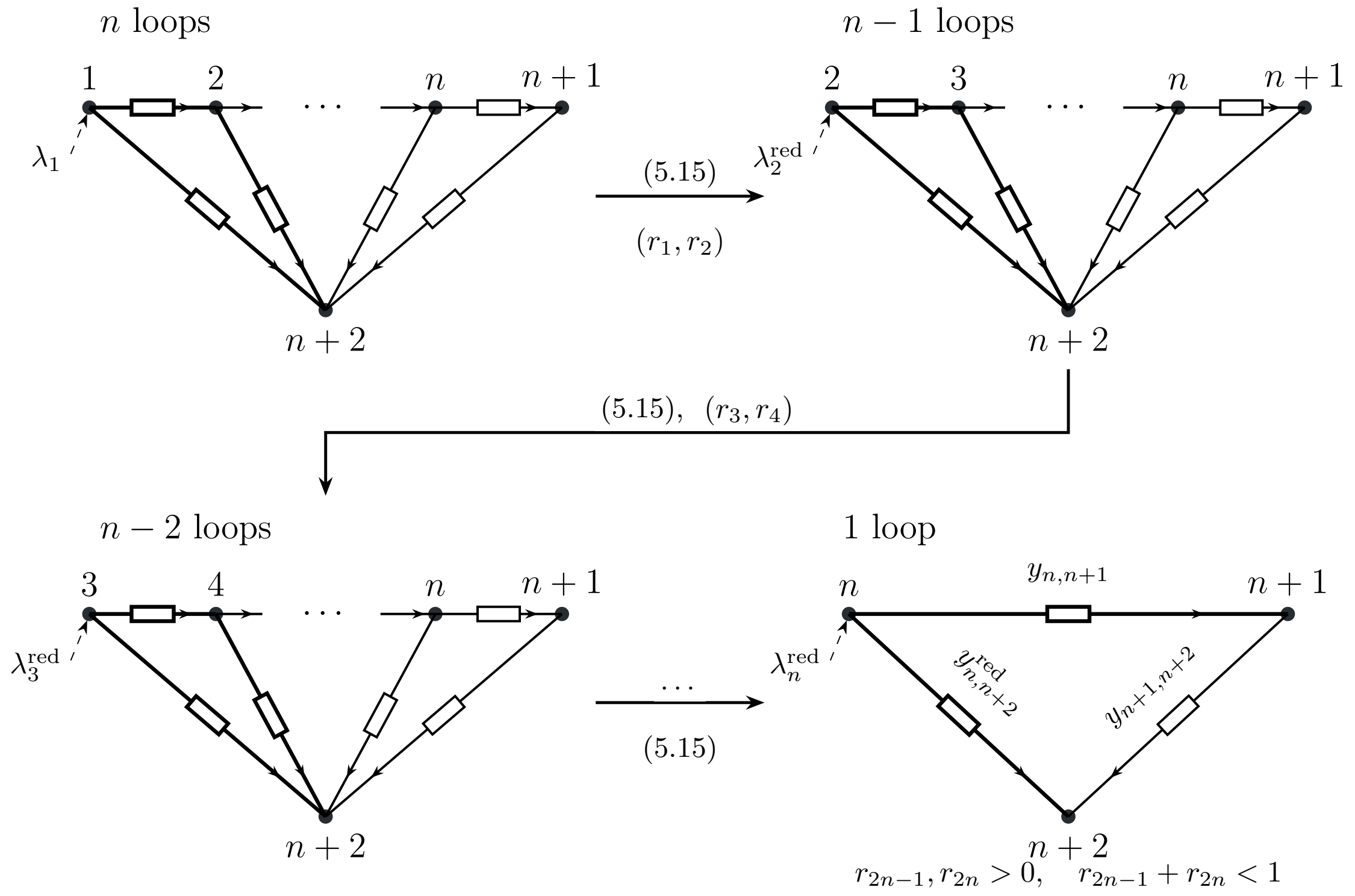

Iterating Proposition~\ref{prop-fan-step} for $i=1,\ldots,n-1$ reduces the $n$-loop fan to the final one-loop triangle on $\{n,n+1,n+2\}$, whose currents reconstruct the entire fan polynomially in reverse order (Figure~\ref{local-B4-fan-chain}).

\begin{proofbox}
\textbf{Proof of Theorem~\ref{thm-fan-kernel} (equations~\eqref{eq-1-40}--\eqref{eq-1-39}).} Solving the forward source recurrence~\eqref{eq-fan-source-transfer} (with empty products equal to $1$) gives
\begin{equation}\label{eq-B-fan-first-cycle}
\lambda_{i,+\dot a}^{\mathrm{red}}
=\sum_{\substack{1\le j\le i\\\mathrm{vertices}}}
\left(\prod_{k=j}^{i-1}r_{2k-1}\right)\lambda_{j,+\dot a},
\qquad 1\le i\le n.
\end{equation}
On the final triangle $\{n,n+1,n+2\}$ in simplex coordinates $(r_{2n-1},r_{2n})\in\Delta^2$, equation~\eqref{eq-triangle-current} determines $y_{n,n+1}$ and initiates the backward recurrence from~\eqref{eq-fan-step-currents}:
\begin{equation}\label{eq-B-fan-general-fractions}
\begin{aligned}
y_{n,n+1,+\dot a}
&=r_{2n-1}\lambda_{n,+\dot a}^{\mathrm{red}}
-r_{2n}\lambda_{n+1,+\dot a},\\
y_{i,i+1,+\dot a}
&=r_{2i-1}r_{2i}\lambda_{i,+\dot a}^{\mathrm{red}}
-(1-r_{2i})\lambda_{i+1,+\dot a}
+(1-r_{2i})y_{i+1,i+2,+\dot a},
\qquad i=n-1,\ldots,1,
\end{aligned}
\end{equation}
which establishes~\eqref{eq-1-38} (and~\eqref{eq-1-39} by vertex conservation). Multiplying the $n-1$ local Jacobians~\eqref{b-00282} leaves the reduced triangle integral
\begin{equation}\label{eq-fan-operator-product}
\begin{aligned}
\mathcal I_\Gamma(\lambda;z)
={}&\int_{(0,1)^{2n-2}}\prod_{j=1}^{2n-2}dr_j
\int_{\substack{y^{\mathrm{red}}\in\Delta_\triangle\\
\text{current conservation at }\lambda^{\mathrm{red}}}}
\prod_{i=1}^{n-1}
\left(r_{2i}\lambda_{i,+\dot a_i}^{\mathrm{red}}
y_{i+1,n+2,+}^{\mathrm{red}}{}^{\dot a_i}\right)\\
&\qquad\times e^{-\sum_{e\in I}y_{e,+\dot b}(r)z_e^{+\dot b}}
\,(-d^2y_{n,n+1}),
\end{aligned}
\end{equation}
where $T_\triangle=\{(n,n+2),(n+1,n+2)\}$ is the spanning tree of the final triangle and its chord two-form~\eqref{b-00267} is
\begin{equation}\label{eq-B-fan-second-merge}
-d^2y_{n,n+1}
=\bigl(\lambda_{n,+\dot a}^{\mathrm{red}}\lambda_{n+1,+}{}^{\dot a}\bigr)
dr_{2n-1}\wedge dr_{2n}.
\end{equation}
Combining $(0,1)^{2n-2}$ with $\Delta^2$ gives the domain~\eqref{eq-1-37}, and substituting~\eqref{b-00282} and~\eqref{eq-B-fan-first-cycle} yields~\eqref{eq-1-40} (reducing at $n=1$ to~\eqref{eq-1-25}).~\hfill\qedsymbol
\end{proofbox}

\subsection{Final Taylor expansion}\label{endpoint-sources-and-taylor-expansion}

We now specialize Theorem~\ref{thm-fan-kernel} to $\lambda_2=\cdots=\lambda_n=0$ and prove~\eqref{eq-1-41}, \eqref{eq-1-42}, and~\eqref{eq-fan-full-taylor}. When $\lambda_2=\cdots=\lambda_n=0$, equation~\eqref{eq-B-fan-first-cycle} collapses to $\lambda_i^{\mathrm{red}}=\bigl(\prod_{k=1}^{i-1}r_{2k-1}\bigr)\lambda_1$. Because $\lambda_{1,+\dot a}\lambda_{1,+}{}^{\dot a}=0$, only the $\lambda_{n+1}$ piece of $y_{i+1,i+2}$ in~\eqref{eq-1-38}---namely $-r_{2n}\bigl(\prod_{k=i+1}^{n-1}(1-r_{2k})\bigr)\lambda_{n+1}$---survives when contracted against $\lambda_i^{\mathrm{red}}$ in~\eqref{eq-1-40}. Multiplying the $n-1$ elimination factors with the final triangle factor $(\prod_{k=1}^{n-1}r_{2k-1})(\lambda_{1,+\dot b}\lambda_{n+1,+}{}^{\dot b})$ collects powers of each $r_j$ and establishes~\eqref{eq-1-41} (where for $Q_n$ in renormalizable theories the interaction vertices carry no derivatives, so $z_{i,i+1}=0$ and $z_{i,n+2}=-z_i$).

In~\eqref{eq-1-41}, the spoke-displacement exponent translates the two endpoint output fields: substituting~\eqref{eq-1-38}--\eqref{eq-1-39} gives
\begin{equation}\label{eq-fan-output-translation}
\sum_{j=1}^{n+1}y_{j,n+2,+\dot a}(r)z_j^{+\dot a}
=\lambda_{1,+\dot a}x_1^{+\dot a}(r,z)
+\lambda_{n+1,+\dot a}x_{n+1}^{+\dot a}(r,z),
\end{equation}
with the explicit polynomial translation vectors
\begin{equation}\label{eq-fan-output-points}
\begin{aligned}
x_1^{+\dot a}(r,z)
={}&\sum_{j=1}^{n}(1-r_{2j-1})
\left(\prod_{k=1}^{j-1}r_{2k-1}\right)\\
&\times\left[
\left(\prod_{k=1}^{j-1}(1-r_{2k})\right)z_1^{+\dot a}
+\sum_{i=2}^{j}r_{2i-2}
\left(\prod_{k=i}^{j-1}(1-r_{2k})\right)z_i^{+\dot a}
\right]\\
&+\left(\prod_{k=1}^{n}r_{2k-1}\right)z_{n+1}^{+\dot a},\\[3pt]
x_{n+1}^{+\dot a}(r,z)
={}&r_{2n}\left[
\left(\prod_{k=1}^{n-1}(1-r_{2k})\right)z_1^{+\dot a}
+\sum_{i=2}^{n}r_{2i-2}
\left(\prod_{k=i}^{n-1}(1-r_{2k})\right)z_i^{+\dot a}
\right]\\
&+(1-r_{2n})z_{n+1}^{+\dot a}.
\end{aligned}
\end{equation}

Every monomial integral over the product domain~\eqref{eq-1-37} factorizes into elementary interval and two-simplex Beta integrals:
\begin{equation}\label{eq-1-44}
\int_{\eqref{eq-1-37}}dr_1\cdots dr_{2n}\,\prod_{j=1}^{2n}r_j^{m_j}=\frac{m_{2n-1}!\,m_{2n}!}{(m_{2n-1}+m_{2n}+2)!}\prod_{j=1}^{2n-2}\frac1{m_j+1}.
\end{equation}
In particular, at zero displacement ($z=0$), integrating~\eqref{eq-1-41} via~\eqref{eq-1-44} (and $\int_0^1 r_{2j}(1-r_{2j})^{j-1}\,dr_{2j}=\frac{1}{j(j+1)}$) yields~\eqref{eq-1-42} immediately:
\begin{equation*}
\begin{aligned}
\left.\mathcal I_\Gamma(\lambda;0)\right|_{\lambda_2=\cdots=\lambda_n=0}
&=(\lambda_{1,+\dot a}\lambda_{n+1,+}{}^{\dot a})^n
\left(\prod_{j=1}^{n-1}\frac{1}{n-j+1}\right)
\left(\prod_{j=1}^{n-1}\frac{1}{j(j+1)}\right)
\frac{0!\,(n-1)!}{(n+1)!}\\
&=(\lambda_{1,+\dot a}\lambda_{n+1,+}{}^{\dot a})^n
\cdot\frac{1}{n!}\cdot\frac{1}{(n-1)!\,n!}\cdot\frac{1}{n(n+1)}
=\frac{(\lambda_{1,+\dot a}\lambda_{n+1,+}{}^{\dot a})^n}{(n!)^3(n+1)}.
\end{aligned}
\end{equation*}

More generally, keeping arbitrary displacements $(z_{i,i+1},z_{j,n+2})$ on all $2n+1$ edges at $\lambda_2=\cdots=\lambda_n=0$, spoke conservation~\eqref{eq-1-39} rewrites the exponent $-\sum_{e\in I}y_{e,+\dot b}z_e^{+\dot b}$ in terms of the path currents $y_{i,i+1}$ and cycle combinations $(z_{i,i+1}+z_{i+1,n+2}-z_{i,n+2})$:
\begin{equation*}
-\sum_{e\in I}y_{e,+\dot b}z_e^{+\dot b}
=-\lambda_{1,+\dot b}z_{1,n+2}^{+\dot b}
-\lambda_{n+1,+\dot b}z_{n+1,n+2}^{+\dot b}
-\sum_{i=1}^n y_{i,i+1,+\dot b}(z_{i,i+1}+z_{i+1,n+2}-z_{i,n+2})^{+\dot b}.
\end{equation*}
By~\eqref{eq-1-38}, each $y_{i,i+1}$ is a sum of $n-i+1$ positive $\lambda_1$ monomials (indexed by $k=i,\ldots,n$) and one negative $\lambda_{n+1}$ monomial. Expanding the exponential with multinomial exponents $r_0,s_0\ge0$, $r_{ik}\ge0$ ($1\le i\le k\le n$), and $s_i\ge0$ ($1\le i\le n$) at total derivative degree $m$, and integrating each factored $(r_1,\ldots,r_{2n})$ Beta/simplex integrand via~\eqref{eq-1-44}, gives the explicit all-order Taylor expansion used in our computer census:
\begin{equation}\label{eq-fan-full-taylor}
\begin{aligned}
&\left.\mathcal I_\Gamma(\lambda;z)
\right|_{\lambda_2=\cdots=\lambda_n=0}\\
&=(\lambda_{1,+\dot a}\lambda_{n+1,+}{}^{\dot a})^n
\sum_{m=0}^{\infty}
\sum_{\substack{r_0,s_0,r_{ik},s_i\ge0\\
r_0+s_0+\sum_{i=1}^ns_i+\sum_{k=1}^n\sum_{i=1}^kr_{ik}=m}}\\
&\quad\times
\frac{[-\lambda_{1,+\dot b}z_{1,n+2}^{+\dot b}]^{r_0}
[-\lambda_{n+1,+\dot c}z_{n+1,n+2}^{+\dot c}]^{s_0}}
{r_0!s_0!}\\
&\quad\times\prod_{1\le i\le k\le n}
\frac{\left[-\lambda_{1,+\dot d}
(z_{i,i+1}+z_{i+1,n+2}-z_{i,n+2})^{+\dot d}\right]^{r_{ik}}}
{r_{ik}!}\\
&\quad\times\prod_{i=1}^n
\frac{\left[\lambda_{n+1,+\dot e}
(z_{i,i+1}+z_{i+1,n+2}-z_{i,n+2})^{+\dot e}\right]^{s_i}}
{s_i!}\\
&\quad\times
\frac{\left(\sum_{i=1}^nr_{in}\right)!
\left(n-1+\sum_{i=1}^ns_i\right)!}
{\left(n+1+\sum_{i=1}^nr_{in}+\sum_{i=1}^ns_i\right)!}\\
&\quad\times\prod_{j=1}^{n-1}
\frac{1}{n-j+1+\sum_{k=j}^{n}\sum_{i=1}^{k}r_{ik}}\\
&\quad\times\prod_{j=1}^{n-1}
\frac{
\left(1+\sum_{i=1}^{j}r_{ij}\right)!
\left(j-1+\sum_{i=1}^{j}\left[s_i+\sum_{k=j+1}^{n}r_{ik}\right]\right)!}
{\left(j+1+\sum_{i=1}^{j}\left[s_i+\sum_{k=j}^{n}r_{ik}\right]\right)!},
\end{aligned}
\end{equation}
whose $m=0$ term ($r_0=s_0=r_{ik}=s_i=0$) reproduces~\eqref{eq-1-42}.

\section{Quantum multiplication and the Leibniz rule}\label{sec-product}

At tree level, $Q_0=\{\mathcal I,\,\cdot\,\}$ contracts a single input letter and therefore acts as a graded derivation of pointwise operator multiplication $AB$. Starting at one loop, however, $Q_n$ ($n\ge1$) contracts $n+1$ input letters simultaneously and breaks the classical Leibniz rule whenever its spokes attach to both factors of a composite product $AB$. To exhibit this failure concretely, recall from Section~\ref{the-all-order-supercharge-of-mathcal-n4-yangmills-theory} (proved at all loop orders in Theorem~\ref{b-00431} of Section~\ref{sec-supercharge}) the one-loop supercharge $Q_1$ of $\mathcal N=4$ Yang--Mills theory~\cite{BK}, written in terms of the superfield $C^A(x,\theta)$ of Appendix~\ref{sec-N4-superfield} ($\theta_I$, $I=1,2,3$) and the displaced functional derivatives~\eqref{eq-sum-deltaC-norm}:
\begin{equation}\label{eq-2-16}
\begin{aligned}
Q_1={}&-\frac{1}{64\pi^2}
\int_0^1dr_2\int_0^{1-r_2}dr_1\int d^3\theta\int d^4z_1\,d^4z_2\,
C_{ARD}C_{BRE}\\
&\times(\partial_{+\dot a}C^D)
\big(x+(1-r_1)z_1+r_1z_2,\theta\big)\,
(\partial_+{}^{\dot a}C^E)
\big(x+r_2z_1+(1-r_2)z_2,\theta\big)\\
&\times\frac{\delta}{\delta C^A(x+z_1,\theta)}
\frac{\delta}{\delta C^B(x+z_2,\theta)}.
\end{aligned}
\end{equation}
When $Q_1$ acts on the pointwise product $AB$ of two operators without holomorphic derivatives, the two equal cross-contraction assignments $(A,B)\leftrightarrow(z_1,z_2)$ and $(z_2,z_1)$ multiply the simplex area $\int_0^1dr_2\int_0^{1-r_2}dr_1=\tfrac12$ by $2$, leaving the finite coincident contact defect
\begin{equation}\label{eq-2-29}
\begin{aligned}
&Q_1(AB)-(Q_1A)B-(-1)^{|A|}A(Q_1B)\\
&\qquad=-\frac1{64\pi^2}\int d^3\theta\,
C_{XRD}C_{YRE}
(\partial_{+\dot a}C^D)(\partial_+{}^{\dot a}C^E)
\frac{\delta A}{\delta C^X(\theta)}\frac{\delta B}{\delta C^Y(\theta)}.
\end{aligned}
\end{equation}
Whereas separated insertions satisfy the graded Leibniz rule by~\eqref{eq-C-separated-insertions}, bringing two one-loop cocycles $A$ and $B$ to the same point produces the nonvanishing collision term~\eqref{eq-2-29} via~\eqref{eq-C-contact-boundary}, so the naive product $AB$ is generally not closed at $O(\hbar)$.

The marked quantum product $A\star B$ of Gaiotto, Kulp, and Wu~\cite[Sections 4.3 and 4.5]{higher}, \cite[footnote 28]{semichiral} restores the Leibniz rule to all loop orders: its $O(\hbar)$ correction cancels~\eqref{eq-2-29}, and its higher-loop terms yield the exact derivation identity $\mathbf Q(A\star B)=(\mathbf QA)\star B+(-1)^{|A|}A\star(\mathbf QB)$. In Section~\ref{sec-dressed-chain} we prove the all-order master equation and $\mathbf Q^2=0$ for gauge-anomaly-free $\mathcal N=1$ matter representations, and derive the all-order Leibniz rule; Section~\ref{sec-order-by-order} unpacks the resulting order-by-order cocycle completions; Section~\ref{sec-ring} proves that $\star$ endows $H(\mathbf Q)$ (and every intermediate space $\mathrm{BPS}_{\hbar^{r-1}}$) with an associative, graded-commutative, unital ring structure; Sections~\ref{the-graph-formula} and~\ref{sec-product-loops} give the operatope graph expansion of $A\star B$, prove its polynomial termination in $\hbar$, and evaluate explicit one- and two-loop products in $\mathcal N=4$ Yang--Mills theory; and Section~\ref{sec-normal-ordering} proves equivalence on cohomology with the Bochner--Martinelli contour normal-ordering product alongside an explicit nonvanishing chain-level associator.

\subsection{The Leibniz rule of the quantum supercharge}\label{sec-dressed-chain}

Here $\{\,\cdot\,\}$ denotes the higher brackets of Definition~\ref{b-00151}. Writing $\{X,Y\}_{\mathcal I}$ for the interaction-dressed binary bracket summing all perturbative insertions of $\mathcal I$, we have
\begin{equation}\label{eq-2-51}
\mathbf Q\mathcal O=\sum_{s\ge1}\frac{\hbar^{s-1}}{s!}\bigl\{\underbrace{\mathcal I,\ldots,\mathcal I}_{s},\mathcal O\bigr\},\qquad
\{X,Y\}_{\mathcal I}=\sum_{r\ge0}\frac{\hbar^r}{r!}\bigl\{\underbrace{\mathcal I,\ldots,\mathcal I}_{r},X,Y\bigr\}.
\end{equation}

\rthm{Lemma}{b-00452}{The quantum master equation; \cite[Theorem 7]{Wang}, \cite[Section 3.2]{operatope}, \cite[Equations (1.12) and (2.25)--(2.26)]{BK}} \emph{Suppose that the interaction satisfies}
\begin{equation}\label{eq-2-40}
\sum_{k\ge2}\frac{\hbar^{k-2}}{k!}\bigl\{\underbrace{\mathcal I,\ldots,\mathcal I}_{k}\bigr\}
=\sum_{n\ge0}\frac{\hbar^n}{n+2}Q_n\mathcal I=0.
\end{equation}
\emph{Then $\mathbf Q^2=0$; at the lowest loop orders,}
\begin{equation}\label{eq-nilpotence}
\begin{aligned}
Q_0^2&=0,& Q_0Q_1+Q_1Q_0&=0,\\
Q_0Q_2+Q_1^2+Q_2Q_0&=0,&
Q_0Q_3+Q_1Q_2+Q_2Q_1+Q_3Q_0&=0.
\end{aligned}
\end{equation}

\begin{proofbox}
\textbf{Proof.} The higher brackets satisfy the quadratic $L_\infty$ (generalized Jacobi) identities~\cite[Section 3.2]{operatope}, \cite[Theorem 7]{Wang}, \cite[Eq.~(4.19)]{higher}, which express $\Delta_{\mathrm{BV}}^2=0$ and compatibility with $Q_{\mathrm{BV}}=\bar\partial$ on UV collision faces independently of the QME. Evaluated on $N$ identical interaction entries $\mathcal I$ and one local operator $\mathcal O$, the quadratic identity reads
\begin{equation}\label{eq-2-46}
\begin{aligned}
0={}&\sum_{r=1}^{N-1}\binom Nr
\bigl\{\underbrace{\mathcal I,\ldots,\mathcal I}_{N-r},\bigl\{\underbrace{\mathcal I,\ldots,\mathcal I}_{r},\mathcal O\bigr\}\bigr\}\\
&-\sum_{k=2}^{N}\binom Nk
\bigl\{\underbrace{\mathcal I,\ldots,\mathcal I}_{N-k},\bigl\{\underbrace{\mathcal I,\ldots,\mathcal I}_{k}\bigr\},\mathcal O\bigr\}.
\end{aligned}
\end{equation}
Multiplying~\eqref{eq-2-46} by $\hbar^{N-2}/N!$ and summing over $N\ge2$, the first sum assembles into $\mathbf Q^2\mathcal O$ while the second factorizes into the QME insertion:
\begin{equation}\label{eq-2-48}
\begin{aligned}
\mathbf Q^2\mathcal O
&=\sum_{\substack{m\ge0\\\text{interaction entries}}}\frac{\hbar^m}{m!}
\Bigg\{\underbrace{\mathcal I,\ldots,\mathcal I}_{m},
\underbrace{\sum_{\substack{k\ge2\\\text{interaction entries}}}
\frac{\hbar^{k-2}}{k!}\bigl\{\mathcal I_1,\ldots,\mathcal I_k\bigr\}}_{0\text{ by \eqref{eq-2-40}}\;(\mathcal I_i=\mathcal I)},
\mathcal O\Bigg\}=0,
\end{aligned}
\end{equation}
whose lowest $\hbar$-powers give~\eqref{eq-nilpotence}.~\hfill\qedsymbol
\end{proofbox}

\rthm{Theorem}{b-00465}{Explicit QME and gauge-anomaly cancellation} ~ \emph{Suppose that $\bar W(\gamma,w)$ is gauge invariant and the chiral matter representation $R$ obeys}
\begin{equation}\label{eq-2-41}
\operatorname{Tr}_R\!\left(t_A\{t_B,t_C\}\right)=0.
\end{equation}
\emph{Then the twisted interaction density $\mathcal I$ of~\eqref{eq-B-10} satisfies, modulo holomorphic total derivatives,}
\begin{equation}\label{b-00468}
\bar\partial\mathcal I=0,\qquad
Q_0\mathcal I=0,\qquad Q_1\mathcal I=0,\qquad
Q_2\mathcal I=Q_3\mathcal I=\cdots=0.
\end{equation}
\emph{Consequently, $\mathcal I$ solves the all-order QME~\eqref{eq-2-40}, and $\mathbf Q^2=0$ by Lemma~\ref{b-00452}.}

\begin{proofbox}
\textbf{Proof.} Applying $\bar\partial$ to $[\mathcal I]_{(0,2)}$ produces an antiholomorphic total derivative. At tree level ($n=0$), the gauge and matter kinetic couplings satisfy $\{\mathcal I_{\mathrm{gauge}}+\mathcal I_{\mathrm{matter}},\mathcal I_{\mathrm{gauge}}+\mathcal I_{\mathrm{matter}}\}=0$ by the Jacobi identity $C^A{}_{B[C}C^B{}_{DE]}=0$ and $[t_A,t_B]=iC^C{}_{AB}t_C$. Because $\bar W(\gamma,w)$ contains no $b$ or $\beta$ antifields, $\{\bar W,\bar W\}=0$, and up to holomorphic total derivatives its cross-bracket with the gauge and matter vertices vanishes by gauge invariance of $\bar W$:
\begin{equation*}
\begin{aligned}
Q_0\mathcal I=\{\mathcal I,\mathcal I\}
={}&-2igc^A(t_A)^j{}_i\gamma_j
\frac{\partial\bar W}{\partial\gamma_i}
-2gc^AC^B{}_{AC}w^C_{\dot a}
\frac{\partial_L\bar W}{\partial w^B_{\dot a}}=0.
\end{aligned}
\end{equation*}

At one loop ($n=1$), $Q_1\mathcal I=\frac12\{\mathcal I,\mathcal I,\mathcal I\}$ is evaluated by the one-loop triangle fan ($\mathcal I_{\triangle}=\frac12$). The pure-gauge $(bcc)^3$ loop vanishes identically because the adjoint generators $(t_A^{\mathrm{adj}})^D{}_E=iC^D{}_{AE}$ are antisymmetric ($\operatorname{Tr}_{\mathrm{adj}}(t_A^{\mathrm{adj}}\{t_B^{\mathrm{adj}},t_C^{\mathrm{adj}}\})=0$), whereas the two opposite cyclic orientations of the matter $(c\gamma\beta)^3$ loop sum to the symmetrized trace~\eqref{eq-2-41}:
\begin{equation}\label{b-00478}
Q_1\mathcal I
=\frac{ig^3}{128\pi^2}
\operatorname{Tr}_R\!\left(t_A\{t_B,t_C\}\right)
c^A(\partial_{+\dot a}c^B)
(\partial_+{}^{\dot a}c^C)=0.
\end{equation}
Finally, every monomial of $\mathcal I$ contains at most one antifield $b$ or $\beta$, so for $n\ge2$ the antifield count~\eqref{eq-antifield-loss} would require
\begin{equation*}
\begin{aligned}
0&\le\#(b,\beta\text{ in output})\\
&=\#(b,\beta\text{ in input monomial})
-n-\#(\bar W\text{ interaction vertices})
\le 1-n<0,
\end{aligned}
\end{equation*}
forcing $Q_n\mathcal I=0$ for all $n\ge2$.~\hfill\qedsymbol
\end{proofbox}

\begin{figure}[H]
\centering
\includegraphics[width=0.86\linewidth]{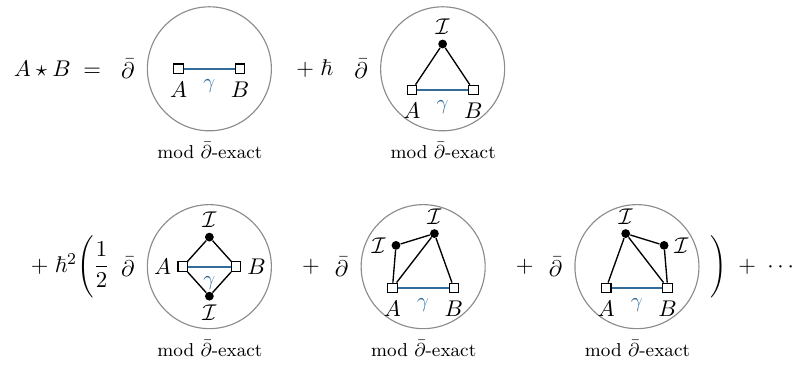}
\caption{The quantum product $A\star B$ through order $\hbar^2$. Squares denote $A,B$, dots the interactions $\mathcal I$, and the blue edge $\gamma$ the auxiliary contraction. Applying $\bar\partial$ gives the heat-edge sum; at order $\hbar^0$ it reduces to $AB$.}
\label{fig-product-graphs}
\end{figure}

Whereas the one-input $L_\infty$ identities give $\mathbf Q^2=0$, their two-input counterparts make $\mathbf Q$ a derivation of the interaction-dressed bracket $\{X,Y\}_{\mathcal I}$ in~\eqref{eq-2-52}. Because $\{X,Y\}_{\mathcal I}$ carries degree $+1$ and vanishes when no fields contract between $X$ and $Y$, Gaiotto, Kulp, and Wu~\cite[Sections 4.3 and 4.5]{higher}, \cite[footnote 28]{semichiral} adjoin a decoupled free auxiliary pair $(a,b_{\mathrm{aux}})$ with free action $\int_{\mathbb C^2}\Omega\wedge[a\bar\partial b_{\mathrm{aux}}]_{(0,2)}$ and degrees $|a|=0$, $|b_{\mathrm{aux}}|=1$. Since $(a,b_{\mathrm{aux}})$ does not appear in $\mathcal I$ and $\bar\partial$ vanishes on holomorphic jets, $Q_na=Q_nb_{\mathrm{aux}}=0$ for all $n\ge0$, so
\begin{equation*}
\mathbf Q(Aa)=(\mathbf QA)a,
\qquad
\mathbf Q(b_{\mathrm{aux}}B)=-b_{\mathrm{aux}}\mathbf QB.
\end{equation*}
Extracting the unique auxiliary Wick contraction $\gamma=\{(A,B)\}$ and setting the remaining auxiliary fields to zero ($\mathrm{aux}=0$) lowers the ghost number by $1$---so $|A\star B|=|A|+|B|$---and recovers ordinary pointwise multiplication $AB$ at $O(\hbar^0)$ (Figure~\ref{fig-product-graphs}):
\begin{equation}\label{b-00504}
\begin{gathered}
A\star B
=\sum_{n\ge0}\frac{\hbar^n}{n!}
\bigl\{\underbrace{\mathcal I,\ldots,\mathcal I}_{n},A,B\bigr\}^{\gamma,0}
=\left.\{Aa,b_{\mathrm{aux}}B\}_{\mathcal I}
\right|_{\mathrm{aux}=0},\\
\{A,B\}^{\gamma,0}
=(-1)^{|A|}\frac{\partial_L}{\partial a}\frac{\partial_L}{\partial b_{\mathrm{aux}}}
(Aa\,b_{\mathrm{aux}}B)=AB.
\end{gathered}
\end{equation}
The equivalence between this marked-edge prescription and Bochner--Martinelli contour normal ordering is proved in Section~\ref{sec-normal-ordering}.

\rthm{Theorem}{thm-leibniz-rule}{Leibniz rule}
\emph{Suppose that the interaction satisfies the QME~\eqref{eq-2-40} (for~\eqref{eq-B-10}, under the anomaly-cancellation condition~\eqref{eq-2-41} of Theorem~\ref{b-00465}). For homogeneous local operators $X,Y$, with $|X|$ the fermion number of $X$,}
\begin{equation}\label{eq-2-52}
\mathbf Q\{X,Y\}_{\mathcal I}
-\{\mathbf QX,Y\}_{\mathcal I}
+(-1)^{|X|}\{X,\mathbf QY\}_{\mathcal I}=0,
\end{equation}
\emph{and the marked quantum product~\eqref{b-00504} obeys the exact graded Leibniz rule}
\begin{equation}\label{eq-chain}
\mathbf Q(A\star B)
=(\mathbf QA)\star B+(-1)^{|A|}A\star(\mathbf QB).
\end{equation}
\emph{In particular, $\star$ maps quantum cocycles to quantum cocycles and descends to $\mathbf Q$-cohomology.}

\begin{proofbox}
\textbf{Proof.} Partitioning the quadratic $L_\infty$ identity~\cite[Section 3.2]{operatope}, \cite[Theorem 7]{Wang}, \cite[Eq.~(4.19)]{higher} on $N$ copies of $\mathcal I$ and two operators $X,Y$ according to whether the inner bracket contains both $X$ and $Y$, only $X$, only $Y$, or only interactions, and multiplying by
\begin{equation}\label{eq-quadratic-two}
\frac{\hbar^{N-1}}{N!}\binom Nr
=\frac{\hbar^{N-r-1}}{(N-r)!}\frac{\hbar^r}{r!},
\end{equation}
summing over $N\ge1$ gives
\begin{equation}\label{eq-quadratic-two-summed}
\begin{aligned}
0={}&\mathbf Q\{X,Y\}_{\mathcal I}
-\{\mathbf QX,Y\}_{\mathcal I}
+(-1)^{|X|}\{X,\mathbf QY\}_{\mathcal I}\\
&+\sum_{\substack{s\ge0\\\text{interaction entries}}}
\frac{\hbar^{s+1}}{s!}
\Bigg\{\underbrace{\mathcal I,\ldots,\mathcal I}_{s},
\underbrace{\sum_{\substack{k\ge2\\\text{interaction entries}}}
\frac{\hbar^{k-2}}{k!}\bigl\{\mathcal I_1,\ldots,\mathcal I_k\bigr\}}_{0\text{ by \eqref{eq-2-40}}\;(\mathcal I_i=\mathcal I)},X,Y\Bigg\},
\end{aligned}
\end{equation}
establishing~\eqref{eq-2-52}. Substituting $X=Aa$ (fermion number $|A|$) and $Y=b_{\mathrm{aux}}B$ (with $\mathbf Q(b_{\mathrm{aux}}B)=-b_{\mathrm{aux}}\mathbf QB$) into~\eqref{eq-2-52} and projecting onto the single auxiliary contraction yields~\eqref{eq-chain}:
\begin{equation*}
0=\mathbf Q(A\star B)-(\mathbf QA)\star B-(-1)^{|A|}A\star(\mathbf QB).
\end{equation*}
Hence $\mathbf QA=\mathbf QB=0$ implies $\mathbf Q(A\star B)=0$, while for coboundary shifts $\mathbf QX,\mathbf QY$ ($|X|=|A|-1$, $|Y|=|B|-1$ mod $2$) equation~\eqref{eq-chain} and $\mathbf Q^2=0$ give
\begin{equation}\label{eq-representatives}
\begin{aligned}
&(A+\mathbf QX)\star(B+\mathbf QY)-A\star B\\
&\qquad=\mathbf Q\bigl(X\star B+(-1)^{|A|}A\star Y
+X\star\mathbf QY\bigr),
\end{aligned}
\end{equation}
proving that $[A\star B]_{\mathbf Q}$ depends only on the cohomology classes $[A]_{\mathbf Q}$ and $[B]_{\mathbf Q}$.~\hfill\qedsymbol
\end{proofbox}

\subsection{Order-by-order representatives}\label{sec-order-by-order}

Expanding the derivation identity~\eqref{eq-chain} in powers of $\hbar$ shows directly how the loop terms of $A\star B$ repair the failure of the pointwise Leibniz rule. For classical cocycles $Q_0A=Q_0B=0$, the $O(\hbar)$ and $O(\hbar^2)$ coefficients of~\eqref{eq-chain} read
\begin{equation}\label{eq-2-55}
\begin{aligned}
Q_0\{\mathcal I,A,B\}^{\gamma,0}+Q_1(AB)
={}&(Q_1A)B+(-1)^{|A|}A(Q_1B),\\[3pt]
\frac12Q_0\{\mathcal I,\mathcal I,A,B\}^{\gamma,0}
+Q_1\{\mathcal I,A,B\}^{\gamma,0}+Q_2(AB)
={}&(Q_2A)B+(-1)^{|A|}A(Q_2B)\\
&+\{\mathcal I,Q_1A,B\}^{\gamma,0}
+(-1)^{|A|}\{\mathcal I,A,Q_1B\}^{\gamma,0}.
\end{aligned}
\end{equation}
At one loop, the classical variation $Q_0\{\mathcal I,A,B\}^{\gamma,0}$ cancels the cross-contraction defect $Q_1(AB)-(Q_1A)B-(-1)^{|A|}A(Q_1B)$ of~\eqref{eq-2-29} identically; at two loops, $\frac12\{\mathcal I,\mathcal I,A,B\}^{\gamma,0}$ simultaneously absorbs the two-loop cross-contractions in $Q_2(AB)$ and the one-loop variation $Q_1\{\mathcal I,A,B\}^{\gamma,0}$ of the one-loop product correction.

More generally, given any two full quantum cocycles $A=\sum_{i\ge0}\hbar^iA_i$ and $B=\sum_{j\ge0}\hbar^jB_j$ ($\mathbf QA=\mathbf QB=0$), expanding $\mathcal O=A\star B=\sum_{r\ge0}\hbar^r\mathcal O_r$ gives
\begin{equation}\label{eq-2-32}
\mathcal O_r=\sum_{i+j+n=r}\frac1{n!}\bigl\{\underbrace{\mathcal I,\ldots,\mathcal I}_{n},A_i,B_j\bigr\}^{\gamma,0},\qquad\mathcal O_0=A_0B_0\ \text{\eqref{b-00504}},
\end{equation}
\begin{equation}\label{eq-2-33}
\begin{aligned}
\mathcal O_1={}&A_1B_0+A_0B_1
+{\{\mathcal I,A_0,B_0\}^{\gamma,0}},\\
\mathcal O_2={}&A_2B_0+A_1B_1+A_0B_2\\
&+\{\mathcal I,A_1,B_0\}^{\gamma,0}
+\{\mathcal I,A_0,B_1\}^{\gamma,0}
+{\frac12\{\mathcal I,\mathcal I,A_0,B_0\}^{\gamma,0}}.
\end{aligned}
\end{equation}
By~\eqref{eq-chain}, these coefficients automatically solve the cocycle-completion equations $\sum_{k=0}^rQ_k\mathcal O_{r-k}=0$~\eqref{eq-liftingequations} at every order $r\ge0$, providing an explicit closed-form completion of $A_0B_0$ without solving linear $Q_0$-inversion systems order by order (as exploited for multi-graviton operators in Section~\ref{sec-lifting}).

\subsection{The quantum BPS ring}\label{sec-ring}

Write $[\cdot]_{\mathbf Q}$ for the cohomology class of a quantum cocycle in $H(\mathbf Q)$.

\rthm{Theorem}{b-00513}{Associativity, graded commutativity, and descent of the quantum BPS ring} \emph{Suppose the interaction satisfies the quantum master equation~\eqref{eq-2-40}. Then the marked product~\eqref{b-00504} induces an associative, graded-commutative, unital multiplication on $H(\mathbf Q)$: for homogeneous quantum cocycles $A,B,D$,}
\begin{equation}\label{eq-ring-associativity}
[(A\star B)\star D]_{\mathbf Q}
=[A\star(B\star D)]_{\mathbf Q},
\end{equation}
\begin{equation}\label{eq-ring-commutativity}
[A\star B]_{\mathbf Q}
=(-1)^{|A||B|}[B\star A]_{\mathbf Q},
\end{equation}
\emph{with strict cochain-level unit}
\begin{equation}\label{eq-ring-unit}
A\star1=1\star A=A.
\end{equation}
\emph{Furthermore, under the successive homotopy transfers of Section~\ref{sec-homotopy-transfer}, $\star$ induces on every intermediate space $\mathrm{BPS}_{\hbar^{r-1}}$ ($r\ge1$) the ordinary product of leading representatives, and each differential $d_r$ is a graded derivation of this product.}

\begin{proofbox}
\textbf{Proof.} \emph{(i) Descent to $\mathrm{BPS}_{\hbar^{r-1}}$ and strict unit.} Equation~\eqref{eq-representatives} ensures that $\star$ is well defined on $H(\mathbf Q)$. Similarly, any class in $\mathrm{BPS}_{\hbar^{r-1}}$ is represented by a partial completion $\tilde A=a+\hbar a_1+\cdots$ whose $\mathbf Q$-variation begins at order $\hbar^r$ with representative $d_ra$; since $\tilde A\star\tilde B=ab+O(\hbar)$, equation~\eqref{eq-chain} implies that $\mathbf Q(\tilde A\star\tilde B)$ also vanishes through $O(\hbar^{r-1})$ and begins at $O(\hbar^r)$ with $(d_ra)b+(-1)^{|a|}a(d_rb)$, proving both induced multiplication and the derivation property of $d_r$. For the unit~\eqref{eq-ring-unit}, the identity operator $1$ carries no fields, so in any marked graph with $n\ge1$ interaction vertices it attaches solely via the single marked edge: as a degree-one leaf in the electrical network, its resistor carries fixed external current regardless of its resistance, so its current differential vanishes (equivalently, deleting that leaf leaves $|I'|=2n$ edges on $|V'|=n+1\ge2$ vertices, violating the Laman bound $2n\le2(n+1)-3$). Thus only the $n=0$ term survives, giving $A\star1=1\star A=A$ identically.

\emph{(ii) Common-root homotopy and vanishing of parallel marks.} For any three $\mathbf Q$-closed insertions $X,Y,Z$, summing the quadratic $L_\infty$ identities over all interaction insertions as in~\cite[Sections 4.3 and 4.5]{higher} (where pure-interaction inner brackets vanish by~\eqref{eq-2-40} and single-operator inner brackets vanish by $\mathbf Q$-closure, while a ternary graph with $n$ interactions has $|V|=n+3$, $|I|=2n+3$, and loop order $n+1$) gives, with the sign conventions of~\eqref{eq-A-28},
\begin{equation}\label{eq-ring-twisted-jacobi}
\begin{aligned}
0={}&\{X,\{Y,Z\}_{\mathcal I}\}_{\mathcal I}
-\{\{X,Y\}_{\mathcal I},Z\}_{\mathcal I}\\
&-(-1)^{(|X|+1)(|Y|+1)}
\{Y,\{X,Z\}_{\mathcal I}\}_{\mathcal I}\\
&+\mathbf Q\sum_{n\ge0}\frac{\hbar^{n+1}}{n!}
\bigl\{\underbrace{\mathcal I,\ldots,\mathcal I}_{n},X,Y,Z\bigr\}.
\end{aligned}
\end{equation}
Introduce two independent decoupled auxiliary pairs $(a_1,b_{\mathrm{aux},1})$ and $(a_2,b_{\mathrm{aux},2})$ ($|a_i|=0$, $|b_{\mathrm{aux},i}|=1$) and set
\begin{equation*}
X=Aa_1,\qquad Y=Da_2,\qquad
Z=b_{\mathrm{aux},1}b_{\mathrm{aux},2}B,
\end{equation*}
placing $Z$ at the root $0$ so that the auxiliary contractions mark both edges $(A,B)$ and $(D,B)$ (Figure~\ref{b-00530}).

\begin{figure}[H]
\centering
\includegraphics[width=0.98\linewidth]{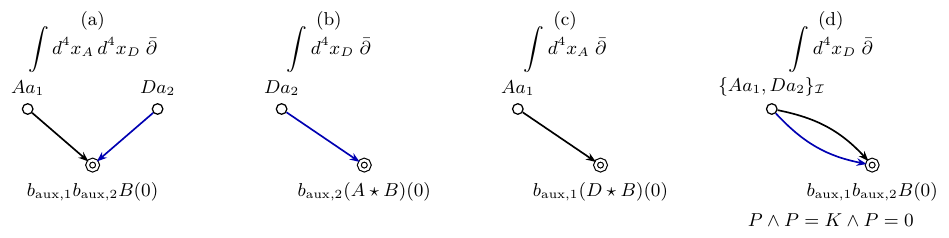}
\caption{Common-root integration and pair-collapse channels. The double circle fixes $B$ at $0$; all other positions range over $\mathbb C^2$. Only auxiliary contractions are drawn; the integrals and the ultraviolet part of $\bar\partial$ act on the full graph. Parallel marks in (d) give zero.}
\label{b-00527}\label{b-00528}\label{b-00529}\label{b-00530}
\end{figure}

Evaluating the three binary channels in~\eqref{eq-ring-twisted-jacobi}: in the first two (panels (b) and (c)), contracting $(a_1,b_{\mathrm{aux},1})$ or $(a_2,b_{\mathrm{aux},2})$ yields
\begin{equation}\label{eq-ring-root-contractions}
\begin{aligned}
\left.\{Aa_1,b_{\mathrm{aux},1}b_{\mathrm{aux},2}B\}_{\mathcal I}
\right|_{a_1=b_{\mathrm{aux},1}=0}
&=(-1)^{|A|}b_{\mathrm{aux},2}(A\star B),\\
\left.\{Da_2,b_{\mathrm{aux},1}b_{\mathrm{aux},2}B\}_{\mathcal I}
\right|_{a_2=b_{\mathrm{aux},2}=0}
&=(-1)^{|D|+1}b_{\mathrm{aux},1}(D\star B),
\end{aligned}
\end{equation}
where in the second line $\partial_L/\partial b_{\mathrm{aux},2}$ passes odd $b_{\mathrm{aux},1}$. In the third channel (panel (d)), $\{Aa_1,Da_2\}_{\mathcal I}$ first collapses $A$ and $D$ to a single vertex, after which both marked edges connect that collapsed vertex in parallel to $B$; by Corollary~\ref{b-00188}~\eqref{eq-1-13}, parallel edges carry collinear currents and annihilate the current measure at every derivative order:
\begin{equation}\label{eq-ring-parallel-channel}
\left.
\{\{Aa_1,Da_2\}_{\mathcal I},
b_{\mathrm{aux},1}b_{\mathrm{aux},2}B\}_{\mathcal I}
\right|_{\mathrm{aux}=0}=0.
\end{equation}
Substituting~\eqref{eq-ring-root-contractions}--\eqref{eq-ring-parallel-channel} into~\eqref{eq-ring-twisted-jacobi} and multiplying by $(-1)^{|D|+1}$ yields the chain-level common-root homotopy
\begin{equation}\label{eq-ring-common-root-homotopy}
\begin{aligned}
&A\star(D\star B)-(-1)^{|A||D|}D\star(A\star B)\\
&\quad=(-1)^{|D|}\mathbf Q
\left[
\left.\sum_{n\ge0}\frac{\hbar^{n+1}}{n!}
\bigl\{\underbrace{\mathcal I,\ldots,\mathcal I}_{n},Aa_1,Da_2,
 b_{\mathrm{aux},1}b_{\mathrm{aux},2}B\bigr\}
\right|_{\mathrm{aux}=0}\right]\\
&\quad=(-1)^{|D|}\mathbf Q\sum_{n\ge0}\frac{\hbar^{n+1}}{n!}
\bigl\{\underbrace{\mathcal I,\ldots,\mathcal I}_{n},A,D,B\bigr\}^{\gamma,\gamma^{\prime}},
\end{aligned}
\end{equation}
where $\gamma=\{(A,B)\}$ and $\gamma'=\{(D,B)\}$ have zero marked displacements (note that the $n$-interaction term has $2n+1$ physical contractions while $\mathcal I_1,\ldots,\mathcal I_n$ supply at most $n$ antifields $b,\beta$, so the sum terminates finitely on polynomial inputs).

\emph{(iii) Graded commutativity and associativity.} Setting $B=1$ in~\eqref{eq-ring-common-root-homotopy} and using $D\star1=D$, $A\star1=A$ immediately proves graded commutativity~\eqref{eq-ring-commutativity} on $H(\mathbf Q)$. Combining~\eqref{eq-ring-commutativity} with~\eqref{eq-ring-common-root-homotopy} on $H(\mathbf Q)$ then gives associativity~\eqref{eq-ring-associativity}:
\begin{equation}\label{eq-ring-associativity-proof}
\begin{aligned}
[(A\star B)\star D]_{\mathbf Q}
&\underset{\eqref{eq-ring-commutativity}}{=}(-1)^{(|A|+|B|)|D|}[D\star(A\star B)]_{\mathbf Q}\\
&\underset{\eqref{eq-ring-common-root-homotopy}}{=}(-1)^{|B||D|}[A\star(D\star B)]_{\mathbf Q}\\
&\underset{\eqref{eq-ring-commutativity}}{=}[A\star(B\star D)]_{\mathbf Q}.\quad\qedsymbol
\end{aligned}
\end{equation}
\end{proofbox}

For fixed quantum classes in $H(\mathbf Q)$, the marked-graph product is independent of cocycle representatives by~\eqref{eq-representatives}, and any alternative local product prescription differing by a $\mathbf Q$-exact homotopy on closed inputs induces the same ring structure on $H(\mathbf Q)$ (distinct from the finite $\hbar$-completion freedom of classical classes classified in Theorem~\ref{product-uniqueness-ch9-p01}).

\subsection{The graph formula of the quantum product}\label{the-graph-formula}

On each physical edge $(v,w)$, $\Delta_{vw}$ contracts paired fields between vertices $v$ and $w$ via~\eqref{eq-A-20}---for instance, in $\mathcal N=4$ superfield variables~\eqref{eq-A-16},
\begin{equation*}
\Delta_{vw}=-\delta^{DE}\int d^3\theta\,
\frac{\delta}{\delta C_v^D(\theta)}
\frac{\delta}{\delta C_w^E(\theta)}.
\end{equation*}
Substituting the operatope representation~\eqref{eq-operatope} into~\eqref{b-00504} expresses the $n$-loop product coefficient as a sum over Laman graphs containing the marked edge $\gamma=\{(n+1,n+2)\}$:
\begin{equation}\label{eq-2-31}
\begin{aligned}
\frac1{n!}\bigl\{\underbrace{\mathcal I,\ldots,\mathcal I}_{n},A,B\bigr\}^{\gamma,0}
={}&\left.\frac1{n!}\bigl\{\underbrace{\mathcal I,\ldots,\mathcal I}_{n},Aa,b_{\mathrm{aux}}B\bigr\}\right|_{\mathrm{aux}=0}\\
={}&\frac{(-1)^n}{n!(8\pi^2)^n}
\sum_{\substack{\Gamma\ \mathrm{labelled,\ Laman}\\
|V|=n+2,\ |I|=2n+1\\(n+1,n+2)\in I}}
\Bigg[\int_{\substack{y\in\Delta_\Gamma\\
\sum_{e:e(0)=v}y_e-\sum_{e:e(1)=v}y_e=\lambda_v}}\\
&\qquad\quad e^{-\sum_{e\in I}y_{e,+\dot a}z_e^{+\dot a}}
\bigwedge_{e\in I\setminus T}(-d^2y_e)\\
&\qquad\times\prod_{(v,w)\in I\setminus\{(n+1,n+2)\}}^{\mathrm{edge\ order}}
\Delta_{vw}(\mathcal I_1\cdots\mathcal I_nA_{n+1}B_{n+2})
\Bigg]_{\mathrm{jets},\,x_v=x,\,z=0}.
\end{aligned}
\end{equation}
Here $T\subset I$ is any spanning tree containing the marked edge $(n+1,n+2)$, the $n=|I|-|T|$ complementary chords carry $-d^2y_e=-dy_{e,+\dot+}\wedge dy_{e,+\dot-}$, and current conservation holds at every vertex. Derivatives on contracted input letters act via~\eqref{eq-sum-spinor-conventions} and edge displacements,
\begin{equation}\label{eq-product-current-insertion}
\frac{\partial}{\partial z_e^{+\dot a}}
e^{-\sum_{f\in I}y_{f,+\dot b}z_f^{+\dot b}}
=-y_{e,+\dot a}
e^{-\sum_{f\in I}y_{f,+\dot b}z_f^{+\dot b}},
\end{equation}
while $\lambda_v$ differentiates surviving output fields at $v$ before setting $z=0$ and $x_v=x$. Section~\ref{sec-forests} classifies the contributing graphs as product forests and reduces their current integrals polynomially.

\rthm{Corollary}{local-ch8-finite-product-corollary}{Finiteness of the quantum product}
\emph{For polynomial local inputs $A,B$ and any polynomial superpotential $\bar W$ in~\eqref{eq-B-10}, the marked product $A\star B$ terminates at a finite power of $\hbar$.}

\begin{proofbox}
\textbf{Proof.} An order-$n$ graph in~\eqref{eq-2-31} has $n$ interaction vertices and $2n$ physical contractions, so Theorem~\ref{b-00249} gives
\begin{equation*}
\begin{aligned}
\#(b,\beta\text{ in output})
&=\#(b,\beta\text{ in }A)+\#(b,\beta\text{ in }B)
+\underbrace{n-\#(\bar W\text{ vertices})}_{\text{interaction terminals}}
-\underbrace{2n}_{\text{physical contractions}}\\
&=\#(b,\beta\text{ in }A)+\#(b,\beta\text{ in }B)-n-\#(\bar W\text{ vertices})\ge0,
\end{aligned}
\end{equation*}
and therefore
\begin{equation}\label{local-ch8-finite-product-bound}
[\hbar^n](A\star B)=0
\quad\text{whenever}\quad
n>\#(b,\beta\text{ in }A)+\#(b,\beta\text{ in }B).
\end{equation}
Taking the maximum over the finitely many monomials of $A$ and $B$ proves polynomial termination in $\hbar$ (and similarly for any iterated product of finitely many polynomial inputs; for a formal power-series superpotential, the bound holds coefficientwise in its couplings).~\hfill\qedsymbol
\end{proofbox}

\subsection{One- and two-loop examples in \texorpdfstring{$\mathcal N=4$}{N=4} Yang--Mills theory}\label{sec-product-loops}

In $\mathcal N=4$ Yang--Mills theory with superfield action~\eqref{eq-A-15}, the one-loop product $(A\star B)_{\mathrm{1\text{-}loop}}(x)$ comes from the single triangle on vertices $(1:\mathcal I,2:A,3:B)$ with marked edge $\gamma=\{(2,3)\}$. Evaluating its triangle current integral on the two-simplex $\Delta^2=\{r_1,r_2>0,\,r_1+r_2<1\}$ and pairing with the displaced functional derivatives~\eqref{eq-sum-deltaC-norm} yields
\begin{equation}\label{eq-2-34}
\begin{aligned}
&(A\star B)_{\mathrm{1\text{-}loop}}(x)
=\hbar\{\mathcal I,A,B\}^{\{(2,3)\},0}(x)
\\
&\quad=\frac{i\hbar}{32\pi^2}\int_0^1dr_1\int_0^{1-r_1}dr_2
\int d^3\theta\,C_X{}^{YZ}
\int d^4z_1\,d^4z_2\\
&\qquad\times(\partial_{+\dot a}C^X)
\big(x+r_1z_1+(1-r_1)z_2,\theta\big)\\
&\qquad\times\left.
\partial_{2,+}{}^{\dot a}
\left(\frac{\delta A(x_2)}{\delta C^Y(x_2+z_1,\theta)}\right)
\right|_{x_2=x+r_2(z_2-z_1)}
\frac{\delta B(x)}{\delta C^Z(x+z_2,\theta)},
\end{aligned}
\end{equation}
where $\partial_{2,+}{}^{\dot a}$ differentiates uncontracted fields in $A(x_2)$ prior to setting $x_2=x+r_2(z_2-z_1)$. In particular, when $A$ and $B$ contain no input holomorphic derivatives ($z_1=z_2=0$), integrating $\int_{\Delta^2}dr_1\,dr_2=\tfrac12$ reduces~\eqref{eq-2-34} immediately to
\begin{equation*}
(A\star B)_{\mathrm{1\text{-}loop}}(x)\big|_{\text{no input derivatives}}
=\frac{i\hbar}{64\pi^2}\int d^3\theta\,C_X{}^{YZ}
(\partial_{+\dot a}C^X)(x,\theta)\,
\partial_+{}^{\dot a}\!\left(\frac{\delta A}{\delta C^Y(\theta)}\right)
\frac{\delta B}{\delta C^Z(\theta)},
\end{equation*}
whose $Q_0$-variation cancels~\eqref{eq-2-29} and evaluates multi-graviton one-loop completions directly (Section~\ref{sec-graviton-example}).

\begin{figure}[H]
\centering
\includegraphics[width=0.96\linewidth]{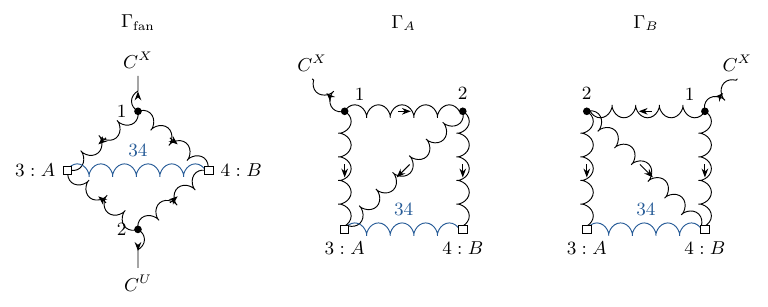}
\caption{The two-loop quantum-product graphs $\Gamma_{\mathrm{fan}}$, $\Gamma_A$ and $\Gamma_B$. Squares are the insertions $3:A$ and $4:B$; dots are cubic interactions. Black arrows orient the physical contractions, and the blue edge $(3,4)$ carries the auxiliary pairing. The fan leaves $C^X,C^U$ uncontracted; each triangle leaves $C^X$.}
\label{fig-two-loop-product}
\end{figure}

At two loops ($|V|=4$ with interactions $1,2$ and insertions $3:A,4:B$, marked edge $\gamma=\{(3,4)\}$), three Laman topologies contribute (Figure~\ref{fig-two-loop-product}), with ordered edge lists
\begin{equation}\label{eq-productedges}
\begin{aligned}
\Gamma_{\mathrm{fan}}&:(13,14,23,24,34),\\
\Gamma_A&:(12,13,23,24,34),\\
\Gamma_B&:(12,14,23,24,34).
\end{aligned}
\end{equation}

All three two-loop topologies reduce via series--parallel moves to fan-type integrals with independent external momenta at both interactions and $A$: in~\eqref{eq-product2A} below, the first integral (carrying $1/2!=1/2$ from exchanging $1\leftrightarrow2$) is $\Gamma_{\mathrm{fan}}$ with two surviving output superfields $C^X(x_1,\theta^{(1)})$ and $C^U(x_2,\theta^{(2)})$, while the second and third integrals (where the two labellings $1\leftrightarrow2$ double $1/2!$ to $1$) are the asymmetric graphs $\Gamma_A$ and $\Gamma_B$ leaving a single twice-differentiated superfield $C^X(x_1,\theta)$. Writing $\partial_1,\partial_2,\partial_3$ for holomorphic derivatives acting on $C^X(x_1)$, $C^U(x_2)$, and the uncontracted letters of $A(x_3)$ respectively before setting $x_1=x_2=x_3=x$, the complete two-loop product is
\begin{align}
&(A\star B)_{\mathrm{2\text{-}loop}}(x)=\frac{\hbar^2}{2}\{\mathcal I,\mathcal I,A,B\}^{\gamma,0}(x)\notag\\*
&=\frac{\hbar^2}{2(32\pi^2)^2}C_X{}^{YZ}C_U{}^{VW}
\int_0^1dr_1\int_0^1dr_3
\int_{\substack{r_2,r_4>0\\r_2+r_4<1}}dr_2\,dr_4\,(1-r_2-r_4)\notag\\*
&\quad\times\int d^3\theta^{(1)}d^3\theta^{(2)}
\int\prod_{e\in\{13,14,23,24\}}d^4z_e\notag\\*
&\quad\times\exp\!\Big\{-\tfrac12[r_1z_{13}+(1-r_1)z_{14}]^{+\dot c}\partial_{1,+\dot c}
-\tfrac12[r_3z_{23}+(1-r_3)z_{24}]^{+\dot d}\partial_{2,+\dot d}\notag\\*
&\hspace{60pt}-\tfrac12[r_2(z_{14}-z_{13})+r_4(z_{24}-z_{23})]^{+\dot e}
(r_1\partial_1+r_3\partial_2+\partial_3)_{+\dot e}\Big\}\notag\\*
&\quad\times\partial_{1,+\dot a}(\partial_3+r_3\partial_2)_+{}^{\dot a}
\partial_{2,+\dot b}(\partial_3+r_1\partial_1)_+{}^{\dot b}\notag\\*
&\quad\times\Bigg[C^X(x_1,\theta^{(1)})C^U(x_2,\theta^{(2)})
\frac{\delta^2 A(x_3)}{\delta C^Y(x_3+z_{13},\theta^{(1)})\,
\delta C^V(x_3+z_{23},\theta^{(2)})}\notag\\*
&\hspace{60pt}\times
\frac{\delta^2 B(x)}{\delta C^Z(x+z_{14},\theta^{(1)})\,
\delta C^W(x+z_{24},\theta^{(2)})}\Bigg]_{x_1=x_2=x_3=x}\notag\\[4pt]
&+\frac{\hbar^2}{(32\pi^2)^2}C_X{}^{YR}C_R{}^{ZW}
\int_0^1dr_1\int_0^1dr_2
\int_{\substack{r_3,r_4>0\\r_3+r_4<1}}dr_3\,dr_4\,
 r_2(1-r_1)(1-r_3-r_4)\notag\\*
&\quad\times\int d^3\theta\int\prod_{e\in\{13,23,24\}}d^4z_e\notag\\*
&\quad\times\exp\!\Big\{-\tfrac12\Big[
\big[r_1r_2+(1-r_2)(r_1r_3+r_4)\big]z_{13}
-r_2\big[r_1(1-r_3)-r_4\big]z_{23}\notag\\*
&\hspace{60pt}+(1-r_1r_3-r_4)z_{24}\Big]^{+\dot c}\partial_{1,+\dot c}
-\tfrac12(1-r_3-r_4)
[z_{24}-(1-r_2)z_{13}-r_2z_{23}]^{+\dot d}\partial_{3,+\dot d}\Big\}\notag\\*
&\quad\times\partial_{1,+\dot a}\partial_{1,+\dot b}
\partial_{3,+}{}^{\dot a}\partial_{3,+}{}^{\dot b}
\Bigg[C^X(x_1,\theta)
\frac{\delta^2 A(x_3)}{\delta C^Y(x_3+z_{13},\theta)\,\delta C^Z(x_3+z_{23},\theta)}\notag\\*
&\hspace{60pt}\times\frac{\delta B(x)}{\delta C^W(x+z_{24},\theta)}
\Bigg]_{x_1=x_3=x}\notag\\[4pt]
&-\frac{\hbar^2}{(32\pi^2)^2}C_X{}^{ZR}C_R{}^{YW}
\int_0^1dr_1\int_0^1dr_2
\int_{\substack{r_3,r_4>0\\r_3+r_4<1}}dr_3\,dr_4\,r_1r_2r_4\notag\\*
&\quad\times\int d^3\theta\int\prod_{e\in\{14,23,24\}}d^4z_e\notag\\*
&\quad\times\exp\!\Big\{-\tfrac12\Big[
\big[1-r_1r_2-r_1(1-r_2)r_3\big]z_{14}
+r_1r_3z_{23}+r_1r_2(1-r_3)z_{24}\Big]^{+\dot c}\partial_{1,+\dot c}\notag\\*
&\hspace{60pt}-\tfrac12r_4[(1-r_2)z_{14}-z_{23}+r_2z_{24}]^{+\dot d}\partial_{3,+\dot d}\Big\}\notag\\*
&\quad\times\partial_{1,+\dot a}\partial_{1,+\dot b}
\partial_{3,+}{}^{\dot a}\partial_{3,+}{}^{\dot b}
\Bigg[C^X(x_1,\theta)
\frac{\delta A(x_3)}{\delta C^Y(x_3+z_{23},\theta)}\notag\\*
&\hspace{60pt}\times\frac{\delta^2 B(x)}{\delta C^Z(x+z_{14},\theta)\,\delta C^W(x+z_{24},\theta)}
\Bigg]_{x_1=x_3=x}.
\label{eq-product2A}
\end{align}

\subsection{Normal ordering and an explicit associator}\label{sec-normal-ordering}

\paragraph{Normal ordering.}\label{quantum-normal-ordering}

In direct analogy with contour normal ordering in two-dimensional CFT, one can also define a geometric product in four-dimensional holomorphic field theory by integrating a separated insertion against the Bochner--Martinelli kernel over a three-sphere~\cite[Section 2.3]{semichiral}: for scale $L>0$ and $S_R^3=\partial B_R$ the three-sphere of radius $R$ around $0$,
\begin{equation}\label{eq-contour-product-definition}
(A\star^{\prime}B)[L]=\int_{S_R^3}\Omega\wedge[P_{0,\infty}(x)]_\Omega\,(A(x)B(0))[L].
\end{equation}
We now prove that $\star'$ and the marked-graph product $\star$ of~\eqref{b-00504} represent the exact same cohomology product on $H(\mathbf Q_L)$ in the scale-$L$ effective complex of Section~\ref{sec-brackets}.

At fixed $L>0$, the joint effective two-operator insertion (including Wick contractions between $A(x)$ and $B(0)$, and requiring no UV counterterms by~\cite{Wang}) is
\begin{equation}\label{eq-product-joint-effective}
(A(x)B(0))[L]e^{I[L]/\hbar}
=\lim_{\varepsilon\downarrow0}
e^{\hbar\partial_{P_{\varepsilon,L}}}
\bigl(A(x)B(0)e^{I/\hbar}\bigr),
\end{equation}
with $I=\int_{\mathbb C^2}\Omega\wedge[\mathcal I]_{(0,2)}$, and local insertions intertwine $\mathbf Q$ and $\mathbf Q_L$ via
\begin{equation}\label{eq-product-local-Ward}
\mathbf Q_L(\mathcal O[L])=(\mathbf Q\mathcal O)[L],
\qquad
\mathbf Q_L=\bar\partial+(I[L],\,\cdot\,)_L+\hbar\Delta_L.
\end{equation}
For separated insertions, holomorphic descent~\cite[Sections 2.2--2.3]{semichiral} provides the antiholomorphic translation homotopy: writing $A(x,\bar\theta)[L]$ for the full Dolbeault multiplet of $A[L]$ in the coordinates~\eqref{eq-B-7}, its $(0,1)$-form descendant satisfies
\begin{equation}\label{eq-ring-descendant}
\begin{aligned}
A^{(1)}(x)[L]
&=\frac14dx^{-\dot a}
\left.\frac{\partial_L A(x,\bar\theta)[L]}{\partial\bar\theta^{\dot a}}
\right|_{\bar\theta=0},\\
\mathbf Q_L A^{(1)}[L]+\bar\partial_x A[L]&=(\mathbf QA)^{(1)}[L].
\end{aligned}
\end{equation}
In particular, when $\mathbf QA=0$, writing $A^{(1)}[L]=dx^{-\dot a}A^{(1)}_{-\dot a}[L]$ expresses antiholomorphic derivatives as $\mathbf Q_L$-exact:
\begin{equation}\label{eq-ring-descendant-component}
A^{(1)}[L]=dx^{-\dot a}A^{(1)}_{-\dot a}[L],\qquad
\frac{\partial A[L]}{\partial x^{-\dot a}}
=\bigl(\bar\partial+(I[L],\,\cdot\,)_L+\hbar\Delta_L\bigr)
A^{(1)}_{-\dot a}[L].
\end{equation}

Recall from~\eqref{eq-C-explicit-propagator} and~\eqref{eq-C-heat-spatial} the Bochner--Martinelli residue normalization on any positively oriented embedded three-sphere $S^3$ linking the origin once:
\begin{equation}\label{eq-ring-propagator-residue}
\begin{gathered}
[P_{0,\infty}(x)]_\Omega
=-\frac{x^-_{\dot a}dx^{-\dot a}}
{8\pi^2|x|_E^4},\qquad
d\bigl(\Omega\wedge[P_{0,\infty}]_\Omega\bigr)
=\delta_E^{(4)}(x)d^4x_E,\\
\int_{S^3}\Omega\wedge[P_{0,\infty}]_\Omega
=\lim_{\varepsilon\downarrow0}\int_{|x|_E\le r}\Omega\wedge[K_\varepsilon]_\Omega
=\lim_{\varepsilon\downarrow0}\left[1-\left(1+\frac{r^2}{2\varepsilon}\right)e^{-r^2/(2\varepsilon)}\right]=1.
\end{gathered}
\end{equation}
Writing $(A(x)B(0))^{(1)}[L]$ for the $(0,1)$-form descendant with respect to $x$, any two homologous contours $S_1^3-S_0^3=\partial V$ disjoint from $0$ define the same class in $H(\mathbf Q_L)$: for $\mathbf QA=\mathbf QB=0$, descent gives $\bar\partial_x(A(x)B(0))[L]=-\mathbf Q_L(A(x)B(0))^{(1)}[L]$, and since $d(\Omega\wedge[P_{0,\infty}]_\Omega)=0$ on $V$, Stokes' theorem yields
\begin{equation}\label{eq-ring-sphere-homotopy}
\begin{aligned}
&\int_{S_1^3}\Omega\wedge[P_{0,\infty}]_\Omega(A(x)B(0))[L]
-\int_{S_0^3}\Omega\wedge[P_{0,\infty}]_\Omega(A(x)B(0))[L]\\
&\quad=-\int_V\Omega\wedge[P_{0,\infty}]_\Omega\wedge\bar\partial_x(A(x)B(0))[L]
=-\mathbf Q_L\int_V\Omega\wedge[P_{0,\infty}]_\Omega
\wedge(A(x)B(0))^{(1)}[L].
\end{aligned}
\end{equation}

\rthm{Theorem}{b-00598}{Normal-ordering comparison} \emph{Let $A$ and $B$ be homogeneous quantum cocycles and $L>0$. In the effective-observable complex at scale $L$ with differential~\eqref{eq-product-local-Ward}, the contour product~\eqref{eq-contour-product-definition} and the marked product~\eqref{b-00504} satisfy}
\begin{equation}\label{eq-normal-ordering-theorem}
\begin{aligned}
(A\star^{\prime}B)[L]
&=(A\star B)[L]
-\mathbf Q_L\int_{B_R}\Omega\wedge[P_{0,\infty}(x)]_\Omega
\wedge(A(x)B(0))^{(1)}[L],\\
[(A\star^{\prime}B)[L]]&=[(A\star B)[L]]\quad\text{in }H(\mathbf Q_L).
\end{aligned}
\end{equation}

\begin{proofbox}
\textbf{Proof.} Shrinking the inner boundary of $V$ to $x=0$ replaces $V$ by the ball $B_R$, where the $(2,2)$-form integrand $\Omega\wedge[P_{0,\infty}(x)]_\Omega\wedge(A(x)B(0))^{(1)}[L]$ is evaluated coefficientwise as a regulated graph limit in the auxiliary extension: physical edges carry $P_{\varepsilon,L}$, while the auxiliary marked edge (ordered last, with zero displacement and its single $\hbar$ factor divided out) carries $P_{\varepsilon,\infty}$. Assigning independent UV cutoffs $\varepsilon_e>0$ to edges produces the same local collision integral once all lower UV facets are summed, because every positive resistance ray intersects their union at
\begin{equation}\label{eq-independent-cutoff-section}
[t]\longmapsto\left(\max_{f\in I}\frac{\varepsilon_f}{t_f}\right)(t_e)_{e\in I},
\end{equation}
and the closed projective current form on $\mathbb{RP}_{>0}^{|I|-1}$ is independent of this radial section as $\varepsilon_e\to0$~\cite[Theorem 6]{Wang}. On a connected cluster with $n$ interactions ($|V|=n+2$, $|I|=2n+1$), edge variations and $\hbar$ powers satisfy
\begin{equation}\label{eq-normal-ordering-J}
\begin{gathered}
\bar\partial P_{\varepsilon,L}=K_\varepsilon-K_L,
\qquad \bar\partial P_{\varepsilon,\infty}=K_\varepsilon,\\
|V|=n+2,\qquad |I|=2n+1,\qquad
\frac1\hbar\frac{\hbar^{2n+1}}{n!\hbar^n}=\frac{\hbar^n}{n!}.
\end{gathered}
\end{equation}
Applying~\eqref{eq-normal-ordering-J} with the signs $(-1)^{j-1}$ of~\eqref{eq-A-28}, the upper $K_L$ faces assemble into $\mathbf Q_L$; pure-interaction UV collision clusters cancel by the QME, and single-insertion UV clusters vanish by $\mathbf QA=\mathbf QB=0$. On the remaining UV clusters collapsing both $A,B$ and $n$ interactions to $0$, rescaling $t_e\mapsto\varepsilon t_e$, $x_v\mapsto\sqrt\varepsilon\,x_v$ sends $L/\varepsilon\to\infty$ and reproduces the exact marked kernel~\eqref{eq-2-31} (with $n=0$ sign $+AB$ fixed by~\eqref{b-00504}), while the external Wick contractions reconstruct the scale-$L$ effective operator:
\begin{equation}\label{eq-normal-ordering-smeared}
\left[
\left.\sum_{n\ge0}\frac{\hbar^n}{n!}
\bigl\{\underbrace{\mathcal I,\ldots,\mathcal I}_{n},Aa,b_{\mathrm{aux}}B\bigr\}
\right|_{\mathrm{aux}=0}
\right][L]
=(A\star B)[L].
\end{equation}
Combining this UV collision residue at $0$ with the outer sphere boundary $S_R^3$ and $\bar\partial_x(A(x)B(0))[L]=-\mathbf Q_L(A(x)B(0))^{(1)}[L]$ away from collisions (noting that $\Omega\wedge[P_{0,\infty}]_\Omega$ has odd total degree $3$) gives the regulated Stokes identity
\begin{equation}\label{eq-normal-ordering-stokes}
\begin{aligned}
&\mathbf Q_L\int_{B_R}\Omega\wedge[P_{0,\infty}]_\Omega\wedge(A(x)B(0))^{(1)}[L]\\
&\quad=-\int_{B_R}\Omega\wedge[P_{0,\infty}]_\Omega\wedge\mathbf Q_L(A(x)B(0))^{(1)}[L]\\
&\quad=\int_{B_R}\Omega\wedge[P_{0,\infty}]_\Omega\wedge\bar\partial_x(A(x)B(0))[L]\\
&\quad=(A\star B)[L]-\int_{S_R^3}\Omega\wedge[P_{0,\infty}]_\Omega\,(A(x)B(0))[L],
\end{aligned}
\end{equation}
which rearranges to~\eqref{eq-normal-ordering-theorem}.~\hfill\qedsymbol
\end{proofbox}

\paragraph{An explicit associator example.}\label{an-explicit-associator}

Although $\star$ is associative on $H(\mathbf Q)$ (Theorem~\ref{b-00513}), it is \emph{not} associative at the cochain level: the ternary homotopy in~\eqref{eq-ring-common-root-homotopy} produces a genuine, nonzero $\mathbf Q$-exact associator even on gauge-invariant $1/16$-BPS cocycles.

\rthm{Example}{b-00610}{Nonvanishing chain-level associator in $\mathcal N=4$ Yang--Mills theory} In $\mathcal N=4$ Yang--Mills theory, let $A$ and $B$ be gauge-invariant polynomials in the undifferentiated scalars $(\gamma^I)_X$ (a subsector of the chiral BMN sector), and set $D=(\gamma^3)_X(\beta_2)^X$. Then $A$, $B$, and $D$ are exact all-loop quantum cocycles ($\mathbf QA=\mathbf QB=\mathbf QD=0$), and their chain-level associator is the nonzero $\mathbf Q$-coboundary
\begin{equation}\label{eq-ring-scalar-associator}
\begin{aligned}
&(A\star B)\star D-A\star(B\star D)\\
&\quad=\mathbf Q\left\{
\frac{\hbar}{64\pi^2}
\left[
(\partial_{+\dot a}B)\partial_+{}^{\dot a}
\frac{\partial A}{\partial(\gamma^2)_X}
+(\partial_{+\dot a}A)\partial_+{}^{\dot a}
\frac{\partial B}{\partial(\gamma^2)_X}
\right](\gamma^3)_X\right\}.
\end{aligned}
\end{equation}

\begin{proofbox}
\textbf{Proof.} Gauge invariance gives $Q_0A=Q_0B=0$, and $Q_0D\propto C^{XYZ}(\gamma^3)_X(\gamma^3)_Y(\gamma^1)_Z=0$ by antisymmetry of $C^{XYZ}$. At loop level, $Q_1D$ vanishes because $(\theta_1\theta_2)\theta_2=0$ in~\eqref{eq-A-16}, while $Q_{n\ge2}D=0$ because $D$ has only two letters; similarly $Q_{n\ge1}A=Q_{n\ge1}B=0$ by Corollary~\ref{b-00235} since $A,B$ contain no $(b,\beta)$ antifields. Thus $\mathbf QA=\mathbf QB=\mathbf QD=0$.

By Corollary~\ref{local-ch8-finite-product-corollary}, $A\star B=AB$ receives no loop corrections ($\#(b,\beta)=0$), whereas any product involving $D$ ($\#(b,\beta)=1$) is one-loop exact. Evaluating~\eqref{eq-2-34} on $(A,D)$ at $z_1=z_2=0$ gives
\begin{equation*}
\{\mathcal I,A,D\}^{\gamma,0}
=\frac{g}{64\pi^2}C_{EX}{}^Y
(\partial_{+\dot a}c^E)
\partial_+{}^{\dot a}
\left(\frac{\partial A}{\partial(\gamma^2)_X}\right)
(\gamma^3)_Y.
\end{equation*}
Subtracting $A\star(B\star D)=A(BD+\hbar\{\mathcal I,B,D\}^{\gamma,0})+\hbar\{\mathcal I,A,BD\}^{\gamma,0}$ from $(A\star B)\star D=(AB)\star D=ABD+\hbar\{\mathcal I,AB,D\}^{\gamma,0}$ leaves the cross-derivative terms of $\partial_{+\dot a}\partial/\partial(\gamma^2)_X$:
\begin{equation*}
\begin{aligned}
&(A\star B)\star D-A\star(B\star D)\\
&\quad=\frac{\hbar g}{64\pi^2}C_{EX}{}^Y
(\partial_{+\dot a}c^E)
\left[
\frac{\partial A}{\partial(\gamma^2)_X}
\partial_+{}^{\dot a}B
+(\partial_+{}^{\dot a}A)
\frac{\partial B}{\partial(\gamma^2)_X}
\right](\gamma^3)_Y.
\end{aligned}
\end{equation*}
On the other hand, since $A$ is gauge invariant ($c^EC_{EY}{}^X(\gamma^I)_X\partial A/\partial(\gamma^I)_Y=0$), differentiating with respect to $(\gamma^2)_X$ and contracting with $(\gamma^3)_X$ gives
\begin{equation*}
Q_0\left[
\partial_+{}^{\dot a}\left(
\frac{\partial A}{\partial(\gamma^2)_X}\right)(\gamma^3)_X
\right]
=-gC_{EX}{}^Y(\partial_+{}^{\dot a}c^E)
\frac{\partial A}{\partial(\gamma^2)_X}(\gamma^3)_Y,
\end{equation*}
and analogously for $B$. Because the primitive inside braces in~\eqref{eq-ring-scalar-associator} is built purely from $\gamma$ and its holomorphic derivatives, $Q_{n\ge1}$ annihilates it identically, so $\mathbf Q$ reduces on it to $Q_0$; using $(\partial_{+\dot a}B)(\partial_+{}^{\dot a}c^E)=-(\partial_{+\dot a}c^E)(\partial_+{}^{\dot a}B)$ yields~\eqref{eq-ring-scalar-associator}.~\hfill\qedsymbol
\end{proofbox}

\section{Spiders and product forests}\label{sec-forests}

In a theory with at most cubic interactions, every Feynman graph contributing to the quantum supercharge $Q_n$ is a fan (Proposition~\ref{b-00210}, Corollary~\ref{b-00222}, and Section~\ref{sec-fans}). Two natural extensions require graph families beyond fans. First, nonrenormalizable or infrared-effective superpotentials $\bar W$ of degree four or higher---such as the quartic superpotential~\eqref{local-ch9-mass-ir-potential} of mass-deformed $\mathcal N=4$ Yang--Mills theory---produce \emph{spider} graphs in $Q_n$, in which several cubic paths meet at the superpotential vertex (Proposition~\ref{b-00240}). Second, even in a purely cubic theory, the marked product $A\star B$ of Section~\ref{sec-product} receives contributions from \emph{product forests} spanning the two insertions $A$ and $B$: through two loops every component tree reduces to a fan, whereas from three loops onward branched cubic trees appear with no degree-two interaction vertices accessible by series--parallel moves alone.

In this section we extend the circuit reduction of Section~\ref{sec-fans}---transferring injected currents $\lambda$ forward as Schwinger resistances $t$ are eliminated, and reconstructing the edge currents $y$ backward as polynomials in dimensionless current-division ratios---to both families. Section~\ref{sec-spiders} evaluates the general spider kernel in closed form and extracts its contribution to $Q_n$. Section~\ref{sec-dressed-definition} proves that every nonzero marked product graph decomposes into rooted trees between $A$ and $B$. Section~\ref{sec-forest-graphs} introduces the single and paired source-free star--triangle moves, proving that every cubic product forest reduces to polynomial interval and simplex integrals, and Section~\ref{sec-product-four-loops} illustrates both star--triangle reductions explicitly on the three-loop product forests.

\subsection{Spider kernels and supercharges}\label{sec-spiders}
\begin{figure}[H]
\centering
\includegraphics[width=.94\linewidth]{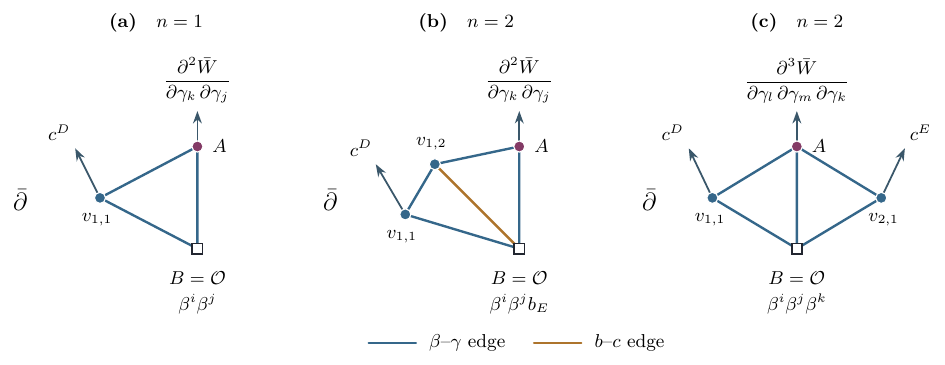}
\caption{One- and two-loop spider contractions. The insertion at $B$ and the surviving vertex factors are displayed; blue and ochre edges contract $\beta$--$\gamma$ and $b$--$c$, respectively.}\label{fig-spider}
\end{figure}

\rthm{Definition}{b-00380}{Spiders}
A \emph{spider} $\Gamma$ with $a\ge 1$ arms of lengths $\ell_1,\ldots,\ell_a\ge 1$ consists of $a$ disjoint paths meeting at a hub vertex $A$, together with a second vertex $B$ joined by a spoke to every arm vertex and to $A$ (Figure~\ref{fig-spider}; at $a=1$ it reduces to an $\ell_1$-loop fan).
For the supercharge $Q_n\mathcal O$, $A$ carries a superpotential vertex $\bar W$ and $B$ carries the local insertion $\mathcal O$.
Arm $i$ has vertices $v_{i,1},\ldots,v_{i,\ell_i}$, ordered towards $v_{i,\ell_i+1}=A$.
Its edge set and loop number are
\begin{equation}\label{eq-1-47}
\begin{gathered}
I=\bigcup_{\substack{i\in\{1,\ldots,a\}\\\mathrm{arms}}}
\bigcup_{j=1}^{\ell_i}\{(v_{i,j},v_{i,j+1}),(v_{i,j},B)\}\cup\{(A,B)\},\\
n=\sum_{\substack{i\in\{1,\ldots,a\}\\\mathrm{arms}}}\ell_i,
\end{gathered}
\end{equation}
so $|V|=n+2$ and $|I|=2n+1=2|V|-3$. Order the edges arm by arm, placing $(v_{i,j},v_{i,j+1})$ before $(v_{i,j},B)$, and $(A,B)$ last.

\rthm{Theorem}{thm-spider-kernel}{Spider kernel}
We keep the sources at $A$ and each of $v_{i,1},\ldots,v_{i,\ell_i}$ arbitrary, with total current conservation fixing $\lambda_B=-\lambda_A-\sum_{i=1}^a\sum_{j=1}^{\ell_i}\lambda_{v_{i,j}}$. The $2n$ current-division ratios range over the product of $(0,1)^{2n-a}$ with the $a$-simplex $\Delta^a$:
\begin{equation}
0<r_{i,j}<1\quad(j<2\ell_i),\qquad
r_{i,2\ell_i}>0,\qquad
\sum_{\substack{i\in\{1,\ldots,a\}\\\mathrm{arms}}}r_{i,2\ell_i}<1.
\label{eq-spiderdomain}
\end{equation}
Here $r_{i,2j-1}$ transfers the source at $v_{i,j}$ towards $A$, while
$r_{i,2j}$ is the current fraction carried by the newly created parallel branch along spoke $(v_{i,j+1},B)$.
The sources collected by the successive eliminations are
\begin{equation}
\begin{aligned}
\lambda_{v_{i,j},+\dot a}^{\mathrm{red}}
&=\sum_{\substack{h\in\{1,\ldots,j\}\\\mathrm{arm\ vertices}}}
\left(\prod_{k=h}^{j-1}r_{i,2k-1}\right)
\lambda_{v_{i,h},+\dot a},\\
y_{AB,+\dot a}^{\mathrm{red}}
&=\lambda_{A,+\dot a}
+\sum_{\substack{i\in\{1,\ldots,a\}\\\mathrm{arms}}}
r_{i,2\ell_i-1}\lambda_{v_{i,\ell_i},+\dot a}^{\mathrm{red}}.
\end{aligned}
\label{eq-spider-reduced}
\end{equation}
Starting from $y_{v_{i,\ell_i+1}B}^{\mathrm{red}}=y_{AB}^{\mathrm{red}}$,
reconstruct the edge currents backward along each arm by
\begin{equation}
\begin{aligned}
y_{v_{i,j}B,+\dot a}^{\mathrm{red}}
&=(1-r_{i,2j-1})\lambda_{v_{i,j},+\dot a}^{\mathrm{red}}
+r_{i,2j}y_{v_{i,j+1}B,+\dot a}^{\mathrm{red}},\\
y_{v_{i,j}v_{i,j+1},+\dot a}
&=\lambda_{v_{i,j},+\dot a}^{\mathrm{red}}
-y_{v_{i,j}B,+\dot a}^{\mathrm{red}},\\
y_{v_{i,1}B,+\dot a}&=y_{v_{i,1}B,+\dot a}^{\mathrm{red}},\\
y_{v_{i,j}B,+\dot a}
&=(1-r_{i,2j-2})y_{v_{i,j}B,+\dot a}^{\mathrm{red}}\quad(j\ge2),\\
y_{AB,+\dot a}
&=\left(1-\sum_{\substack{i\in\{1,\ldots,a\}\\\mathrm{arms}}}
r_{i,2\ell_i}\right)y_{AB,+\dot a}^{\mathrm{red}}.
\end{aligned}
\label{eq-1-51}
\end{equation}
With the spokes and $(A,B)$ chosen as spanning tree, the spider kernel is
\begin{equation}
\begin{aligned}
\mathcal I_\Gamma(\lambda;z)
&=\int_{\eqref{eq-spiderdomain}}
e^{-\sum_{e\in I\,\mathrm{(edges)}}y_{e,+\dot a}z_e^{+\dot a}}
\bigwedge_{i=1}^a\bigwedge_{j=1}^{\ell_i}
(-d^2y_{v_{i,j}v_{i,j+1}})\\
&=\int_{\eqref{eq-spiderdomain}}
\prod_{i=1}^a\prod_{j=1}^{2\ell_i}dr_{i,j}\,
\left(1-\sum_{\substack{i\in\{1,\ldots,a\}\\\mathrm{arms}}}r_{i,2\ell_i}\right)\\
&\quad\times\prod_{i=1}^a
\left[\left(\lambda_{v_{i,\ell_i},+\dot a}^{\mathrm{red}}
y_{AB,+}^{\mathrm{red}}{}^{\dot a}\right)
\prod_{j=1}^{\ell_i-1}
\left((1-r_{i,2j})\lambda_{v_{i,j},+\dot b}^{\mathrm{red}}
y_{v_{i,j+1}B,+}^{\mathrm{red}}{}^{\dot b}\right)\right]\\
&\quad\times
e^{-\sum_{e\in I\,\mathrm{(edges)}}y_{e,+\dot c}(r)z_e^{+\dot c}}.
\end{aligned}
\label{eq-1-54-general}
\end{equation}

\begin{figure}[ht!]
\centering
\includegraphics[width=.96\linewidth]{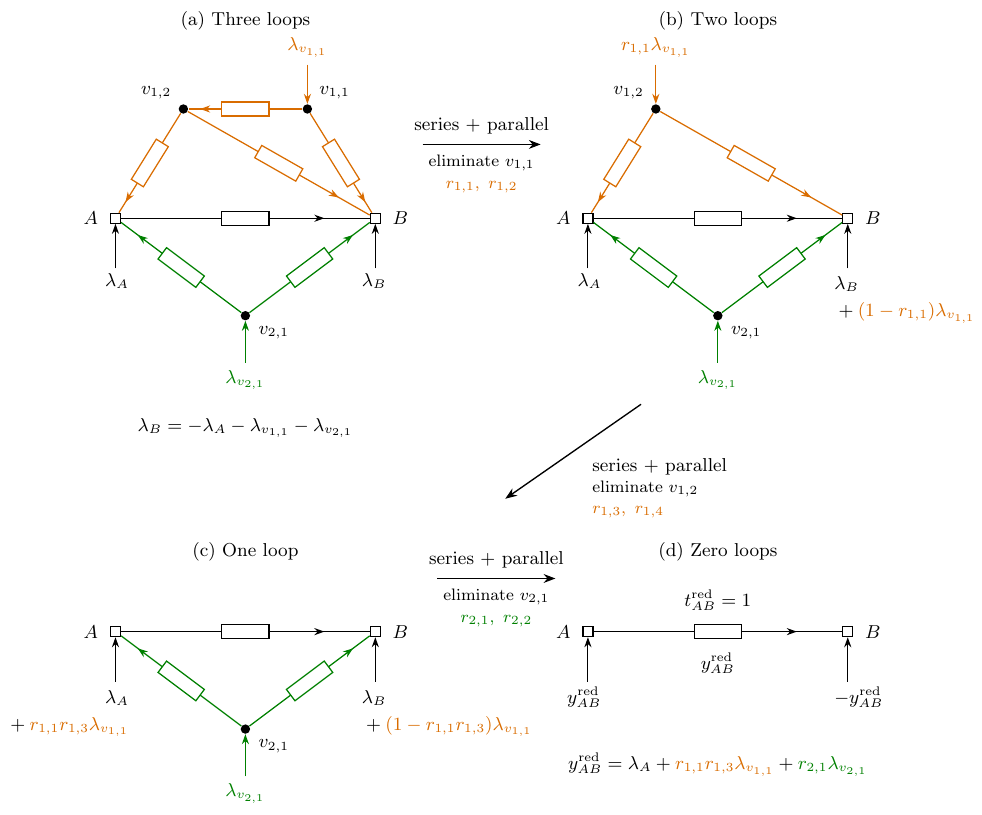}
\caption{Step-by-step series--parallel circuit reduction of a three-loop, two-arm spider ($\ell_1=2,\ell_2=1$), illustrating forward source transfer to $y_{AB}^{\mathrm{red}}=\lambda_A+r_{1,1}r_{1,3}\lambda_{v_{1,1}}+r_{2,1}\lambda_{v_{2,1}}$ in~\eqref{eq-spider-reduced}.}\label{fig-spider-reduction}
\end{figure}

\begin{proofbox}
\textbf{Proof.} Apply the series--parallel move~\eqref{eq-kernel-series-parallel} inward along each arm (Figure~\ref{fig-spider-reduction}), setting $(r_1,r_2)=(r_{i,2j-1},1-r_{i,2j})$ at step $j<\ell_i$.
At a nonterminal vertex $v_{i,j}$ ($1\le j<\ell_i$), keep the earlier fractions fixed; $\equiv$ denotes equality modulo the retained loop-current differentials:
\begin{equation}\label{eq-spider-arm-step}
\begin{aligned}
\lambda_{v_{i,j+1}}^{\mathrm{red}}
&=\lambda_{v_{i,j+1}}+r_{i,2j-1}\lambda_{v_{i,j}}^{\mathrm{red}},\\
dy_{v_{i,j+1}B,+\dot a}^{\mathrm{red}}
&\equiv\lambda_{v_{i,j},+\dot a}^{\mathrm{red}}\,dr_{i,2j-1},\\
dy_{v_{i,j}v_{i,j+1},+\dot a}
&\equiv(1-r_{i,2j})\lambda_{v_{i,j},+\dot a}^{\mathrm{red}}\,dr_{i,2j-1}
-y_{v_{i,j+1}B,+\dot a}^{\mathrm{red}}\,dr_{i,2j},\\
-d^2y_{v_{i,j}v_{i,j+1}}
&\equiv(1-r_{i,2j})
\bigl(\lambda_{v_{i,j},+\dot a}^{\mathrm{red}}
y_{v_{i,j+1}B,+}^{\mathrm{red}}{}^{\dot a}\bigr)
dr_{i,2j-1}\wedge dr_{i,2j}.
\end{aligned}
\end{equation}
The last line is the current-area Jacobian of this elimination, with $d^2y=dy_{+\dot+}\wedge dy_{+\dot-}$.

After these $n-a$ eliminations, a single arm edge $(v_{i,\ell_i},A)$ remains on each arm, forming $a$ series branches in parallel with $(A,B)$.
With all earlier fractions fixed, \eqref{eq-spider-reduced} gives
\begin{equation}\label{eq-spider-terminal-arms}
\begin{aligned}
y_{v_{i,\ell_i}A,+\dot a}
&=r_{i,2\ell_i-1}\lambda_{v_{i,\ell_i},+\dot a}^{\mathrm{red}}
-r_{i,2\ell_i}y_{AB,+\dot a}^{\mathrm{red}},\\
dy_{AB,+\dot a}^{\mathrm{red}}
&=\sum_{\substack{k\in\{1,\ldots,a\}\\\mathrm{arms}}}
\lambda_{v_{k,\ell_k},+\dot a}^{\mathrm{red}}\,dr_{k,2\ell_k-1},\\
dy_{v_{i,\ell_i}A,+\dot a}
&=\lambda_{v_{i,\ell_i},+\dot a}^{\mathrm{red}}\,dr_{i,2\ell_i-1}
-y_{AB,+\dot a}^{\mathrm{red}}\,dr_{i,2\ell_i}\\
&\quad-r_{i,2\ell_i}
\sum_{\substack{k\in\{1,\ldots,a\}\\\mathrm{arms}}}
\lambda_{v_{k,\ell_k},+\dot a}^{\mathrm{red}}\,dr_{k,2\ell_k-1}.
\end{aligned}
\end{equation}
Each $dr_{i,2\ell_i}$ occurs only in its own arm's differential, so the top-degree wedge is
\begin{equation}\label{eq-spider-terminal-wedge}
\begin{aligned}
&\bigwedge_{i=1}^a(-d^2y_{v_{i,\ell_i}A})\\
&=\bigwedge_{i=1}^a\Bigg\{\Bigg[
\bigl(\lambda_{v_{i,\ell_i},+\dot a}^{\mathrm{red}}
y_{AB,+}^{\mathrm{red}}{}^{\dot a}\bigr)dr_{i,2\ell_i-1}\\
&\hspace{65pt}-r_{i,2\ell_i}
\sum_{\substack{k\in\{1,\ldots,a\}\\\mathrm{arms}}}
\bigl(\lambda_{v_{k,\ell_k},+\dot b}^{\mathrm{red}}
y_{AB,+}^{\mathrm{red}}{}^{\dot b}\bigr)dr_{k,2\ell_k-1}
\Bigg]\wedge dr_{i,2\ell_i}\Bigg\}\\
&=\Bigg[
\prod_{i=1}^a\bigl(\lambda_{v_{i,\ell_i},+\dot a}^{\mathrm{red}}
y_{AB,+}^{\mathrm{red}}{}^{\dot a}\bigr)\\
&\hspace{28pt}-\sum_{\substack{i\in\{1,\ldots,a\}\\\mathrm{arms}}}
r_{i,2\ell_i}
\prod_{k=1}^a\bigl(\lambda_{v_{k,\ell_k},+\dot b}^{\mathrm{red}}
y_{AB,+}^{\mathrm{red}}{}^{\dot b}\bigr)
\Bigg]\bigwedge_{i=1}^a(dr_{i,2\ell_i-1}\wedge dr_{i,2\ell_i})\\
&=\left(1-\sum_{\substack{i\in\{1,\ldots,a\}\\\mathrm{arms}}}
r_{i,2\ell_i}\right)
\prod_{i=1}^a\bigl(\lambda_{v_{i,\ell_i},+\dot a}^{\mathrm{red}}
y_{AB,+}^{\mathrm{red}}{}^{\dot a}\bigr)
\bigwedge_{i=1}^a(dr_{i,2\ell_i-1}\wedge dr_{i,2\ell_i}).
\end{aligned}
\end{equation}
Terms selecting the summed one-form from two different arms vanish because that one-form is repeated in the wedge; for a term selecting it only from arm $i$, only its $k=i$ summand survives, giving the factor $-r_{i,2\ell_i}$.

Equation~\eqref{eq-spider-terminal-wedge} supplies the measure of the $a$ terminal arm edges, while the full measure in~\eqref{eq-1-54-general} combines it with the $n-a$ preceding eliminations:
\begin{equation}\label{eq-spider-full-measure}
\begin{aligned}
&\bigwedge_{i=1}^a\bigwedge_{j=1}^{\ell_i}
(-d^2y_{v_{i,j}v_{i,j+1}})\\
&=\prod_{i=1}^a\prod_{j=1}^{\ell_i-1}
\Bigl[(1-r_{i,2j})\lambda_{v_{i,j},+\dot a}^{\mathrm{red}}
y_{v_{i,j+1}B,+}^{\mathrm{red}}{}^{\dot a}\Bigr]\\
&\quad\times\left[\bigwedge_{i=1}^a\bigwedge_{j=1}^{\ell_i-1}
(dr_{i,2j-1}\wedge dr_{i,2j})\right]
\wedge\left.\bigwedge_{i=1}^a(-d^2y_{v_{i,\ell_i}A})\right|_{\substack{\text{earlier fractions fixed}}}.
\end{aligned}
\end{equation}
Substituting~\eqref{eq-spider-terminal-wedge} yields~\eqref{eq-1-54-general}; reordering the two-form factors introduces no sign.~\hfill\qedsymbol
\end{proofbox}

In a spider diagram contributing to $Q_n\mathcal O$ (Proposition~\ref{b-00240}), every inner arm vertex $v_{i,j}$ ($j>1$) is cubic of full graph degree three and therefore has no uncontracted field, whereas each outer arm vertex $v_{i,1}$ has degree two and retains one surviving field (and $A:\bar W$ retains $\deg\bar W-a-1$ surviving fields):
\begin{equation}\label{eq-spider-qn-sources}
\begin{gathered}
\lambda_{v_{i,j},+\dot a}=0\quad(j>1),\qquad
\lambda_{v_{i,j},+\dot a}^{\mathrm{red}}
=\left(\prod_{k=1}^{j-1}r_{i,2k-1}\right)\lambda_{v_{i,1},+\dot a},\\
y_{AB,+\dot a}^{\mathrm{red}}
=\lambda_{A,+\dot a}+\sum_{i=1}^a\left(\prod_{k=1}^{\ell_i}r_{i,2k-1}\right)\lambda_{v_{i,1},+\dot a}.
\end{gathered}
\end{equation}
Because $\lambda_{v_{i,j},+\dot a}^{\mathrm{red}}$ is then proportional to $\lambda_{v_{i,1},+\dot a}$ along arm $i$, only the $(\prod_{m=j+1}^{\ell_i}r_{i,2m})y_{AB,+\dot a}^{\mathrm{red}}$ part of $y_{v_{i,j+1}B,+\dot a}^{\mathrm{red}}$ survives its antisymmetric spinor contraction in~\eqref{eq-1-54-general}. Consequently, the specialized spider kernel takes the factorized form
\begin{equation}
\begin{aligned}
\left.\mathcal I_\Gamma(\lambda;z)\right|_{\lambda_{v_{i,j}}=0\,(j>1)}
={}&\int_{\eqref{eq-spiderdomain}}
\prod_{i=1}^a\prod_{j=1}^{2\ell_i}dr_{i,j}\,
\left(1-\sum_{\substack{i\in\{1,\ldots,a\}\\\mathrm{arms}}}
r_{i,2\ell_i}\right)\\
&\times\prod_{i=1}^a\left[
r_{i,2\ell_i}^{\ell_i-1}
\prod_{j=1}^{\ell_i-1}
r_{i,2j-1}^{\ell_i-j}r_{i,2j}^{j-1}(1-r_{i,2j})
\left(\lambda_{v_{i,1},+\dot a}
y_{AB,+}^{\mathrm{red}}{}^{\dot a}\right)^{\ell_i}
\right]\\
&\times e^{-\sum_{e\in I\,\mathrm{(edges)}}y_{e,+\dot c}(r)z_e^{+\dot c}}.
\end{aligned}
\label{eq-1-54}
\end{equation}
At zero displacement ($z=0$), $y_{AB}^{\mathrm{red}}$ depends only on the $n$ odd-indexed source-transfer ratios $r_{i,2j-1}$. Integrating out the parallel fractions $r_{i,2j}\in(0,1)$ ($1\le j<\ell_i$) via $\int_0^1 r^{j-1}(1-r)\,dr=\frac{1}{j(j+1)}$ gives $\prod_{j=1}^{\ell_i-1}\frac{1}{j(j+1)}=\frac{1}{(\ell_i-1)!\,\ell_i!}$ on each arm, while integrating $(r_{1,2\ell_1},\ldots,r_{a,2\ell_a})$ over $\Delta^a$ via~\eqref{eq-Dirichlet} gives $\frac{\prod_{i=1}^a(\ell_i-1)!}{(n+1)!}$. Cancelling $\prod_{i=1}^a(\ell_i-1)!$ leaves an $n$-cube integral over the source-transfer ratios alone:
\begin{equation}\label{eq-spider-zero-displacement}
\left.\mathcal I_\Gamma(\lambda;0)\right|_{\lambda_{v_{i,j}}=0\,(j>1)}
=\frac{1}{(n+1)!\,\prod_{i=1}^a\ell_i!}
\int_{(0,1)^n}\prod_{i=1}^a\left[
\left(\prod_{j=1}^{\ell_i}dr_{i,2j-1}\,r_{i,2j-1}^{\ell_i-j}\right)
\bigl(\lambda_{v_{i,1},+\dot a}y_{AB,+}^{\mathrm{red}}{}^{\dot a}\bigr)^{\ell_i}
\right].
\end{equation}
At $a=1$ and $\ell_1=n$, the pairing $\lambda_{v_{1,1},+\dot a}y_{AB,+}^{\mathrm{red}}{}^{\dot a}=\lambda_{v_{1,1},+\dot a}\lambda_{A,+}{}^{\dot a}$ is independent of $r$, and the remaining cube integral $\prod_{j=1}^n\int_0^1 r^{n-j}\,dr=\frac{1}{n!}$ reproduces the all-loop fan value~\eqref{eq-1-42}, $\frac{(\lambda_{v_{1,1},+\dot a}\lambda_{A,+}{}^{\dot a})^n}{(n!)^3(n+1)}$. At $a=2$ and $\ell_1=\ell_2=1$ ($n=2$), \eqref{eq-spider-zero-displacement} evaluates directly to
\begin{equation}\label{eq-spider-two-arm-value}
\left.\mathcal I_\Gamma(\lambda;0)\right|_{a=2,\,\ell_1=\ell_2=1}
=\frac{1}{24}
\bigl(\lambda_{v_{1,1},+\dot a}(\lambda_{v_{2,1}}+2\lambda_A)_+{}^{\dot a}\bigr)
\bigl(\lambda_{v_{2,1},+\dot b}(\lambda_{v_{1,1}}+2\lambda_A)_+{}^{\dot b}\bigr),
\end{equation}
which simplifies to $-\frac{1}{24}(\lambda_{v_{1,1},+\dot a}\lambda_{v_{2,1},+}{}^{\dot a})^2$ when all four legs of a quartic superpotential vertex at $A$ are contracted ($\lambda_A=0$).

Inserting~\eqref{eq-1-54} into the general supercharge formula~\eqref{eq-2-3}, a labelled spider $\Gamma$ of arm lengths $(\ell_1,\ldots,\ell_a)$ with $\sum_{i=1}^a\ell_i=n$ contributes to $Q_n\mathcal O(x)$ the local operator
\begin{equation}\label{eq-spider-supercharge}
(Q_n\mathcal O)_{\mathrm{spider},\,\Gamma}(x)
=\frac{(-1)^n}{(n+1)!\,(8\pi^2)^n}
\left.\Bigl[\mathcal I_\Gamma(\lambda;z)\right|_{\lambda_{v_{i,j}}=0\,(j>1)}
W_\Gamma[\mathcal I_1,\ldots,\mathcal I_{n+1},\mathcal O]\Bigr]_{\mathrm{jets},\,x_v=x,\,z=0},
\end{equation}
where only the spoke displacements $z_{v_{i,j}B}$ and $z_{AB}$ act on the differentiated letters of $\mathcal O$, while $\lambda_{v_{i,1}}$ and $\lambda_A$ act as holomorphic derivatives on the surviving interaction fields at $v_{i,1}$ ($1\le i\le a$) and $A$.

\subsection{Product forests}\label{sec-dressed-definition}
\rthm{Theorem}{b-00249}{Product forests}
For the interaction~\eqref{eq-B-10}, a nonzero connected marked graph $\Gamma$ contributing
to $A\star B$ at order $\hbar^n$ has $n$ interaction vertices,
$2n$ physical edges and the marked edge $(A,B)$. Its interaction subgraph decomposes into trees,
\begin{equation}
\begin{gathered}
\Gamma\setminus\{A,B\}=\coprod_jT_j,\qquad
T_j\text{ a tree},\qquad
\#(\bar W\text{ vertices in }T_j)\le1,\\
|I_j|=|T_j|-1,\qquad |E_j|=|T_j|+1.
\end{gathered}
\label{eq-1-19}
\end{equation}
Here $I_j$ denotes the internal edges of $T_j$ and $E_j$ the spokes connecting $T_j$ to
the insertions $A,B$. Every tree $T_j$ meets both insertions and carries at least one nonzero source.
The antifield count is
\begin{equation}\label{eq-product-antifield-loss}
\#(b,\beta)_{\mathrm{out}}
=\#(b,\beta)_A+\#(b,\beta)_B
-n-\#(\bar W\text{ vertices}).
\end{equation}
In particular, a tree composed solely of cubic vertices retains exactly one uncontracted field, located at its unique vertex of full graph degree two, while all its remaining vertices have full degree three.
For a tree containing one superpotential monomial $\bar W$, where $\deg_\Gamma\bar W$ counts its contracted fields and $\deg\bar W$ counts all its fields,
\begin{equation}
\begin{aligned}
\#(\text{degree-two cubic vertices in }T_j)&=\deg_\Gamma\bar W-2,\\
\#(\text{fields surviving at }\bar W)&=\deg\bar W-\deg_\Gamma\bar W,\\
\#(\text{fields surviving in }T_j)&=\deg\bar W-2.
\end{aligned}
\label{eq-1-23}
\end{equation}
Quadratic superpotential terms give no loop graph.

\begin{proofbox}
\textbf{Proof.} Connectedness of each component $T_j$ of $\Gamma\setminus\{A,B\}$ and the antifield-terminal bound give
\begin{equation}
|T_j|-1\le|I_j|\le|T_j|-\#(\bar W\text{ vertices in }T_j).
\end{equation}
A physical edge directly between $A$ and $B$ would be parallel to the marked edge $(A,B)$ and hence vanishes by Corollary~\ref{b-00188}.
Applying the subgraph Laman bound to $T_j\cup\{A,B\}$ together with the total physical edge count $2n$ gives
\begin{equation}
\begin{aligned}
|I_j|+|E_j|+1&\le2(|T_j|+2)-3,\\
\sum_{j\,\mathrm{(components)}}(|I_j|+|E_j|)
&=2n=2\sum_{j\,\mathrm{(components)}}|T_j|,
\end{aligned}
\qquad\Longrightarrow\qquad |I_j|+|E_j|=2|T_j|.
\end{equation}
Suppose a component $T_j$ carries no external source ($\lambda_v=0$ for all $v\in T_j$). Then $T_j$ is a passive resistor network driven solely by the total current $J$ flowing through it from $A$ to $B$, so every edge current in $T_j\cup E_j$ is proportional to $J$, $y_e=c_e(t)\,J$ (Figure~\ref{fig-sourceless}). Its two-forms therefore satisfy
\begin{equation}\label{eq-sourceless-component}
\begin{aligned}
d^2y_e
&=c_e^2\,dJ_{+\dot+}\wedge dJ_{+\dot-}
+c_e\,dc_e\wedge
(J_{+\dot+}dJ_{+\dot-}-J_{+\dot-}dJ_{+\dot+}),\\
d^2y_e\wedge d^2y_f&=0.
\end{aligned}
\end{equation}
\begin{figure}[H]
\centering
\includegraphics[width=.94\linewidth]{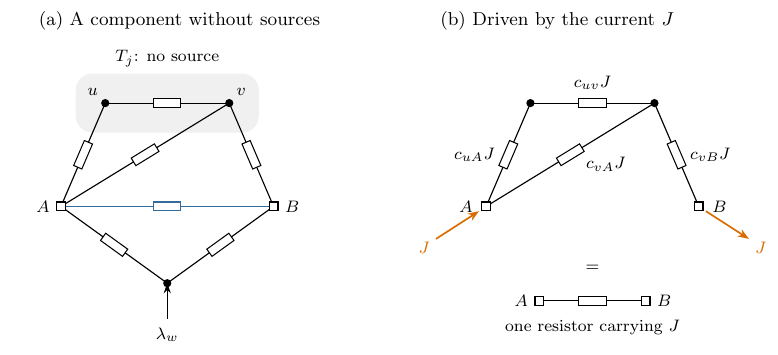}
\caption{A source-free component is driven by the voltage between $A$ and $B$; its currents are collinear, so its current-area form vanishes.}\label{fig-sourceless}
\end{figure}

For $|T_j|\ge2$, choosing a spanning tree assigns $|I_j|+|E_j|-|T_j|=|T_j|\ge2$ chord currents to this component, whose wedge product vanishes by~\eqref{eq-sourceless-component}; for $|T_j|=1$, the series--parallel reduction~\eqref{eq-kernel-series-parallel} gives $r_2\lambda_{v,+\dot a}y_{AB,+}^{\mathrm{red}}{}^{\dot a}=0$. Every nonzero component must therefore carry a source.

An internal cycle in $T_j$ would force $|I_j|=|T_j|$, no $\bar W$ vertex, and $|E_j|=|T_j|$; its cubic vertices would then have $2|I_j|+|E_j|=3|T_j|$ contracted terminals and hence no surviving field to supply a source. Thus every nonzero component $T_j$ is a tree, proving~\eqref{eq-1-19}. Since no vertex can carry two parallel spokes to the same insertion, $|E_j|=|T_j|+1$ implies that $T_j$ meets both $A$ and $B$.

Each physical contraction consumes one antifield. There are $n-\#(\bar W\text{ vertices})$ interaction antifields and $2n$ physical contractions, giving~\eqref{eq-product-antifield-loss}. Finally, summing full graph degrees over $T_j$ gives
\begin{equation}
\sum_{v\in T_j\,\mathrm{(vertices)}}\deg_\Gamma v
=2|I_j|+|E_j|=2(|T_j|-1)+(|T_j|+1)=3|T_j|-1.
\end{equation}
Every surviving vertex has $\deg_\Gamma v\ge2$. When all vertices of $T_j$ are cubic ($\deg_\Gamma v\le3$), exactly one vertex has $\deg_\Gamma v=2$ and all others have $\deg_\Gamma v=3$, leaving a single surviving field at the degree-two vertex. With one $\bar W$ vertex, the same degree sum gives~\eqref{eq-1-23}; for $\deg\bar W=2$, all terminals are contracted and the passive-network vanishing applies.~\hfill\qedsymbol
\end{proofbox}

\begin{figure}[H]
\centering
\includegraphics[width=.94\linewidth]{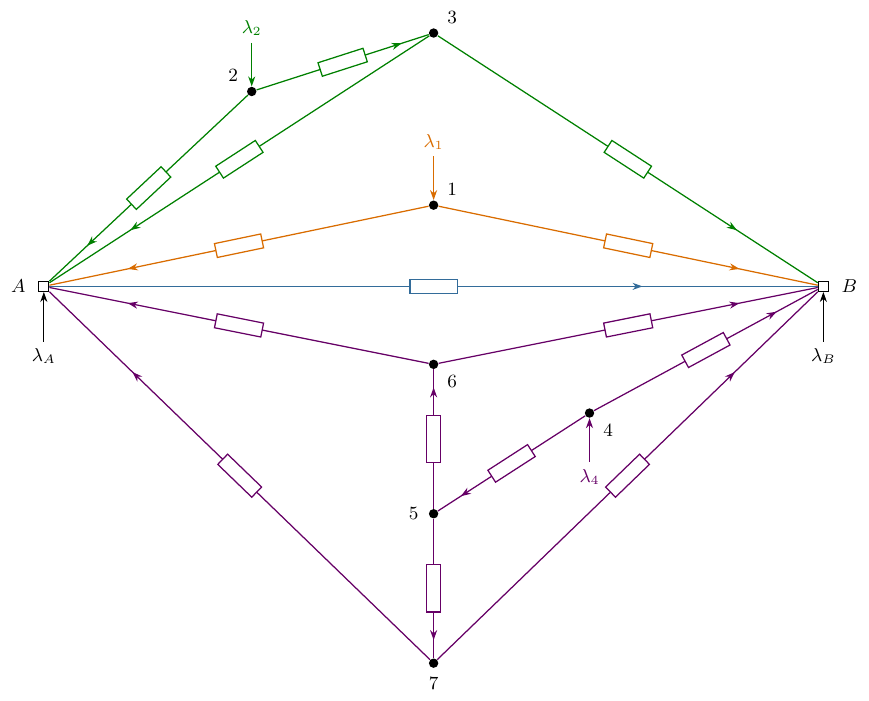}
\caption{A seven-loop product graph with three trees between $A$ and $B$; the marked edge $AB$ is blue. Each tree has one source, at its unique vertex of degree two.}\label{fig-forest-components}
\end{figure}

\subsection{Reduction of product forests}\label{sec-forest-graphs}

\begin{figure}[H]
\centering
\includegraphics[width=.98\linewidth]{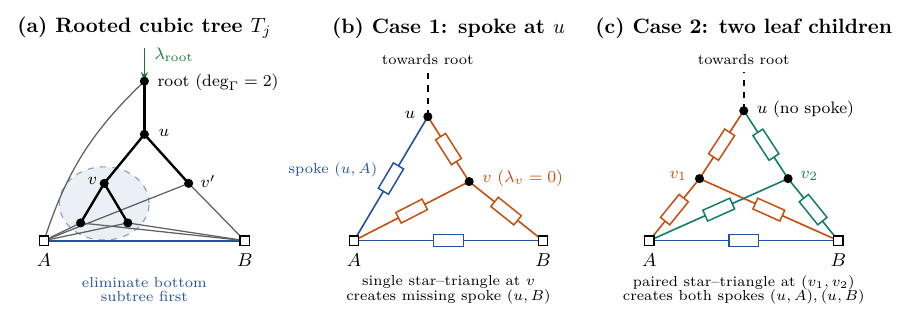}
\caption{Bottom-up reduction of a rooted cubic product tree $T_j$. (a) Rooting $T_j$ at its unique degree-two vertex ($\lambda_{\mathrm{root}}$) and eliminating descendants upward leaves source-free degree-three leaves joined to $A$ and $B$. (b) When the parent $u$ carries a spoke, a single star--triangle move at $v$ (Theorem~\ref{b-01159}) creates the missing spoke on $u$. (c) When $u$ carries no spoke, a paired star--triangle move on its two leaf children $(v_1,v_2)$ (Theorem~\ref{b-01173}) creates both spokes on $u$.}\label{fig-ch7-rooted-tree-idea}
\end{figure}

In a cubic product forest (Figure~\ref{fig-forest-components}), root each component tree $T_j$ at its unique vertex of full graph degree two (which carries the tree's surviving source $\lambda_{\mathrm{root}}$). Every non-root leaf $v$ of $T_j$ has one internal edge $(v,u)$ to its parent $u$ and two spokes $(v,A)$ and $(v,B)$, so $v$ is source-free of full graph degree three. Consequently, $v$ cannot be removed by a series--parallel move, which requires degree two. Instead, two local configurations arise when eliminating leaves from the bottom up (Figure~\ref{fig-ch7-rooted-tree-idea}):
\begin{enumerate}
\item if the parent $u$ has a single child $v$ in $T_j$, then $u$ already carries one insertion spoke $(u,x)$ ($x\in\{A,B\}$), so eliminating the source-free star at $v$ merges two triangle edges in parallel with $(u,x)$ and $(A,B)$ and creates the missing spoke $(u,w)$ (Theorem~\ref{b-01159});
\item if $u$ has two leaf children $v_1,v_2$ in $T_j$ and hence no insertion spoke, eliminating both stars at $v_1,v_2$ simultaneously creates both spokes $(u,A)$ and $(u,B)$ and merges their $(A,B)$ branches in parallel with $(A,B)$ (Theorem~\ref{b-01173}).
\end{enumerate}
In both reductions below, $T^{\mathrm{red}}$ denotes a chosen spanning tree of the reduced graph $\Gamma^{\mathrm{red}}$.

\begin{figure}[H]
\centering
\includegraphics[width=.98\linewidth]{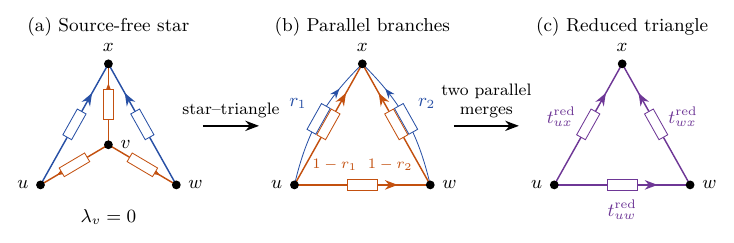}
\caption{A source-free star followed by two parallel merges; $r_1,r_2$ are the original-edge current fractions.}\label{fig-star-triangle-free}
\end{figure}

\rthm{Theorem}{b-01159}{Star--triangle}
In a Laman graph, let $v$ be a source-free vertex of degree three with neighbours $u,w,x$, with $ux,wx$ present
and $uw$ absent. The star--triangle move and the two parallel merges
remove one vertex and one loop, leaving a Laman graph (Figure~\ref{fig-star-triangle-free}). Order the local edges as $(vu,vw,vx,ux,wx)$ before the move and $(uw,ux,wx)$ after it, followed by the other edges in their original order.
If $r_1,r_2\in(0,1)$ are the current
fractions carried by the original edges $ux,wx$, the reverse current map is
\begin{equation}
\begin{aligned}
y_{vu}&=-y_{uw}^{\mathrm{red}}-(1-r_1)y_{ux}^{\mathrm{red}},\\
y_{vw}&=y_{uw}^{\mathrm{red}}-(1-r_2)y_{wx}^{\mathrm{red}},\\
y_{vx}&=(1-r_1)y_{ux}^{\mathrm{red}}+(1-r_2)y_{wx}^{\mathrm{red}},\\
y_{ux}&=r_1y_{ux}^{\mathrm{red}},\qquad
y_{wx}=r_2y_{wx}^{\mathrm{red}}.
\end{aligned}
\label{b-01166}
\end{equation}
Choose $T^{\mathrm{red}}$ containing $ux,wx$. Then
\begin{equation}
\begin{aligned}
\mathcal I_\Gamma(\lambda;z)
={}&\int_0^1dr_1\int_0^1dr_2
\int_{\substack{y^{\mathrm{red}}\in\Delta_{\Gamma^{\mathrm{red}}}\\
\mathrm{current\ conservation\ at}\ \lambda}}
r_1r_2y_{ux,+\dot a}^{\mathrm{red}}y_{wx,+}^{\mathrm{red}}{}^{\dot a}\\
&\quad\times
e^{-\sum_{e\in I\,\mathrm{(edges)}}y_{e,+\dot b}z_e^{+\dot b}}
\bigwedge_{e\in I^{\mathrm{red}}\setminus T^{\mathrm{red}}}
(-d^2y_e^{\mathrm{red}}).
\end{aligned}
\label{eq-kernel-star-triangle}
\end{equation}

\begin{proofbox}
\textbf{Proof.} Only vertex subsets $S$ containing $u,w$ gain an edge, and
\begin{equation}\label{eq-star-laman-count}
|I_S^{\mathrm{red}}|
\le |I_{S\cup\{v,x\}}|-2|\{v,x\}\setminus S|
\le2|S|-3.
\end{equation}
The total counts decrease by one vertex and two edges.
With $S_v=t_{vu}t_{vw}+t_{vu}t_{vx}+t_{vw}t_{vx}$, the star--triangle resistance transformation replaces the star $(t_{vu},t_{vw},t_{vx})$ by triangle resistances $(S_v/t_{vx},S_v/t_{vw},S_v/t_{vu})$ on $(uw,ux,wx)$, and the parallel merges~\eqref{eq-1-63} with the original edges $ux,wx$ yield~\eqref{b-01166}. Taking $T=(T^{\mathrm{red}}\setminus\{ux,wx\})\cup\{vu,vw,vx\}$ as spanning tree before the move, current conservation at fixed sources implies
\begin{equation}
dy_{ux,+\dot a}
\equiv y_{ux,+\dot a}^{\mathrm{red}}dr_1
-r_1dy_{uw,+\dot a}^{\mathrm{red}},\qquad
dy_{wx,+\dot a}
\equiv y_{wx,+\dot a}^{\mathrm{red}}dr_2
+r_2dy_{uw,+\dot a}^{\mathrm{red}},
\end{equation}
modulo the other retained chord differentials. Hence
\begin{equation}
\begin{aligned}
(-d^2y_{ux})\wedge(-d^2y_{wx})
\equiv{}&-r_1r_2
(y_{ux,+\dot a}^{\mathrm{red}}y_{wx,+}^{\mathrm{red}}{}^{\dot a})\\
&\quad\times dr_1\wedge dr_2\wedge(-d^2y_{uw}^{\mathrm{red}}).
\end{aligned}
\label{eq-star-current-wedge}
\end{equation}
In projective Schwinger gauge $t_{wx}=1$, the forward resistance map $(t_{vu},t_{vw},t_{vx},t_{ux})\mapsto(r_1,r_2,t_{uw}^{\mathrm{red}}/t_{wx}^{\mathrm{red}},t_{ux}^{\mathrm{red}}/t_{wx}^{\mathrm{red}})$---with $r_1=S_v/(S_v+t_{vw}t_{ux})$, $r_2=S_v/(S_v+t_{vu})$, $t_{uw}^{\mathrm{red}}/t_{wx}^{\mathrm{red}}=(S_v+t_{vu})/t_{vx}$, and $t_{ux}^{\mathrm{red}}/t_{wx}^{\mathrm{red}}=t_{ux}(S_v+t_{vu})/(S_v+t_{vw}t_{ux})$---has Jacobian determinant
\begin{equation*}
\det\frac{\partial\bigl(r_1,r_2,t_{uw}^{\mathrm{red}}/t_{wx}^{\mathrm{red}},t_{ux}^{\mathrm{red}}/t_{wx}^{\mathrm{red}}\bigr)}{\partial(t_{vu},t_{vw},t_{vx},t_{ux})}=-\frac{t_{ux}S_v^2}{t_{vx}^2(S_v+t_{vw}t_{ux})^3}<0\qquad\text{on }\mathbb R_{>0}^4,
\end{equation*}
so the resistance change reverses orientation and cancels the minus sign in~\eqref{eq-star-current-wedge}.~\hfill\qedsymbol
\end{proofbox}

\begin{figure}[H]
\centering
\includegraphics[width=.98\linewidth]{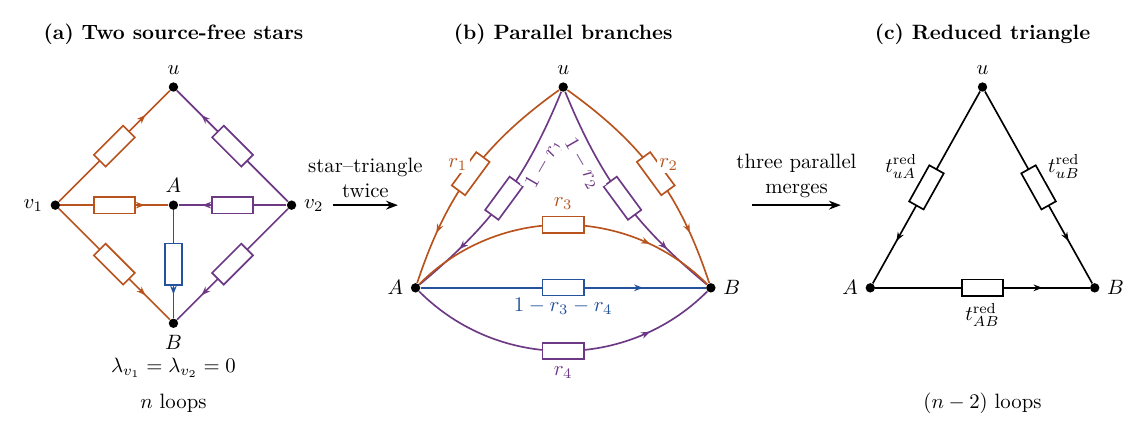}
\caption{Two source-free stars followed by three parallel merges.}\label{fig-ch7-paired-reduction}
\end{figure}

\rthm{Theorem}{b-01173}{Paired star--triangle}
In a Laman graph, let source-free vertices $v_1,v_2$ each have degree three and be joined to $u,A,B$, with $AB$
present and $uA,uB$ absent. The two star--triangle moves and three parallel
merges remove two vertices and two loops, leaving a Laman graph (Figure~\ref{fig-ch7-paired-reduction}). The local edge orders are $(v_1u,v_1A,v_1B,v_2u,v_2A,v_2B,AB)$ and $(uA,uB,AB)$ before and after the move, followed by the other edges in their original order.
Let $r_1,r_2$ be the current
fractions contributed by the $v_1$ star to $uA,uB$, and $r_3,r_4$ the
fractions contributed by $v_1,v_2$ to $AB$. Their domain is
$0<r_1,r_2<1$, $r_3,r_4>0$, $r_3+r_4<1$, and
\begin{equation}
\begin{aligned}
y_{v_1u}&=-r_1y_{uA}^{\mathrm{red}}-r_2y_{uB}^{\mathrm{red}},&
y_{v_2u}&=-(1-r_1)y_{uA}^{\mathrm{red}}-(1-r_2)y_{uB}^{\mathrm{red}},\\
y_{v_1A}&=r_1y_{uA}^{\mathrm{red}}-r_3y_{AB}^{\mathrm{red}},&
y_{v_2A}&=(1-r_1)y_{uA}^{\mathrm{red}}-r_4y_{AB}^{\mathrm{red}},\\
y_{v_1B}&=r_2y_{uB}^{\mathrm{red}}+r_3y_{AB}^{\mathrm{red}},&
y_{v_2B}&=(1-r_2)y_{uB}^{\mathrm{red}}+r_4y_{AB}^{\mathrm{red}},\\
y_{AB}&=(1-r_3-r_4)y_{AB}^{\mathrm{red}}.
\end{aligned}
\label{b-01180}
\end{equation}
Choose $T^{\mathrm{red}}$ containing $uA,uB$. The reduced chord $AB$
remains in the area wedge, and
\begin{equation}
\begin{aligned}
\mathcal I_\Gamma(\lambda;z)
={}&\int_0^1dr_1\int_0^1dr_2
\int_{\substack{r_3,r_4>0\\r_3+r_4<1}}dr_3\,dr_4
\int_{\substack{y^{\mathrm{red}}\in\Delta_{\Gamma^{\mathrm{red}}}\\
\mathrm{current\ conservation\ at}\ \lambda}}
(r_1-r_2)(1-r_3-r_4)\\
&\quad\times
(y_{uA,+\dot a}^{\mathrm{red}}y_{AB,+}^{\mathrm{red}}{}^{\dot a})
(y_{uB,+\dot b}^{\mathrm{red}}y_{AB,+}^{\mathrm{red}}{}^{\dot b})\\
&\quad\times
e^{-\sum_{e\in I\,\mathrm{(edges)}}y_{e,+\dot c}z_e^{+\dot c}}
\bigwedge_{e\in I^{\mathrm{red}}\setminus T^{\mathrm{red}}}
(-d^2y_e^{\mathrm{red}}).
\end{aligned}
\label{eq-kernel-two-leaves}
\end{equation}

\begin{proofbox}
\textbf{Proof.} Every vertex subset $S$ containing a newly created edge satisfies
\begin{equation}\label{eq-paired-laman-count}
|I_S^{\mathrm{red}}|
\le |I_{S\cup\{v_1,v_2,A,B\}}|-2|\{v_1,v_2,A,B\}\setminus S|
\le2|S|-3.
\end{equation}
Other subsets are unchanged, and the total counts decrease by two vertices and four edges.
Applying the star--triangle transformation independently at $v_1$ and $v_2$ followed by the parallel merges on $(uA,uB,AB)$ gives~\eqref{b-01180}.
Taking $T=(T^{\mathrm{red}}\setminus\{uA,uB\})\cup\{v_1u,v_1A,v_2u,v_2B\}$ before the move leaves local chords $(v_1B,v_2A,AB)$; modulo the other retained chord differentials, conservation gives
\begin{equation}
\begin{aligned}
dy_{AB}
&=(1-r_3-r_4)dy_{AB}^{\mathrm{red}}
-y_{AB}^{\mathrm{red}}(dr_3+dr_4),\\
dy_{v_1B}
&\equiv y_{uB}^{\mathrm{red}}dr_2+y_{AB}^{\mathrm{red}}dr_3
+(r_3-r_2)dy_{AB}^{\mathrm{red}},\\
dy_{v_2A}
&\equiv-y_{uA}^{\mathrm{red}}dr_1-y_{AB}^{\mathrm{red}}dr_4
+(1-r_1-r_4)dy_{AB}^{\mathrm{red}}.
\end{aligned}
\end{equation}
Wedging $(-d^2y_{v_1B})\wedge(-d^2y_{v_2A})\wedge(-d^2y_{AB})$ produces $(y_{uA,+\dot a}^{\mathrm{red}}y_{AB,+}^{\mathrm{red}}{}^{\dot a})(y_{uB,+\dot b}^{\mathrm{red}}y_{AB,+}^{\mathrm{red}}{}^{\dot b})\,dr_1\wedge dr_2\wedge dr_3\wedge dr_4\wedge(-d^2y_{AB}^{\mathrm{red}})$ multiplied by the scalar factor
\begin{equation}
\begin{aligned}
&(1-r_3-r_4)^2
-(1-r_3-r_4)(1-r_1-r_4)
+(1-r_3-r_4)(r_3-r_2)\\
&\qquad=(r_1-r_2)(1-r_3-r_4).
\end{aligned}
\end{equation}
Since the two star--triangle maps on $(v_1u,v_1A,v_1B)$ and $(v_2u,v_2A,v_2B)$ each reverse orientation once and the three parallel merges preserve orientation, the composite resistance map $(t_{v_1u},t_{v_1A},t_{v_1B},t_{v_2u},t_{v_2A},t_{v_2B})\mapsto(r_1,r_2,r_3,r_4,t_{uA}^{\mathrm{red}}/t_{AB}^{\mathrm{red}},t_{uB}^{\mathrm{red}}/t_{AB}^{\mathrm{red}})$ at $t_{AB}=1$ has strictly positive Jacobian determinant, preserving orientation.~\hfill\qedsymbol
\end{proofbox}

\rthm{Theorem}{b-00405}{Cubic product kernels}
Every cubic product forest reduces to the marked resistor by the
series--parallel move~\eqref{eq-kernel-series-parallel} and the two source-free star--triangle moves~\eqref{eq-kernel-star-triangle} and~\eqref{eq-kernel-two-leaves}.
The inverse maps reconstruct all currents polynomially in $2n$ interval
and simplex coordinates. The current-area density is polynomial, and
every displacement coefficient is a finite sum of interval and simplex
moments.

\begin{proofbox}
\textbf{Proof.} Root each tree $T_j$ at its unique vertex of full degree two and eliminate descendants from the leaves upward. A non-root vertex $v$, once its descendants have been removed, is source-free and joined to its parent $u$ and to both insertions $A,B$.
If $u$ already carries an insertion spoke, the single star--triangle move (Theorem~\ref{b-01159}) removes $v$ and supplies the second spoke on $u$. If $u$ has no insertion spoke, it has two children in $T_j$; once their descendants are removed, the paired star--triangle move (Theorem~\ref{b-01173}) removes both children at once and supplies $(u,A)$ and $(u,B)$. In either case $u$ becomes joined to both $A$ and $B$. Once all non-root vertices of $T_j$ are removed, the root is joined to $A$ and $B$ and is eliminated by~\eqref{eq-kernel-series-parallel} (with the terminal parallel merge across all trees performed as in~\eqref{eq-spider-terminal-wedge}).

Each move removes one loop and introduces two coordinates, or removes two loops and introduces four coordinates, with inverse current map linear in the retained currents and polynomial in the new fractions. At the final marked edge $(A,B)$, total current conservation fixes the current in terms of the sources. Composing the inverse maps therefore expresses every original current and the full Jacobian density as polynomials in $2n$ interval and simplex variables. Expanding the displacement exponential reduces every jet coefficient to standard simplex integrals, evaluated for $k_j\in\mathbb Z_{\ge0}$ by
\begin{equation}\label{eq-Dirichlet}
\int_{\substack{r_j>0\\\sum_{j\,\mathrm{(coordinates)}}r_j<1}}
\prod_{j=1}^{\dim\Delta}r_j^{k_j}\,dr_j
=\frac{\prod_{j=1}^{\dim\Delta}k_j!}
{\left(\dim\Delta+\sum_{j\,\mathrm{(coordinates)}}k_j\right)!}.
\end{equation}
This completes the reduction.~\hfill\qedsymbol
\end{proofbox}

Higher-degree superpotentials can leave several sources in one tree, as counted in~\eqref{eq-1-23}.
Appendix~\ref{app-sourced-star} gives the sourced star--triangle move; it reconstructs the currents rationally in the remaining resistances.

\subsection{Three-loop product forests and star--triangle examples}\label{sec-product-four-loops}

Through two loops, every tree in $\Gamma\setminus\{A,B\}$ has at most two interaction vertices and therefore reduces to a fan kernel: the resulting one-loop triangle and three two-loop graphs ($\Gamma_{\mathrm{fan}}$, $\Gamma_A$, $\Gamma_B$) for $A\star B$ in $\mathcal N=4$ Yang--Mills theory are displayed in~\eqref{eq-2-34} and~\eqref{eq-product2A} of Section~\ref{sec-product-loops}.

Three loops ($n=3$) is the first order at which Theorems~\ref{b-01159} and~\ref{b-01173} become essential. Ordering the interaction vertices $1,2,3$ depth-first and setting $A=4$, $B=5$ with lexicographic edge order (and marked edge $(4,5)$ last, matching the companion tables~\cite{Repo}), there are eight connected three-loop product-forest topologies: three multi-tree forests assembled from one- and two-loop fan trees, and five single-tree graphs with a single surviving source $\lambda_1$ at the degree-two root $1$ (Table~\ref{tab-ch7-three-loop-forests}).

{\footnotesize\renewcommand{\arraystretch}{1.35}\setlength{\tabcolsep}{3.8pt}
\newcommand{\forestcell}[1]{\rule{0pt}{10.5mm}\raisebox{-.45\height}{#1}\rule[-9.5mm]{0pt}{0pt}}
\begin{xltabular}{\linewidth}{|l|c|l|c|>{\raggedright\arraybackslash}X|>{\raggedright\arraybackslash}p{3.6cm}|}
\caption{The eight three-loop cubic product-forest topologies (bold black: internal tree edges; grey: spokes to $A,B$; blue: marked edge $(A,B)$; orange arrows: surviving root sources), their symmetry factors $|\operatorname{Aut}F|$ fixing $A,B$, reduction moves, and zero-displacement kernels $\mathcal I_\Gamma(\lambda;0)$ in lexicographic edge order (with $\langle u,v\rangle=u_{+\dot a}v_+^{\dot a}$).}\label{tab-ch7-three-loop-forests}\\
\hline
$F$ & Diagram & Trees in $\Gamma\setminus\{A,B\}$ & $|\operatorname{Aut}F|$ & Reduction moves & Zero-displacement $\mathcal I_\Gamma(\lambda;0)$\\
\hline\endfirsthead
\hline
$F$ & Diagram & Trees in $\Gamma\setminus\{A,B\}$ & $|\operatorname{Aut}F|$ & Reduction moves & Zero-displacement $\mathcal I_\Gamma(\lambda;0)$\\
\hline\endhead
$3.1$ & \forestcell{\includegraphics[width=2.55cm]{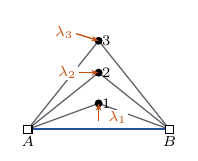}} & $(A,B),\,(A,B),\,(A,B)$ & $6$ & Three series--parallel moves~\eqref{eq-kernel-series-parallel} & Three-root polynomial in $\lambda_1,\lambda_2,\lambda_3,\lambda_A$\\
\hline
$3.2$ & \forestcell{\includegraphics[width=2.55cm]{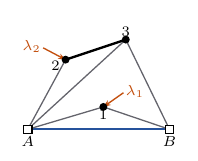}} & $(A,B),\,(A,(A,B))$ & $1$ & Fan + series--parallel~\eqref{eq-kernel-series-parallel} & Two-root polynomial in $\lambda_1,\lambda_2,\lambda_A$\\
\hline
$3.3$ & \forestcell{\includegraphics[width=2.55cm]{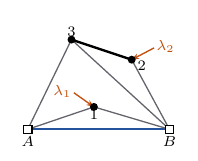}} & $(A,B),\,(B,(A,B))$ & $1$ & Fan + series--parallel~\eqref{eq-kernel-series-parallel} & Two-root polynomial in $\lambda_1,\lambda_2,\lambda_A$\\
\hline
$3.4$ & \forestcell{\includegraphics[width=2.55cm]{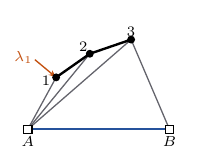}} & $(A,(A,(A,B)))$ & $1$ & Three-loop fan (Theorem~\ref{thm-spider-kernel}, $a=1$) & $+\frac{1}{864}\langle\lambda_1,\lambda_A\rangle^3$\\
\hline
$3.5$ & \forestcell{\includegraphics[width=2.55cm]{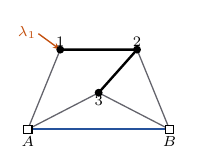}} & $(A,(B,(A,B)))$ & $1$ & Two single star--triangle moves (Theorem~\ref{b-01159}) & $-\frac{1}{576}\langle\lambda_1,\lambda_A\rangle^3$\\
\hline
$3.6$ & \forestcell{\includegraphics[width=2.55cm]{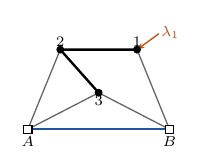}} & $(B,(A,(A,B)))$ & $1$ & Two single star--triangle moves (Theorem~\ref{b-01159}) & $-\frac{1}{576}\langle\lambda_1,\lambda_A\rangle^3$\\
\hline
$3.7$ & \forestcell{\includegraphics[width=2.55cm]{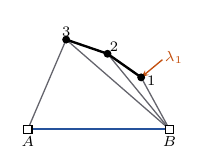}} & $(B,(B,(A,B)))$ & $1$ & Three-loop fan (Theorem~\ref{thm-spider-kernel}, $a=1$) & $+\frac{1}{864}\langle\lambda_1,\lambda_A\rangle^3$\\
\hline
$3.8$ & \forestcell{\includegraphics[width=2.55cm]{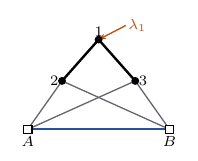}} & $((A,B),(A,B))$ & $2$ & Paired star--triangle move (Theorem~\ref{b-01173}) & $0$ (first nonvanishing at $O(z)$)\\
\hline
\end{xltabular}}

We evaluate the two genuinely non-fan single-tree topologies explicitly, together with their step-by-step circuit reductions, to illustrate Theorems~\ref{b-01159} and~\ref{b-01173}:
\begin{enumerate}
\item \emph{Alternating three-vertex chain $(A,(B,(A,B)))$ via Theorem~\ref{b-01159} (Figure~\ref{fig-ch7-case1-circuit}).} The ordered edge list is $(12,1A,23,2B,3A,3B,AB)$, with root $1$ carrying $\lambda_1$ and vertices $2,3$ source-free. Eliminating leaf $3$ (joined to $u=2$, $w=A$, $x=B$) by Theorem~\ref{b-01159} with fractions $(r_1,r_2)\in(0,1)^2$ creates edge $2A$, and eliminating vertex $2$ (now joined to $u=1$, $w=B$, $x=A$) by Theorem~\ref{b-01159} with fractions $(r_3,r_4)\in(0,1)^2$ reduces the graph to the one-loop triangle $(1A,1B,AB)$ with simplex coordinates $(r_5,r_6)\in\Delta^2$ as in~\eqref{eq-triangle-current}--\eqref{eq-1-25} ($y_{1A}^{(2)}=r_5\lambda_1-r_6\lambda_A$, $y_{1B}^{(2)}=(1-r_5)\lambda_1+r_6\lambda_A$, $y_{AB}^{(2)}=r_5\lambda_1+(1-r_6)\lambda_A$).

\begin{figure}[H]
\centering
\includegraphics[width=.96\linewidth]{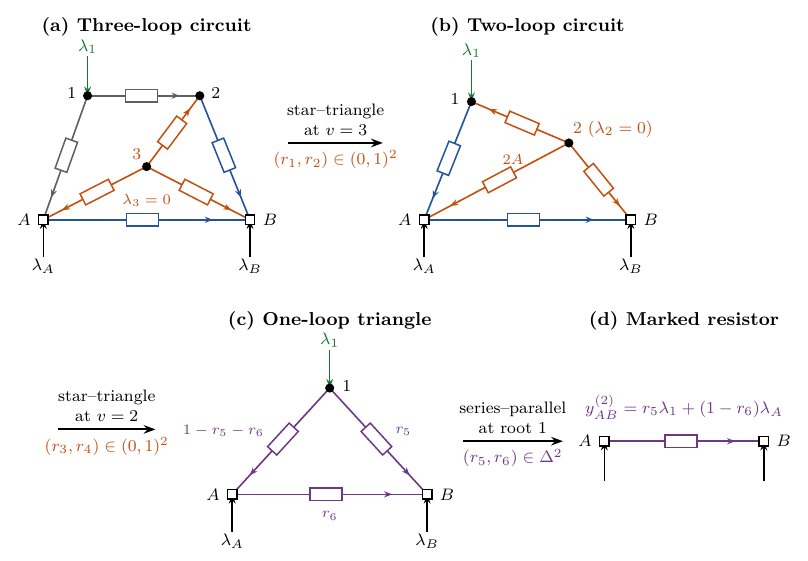}
\caption{Circuit reduction of the alternating three-loop chain $(A,(B,(A,B)))$ (forest $3.5$): two successive source-free star--triangle moves at $v=3$ and $v=2$ (orange stars, merged against blue edges) leave the reduced one-loop triangle $(1,A,B)$ (purple), followed by series--parallel elimination of root $1$.}\label{fig-ch7-case1-circuit}
\end{figure}

Tracking the two edge-order transpositions (whose minus signs cancel) and multiplying the two star--triangle Jacobians $r_1r_2(y_{2B,+\dot a}^{(1)}y_{AB,+}^{(1)\dot a})=r_1r_2r_4(1-r_5-r_6)(\lambda_{1,+\dot a}\lambda_{A,+}{}^{\dot a})$ and $r_3r_4(y_{1A,+\dot b}^{(2)}y_{BA,+}^{(2)\dot b})=-r_3r_4r_5(\lambda_{1,+\dot b}\lambda_{A,+}{}^{\dot b})$ with the triangle area $(\lambda_{1,+\dot c}\lambda_{A,+}{}^{\dot c})\,dr_5\wedge dr_6$ yields
\begin{equation}\label{eq-ch7-alt-chain-kernel}
\begin{aligned}
\mathcal I_{(A,(B,(A,B)))}(\lambda;z)
={}&-(\lambda_{1,+\dot a}\lambda_{A,+}{}^{\dot a})^3
\int_{(0,1)^4}dr_1\,dr_2\,dr_3\,dr_4
\int_{\substack{r_5,r_6>0\\r_5+r_6<1}}dr_5\,dr_6\\
&\quad\times r_1r_2r_3r_4^2\,r_5(1-r_5-r_6)\,
e^{-\sum_{e\in I}y_{e,+\dot d}(r)z_e^{+\dot d}},
\end{aligned}
\end{equation}
where the external spoke currents reconstructed by~\eqref{b-01166} are $y_{1A}=r_3y_{1A}^{(2)}$, $y_{2B}=r_1\bigl(y_{1B}^{(2)}+(1-r_4)y_{AB}^{(2)}\bigr)$, $y_{3A}=(1-r_3)y_{1A}^{(2)}-(1-r_2r_4)y_{AB}^{(2)}$, and $y_{3B}=(1-r_1)y_{1B}^{(2)}+\bigl(1-r_1-(1-r_1-r_2)r_4\bigr)y_{AB}^{(2)}$. At $z=0$, integrating $(r_1,r_2,r_3,r_4)$ gives $\frac12\cdot\frac12\cdot\frac12\cdot\frac13=\frac{1}{24}$ and~\eqref{eq-Dirichlet} gives $\int_{\Delta^2}r_5(1-r_5-r_6)\,dr_5\,dr_6=\frac{1}{24}$, producing a rational denominator outside the fan sequence $\frac{1}{(n!)^3(n+1)}$:
\begin{equation}\label{eq-product-four-loop-evaluated}
\mathcal I_{(A,(B,(A,B)))}(\lambda;0)
=-\frac{1}{576}(\lambda_{1,+\dot a}\lambda_{A,+}{}^{\dot a})^3,\qquad
\mathcal I_{((A,B),(A,B))}(\lambda;0)=0.
\end{equation}

\item \emph{Symmetric three-vertex fork $((A,B),(A,B))$ via Theorem~\ref{b-01173} (Figure~\ref{fig-ch7-case2-circuit}).} Here the root $1$ (carrying $\lambda_1$) has two source-free leaf children $2,3$, with ordered edge list $(12,13,2A,2B,3A,3B,AB)$. Moving $13$ past $(2A,2B)$ is an even permutation and brings the graph directly into the local edge order of Theorem~\ref{b-01173} at $(v_1,v_2,u)=(2,3,1)$, reducing it in a single step to the triangle $(1A,1B,AB)$ with simplex coordinates $(r_5,r_6)\in\Delta^2$.

\begin{figure}[H]
\centering
\includegraphics[width=.98\linewidth]{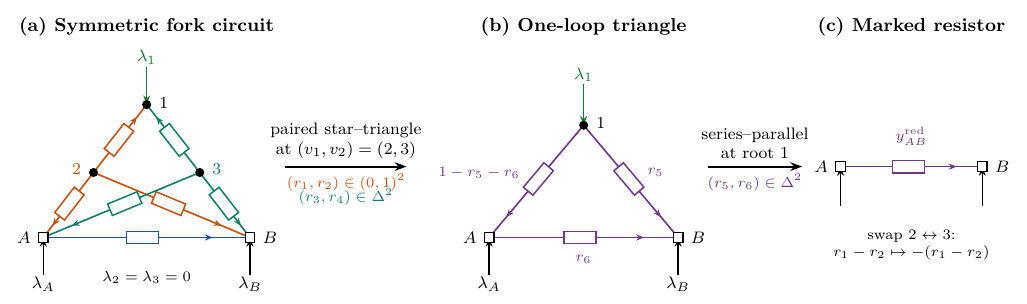}
\caption{Circuit reduction of the symmetric three-loop fork $((A,B),(A,B))$ (forest $3.8$): the paired star--triangle move eliminates both source-free leaf stars at $2$ (orange) and $3$ (teal) simultaneously to form the one-loop triangle $(1,A,B)$ (purple), followed by series--parallel elimination of root $1$.}\label{fig-ch7-case2-circuit}
\end{figure}

Since $y_{1A,+\dot a}^{\mathrm{red}}y_{AB,+}^{\mathrm{red}}{}^{\dot a}=r_5(\lambda_{1,+\dot a}\lambda_{A,+}{}^{\dot a})$ and $y_{1B,+\dot b}^{\mathrm{red}}y_{AB,+}^{\mathrm{red}}{}^{\dot b}=(1-r_5-r_6)(\lambda_{1,+\dot b}\lambda_{A,+}{}^{\dot b})$, equation~\eqref{eq-kernel-two-leaves} gives the six-parameter representation
\begin{equation}\label{eq-ch7-fork-kernel}
\begin{aligned}
\mathcal I_{((A,B),(A,B))}(\lambda;z)
={}&(\lambda_{1,+\dot a}\lambda_{A,+}{}^{\dot a})^3
\int_{(0,1)^2}dr_1\,dr_2
\int_{\substack{r_3,r_4>0\\r_3+r_4<1}}dr_3\,dr_4
\int_{\substack{r_5,r_6>0\\r_5+r_6<1}}dr_5\,dr_6\\
&\quad\times (r_1-r_2)(1-r_3-r_4)\,r_5(1-r_5-r_6)\,
e^{-\sum_{e\in I}y_{e,+\dot d}(r)z_e^{+\dot d}},
\end{aligned}
\end{equation}
with spoke currents $y_{2A}=r_1y_{1A}^{\mathrm{red}}-r_3y_{AB}^{\mathrm{red}}$, $y_{2B}=r_2y_{1B}^{\mathrm{red}}+r_3y_{AB}^{\mathrm{red}}$, $y_{3A}=(1-r_1)y_{1A}^{\mathrm{red}}-r_4y_{AB}^{\mathrm{red}}$, and $y_{3B}=(1-r_2)y_{1B}^{\mathrm{red}}+r_4y_{AB}^{\mathrm{red}}$. At zero displacement ($z=0$), the factor $(r_1-r_2)$ is odd under $r_1\leftrightarrow r_2$ (reflecting the orientation reversal under exchanging the two identical leaves $2\leftrightarrow 3$), so $\mathcal I_{((A,B),(A,B))}(\lambda;0)=0$ identically as recorded in~\eqref{eq-product-four-loop-evaluated}. The symmetric fork first contributes when the inputs carry at least one holomorphic derivative; for example, differentiating with respect to $z_{2A}^{+\dot c}$ at $z=0$ inserts $-y_{2A,+\dot c}$ and gives
\begin{equation}\label{eq-ch7-fork-first-jet}
\left.-\frac{\partial}{\partial z_{2A}^{+\dot c}}
\mathcal I_{((A,B),(A,B))}(\lambda;z)\right|_{z=0}
=\frac{1}{8640}(\lambda_{1,+\dot a}\lambda_{A,+}{}^{\dot a})^3\,
\bigl(2\lambda_{1,+\dot c}-\lambda_{A,+\dot c}\bigr).
\end{equation}
\end{enumerate}
The complete polynomial tree charts and assembled forest moments for $A\star B$ through seven loops ($1,3,8,23,65,194,576$ product forests at $n=1,\ldots,7$), together with explicit Leibniz-rule verifications of~\eqref{eq-chain} through four loops, are archived in the companion repository~\cite{KernelRepo}.

\section{Quantum supercharges and \texorpdfstring{$\mathcal N=4$}{N=4} Yang--Mills theory}\label{sec-supercharge}

Combining the scalar graph kernels of Sections~\ref{sec-BV}--\ref{sec-fans} with the BV fibre contractions yields the loop-corrected supersymmetry transformation in any holomorphically twisted $\mathcal N=1$ gauge theory. Let $W_\Gamma[\mathcal I_1,\ldots,\mathcal I_{n+1},\mathcal O]$ denote the ordered Wick contraction~\eqref{eq-W-fibre-explicit} of $n+1$ interaction vertices $\mathcal I_1,\ldots,\mathcal I_{n+1}$ with a local operator $\mathcal O(x)$, including Koszul signs and with edge displacements $z_e$ acting on input letters as in~\eqref{eq-W-jet-edge}. Restoring the scalar propagator normalization~\eqref{eq-1-7} gives
\begin{equation}\label{eq-2-3}
\begin{aligned}
Q_n\mathcal O
={}&\frac{(-1)^n}{(n+1)!\,(8\pi^2)^n}
\sum_{\substack{\Gamma\ \mathrm{labelled,\ rooted,\ Laman}}}
\Bigl[\mathcal I_\Gamma(\lambda;z)\,
W_\Gamma[\mathcal I_1,\ldots,\mathcal I_{n+1},\mathcal O]\Bigr].
\end{aligned}
\end{equation}
At $n=0$, the single-edge tree gives $Q_0\mathcal O=\{\mathcal I,\mathcal O\}$, reproducing the classical supersymmetry transformation.

Whenever the interaction $\mathcal I$ contains no holomorphic derivatives (including nonrenormalizable higher-degree superpotentials $\bar W$), $Q_n$ obeys a sharp derivative selection rule. Under the independent rescaling $\lambda_{v,+\dot a}\mapsto \kappa\,\lambda_{v,+\dot a}$ and $z_e^{+\dot a}\mapsto \kappa^{-1}z_e^{+\dot a}$ ($\kappa>0$) at fixed $\dot a\in\{\dot+,\dot-\}$, the displacement exponential in~\eqref{eq-operatope} is invariant while the current $2n$-form scales by $\kappa^n$ in each dotted spinor index~\eqref{eq-derivative-homogeneity}. Because each factor of $\lambda_{v,+\dot a}$ differentiates an output field and each $-\partial/\partial z_e^{+\dot a}$ removes a derivative from an input letter, $Q_n$ increases the counts $(d_{\dot+},d_{\dot-})$ of $(\partial_{+\dot+},\partial_{+\dot-})$ derivatives by $(n,n)$, and hence raises the total holomorphic derivative count by $2n$ (an analogous homogeneity argument holds in arbitrary even dimension). Consequently, a local operator carrying $(d_{\dot+},d_{\dot-})$ holomorphic derivatives can lie in the image of $Q_n$ only if $n\le\min(d_{\dot+},d_{\dot-})$; when the homotopy retracts preserve both derivative counts, Lemma~\ref{b-00094} imposes the same bound $n\le\min(d_{\dot+},d_{\dot-})$ for a class in $\mathrm{BPS}_{\hbar^{n-1}}$ to be $d_n$-exact (Corollary~\ref{b-00758}).

We now specialize~\eqref{eq-2-3} to our primary non-abelian application: the all-order anomalous supersymmetry transformation in four-dimensional $\mathcal N=4$ Yang--Mills theory. For the holomorphic twist with respect to $Q_-^4$, the gauge fields and the three adjoint chiral multiplets assemble into a single superfield $C^A(x,\theta)$ on $\mathbb C^{2|3}$~\cite[Section 2]{ChangYin}, \cite[Eq.~(3.52)]{BK}, whose component expansion~\eqref{eq-A-13} and cubic Chern--Simons-type twisted action~\eqref{eq-A-15} are reviewed in Appendix~\ref{sec-N4-superfield}. Functional derivatives at displaced insertion points are normalized by
\begin{equation}\label{eq-2-10}
\frac{\delta C^B(y,\theta')}{\delta C^A(y',\theta)}=\delta_A{}^B\,\delta^{(4)}(y-y')\prod_{I=1}^3(\theta_I-\theta'_I).
\end{equation}

\subsection{The all-order anomalous SUSY transformation}\label{the-all-order-operator}

\rthm{Theorem}{b-00431}{All-loop quantum supercharge in $\mathcal N=4$ Yang--Mills theory} \emph{For the twisted $\mathcal N=4$ superfield action~\eqref{eq-A-15} and any local operator inserted at $x$, the $n$-loop supercharge for $n\ge1$ is}
\begin{equation}\label{eq-2-8}
\begin{aligned}
Q_n={}&\frac{(-1)^ni^{n+1}}{2(32\pi^2)^n}
\int_{\eqref{eq-1-37}}dr_1\cdots dr_{2n}\,
\left[\prod_{j=1}^{n-1}r_{2j-1}^{n-j}r_{2j}(1-r_{2j})^{j-1}\right]r_{2n}^{n-1}
\int d^3\theta\int\prod_{j=1}^{n+1}d^4z_j\\
&\times C_{A_1DR_1}
\left(\prod_{j=2}^{n}C_{A_jR_{j-1}R_j}\right)C_{A_{n+1}R_nE}\\
&\times\big(\partial_{+\dot a_1}\cdots\partial_{+\dot a_n}C^D\big)(x+x_1(r,z),\theta)
\big(\partial_+{}^{\dot a_1}\cdots\partial_+{}^{\dot a_n}C^E\big)(x+x_{n+1}(r,z),\theta)\\
&\times\frac{\delta}{\delta C^{A_1}(x+z_1,\theta)}\cdots
\frac{\delta}{\delta C^{A_{n+1}}(x+z_{n+1},\theta)},
\end{aligned}
\end{equation}
\emph{where the endpoint superfields $C^D,C^E$ are differentiated before evaluation at the translated positions $x+x_1(r,z)$ and $x+x_{n+1}(r,z)$ given in~\eqref{eq-fan-output-points} (and Taylor-expanded to all derivative orders in~\eqref{eq-fan-full-taylor}).}

\begin{figure}[ht!]
\centering
\includegraphics[width=0.96\linewidth]{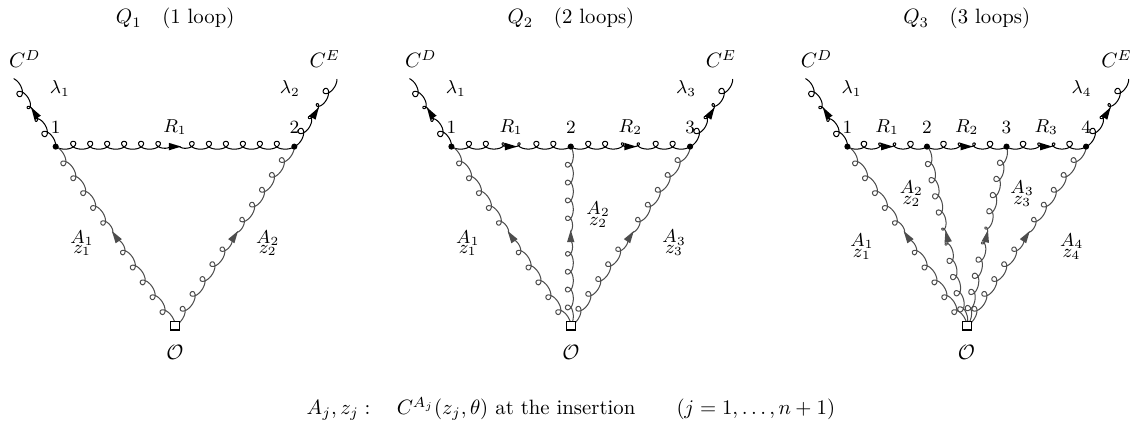}
\caption{One-, two- and three-loop supercharge fans in $\mathcal N=4$ Yang--Mills theory. The spoke labelled $A_j,z_j$ contracts an input $C^{A_j}(x+z_j,\theta)$ at $\mathcal O(x)$. The interaction path contracts the adjoint colour indices $R_1,\ldots,R_n$ and leaves the two endpoint superfields $C^D,C^E$ carrying output momenta $\lambda_1,\lambda_{n+1}$.}
\label{fig-n4-supercharge-fans}
\end{figure}

\begin{proofbox}
\textbf{Proof.} Because the interaction~\eqref{eq-A-15} is strictly cubic, Corollary~\ref{b-00247} restricts the connected $n$-loop graphs of $Q_n$ in~\eqref{eq-2-3} to rooted $n$-loop fans (Figure~\ref{fig-n4-supercharge-fans}). Each cubic vertex contributes $iC_{ABC}$ (summing the $3!$ leg assignments), the $(n+1)!$ vertex relabelings cancel the prefactor $1/(n+1)!$ in~\eqref{eq-2-3}, and reversing the interaction path $1\leftrightarrow n+1$ divides by the fan automorphism factor $2$.

The $n+1$ spokes attach to $\mathcal O(x)$ through the displaced functional derivatives in~\eqref{eq-2-8}, while the $n$ path edges contract the internal colour indices $R_1,\ldots,R_n$ in the order $C_{A_1DR_1}(\prod_{j=2}^n C_{A_jR_{j-1}R_j})C_{A_{n+1}R_nE}$. Only the two endpoint vertices $1$ and $n+1$ carry uncontracted output superfields $C^D$ and $C^E$, so $\lambda_2=\cdots=\lambda_n=0$. Substituting the operator rule~\eqref{eq-D-lambda} into the endpoint fan kernel~\eqref{eq-1-41}--\eqref{eq-fan-output-translation}, with $\partial_{1,+\dot a}$ and $\partial_{n+1,+\dot a}$ acting on $C^D(x,\theta)$ and $C^E(x,\theta)$ respectively and $\partial/\partial x^{+\dot a}=-\tfrac12\partial_{+\dot a}$, gives
\begin{equation*}
\begin{aligned}
(\lambda_{1,+\dot a}\lambda_{n+1,+}{}^{\dot a})^n
&\longmapsto 4^{-n}
\partial_{1,+\dot a_1}\cdots\partial_{1,+\dot a_n}
\partial_{n+1,+}{}^{\dot a_1}\cdots\partial_{n+1,+}{}^{\dot a_n},\\
e^{\lambda_{1,+\dot a}x_1^{+\dot a}(r,z)+
\lambda_{n+1,+\dot a}x_{n+1}^{+\dot a}(r,z)}
C^D(x,\theta)C^E(x,\theta)
&\longmapsto
C^D(x+x_1(r,z),\theta)C^E(x+x_{n+1}(r,z),\theta).
\end{aligned}
\end{equation*}
Collecting the numerical prefactors,
\begin{equation*}
\underbrace{\frac12}_{\text{path reversal}}\quad
\underbrace{i^{n+1}}_{\text{cubic vertices}}\quad
\underbrace{(-8\pi^2)^{-n}}_{\text{scalar-kernel normalization}}\quad
\underbrace{4^{-n}}_{\text{output derivatives}}
=\frac{(-1)^ni^{n+1}}{2(32\pi^2)^n},
\end{equation*}
and inserting the reduced fan density~\eqref{eq-1-41} yields~\eqref{eq-2-8}.\hfill\qedsymbol
\end{proofbox}

In explicit gauge groups such as $SU(N)$, $SO(N)$, and $Sp(N)$, the adjoint colour chain in~\eqref{eq-2-8} simplifies further via trace and completeness identities~\cite{Repo}.

When $\mathcal O(x)$ is built purely from undifferentiated letters ($\partial_{+\dot a}=0$ on all inputs), only the $z=0$ term of~\eqref{eq-2-8} contributes; substituting the zero-displacement fan integral~\eqref{eq-1-42} reduces~\eqref{eq-2-8} to the purely algebraic differential operator
\begin{equation}\label{eq-Qnzero}
\begin{aligned}
Q_n\big|_{\text{no input derivatives}}
={}&\frac{(-1)^ni^{n+1}}{2(32\pi^2)^n(n!)^3(n+1)}
\int d^3\theta\,
C_{A_1DR_1}\left(\prod_{j=2}^nC_{A_jR_{j-1}R_j}\right)
C_{A_{n+1}R_nE}\\
&\times(\partial_{+\dot a_1}\cdots\partial_{+\dot a_n}C^D)
(\partial_+{}^{\dot a_1}\cdots\partial_+{}^{\dot a_n}C^E)
\prod_{j=1}^{n+1}\frac{\delta}{\delta C^{A_j}(\theta)}.
\end{aligned}
\end{equation}
At $n=1$, rewriting $C_{A_1DR}=-C_{A_1RD}$ in~\eqref{eq-2-8} directly reproduces the one-loop operator $Q_1$ displayed in~\eqref{eq-2-16}, while~\eqref{eq-Qnzero} gives its zero-displacement prefactor $-1/(128\pi^2)\,C_{A_1RD}C_{A_2RE}$.

\subsection{One-, two- and three-loop examples in \texorpdfstring{$\mathcal N=4$}{N=4} Yang--Mills theory}\label{examples-in-mathcal-n4-yangmills-theory}

To make the general-$n$ operator~\eqref{eq-2-8} concrete, we display its two-loop specialization $Q_2$---whose explicit form has recently been sought in the holomorphic-twist literature~\cite[Section 3.2.3]{semichiral}---and its three-loop specialization $Q_3$, and evaluate representative component transformations at one, two and three loops. Setting $n=2$ in~\eqref{eq-2-8} and~\eqref{eq-fan-output-points} gives
\begin{equation}\label{eq-2-17}
\begin{aligned}
Q_2={}&-\frac{i}{2(32\pi^2)^2}
\int_0^1dr_1\int_0^1dr_2\int_0^1dr_4
\int_0^{1-r_4}dr_3\,r_1r_2r_4\int d^3\theta\int\prod_{j=1}^{3}d^4z_j\\
&\times C_{A_1DR}C_{A_2RS}C_{A_3SE}\,
(\partial_{+\dot a}\partial_{+\dot b}C^D)
\big(x+x_1(r,z),\theta\big)
(\partial_+{}^{\dot a}\partial_+{}^{\dot b}C^E)
\big(x+x_3(r,z),\theta\big)\\
&\times\frac{\delta}{\delta C^{A_1}(x+z_1,\theta)}
\frac{\delta}{\delta C^{A_2}(x+z_2,\theta)}
\frac{\delta}{\delta C^{A_3}(x+z_3,\theta)},
\end{aligned}
\end{equation}
where
\begin{equation*}
\begin{aligned}
x_1^{+\dot a}(r,z)
&=(1-r_1)z_1^{+\dot a}
+r_1(1-r_3)\big[(1-r_2)z_1^{+\dot a}+r_2z_2^{+\dot a}\big]
+r_1r_3z_3^{+\dot a},\\
x_3^{+\dot a}(r,z)
&=r_4\big[(1-r_2)z_1^{+\dot a}+r_2z_2^{+\dot a}\big]
+(1-r_4)z_3^{+\dot a}.
\end{aligned}
\end{equation*}
Likewise, setting $n=3$ in~\eqref{eq-2-8} and~\eqref{eq-fan-output-points} gives the explicit three-loop supercharge
\begin{equation}\label{eq-Q3-explicit}
\begin{aligned}
Q_3={}&-\frac{1}{2(32\pi^2)^3}
\int_0^1dr_1\int_0^1dr_2\int_0^1dr_3\int_0^1dr_4\int_0^1dr_6
\int_0^{1-r_6}dr_5\,r_1^2r_2r_3r_4(1-r_4)r_6^2\\
&\times\int d^3\theta\int\prod_{j=1}^{4}d^4z_j\,
C_{A_1DR}C_{A_2RS}C_{A_3ST}C_{A_4TE}\\
&\times(\partial_{+\dot a}\partial_{+\dot b}\partial_{+\dot c}C^D)
\big(x+x_1(r,z),\theta\big)
(\partial_+{}^{\dot a}\partial_+{}^{\dot b}\partial_+{}^{\dot c}C^E)
\big(x+x_4(r,z),\theta\big)\\
&\times\frac{\delta}{\delta C^{A_1}(x+z_1,\theta)}
\frac{\delta}{\delta C^{A_2}(x+z_2,\theta)}
\frac{\delta}{\delta C^{A_3}(x+z_3,\theta)}
\frac{\delta}{\delta C^{A_4}(x+z_4,\theta)},
\end{aligned}
\end{equation}
with translated output positions
\begin{equation*}
\begin{aligned}
x_1^{+\dot a}(r,z)
={}&(1-r_1)z_1^{+\dot a}
+r_1(1-r_3)\big[(1-r_2)z_1^{+\dot a}+r_2z_2^{+\dot a}\big]\\
&+r_1r_3(1-r_5)\Bigl\{(1-r_4)\big[(1-r_2)z_1^{+\dot a}+r_2z_2^{+\dot a}\big]+r_4z_3^{+\dot a}\Bigr\}
+r_1r_3r_5z_4^{+\dot a},\\
x_4^{+\dot a}(r,z)
={}&r_6\Bigl\{(1-r_4)\big[(1-r_2)z_1^{+\dot a}+r_2z_2^{+\dot a}\big]+r_4z_3^{+\dot a}\Bigr\}
+(1-r_6)z_4^{+\dot a}.
\end{aligned}
\end{equation*}

On operators built from undifferentiated component letters, substituting the superfield expansion~\eqref{eq-A-13} and component functional derivative~\eqref{eq-A-16} into~\eqref{eq-Qnzero} immediately yields explicit loop anomalies. At one loop ($n=1$), $(\partial_{+\dot a}C^D)(\partial_+{}^{\dot a}C^E)$ is symmetric in $D\leftrightarrow E$:
\begin{itemize}
\item Acting on $\delta^{AA'}(\gamma^1)_{A'}(\beta_1)^B$, the $\gamma^1$- and $\beta_1$-derivatives in~\eqref{eq-A-16} supply $(-\theta_2\theta_3)(-\theta_1)=+\theta_1\theta_2\theta_3$ for both assignments $(A_1,A_2)=(A,B)$ and $(B,A)$, saturating $\int d^3\theta$ and selecting the $\theta=0$ component $C^D|_{\theta=0}=ig\,c^D$ on both output legs:
\begin{equation}\label{eq-2-26}
Q_1\big(\delta^{AA'}(\gamma^1)_{A'}(\beta_1)^B\big)=\frac{g^2}{64\pi^2}C_{ARD}C_{BRE}(\partial_{+\dot a}c^D)(\partial_+{}^{\dot a}c^E).
\end{equation}
\item Acting on $(\beta_1)^A(\beta_2)^B$, both assignments $(A_1,A_2)=(A,B)$ and $(B,A)$ produce $+\theta_1\theta_2$, so $\int d^3\theta$ selects one $\theta=0$ factor $ig\,\partial_{+\dot a}c$ and one $\theta_3$ factor $\theta_3\delta^{EF}\partial_+{}^{\dot a}(\gamma^3)_F$ from the two symmetric output legs:
\begin{equation}\label{eq-2-27}
Q_1\big((\beta_1)^A(\beta_2)^B\big)=\frac{ig}{64\pi^2}(C_{ARD}C_{BRE}+C_{BRD}C_{ARE})(\partial_{+\dot a}c^D)\,\delta^{EF}\partial_+{}^{\dot a}(\gamma^3)_F.
\end{equation}
\end{itemize}
Equations~\eqref{eq-2-26} and~\eqref{eq-2-27} agree, up to field normalization, with equations~(3.66) and~(3.67) of Budzik and Kulp~\cite{BK}.\footnote{On the physical Lagrangian subspace, with $\delta^{AB}(\gamma^I)_B\leftrightarrow\bar\phi_I{}^A$, $(\beta_I)^A\leftrightarrow-\frac{i}{\sqrt2}\psi_+^{IA}$, $\delta^{AB}b_B\leftrightarrow\frac14f_{++}^A$, $\partial_{+\dot a}c^A\leftrightarrow2\bar\lambda_{\dot a}^A$, and $\partial_{+\dot a}\partial_{+\dot b}c^A\leftrightarrow2\nabla_{+(\dot a}\bar\lambda_{\dot b)}^A$, equation~\eqref{eq-2-26} reads $Q_1(\bar\phi_1{}^A\psi_+^{1B})=\frac{i\sqrt2\,g^2}{16\pi^2}C^A{}_{RD}C^{BR}{}_E\,\bar\lambda_{\dot a}^D\bar\lambda^{E\dot a}$, the component form of the Konishi anomaly~\cite{Konishi}.}

At two loops ($n=2$),~\eqref{eq-Qnzero} carries prefactor $-i/[48(32\pi^2)^2]$ and three Grassmann-even functional derivatives $\delta/\delta C^{A_j}(\theta)$. Acting on the three-fermion composite $(\beta_1)^{A_1}(\beta_2)^{A_2}(\beta_3)^{A_3}$---which has $\deg_\theta=6$ and can receive a two-loop ghost-descendant variation because $\deg_\theta(Q_2\mathcal O)=\deg_\theta\mathcal O-6=0$ by~\eqref{eq-graviton-differential-degree}---each permutation $\sigma\in S_3$ of the three spokes contributes $-\theta_1\theta_2\theta_3$, selecting $C^D|_{\theta=0}=ig\,c^D$ and $C^E|_{\theta=0}=ig\,c^E$ on the twice-differentiated endpoints:
\begin{equation}\label{eq-2-28}
Q_2\big((\beta_1)^{A_1}(\beta_2)^{A_2}(\beta_3)^{A_3}\big)=-\frac{ig^2}{48(32\pi^2)^2}\sum_{\sigma\in S_3}C_{A_{\sigma(1)}DR}C_{A_{\sigma(2)}RS}C_{A_{\sigma(3)}SE}(\partial_{+\dot a}\partial_{+\dot b}c^D)(\partial_+{}^{\dot a}\partial_+{}^{\dot b}c^E).
\end{equation}
In physical BPS letters,~\eqref{eq-2-28} gives the explicit two-loop anomalous supersymmetry variation
\begin{equation*}
Q_2\bigl(\psi_+^{1A_1}\psi_+^{2A_2}\psi_+^{3A_3}\bigr)
=-\frac{\sqrt2\,g^2}{6(32\pi^2)^2}\sum_{\sigma\in S_3}
C_{A_{\sigma(1)}DR}C_{A_{\sigma(2)}RS}C_{A_{\sigma(3)}SE}
\bigl(\nabla_{+(\dot a}\bar\lambda_{\dot b)}^D\bigr)
\bigl(\nabla_+^{(\dot a}\bar\lambda^{\dot b)E}\bigr),
\end{equation*}
mapping three chiral fermions directly into a pair of covariantly differentiated antichiral gauginos.

At three loops ($n=3$),~\eqref{eq-Qnzero} carries prefactor $-1/[1728(32\pi^2)^3]$ and four Grassmann-even functional derivatives. Acting on $\delta^{A_1B}b_B(\beta_1)^{A_2}(\beta_2)^{A_3}(\beta_3)^{A_4}$ ($\deg_\theta=9$), the $b$-derivative supplies $-ig$ and the three $\beta_I$-derivatives supply $-\theta_1\theta_2\theta_3$ for each permutation $\sigma\in S_4$, selecting $C^D|_{\theta=0}=ig\,c^D$ and $C^E|_{\theta=0}=ig\,c^E$ on the thrice-differentiated endpoints:
\begin{equation}\label{eq-Q3-component}
\begin{aligned}
Q_3\bigl(\delta^{A_1B}b_B(\beta_1)^{A_2}(\beta_2)^{A_3}(\beta_3)^{A_4}\bigr)
={}&\frac{ig^3}{1728(32\pi^2)^3}\sum_{\sigma\in S_4}
C_{A_{\sigma(1)}DR}C_{A_{\sigma(2)}RS}C_{A_{\sigma(3)}ST}C_{A_{\sigma(4)}TE}\\
&\times(\partial_{+\dot a}\partial_{+\dot b}\partial_{+\dot c}c^D)
(\partial_+{}^{\dot a}\partial_+{}^{\dot b}\partial_+{}^{\dot c}c^E).
\end{aligned}
\end{equation}
In physical BPS letters, with $\partial_{+\dot a}\partial_{+\dot b}\partial_{+\dot c}c^D\leftrightarrow2\nabla_{+(\dot a}\nabla_{+\dot b}\bar\lambda_{\dot c)}^D$, equation~\eqref{eq-Q3-component} reads
\begin{equation*}
\begin{aligned}
Q_3\bigl(f_{++}^{A_1}\psi_+^{1A_2}\psi_+^{2A_3}\psi_+^{3A_4}\bigr)
={}&\frac{\sqrt2\,g^3}{54(32\pi^2)^3}\sum_{\sigma\in S_4}
C_{A_{\sigma(1)}DR}C_{A_{\sigma(2)}RS}C_{A_{\sigma(3)}ST}C_{A_{\sigma(4)}TE}\\
&\times\bigl(\nabla_{+(\dot a}\nabla_{+\dot b}\bar\lambda_{\dot c)}^D\bigr)
\bigl(\nabla_+^{(\dot a}\nabla_+^{\dot b}\bar\lambda^{\dot c)E}\bigr),
\end{aligned}
\end{equation*}
mapping a self-dual field strength and three chiral fermions into a pair of twice-differentiated antichiral gauginos.

\section{Multi-graviton completions and obstruction equations}\label{sec-lifting}

Quantum multiplication constructs perturbatively closed multi-graviton operators in $\mathcal N=4$ Yang--Mills theory. In Section~\ref{sec-graviton-closure} we prove that every iterated marked product of single-graviton representatives is $\mathbf Q$-closed to all loop orders and terminates at a finite order in $\hbar$ (Theorem~\ref{b-00630}), and in Section~\ref{sec-graviton-example} we evaluate an explicit two-graviton product to illustrate how the one-loop contact term of $A\star B$ cancels the anomalous variation $Q_1(AB)$.\footnote{A closed multi-graviton can be the target of an incoming differential. If $\mathbf QX=\hbar^r\mathcal O$, its completed class is $\hbar^r$-torsion and vanishes after $\hbar$ is inverted. At one loop in $SO(7)$, the source is a fortuitous black-hole class and the target is a multi-graviton~\cite[Section 3]{KonishiBH}. Section~\ref{sec-results-SOSp} gives further examples at one and three loops.}

Section~\ref{sec-homotopy-transfer} then develops the successive homotopy-transfer framework that uses the loop operators $Q_n$ to compute the higher-loop differentials $d_r$ on $\mathrm{BPS}_{\hbar^{r-1}}$, establishes the choice-independence and rational dual-functional certification of $d_r\ne 0$ (Proposition~\ref{prop-ch9-subspace-certificate}), proves the fortuitous bound~\eqref{eq-sum-rank} on the total lifting rank across all loop orders (Corollary~\ref{cor-ch9-fortuitous-bound}), and characterizes operator-mixing and renormalization ambiguities (Theorem~\ref{product-uniqueness-ch9-p01}). Because every multi-graviton class admits a closed completion, no differential $d_r$ can map one multi-graviton class to another: every perturbative lift pairs a fortuitous class either with a multi-graviton class or with another fortuitous class, reducing the lifting census from the full classical cohomology to its fortuitous quotient. We remind the reader that quantum closure of a classical representative $\mathcal O_0$ means the existence of loop corrections $\mathcal O_1,\mathcal O_2,\ldots$ satisfying $\mathbf Q(\mathcal O_0+\hbar\mathcal O_1+\hbar^2\mathcal O_2+\cdots)=0$ order by order in $\hbar$, whereas $\mathcal O_0$ itself is in general not annihilated by $\mathbf Q$; nonperturbative instanton effects can modify the cohomology further. The computer implementation and complete sector tables are archived in~\cite{Repo}.

\subsection{Perturbative closure of multi-gravitons}\label{sec-graviton-closure}

The single-graviton representatives of Chang and Yin~\cite[Section 4]{ChangYin} are components of $\operatorname{Tr}(dC)^p$, where
\begin{equation}\label{b-00629}
dC=dx^{+\dot a}\frac{\partial C}{\partial x^{+\dot a}}
+d\theta_I\frac{\partial_LC}{\partial\theta_I}.
\end{equation}

\rthm{Theorem}{b-00630}{All-order perturbative closure of multi-gravitons} \emph{Let $G_1,\ldots,G_m$ be single-graviton representatives. Every iterated quantum product $\mathcal O$ of these operators, with any fixed bracketing, satisfies}
\begin{equation}\label{eq-multigraviton-closure}
\mathbf Q\mathcal O=0,\qquad
\left.\mathcal O\right|_{\hbar=0}=G_1\cdots G_m.
\end{equation}
\emph{For each fixed component, its perturbative completion has finitely many terms in $\hbar$.}

\begin{proofbox}
\textbf{Proof.} The classical differential~\eqref{eq-A-15} acts on $dC$ and $\operatorname{Tr}(dC)^p$ by
\begin{equation}\label{eq-2-58}
Q_0(dC)=[C,dC],\qquad Q_0\operatorname{Tr}(dC)^p=\operatorname{Tr}[C,(dC)^p]=0.
\end{equation}
Consider first a single graviton $G$. At loop order $n\ge1$, the auxiliary Grassmann factor of the $Q_n$ fan kernel reads
\begin{equation}\label{eq-graviton-odd-factor}
\prod_{j=1}^{n}\prod_{I=1}^{3}
\bigl(\theta_I^{(j)}-\theta_I^{(j+1)}\bigr).
\end{equation}
A nonzero coincident value requires at least $3n$ odd derivatives, whereas each of the $n+1$ contracted factors of $dC$ supplies at most one, leaving positive odd degree $3n-(n+1)=2n-1>0$ in the relative Grassmann coordinates. Equivalently, in~\eqref{eq-2-8} all $n+1$ functional derivatives carry the same odd argument $\theta$: each contraction with $dC(\theta')$ leaves at least two of the three factors $\theta_I-\theta'_I$, so the product of the $n+1$ contractions has degree at least $2(n+1)>3$ in $\theta-\theta'$ and vanishes identically before any component of $\operatorname{Tr}(dC)^p$ is extracted:
\begin{equation}\label{eq-2-60}
Q_0G=0,\qquad Q_nG=0\quad(n\geq1).
\end{equation}
Applying the Leibniz identity~\eqref{eq-chain} at each binary vertex of an iterated marked product gives
\begin{equation}\label{b-00642}
\mathbf Q(A\star B)
=(\mathbf QA)\star B+(-1)^{|A|}A\star(\mathbf QB)=0
\qquad(\mathbf QA=\mathbf QB=0),
\end{equation}
so induction on $m$ proves $\mathbf Q\mathcal O=0$, while the leading term of each binary product is ordinary multiplication~\cite[Section 5]{ChangYin}. Finiteness in $\hbar$ follows from Corollary~\ref{local-ch8-finite-product-corollary} in Section~\ref{the-graph-formula}.~\hfill\qedsymbol
\end{proofbox}

For example, for two single gravitons $G_1$ and $G_2$, the marked product $G_1\star G_2$ furnishes an explicit quantum completion of $G_1G_2$ (up to the finite $\hbar$-mixing freedom classified in Theorem~\ref{product-uniqueness-ch9-p01}); equivalently, $G_1G_2$ is never lifted by a failure of $\mathbf Q$-closure.

\paragraph{The graviton part of the cohomology.}\label{the-graviton-part-of-the-cohomology}
At finite rank $N$, the classical cohomology $\mathrm{BPS}_{\hbar^0}=H(Q_0)$ splits into the graviton subalgebra, spanned by polynomials in single-graviton classes, and the remaining fortuitous classes~\cite{microstates}. Theorem~\ref{b-00630} equips every class in the graviton part with an explicit all-order closed completion obtained by replacing ordinary products of single-graviton representatives by marked products $\star$. By Theorem~\ref{b-00513}, the resulting $\mathbf Q$-cohomology class is independent of bracketing, and exchanging adjacent homogeneous factors $G_i,G_j$ multiplies the class by $(-1)^{|G_i||G_j|}$. The same construction dresses any closed completion $\mathcal O[\mathcal O_0]$ of a fortuitous core $\mathcal O_0$: the product $G\star\mathcal O[\mathcal O_0]$ is again $\mathbf Q$-closed.

\subsection{An explicit example}\label{sec-graviton-example}

Take the two single-graviton representatives
\begin{equation}\label{eq-gravitoninputs}
A=\delta^{XY}(\gamma^1)_X(\gamma^1)_Y,
\qquad B=(\gamma^2)_X(\beta_1)^X,
\qquad\mathcal O_0=AB.
\end{equation}
Equation~\eqref{eq-A-13}, with left Grassmann differentiation, identifies both operators as components of $\operatorname{Tr}(dC)^2$:
\begin{equation}\label{local-ch8-graviton-component-extraction}
\begin{aligned}
A&=\left.\delta_{AB}\frac{\partial_L C^A}{\partial\theta_1}
\frac{\partial_L C^B}{\partial\theta_1}\right|_{\theta=0},\\
B&=-\frac12\left.\frac{\partial_L}{\partial\theta_3}
\left(\delta_{AB}\frac{\partial_L C^A}{\partial\theta_2}
\frac{\partial_L C^B}{\partial\theta_2}\right)\right|_{\theta=0}.
\end{aligned}
\end{equation}
Gauge invariance gives $Q_0A=0$, and for $B$ the superpotential term gives
\begin{equation}\label{b-00658}
Q_0B=-iC^{XYZ}(\gamma^2)_X(\gamma^2)_Y(\gamma^3)_Z=0,
\end{equation}
while $Q_nA=Q_nB=0$ ($n\ge1$) by Theorem~\ref{b-00630}. By contrast, the classical product $\mathcal O_0=AB$ is not $Q_1$-closed; evaluating~\eqref{eq-2-8} at $n=1$ gives
\begin{equation}\label{eq-scalarfermionQ1}
Q_1\mathcal O_0
=\frac{g^2}{32\pi^2}
C^X{}_{RD}C^{YR}{}_E
(\gamma^1)_X(\gamma^2)_Y
(\partial_{+\dot a}c^D)(\partial_+{}^{\dot a}c^E).
\end{equation}
Without the quantum product, deciding whether $[\mathcal O_0]_{\hbar^0}$ survives at one loop requires testing linear-algebraically whether the right-hand side of~\eqref{eq-scalarfermionQ1} is $Q_0$-exact: if $\mathrm{RHS}=Q_0\mathcal O'$, then $\mathcal O_0-\hbar\mathcal O'$ is closed through $O(\hbar)$ (and cannot be $\mathbf Q$-exact because $\mathcal O_0$ carries no derivatives), whereas otherwise $[\mathcal O_0]_{\hbar^0}$ is lifted together with another class in $\mathrm{BPS}_{\hbar^0}$.

With the quantum product, no linear-system search is needed: Theorem~\ref{b-00630} guarantees that $\mathcal O_1=[\hbar](A\star B)$ solves $Q_0\mathcal O_1=-Q_1\mathcal O_0$ directly:
\begin{equation}\label{eq-2-65}
\begin{aligned}
\mathcal O_1
&=\{\mathcal I,A,B\}^{\gamma,0}\\
&=\underbrace{\frac{i}{32\pi^2}}_{\text{prefactor \eqref{eq-2-34}}}
\cdot\underbrace{\frac12}_{\int_{\Delta^2}dr_1dr_2}
\cdot\underbrace{(ig)}_{\partial C^D|_{\theta=0}}
\cdot\underbrace{(-2)}_{\delta A/\delta C^Y}
\cdot\underbrace{(-1)}_{\delta B/\delta C^Z}
\cdot\underbrace{(-1)}_{\text{Koszul sign}}\,
C_D{}^{XY}(\partial_{+\dot a}c^D)
(\partial_+{}^{\dot a}(\gamma^1)_X)(\gamma^2)_Y\\
&=\frac{g}{32\pi^2}C_D{}^{XY}(\partial_{+\dot a}c^D)
(\partial_+{}^{\dot a}(\gamma^1)_X)(\gamma^2)_Y.
\end{aligned}
\end{equation}
Let us verify explicitly that $Q_0\mathcal O_1+Q_1\mathcal O_0=0$. Expanding $Q_0\mathcal O_1$ gives
\begin{equation}\label{eq-scalarfermionexpand}
\begin{aligned}
Q_0\mathcal O_1={}&\frac{g^2}{32\pi^2}C_D{}^{XY}\Bigl[
-C^D{}_{EF}(\partial_{+\dot a}c^E)c^F
(\partial_+{}^{\dot a}(\gamma^1)_X)(\gamma^2)_Y\\
&+C_{XE}{}^F(\partial_{+\dot a}c^D)
(\partial_+{}^{\dot a}c^E)(\gamma^1)_F(\gamma^2)_Y\\
&+C_{XE}{}^F(\partial_{+\dot a}c^D)c^E
(\partial_+{}^{\dot a}(\gamma^1)_F)(\gamma^2)_Y\\
&+C_{YE}{}^F(\partial_{+\dot a}c^D)
(\partial_+{}^{\dot a}(\gamma^1)_X)c^E(\gamma^2)_F
\Bigr].
\end{aligned}
\end{equation}
The first, third, and fourth lines cancel by the Jacobi identity for $C_{DXY}$,
\begin{equation}\label{b-00670}
C_{RX}{}^TC_{TYZ}+C_{RY}{}^TC_{XTZ}+C_{RZ}{}^TC_{XYT}=0,
\end{equation}
leaving the second line:
\begin{equation}\label{eq-2-66}
\begin{aligned}
Q_0\mathcal O_1
&=\frac{g^2}{32\pi^2}
C_D{}^{XY}C_{XE}{}^F(\gamma^1)_F(\gamma^2)_Y
(\partial_{+\dot a}c^D)(\partial_+{}^{\dot a}c^E)\\
&=-\frac{g^2}{32\pi^2}
C^X{}_{RD}C^{YR}{}_{E}(\gamma^1)_X(\gamma^2)_Y
(\partial_{+\dot a}c^D)(\partial_+{}^{\dot a}c^E)
=-Q_1\mathcal O_0.
\end{aligned}
\end{equation}
The one-loop contact term of $A\star B$ thus cancels the anomalous one-loop variation of $AB$. Moreover, since $\deg_\theta\mathcal O_0=5$ and $\deg_\theta\mathcal O_1=2$, the component-degree bound~\eqref{eq-2-63}--\eqref{eq-graviton-differential-degree} forces all higher-loop contributions to vanish:
\begin{equation}\label{eq-finiteexample}
\begin{gathered}
\mathcal O=A\star B=\mathcal O_0+\hbar\mathcal O_1,\qquad
Q_{n\ge2}\mathcal O_0=Q_{n\ge1}\mathcal O_1=0,\\
\mathbf Q\mathcal O
=\hbar(Q_1\mathcal O_0+Q_0\mathcal O_1)=0.
\end{gathered}
\end{equation}
Here $\mathcal O_0+\hbar\mathcal O_1$ is already the exact all-order perturbative completion; for higher-charge multi-graviton words where $Q_2\mathcal O_0\ne0$, the higher orders of $A\star B$ supply the required completions systematically.

\subsection{Homotopy transfer and successive obstructions}\label{sec-homotopy-transfer}

The loop operators $Q_n$ organize the perturbative cocycle condition $\mathbf Q\mathcal O=0$ into a hierarchy of linear obstruction equations:
\begin{equation}\label{eq-liftingequations}
\mathbf Q\bigl(\mathcal O_0+\hbar\mathcal O_1+\hbar^2\mathcal O_2+\cdots\bigr)=0
\quad\Longleftrightarrow\quad
\sum_{\substack{j=0\\\text{loop orders}}}^{s}Q_j\mathcal O_{s-j}=0
\qquad(s\ge0).
\end{equation}
We write $\mathrm{BPS}_{\hbar^r}$ for the space of cohomology classes surviving the obstruction differentials $d_1,\ldots,d_r$, after quotienting at each step by the coboundaries in $\operatorname{im}d_r$:
\begin{equation}\label{eq-page-iterated}
\begin{gathered}
\mathrm{BPS}_{\hbar^0}=H(Q_0),\qquad
\mathrm{BPS}_{\hbar}=H(\mathrm{BPS}_{\hbar^0},d_1),\qquad
\mathrm{BPS}_{\hbar^2}=H(\mathrm{BPS}_{\hbar},d_2),\\
\mathrm{BPS}_{\hbar^r}^{k}
=\frac{\ker\bigl(d_r:\mathrm{BPS}_{\hbar^{r-1}}^{k}\to\mathrm{BPS}_{\hbar^{r-1}}^{k+1}\bigr)}
{\operatorname{im}\bigl(d_r:\mathrm{BPS}_{\hbar^{r-1}}^{k-1}\to\mathrm{BPS}_{\hbar^{r-1}}^{k}\bigr)}
\qquad(r\ge1).
\end{gathered}
\end{equation}
Here $k=\mathrm{gh}-2J_L$ is the cochain degree, combining the BV ghost number $\mathrm{gh}$ and the undotted spin $J_L$: the BRST differential carries $(\Delta\mathrm{gh},\Delta J_L)=(1,0)$ and the supercharge carries $(0,-\tfrac12)$, so every $Q_n$ and every $d_r$ raises $k$ by $+1$. On physical ghost-number-zero operators $k=-2J_L$, while Koszul signs $(-1)^{|A|}$ continue to use the fermion number $|A|$ (modulo $2$).
Writing $[\mathcal O_0]_{\hbar^{r-1}}$ for the class of a survivable representative in $\mathrm{BPS}_{\hbar^{r-1}}$, the one-loop differential is $d_1[\mathcal O_0]_{\hbar^0}=[Q_1\mathcal O_0]_{\hbar^0}$, and the $r$-loop obstruction differential acts on a representative admitting a partial completion $\mathcal O_0+\hbar\mathcal O_1+\cdots+\hbar^{r-1}\mathcal O_{r-1}$ by
\begin{equation}\label{local-ch8-obstruction-definition}
d_r[\mathcal O_0]_{\hbar^{r-1}}=
\left[\sum_{\substack{j=1\\\text{loop orders}}}^{r}Q_j\mathcal O_{r-j}\right]_{\hbar^{r-1}},
\qquad
\sum_{\substack{j=0\\\text{loop orders}}}^{s}Q_j\mathcal O_{s-j}=0\quad(0\le s<r).
\end{equation}
To compute the successive differentials systematically, choose at each stage $r\ge0$ an inclusion of representatives $i_r:\mathrm{BPS}_{\hbar^r}\to\mathrm{BPS}_{\hbar^{r-1}}$ (where $\mathrm{BPS}_{\hbar^{-1}}$ denotes the local-operator cochain complex with $d_0=Q_0$, so $i_0:H(Q_0)\to\ker Q_0$), a projection $\pi_r$ in the reverse direction, and a degree-$-1$ contracting homotopy $h_r$ satisfying the standard SDR (strong deformation retraction) identities
\begin{equation}\label{eq-page-contraction}
\begin{gathered}
\pi_ri_r=1,\qquad d_ri_r=\pi_rd_r=0,\qquad
d_rh_r+h_rd_r=1-i_r\pi_r,\\
h_ri_r=\pi_rh_r=h_r^2=0.
\end{gathered}
\end{equation}

\begin{figure}[H]\centering
\includegraphics[width=.98\linewidth]{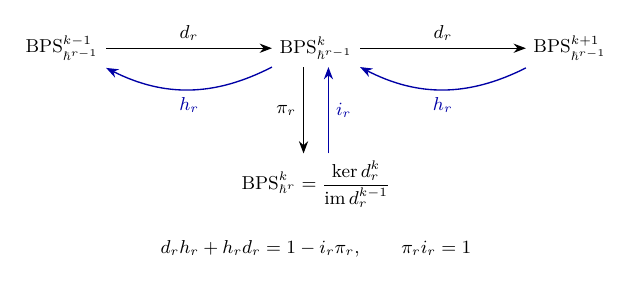}
\caption{The contraction of $d_r$, $r\ge1$: $i_r$ selects cohomology representatives and $h_r$ returns a primitive of each boundary.}\label{b-00686}
\end{figure}
Explicitly (Figure~\ref{b-00686}), choosing a subspace of cocycle representatives for $\ker d_r/\operatorname{im}d_r$ and a direct-sum complement to $\ker d_r$, $h_r$ inverts $d_r$ from $\operatorname{im}d_r$ onto that complement and vanishes on the remaining summands; on any $d_r$-cocycle $a$, the retraction separates its chosen representative from its $d_r$-exact piece:
\begin{equation}\label{local-ch8-representative-decomposition}
d_ra=0\quad\Longrightarrow\quad
a=i_r\pi_ra+\underbrace{d_rh_ra}_{\text{exact part}}.
\end{equation}

\begin{figure}[H]\centering
\includegraphics[width=.98\linewidth]{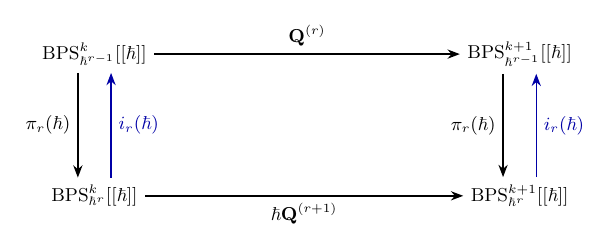}
\caption{Transfer of the full differential, $r\ge1$; the lower arrow retains the factor $\hbar$.}\label{fig-full-transfer}
\end{figure}

\rthm{Proposition}{prop-ch8-transfer-validity}{Successive homotopy transfer}
Let $\mathbf Q^{(0)}=\mathbf Q$, and suppose at stage $r\ge0$ that $(\mathbf Q^{(r)})^2=0$ on $\mathrm{BPS}_{\hbar^{r-1}}[[\hbar]]$ (the formal $\hbar$-series with coefficients in $\mathrm{BPS}_{\hbar^{r-1}}$, or in the local-operator complex at $r=0$) with $\hbar^0$ term $d_r$:
\begin{equation}\label{eq-splitting-relations}
\begin{gathered}
\mathbf Q^{(0)}=Q_0+\hbar Q_1+\hbar^2Q_2+\cdots,\\
\mathbf Q^{(r)}=d_r+\hbar\bigl([\hbar]\mathbf Q^{(r)}\bigr)
+\hbar^2\bigl([\hbar^2]\mathbf Q^{(r)}\bigr)+\cdots,
\qquad d_r=[\hbar^0]\mathbf Q^{(r)}.
\end{gathered}
\end{equation}
Then the homological perturbation formulas \cite[Lemma 2.4]{Crainic} define a deformed contraction $(i_r(\hbar),\pi_r(\hbar),h_r(\hbar))$ and a transferred differential $\mathbf Q^{(r+1)}$ on $\mathrm{BPS}_{\hbar^r}[[\hbar]]$:
\begin{equation}\label{eq-page-transfer}
\begin{aligned}
i_r(\hbar)&=\bigl[1+h_r(\mathbf Q^{(r)}-d_r)\bigr]^{-1}i_r,\\
\pi_r(\hbar)&=\pi_r\bigl[1+(\mathbf Q^{(r)}-d_r)h_r\bigr]^{-1},\\
h_r(\hbar)&=\bigl[1+h_r(\mathbf Q^{(r)}-d_r)\bigr]^{-1}h_r,\\
\hbar\mathbf Q^{(r+1)}&=\pi_r(\mathbf Q^{(r)}-d_r)i_r(\hbar),\\
d_{r+1}&=[\hbar^0]\mathbf Q^{(r+1)}
=\pi_r\bigl([\hbar]\mathbf Q^{(r)}\bigr)i_r.
\end{aligned}
\end{equation}
Passing to $d_r$-cohomology eliminates the constant-order differential $d_r$ and transports all higher orders into $\hbar\mathbf Q^{(r+1)}$ (Figure~\ref{fig-full-transfer}), satisfying
\begin{equation}\label{eq-page-transfer-identities}
\begin{aligned}
\mathbf Q^{(r)}i_r(\hbar)&=\hbar i_r(\hbar)\mathbf Q^{(r+1)},\\
\pi_r(\hbar)\mathbf Q^{(r)}&=\hbar\mathbf Q^{(r+1)}\pi_r(\hbar),\\
\mathbf Q^{(r)}h_r(\hbar)+h_r(\hbar)\mathbf Q^{(r)}&=1-i_r(\hbar)\pi_r(\hbar),\\
\pi_r(\hbar)i_r(\hbar)&=1,\qquad (\mathbf Q^{(r+1)})^2=d_{r+1}^2=0.
\end{aligned}
\end{equation}

\begin{proofbox}
\textbf{Proof.} Because $\mathbf Q^{(r)}-d_r$ has zero $\hbar^0$ term, the geometric series for the resolvents truncate at finite length at every fixed power $\hbar^n$:
\begin{equation}\label{eq-ch8-transfer-inverses}
\begin{aligned}
[\hbar^n]\bigl[1+h_r(\mathbf Q^{(r)}-d_r)\bigr]^{-1}
&=[\hbar^n]\sum_{\substack{\ell=0\\\text{operator powers}}}^{n}
(-1)^\ell\bigl[h_r(\mathbf Q^{(r)}-d_r)\bigr]^\ell,\\
[\hbar^n]\bigl[1+(\mathbf Q^{(r)}-d_r)h_r\bigr]^{-1}
&=[\hbar^n]\sum_{\substack{\ell=0\\\text{operator powers}}}^{n}
(-1)^\ell\bigl[(\mathbf Q^{(r)}-d_r)h_r\bigr]^\ell.
\end{aligned}
\end{equation}
Clearing denominators gives the resolvent relations
\begin{equation}\label{eq-ch8-transfer-resolvent-identities}
\begin{aligned}
i_r(\hbar)&=i_r-h_r(\hbar)(\mathbf Q^{(r)}-d_r)i_r,\\
\pi_r(\hbar)&=\pi_r-\pi_r(\mathbf Q^{(r)}-d_r)h_r(\hbar),\\
h_r(\hbar)&=h_r\bigl[1+(\mathbf Q^{(r)}-d_r)h_r\bigr]^{-1},
\end{aligned}
\end{equation}
which together with the side conditions $h_r^2=h_ri_r=\pi_rh_r=0$ imply
\begin{equation}\label{eq-ch8-transfer-retraction-proof}
\begin{gathered}
h_r(\hbar)^2=h_r(\hbar)i_r=\pi_rh_r(\hbar)=0,\\
\begin{aligned}
\pi_r(\hbar)i_r(\hbar)
&=\pi_ri_r-\pi_rh_r(\hbar)(\mathbf Q^{(r)}-d_r)i_r\\
&\quad-\pi_r(\mathbf Q^{(r)}-d_r)h_r(\hbar)i_r
+\pi_r(\mathbf Q^{(r)}-d_r)h_r(\hbar)^2(\mathbf Q^{(r)}-d_r)i_r=1,\\
h_r(\hbar)i_r(\hbar)&=\pi_r(\hbar)h_r(\hbar)=0.
\end{aligned}
\end{gathered}
\end{equation}
Using $(\mathbf Q^{(r)})^2=d_r^2=0$, we have
\begin{equation}\label{eq-ch8-transfer-perturbation-algebra}
(\mathbf Q^{(r)}-d_r)\mathbf Q^{(r)}
+\mathbf Q^{(r)}(\mathbf Q^{(r)}-d_r)
=(\mathbf Q^{(r)}-d_r)^2,
\end{equation}
and multiplying $\mathbf Q^{(r)}h_r(\hbar)+h_r(\hbar)\mathbf Q^{(r)}$ on both sides by the resolvent denominators yields
\begin{equation}\label{eq-ch8-transfer-homotopy-calculation}
\begin{aligned}
&\bigl[1+h_r(\mathbf Q^{(r)}-d_r)\bigr]
\bigl(\mathbf Q^{(r)}h_r(\hbar)+h_r(\hbar)\mathbf Q^{(r)}\bigr)
\bigl[1+(\mathbf Q^{(r)}-d_r)h_r\bigr]\\
&=\mathbf Q^{(r)}h_r+h_r\mathbf Q^{(r)}
+h_r\bigl[(\mathbf Q^{(r)}-d_r)\mathbf Q^{(r)}
+\mathbf Q^{(r)}(\mathbf Q^{(r)}-d_r)\bigr]h_r\\
&=1-i_r\pi_r+(\mathbf Q^{(r)}-d_r)h_r
+h_r(\mathbf Q^{(r)}-d_r)+h_r(\mathbf Q^{(r)}-d_r)^2h_r\\
&=\bigl[1+h_r(\mathbf Q^{(r)}-d_r)\bigr]
\bigl[1+(\mathbf Q^{(r)}-d_r)h_r\bigr]-i_r\pi_r.
\end{aligned}
\end{equation}
Applying the two inverses establishes the third identity of~\eqref{eq-page-transfer-identities}; commuting it with $\mathbf Q^{(r)}$ and acting on $i_r(\hbar)$ then gives
\begin{equation}\label{eq-ch8-transfer-invariance}
\begin{gathered}
\mathbf Q^{(r)}i_r(\hbar)\pi_r(\hbar)
=i_r(\hbar)\pi_r(\hbar)\mathbf Q^{(r)},\qquad
h_r(\hbar)\mathbf Q^{(r)}i_r(\hbar)=0,\\
\begin{aligned}
\pi_r(\hbar)\mathbf Q^{(r)}i_r(\hbar)
&=\pi_r\mathbf Q^{(r)}i_r(\hbar)
-\pi_r(\mathbf Q^{(r)}-d_r)h_r(\hbar)\mathbf Q^{(r)}i_r(\hbar)\\
&=\pi_r(\mathbf Q^{(r)}-d_r)i_r(\hbar)
=\hbar\mathbf Q^{(r+1)}.
\end{aligned}
\end{gathered}
\end{equation}
The retraction identities therefore imply the chain-map relations and nilpotence:
\begin{equation}\label{eq-ch8-transfer-chain-maps-proved}
\begin{aligned}
\mathbf Q^{(r)}i_r(\hbar)
&=i_r(\hbar)\pi_r(\hbar)\mathbf Q^{(r)}i_r(\hbar)
=\hbar i_r(\hbar)\mathbf Q^{(r+1)},\\
\pi_r(\hbar)\mathbf Q^{(r)}
&=\pi_r(\hbar)\mathbf Q^{(r)}i_r(\hbar)\pi_r(\hbar)
=\hbar\mathbf Q^{(r+1)}\pi_r(\hbar),\\
0&=\pi_r(\hbar)(\mathbf Q^{(r)})^2i_r(\hbar)
=\hbar^2(\mathbf Q^{(r+1)})^2,\\
(\mathbf Q^{(r+1)})^2&=0,\qquad
d_{r+1}^2=[\hbar^0](\mathbf Q^{(r+1)})^2=0,
\end{aligned}
\end{equation}
since multiplication by $\hbar^2$ is injective on formal power series, and extracting $[\hbar]$ in~\eqref{eq-page-transfer} gives the formula for $d_{r+1}$.~\hfill\qedsymbol
\end{proofbox}
Over $\mathbb C[[\hbar]]$, the contraction $(i_r(\hbar),\pi_r(\hbar),h_r(\hbar))$ identifies $H(\mathbf Q^{(r)})$ with $H(\hbar\mathbf Q^{(r+1)})$. Dividing $\hbar\mathbf Q^{(r+1)}$ by $\hbar$ preserves the cocycle space while enlarging the coboundary space by $\hbar$-torsion classes removed at higher loop orders:
\begin{equation}\label{ch9-rescaling-cohomology}
\begin{gathered}
\ker(\hbar\mathbf Q^{(r+1)})=\ker\mathbf Q^{(r+1)},\qquad
\operatorname{im}(\hbar\mathbf Q^{(r+1)})
=\hbar\operatorname{im}\mathbf Q^{(r+1)},\\
H(\hbar\mathbf Q^{(r+1)})
=\frac{\ker\mathbf Q^{(r+1)}}{\hbar\operatorname{im}\mathbf Q^{(r+1)}},\qquad
H^k(\mathbf Q^{(r)})\cong\frac{H^k(\mathbf Q)}{\{u\in H^k(\mathbf Q):\hbar^ru=0\}}.
\end{gathered}
\end{equation}
Indeed, because $C^{(r+1)}$ is $\hbar$-torsion-free, the surjective projection $\pi_r(\hbar)_*:H^k(\mathbf Q^{(r)})\twoheadrightarrow H^k(\mathbf Q^{(r+1)})$ kills precisely the classes annihilated by $\hbar$, and iterating through $r$ stages quotients $H^k(\mathbf Q)$ by its $\hbar^r$-torsion submodule. Once $\hbar$ is inverted over the Laurent field $\mathbb C((\hbar))$, rescaling by $\hbar$ is an isomorphism and all $H(\mathbf Q^{(r)})\otimes_{\mathbb C[[\hbar]]}\mathbb C((\hbar))$ are isomorphic via the chosen transfer maps $i_0(\hbar)\cdots i_{r-1}(\hbar)$ to $H(\mathbf Q)\otimes_{\mathbb C[[\hbar]]}\mathbb C((\hbar))$.

\paragraph{Obstructions and corrected representatives.}\label{the-first-obstruction-maps}
\label{b-00695}\label{local-ch8-transfer-coefficients-intro}%
Expanding the first transfer $\mathbf Q^{(1)}$ in terms of the loop operators $Q_n$ gives
\begin{equation}\label{eq-first-page-series}
\begin{aligned}
[\hbar^0]\mathbf Q^{(1)}&=\pi_0Q_1i_0,\\
[\hbar]\mathbf Q^{(1)}&=\pi_0(Q_2-Q_1h_0Q_1)i_0,\\
[\hbar^2]\mathbf Q^{(1)}&=\pi_0\bigl(Q_3-Q_2h_0Q_1-Q_1h_0Q_2
+Q_1h_0Q_1h_0Q_1\bigr)i_0,
\end{aligned}
\end{equation}
while the first three obstruction differentials $d_1,d_2,d_3$---including at three loops the correction of one-loop representatives in $\mathrm{BPS}_{\hbar^0}$ generated by $h_1$---read
\begin{equation}\label{b-00699}
\begin{aligned}
d_1&=\pi_0Q_1i_0,\\
d_2&=\pi_1\pi_0(Q_2-Q_1h_0Q_1)i_0i_1,\\
d_3&=\pi_2\pi_1\pi_0\Bigl[Q_3-Q_2h_0Q_1-Q_1h_0Q_2+Q_1h_0Q_1h_0Q_1\\
&\hspace{32mm}-(Q_2-Q_1h_0Q_1)i_0h_1\pi_0(Q_2-Q_1h_0Q_1)\Bigr]i_0i_1i_2.
\end{aligned}
\end{equation}
More generally, at every stage $r\ge0$,
\begin{equation}\label{local-ch8-transfer-coefficients}
\begin{aligned}
[\hbar]\mathbf Q^{(r+1)}
&=\pi_r\Bigl([\hbar^2]\mathbf Q^{(r)}-
\bigl([\hbar]\mathbf Q^{(r)}\bigr)h_r\bigl([\hbar]\mathbf Q^{(r)}\bigr)\Bigr)i_r,\\
d_{r+2}&=\pi_{r+1}\bigl([\hbar]\mathbf Q^{(r+1)}\bigr)i_{r+1}.
\end{aligned}
\end{equation}
For a class $a\in\mathrm{BPS}_{\hbar^2}$ (so $d_2i_2a=0$), the transferred inclusion constructs an explicit completion through two loops:
\begin{equation}\label{eq-two-loop-representative-transfer}
\begin{aligned}
\mathcal O_0&=i_0i_1i_2a,\\
\mathcal O_1&=-h_0Q_1\mathcal O_0
-i_0h_1\pi_0(Q_2-Q_1h_0Q_1)\mathcal O_0,\\
\mathcal O_2&=-h_0(Q_2\mathcal O_0+Q_1\mathcal O_1).
\end{aligned}
\end{equation}
Here the second term of $\mathcal O_1$ is a $Q_0$-closed adjustment whose $Q_1$-action cancels the $d_1$-exact projection of $(Q_2-Q_1h_0Q_1)\mathcal O_0$, as verified using $[\hbar](\mathbf Q^{(1)})^2=0$:
\begin{equation}\label{eq-two-loop-representative-projection}
\begin{aligned}
d_1\pi_0(Q_2-Q_1h_0Q_1)\mathcal O_0
&=-\pi_0(Q_2-Q_1h_0Q_1)i_0d_1i_1i_2a=0,\\
Q_0\mathcal O_1&=-Q_1\mathcal O_0,\\
\pi_0(Q_2\mathcal O_0+Q_1\mathcal O_1)
&=(1-d_1h_1)\pi_0(Q_2-Q_1h_0Q_1)\mathcal O_0\\
&=i_1\pi_1\pi_0(Q_2-Q_1h_0Q_1)\mathcal O_0=i_1d_2i_2a=0,\\
Q_0(Q_2\mathcal O_0+Q_1\mathcal O_1)
&=-Q_1^2\mathcal O_0-Q_1Q_0\mathcal O_1=0,\\
Q_0\mathcal O_2&=-Q_2\mathcal O_0-Q_1\mathcal O_1.
\end{aligned}
\end{equation}
The $h_1$ adjustment vanishes if the chosen $h_0$ satisfies $\pi_0(Q_2-Q_1h_0Q_1)\mathcal O_0=0$. For the orthogonal chains in~\eqref{eq-orthogonal-d3-chain}, choose the compatible primitives $Q_0\mathcal O_1=-Q_1\mathcal O_0$ and $Q_0\mathcal O_2=-Q_1\mathcal O_1$ given there. Degree counting gives $Q_2\mathcal O_0=Q_2\mathcal O_1=Q_3\mathcal O_0=0$, hence $d_3[\mathcal O_0]_{\hbar^2}=[Q_1\mathcal O_2]_{\hbar^2}$.
At arbitrary order $r\ge1$, iterating the first identity of~\eqref{eq-page-transfer-identities} gives
\begin{equation}\label{eq-page-representative-chain}
\mathbf Q\,i_0(\hbar)\cdots i_{r-1}(\hbar)
=\hbar^r i_0(\hbar)\cdots i_{r-1}(\hbar)\mathbf Q^{(r)}.
\end{equation}
Setting $\mathcal O_s=[\hbar^s]\bigl(i_0(\hbar)\cdots i_{r-1}(\hbar)i_ra\bigr)$ for $a\in\mathrm{BPS}_{\hbar^r}$ therefore solves the cocycle equations through order $r$ and expresses the next differential $d_{r+1}$ on $\mathrm{BPS}_{\hbar^r}$ in the form~\eqref{local-ch8-obstruction-definition}:
\begin{equation}\label{eq-2-75}
\begin{gathered}
\sum_{\substack{j=0\\\text{loop orders}}}^{s}Q_j\mathcal O_{s-j}=0
\qquad(0\le s\le r),\\
d_{r+1}[\mathcal O_0]_{\hbar^r}
=\left[\sum_{\substack{j=1\\\text{loop orders}}}^{r+1}Q_j\mathcal O_{r+1-j}\right]_{\hbar^r}.
\end{gathered}
\end{equation}
In particular, any classical representative $\mathcal O_0$ admitting an all-order completion $\mathcal O=\sum_{s\ge0}\hbar^s\mathcal O_s$ with $\mathbf Q\mathcal O=0$ satisfies $d_{r+1}[\mathcal O_0]_{\hbar^r}=0$ on every successive cohomology space $\mathrm{BPS}_{\hbar^r}$; by Theorem~\ref{b-00630}, this holds for every multi-graviton representative in $\mathcal N=4$ Yang--Mills theory.

\rthm{Proposition}{prop-ch9-subspace-certificate}{Choice-independence and rational obstruction certificates}
For each cochain degree $k$ and integer $r\ge0$, define canonical subspaces $B_r^k\subseteq Z_r^k\subseteq\mathrm{BPS}_{\hbar^0}^k=H^k(Q_0)$ directly from the loop operators $Q_0,Q_1,\ldots,Q_r$ by setting $Z_0^k=\mathrm{BPS}_{\hbar^0}^k$, $B_0^k=0$, and for $r\ge1$:
\begin{equation}\label{eq-ch9-canonical-filtration}
\begin{aligned}
Z_r^k&=\Bigl\{[\mathcal O_0]_{\hbar^0}\in H^k(Q_0):\exists\,\mathcal O_1,\ldots,\mathcal O_r\text{ with }
\sum_{\substack{j=0\\\text{loop orders}}}^{s}Q_j\mathcal O_{s-j}=0\;\;(0\le s\le r)\Bigr\},\\
B_r^k&=\Bigl\{\Bigl[-\sum_{\substack{j=1\\\text{loop orders}}}^{m}Q_j\mathcal X_{m-j}\Bigr]_{\hbar^0}\in H^k(Q_0):
1\le m\le r,\;\sum_{\substack{j=0\\\text{loop orders}}}^{s}Q_j\mathcal X_{s-j}=0\;\;(0\le s<m)\Bigr\}.
\end{aligned}
\end{equation}
Then:
\begin{enumerate}
\item For every choice of contractions $(i_m,\pi_m,h_m)_{0\le m\le r}$, sending $a\in\mathrm{BPS}_{\hbar^r}^k$ to the classical class $[\,[\hbar^0](i_0(\hbar)\cdots i_{r-1}(\hbar)i_ra)\,]_{\hbar^0}$ induces an isomorphism onto the choice-independent subquotient
\begin{equation}\label{eq-ch9-canonical-subquotient}
\begin{gathered}
\mathrm{BPS}_{\hbar^r}^k\;\cong\;\frac{Z_r^k}{B_r^k},\\
\operatorname{rank}\bigl(d_r:\mathrm{BPS}_{\hbar^{r-1}}^k\to\mathrm{BPS}_{\hbar^{r-1}}^{k+1}\bigr)
=\dim Z_{r-1}^k-\dim Z_r^k=\dim B_r^{k+1}-\dim B_{r-1}^{k+1}.
\end{gathered}
\end{equation}
In particular, $\dim\mathrm{BPS}_{\hbar^r}^k$ and $\operatorname{rank}d_r^k$ are independent of all choices of representatives and homotopies.
\item Let $[\mathcal O_0]_{\hbar^0}\in Z_{r-1}^k$ ($r\ge1$) with any partial completion $\mathcal O_0+\hbar\mathcal O_1+\cdots+\hbar^{r-1}\mathcal O_{r-1}$ solving $\sum_{j=0}^s Q_j\mathcal O_{s-j}=0$ for $0\le s<r$, and define the indeterminate subspace in degree $k+1$:
\begin{equation}\label{eq-ch9-indeterminate-subspace}
\begin{gathered}
W_r^{k+1}=\Bigl\{\sum_{\substack{j=0\\\text{loop orders}}}^{r-1}Q_j\mathcal Y_{r-1-j}:
\sum_{\substack{j=0\\\text{loop orders}}}^{s}Q_j\mathcal Y_{s-j}=0\;\;(0\le s\le r-2)\Bigr\},\\
W_1^{k+1}=\operatorname{im}Q_0,\qquad
W_2^{k+1}=\operatorname{im}Q_0+Q_1(\ker Q_0).
\end{gathered}
\end{equation}
Then $d_r[\mathcal O_0]_{\hbar^{r-1}}\ne0$ in $\mathrm{BPS}_{\hbar^{r-1}}^{k+1}$ if and only if $\sum_{j=1}^r Q_j\mathcal O_{r-j}\notin W_r^{k+1}$. When the operators $Q_0,\ldots,Q_r$ are defined over $\mathbb Q$ in a finite-dimensional sector basis, any rational linear functional $\ell$ vanishing on $W_r^{k+1}$ with $\ell\bigl(\sum_{j=1}^r Q_j\mathcal O_{r-j}\bigr)\ne0$ furnishes an exact, choice-independent certificate over $\mathbb Q$ that $d_r[\mathcal O_0]_{\hbar^{r-1}}\ne0$.
\end{enumerate}

\begin{proofbox}
\textbf{Proof.} For any tuple $(\mathcal X_0,\ldots,\mathcal X_{m-1})$ solving $\sum_{j=0}^s Q_j\mathcal X_{s-j}=0$ for $0\le s<m$, all negative powers of $\hbar$ cancel in $\mathcal O(\hbar)=-\hbar^{-m}\mathbf Q(\sum_{s=0}^{m-1}\hbar^s\mathcal X_s)\in\mathcal A_{\mathrm{loc}}[[\hbar]]$; its leading term is $\mathcal O_0=-\sum_{j=1}^m Q_j\mathcal X_{m-j}$, and nilpotence $\mathbf Q^2=0$ gives $\mathbf Q\mathcal O(\hbar)=0$ to all orders (with $\hbar^m[\mathcal O]_{\mathbf Q}=0$), proving $0=B_0^k\subseteq B_1^k\subseteq\cdots\subseteq B_r^k\subseteq Z_r^k\subseteq\cdots\subseteq Z_0^k$. We prove $\mathrm{BPS}_{\hbar^r}^k\cong Z_r^k/B_r^k$ by induction on $r$, the base case $r=0$ being $\mathrm{BPS}_{\hbar^0}^k=H^k(Q_0)$. Assuming the isomorphism through $r-1$, equation~\eqref{eq-2-75} shows that every $a\in\mathrm{BPS}_{\hbar^{r-1}}^k$ is represented by a tuple $(\mathcal O_0,\ldots,\mathcal O_{r-1})$ solving~\eqref{eq-liftingequations} through $r-1$, with $[\mathcal O_0]_{\hbar^0}\in Z_{r-1}^k$ well defined modulo $B_{r-1}^k$. Under the induction hypothesis on degree $k+1$, the class $d_r a=\bigl[\sum_{j=1}^r Q_j\mathcal O_{r-j}\bigr]_{\hbar^{r-1}}\in\mathrm{BPS}_{\hbar^{r-1}}^{k+1}\cong Z_{r-1}^{k+1}/B_{r-1}^{k+1}$ vanishes if and only if $\sum_{j=1}^r Q_j\mathcal O_{r-j}$ is $Q_0$-cohomologous to $-\sum_{j=1}^m Q_j\mathcal Y_{m-j}$ ($0\le m\le r-1$, with the empty sum at $m=0$) for some lower-order cocycle tuple $\mathcal Y$, i.e., if and only if there exist adjustments $(\widetilde{\mathcal Y}_1,\ldots,\widetilde{\mathcal Y}_{r-1})$ of $(\mathcal O_1,\ldots,\mathcal O_{r-1})$ with $\widetilde{\mathcal Y}_0=0$ and an $\mathcal O_r$ such that $Q_0\mathcal O_r+\sum_{j=1}^r Q_j(\mathcal O_{r-j}-\widetilde{\mathcal Y}_{r-j})=0$, which is precisely $[\mathcal O_0]_{\hbar^0}\in Z_r^k$ and $\sum_{j=1}^r Q_j\mathcal O_{r-j}\in W_r^{k+1}$. Likewise, the image of $d_r:\mathrm{BPS}_{\hbar^{r-1}}^{k-1}\to\mathrm{BPS}_{\hbar^{r-1}}^k$ corresponds under $Z_{r-1}^k/B_{r-1}^k$ precisely to $B_r^k/B_{r-1}^k$. Taking $(\ker d_r)/(\operatorname{im}d_r)=(Z_r^k/B_{r-1}^k)/(B_r^k/B_{r-1}^k)\cong Z_r^k/B_r^k$ proves both~(1) and the subspace criterion $\sum_{j=1}^r Q_j\mathcal O_{r-j}\notin W_r^{k+1}$ in~(2). Over $\mathbb Q$, a vector $v$ lies outside a subspace $W$ if and only if there exists a linear functional $\ell\in W^\perp$ with $\ell(v)\ne0$.~\hfill\qedsymbol
\end{proofbox}

\rthm{Corollary}{cor-ch9-fortuitous-bound}{Fortuitous bound on quantum lifting}
In $\mathcal N=4$ Yang--Mills theory at fixed gauge group and conserved charge sector, let $M^k\subseteq\mathrm{BPS}_{\hbar^0}^k$ be the multi-graviton subspace and $F^k=\mathrm{BPS}_{\hbar^0}^k/M^k$ the fortuitous quotient. Then for every $R\ge1$,
\begin{equation}\label{eq-ch9-fortuitous-rank-bound}
\begin{aligned}
\sum_{\substack{r=1\\\text{obstruction orders}}}^{R}
\operatorname{rank}\bigl(d_r:\mathrm{BPS}_{\hbar^{r-1}}^k\to\mathrm{BPS}_{\hbar^{r-1}}^{k+1}\bigr)
&\le\dim F^k,\\
\dim(M^k\cap B_R^k)\le\sum_{\substack{r=1\\\text{obstruction orders}}}^{R}
\operatorname{rank}\bigl(d_r:\mathrm{BPS}_{\hbar^{r-1}}^{k-1}\to\mathrm{BPS}_{\hbar^{r-1}}^k\bigr)
&\le\dim F^{k-1}.
\end{aligned}
\end{equation}

\begin{proofbox}
\textbf{Proof.} By Theorem~\ref{b-00630}, every class in $M^k$ admits an all-order $\mathbf Q$-closed completion, so $M^k\subseteq Z_R^k$ for every $R\ge1$. Telescoping~\eqref{eq-ch9-canonical-subquotient} gives
\begin{equation*}
\begin{aligned}
\sum_{r=1}^R\operatorname{rank}d_r^k
&=\sum_{r=1}^R(\dim Z_{r-1}^k-\dim Z_r^k)
=\dim\mathrm{BPS}_{\hbar^0}^k-\dim Z_R^k\\
&\le\dim\mathrm{BPS}_{\hbar^0}^k-\dim M^k=\dim F^k.
\end{aligned}
\end{equation*}
Similarly, $\dim(M^k\cap B_R^k)\le\dim B_R^k=\sum_{r=1}^R(\dim B_r^k-\dim B_{r-1}^k)=\sum_{r=1}^R\operatorname{rank}d_r^{k-1}\le\dim F^{k-1}$, bounding the number of linearly independent multi-graviton classes in degree $k$ that can be hit by incoming differentials through order $R$.~\hfill\qedsymbol
\end{proofbox}

\rthm{Proposition}{ch9-finiteness}{Finite graph operations and bounded sectors}
For the interaction~\eqref{eq-B-10} with a polynomial superpotential, the antifield counts give for every polynomial local operator $\mathcal O,A,B$:
\begin{equation}\label{ch9-general-graph-bound}
\begin{aligned}
Q_n\mathcal O&=0
&&\bigl(n>\#(b,\beta\text{ in }\mathcal O)\bigr),\\
[\hbar^n](A\star B)&=0
&&\bigl(n>\#(b,\beta\text{ in }A)+\#(b,\beta\text{ in }B)\bigr).
\end{aligned}
\end{equation}
The homotopy transfer~\eqref{eq-page-transfer} is defined coefficientwise for every formal differential; on any sector with bounded holomorphic derivative totals $\max(n_+,n_-)<\infty$, its resolvent series and reconstructed completions $i_0(\hbar)\cdots i_{r-1}(\hbar)i_ra$ terminate as finite polynomials in $\hbar$ whenever $Q_0$ preserves $n_\pm$ and $Q_n$ raises both $n_+$ and $n_-$ by $n$.

\begin{proofbox}
\textbf{Proof.} Every physical propagator consumes one $b$ or $\beta$, every gauge or matter interaction vertex supplies one, and a $\bar W$ vertex supplies none, so~\eqref{eq-antifield-loss} and~\eqref{eq-product-antifield-loss} give
\begin{equation}\label{ch9-antifield-count}
\begin{aligned}
\#(b,\beta)_{\mathrm{out}}
&=\#(b,\beta)_{\mathcal O}-n-\#(\bar W\text{ vertices}),\\
\#(b,\beta)_{\mathrm{out}}
&=\#(b,\beta)_A+\#(b,\beta)_B-n-\#(\bar W\text{ vertices}).
\end{aligned}
\end{equation}
Nonnegativity of the surviving antifield count proves~\eqref{ch9-general-graph-bound} (with formal superpotentials read coefficientwise).
For interaction vertices without holomorphic derivatives, the current-form homogeneity~\eqref{eq-derivative-homogeneity} gives $n_\pm(Q_n a)=n_\pm(a)+n$, where $n_\pm$ count $\partial_{+\dot\pm}$ derivatives. Choosing homogeneous contractions $(i_r,\pi_r,h_r)$ so that $i_r,\pi_r$ preserve $n_\pm$ and $h_r$ inverts the shift of $d_r$, any nonzero output obeys by induction on $r$:
\begin{equation}\label{eq-ch8-transfer-lowering}
\begin{aligned}
n_\pm(h_ra)&=n_\pm(a)-r,\\
n_\pm\bigl(([\hbar^m]\mathbf Q^{(r)})a\bigr)&=n_\pm(a)+r+m,\\
n_\pm\bigl(h_r([\hbar^m]\mathbf Q^{(r)})a\bigr)
&=n_\pm\bigl(([\hbar^m]\mathbf Q^{(r)})h_ra\bigr)=n_\pm(a)+m,
\end{aligned}
\end{equation}
since a transfer word with $\ell$ perturbation factors of total $\hbar$-excess $m+1$ and $\ell-1$ homotopies shifts $n_\pm$ by $\ell r+(m+1)-(\ell-1)r=r+m+1$. Consequently,
\begin{equation}\label{eq-ch8-transfer-finite-inverses}
\begin{gathered}
[\hbar^s]\bigl[1+h_r(\mathbf Q^{(r)}-d_r)\bigr]^{-1}a
=[\hbar^s]\bigl[1+(\mathbf Q^{(r)}-d_r)h_r\bigr]^{-1}a=0,\\
s>\min_{\pm}\bigl(\max_{\text{fixed sector}}n_\pm-n_\pm(a)\bigr),
\end{gathered}
\end{equation}
which proves finite termination in $\hbar$ whenever the sector maximum of $n_+$ or $n_-$ is finite (including any effective-theory truncation with bounded derivative order).~\hfill\qedsymbol
\end{proofbox}
In particular, in all theories studied in Section~\ref{sec-applications} where a positive conserved weight bounds $n_++n_-$ at fixed quantum numbers---including $\mathcal N=4$ Yang--Mills theory and its Leigh--Strassler and $\beta$-deformations (where $2(E+J_L)=2\Theta+3(n_++n_-)$ is conserved), $\mathcal N=2$ gauge theories, and the mass-deformed theories ordered by $w_{\mathrm{IR}}=q_A+q_B+2(n_++n_-)$---every survivable polynomial local operator $\mathcal O_0$ in the original fields and their derivatives admits a quantum completion $\mathcal O=\sum_{s=0}^S\hbar^s\mathcal O_s$ that is a finite sum in $\hbar$. In $\mathcal N=4$ SYM and its marginal deformations, the component degree $\Theta=\deg_\theta$ of Appendix~\ref{sec-E-sectors} gives the explicit degree drop
\begin{equation}\label{local-ch8-finite-correction-degree}
\begin{gathered}
\deg_\theta\mathcal O_s=\deg_\theta\mathcal O_0-3s,
\qquad \mathcal O_s=0\quad(3s>\deg_\theta\mathcal O_0),\\
\deg_\theta\bigl(h_r([\hbar^m]\mathbf Q^{(r)})a\bigr)
=\deg_\theta\bigl(([\hbar^m]\mathbf Q^{(r)})h_ra\bigr)
=\deg_\theta a-3m,
\end{gathered}
\end{equation}
while for unrestricted couplings without a conserved positive weight, the resolvents in~\eqref{eq-ch8-transfer-inverses} remain well defined coefficientwise as formal series in $\hbar$.

\paragraph{Products and operator mixing.}\label{products-and-classical-representatives}
\label{b-00716}\label{b-00718}\label{b-00720}%
Suppose $A=\sum_{s<r}\hbar^sA_s$ and $B=\sum_{s<r}\hbar^sB_s$ solve the closure equations~\eqref{eq-liftingequations} through order $r-1$, and put $\mathcal O_s=[\hbar^s](A\star B)$ with $\mathcal O_0=A_0B_0$. Expanding the Leibniz identity~\eqref{eq-chain} through order $r$ gives
\begin{equation}\label{eq-2-77}
\begin{gathered}
\sum_{\substack{j=0\\\text{loop orders}}}^{s}Q_j\mathcal O_{s-j}=0\qquad(0\le s<r),\\
\begin{aligned}
Q_0\mathcal O_r+\sum_{\substack{j=1\\\text{loop orders}}}^{r}Q_j\mathcal O_{r-j}
={}&\Bigl(\sum_{\substack{j=1\\\text{loop orders}}}^{r}Q_jA_{r-j}\Bigr)B_0\\
&+(-1)^{|A_0|}A_0\Bigl(\sum_{\substack{j=1\\\text{loop orders}}}^{r}Q_jB_{r-j}\Bigr).
\end{aligned}
\end{gathered}
\end{equation}
Passing to $\mathrm{BPS}_{\hbar^{r-1}}$ shows that the classical product descends to each successive cohomology space and that $d_r$ acts as a graded derivation:
\begin{equation}\label{eq-2-78}
\begin{aligned}
[A_0]_{\hbar^{r-1}}[B_0]_{\hbar^{r-1}}&=[A_0B_0]_{\hbar^{r-1}},\\
d_r([A_0]_{\hbar^{r-1}}[B_0]_{\hbar^{r-1}})
&=(d_r[A_0]_{\hbar^{r-1}})[B_0]_{\hbar^{r-1}}+(-1)^{|A_0|}[A_0]_{\hbar^{r-1}}(d_r[B_0]_{\hbar^{r-1}}),\\
[A_0]_{\hbar^{r-1}}=d_ra,\quad d_r[B_0]_{\hbar^{r-1}}=0
&\quad\Longrightarrow\quad [A_0B_0]_{\hbar^{r-1}}=d_r\bigl(a[B_0]_{\hbar^{r-1}}\bigr),
\end{aligned}
\end{equation}
inheriting associativity, graded commutativity, and the unit from $\mathrm{BPS}_{\hbar^0}=H(Q_0)$ (for $r=3$, the third line of~\eqref{eq-2-78} yields the level-$48$ three-loop lift~\eqref{eq-so13-s8-product-lift} of $[S^8]$ in $SO(13)$ from $[S^7]$).

\rthm{Proposition}{prop-cartan-image-vanishing}{All-order Cartan invisibility of perturbative differential images}
\emph{In $\mathcal N=4$ Yang--Mills theory with classical gauge group $SU(N)$, $SO(N)$, or $Sp(n)$, restrict to the multi-trace subring $\mathcal R_{\mathrm{mt}}\subseteq\mathcal A_{\mathrm{loc}}$ (which equals all of $\mathcal A_{\mathrm{loc}}$ for $SU(N)$, $SO(2n+1)$, and $Sp(n)$, and the $O(2n)$-even direct summand $\mathcal A_{\mathrm{loc}}^+$ for $SO(2n)$), and let $\rho_0:H(Q_0)\to\mathcal A_{\mathrm{Cartan}}$ denote Cartan restriction. Then for every $r\ge1$, $\rho_0$ descends to an algebra homomorphism $\rho_{r-1}:\mathrm{BPS}_{\hbar^{r-1}}(\mathcal R_{\mathrm{mt}})\to\mathcal A_{\mathrm{Cartan}}$ satisfying}
\begin{equation}\label{eq-cartan-image-zero}
\rho_{r-1}(d_ry)=0\qquad\forall\,y\in\mathrm{BPS}_{\hbar^{r-1}}(\mathcal R_{\mathrm{mt}}).
\end{equation}
\emph{In particular, every residual $R_r=\sum_{j=1}^r Q_j\mathcal O_{r-j}$ of a multi-trace partial completion obeys $\rho_0([R_r]_{\hbar^0})=0$, no Cartan-visible multi-trace class is $d_r$-exact for any $r\ge1$, and $d_r=0$ into every multi-trace target sector satisfying $\HQ=\HQC$.}

\begin{proofbox}
\textbf{Proof.} Let $\Delta\in\mathcal R_{\mathrm{mt}}$ be the characteristic-polynomial discriminant of the single scalar $\gamma^1$ in the defining representation. Gauge invariance gives $Q_0\Delta=0$. Since $\Delta$ carries $\theta_2=\theta_3=n_+=n_-=0$, degree counting~\eqref{eq-graviton-differential-degree} gives $Q_n\Delta=0$ for all $n\ge1$, so $\mathbf Q\Delta=0$ and $d_r[\Delta]_{\hbar^{r-1}}=0$ for every $r\ge1$; moreover, a generic Cartan element in the defining representation of $SU(N)$, $SO(N)$ (both $N=2n$ and $N=2n+1$), or $Sp(n)$ has pairwise distinct eigenvalues and $\Delta$ vanishes on every root hyperplane, so $\rho_0([\Delta]_{\hbar^0})$ is a nonzero even polynomial and hence a non-zero-divisor in the Cartan super-polynomial algebra $\mathcal A_{\mathrm{Cartan}}$. Assume inductively that $\rho_{r-1}$ is well defined on $\mathrm{BPS}_{\hbar^{r-1}}(\mathcal R_{\mathrm{mt}})\cong Z_{r-1}/B_{r-1}$ (trivial for $r=1$). For any $y=[x_0]_{\hbar^{r-1}}$, the classical discriminant lemma~\cite[Section 6.2]{KvLM} supplies $m\ge0$ and a multi-graviton polynomial $G$ such that $[\Delta]_{\hbar^0}^m[x_0]_{\hbar^0}=[G]_{\hbar^0}$ in $H(Q_0)$. Since multiplication on $\mathrm{BPS}_{\hbar^{r-1}}$ is inherited from $H(Q_0)$ by~\eqref{eq-2-78} and $[G]_{\hbar^0}\in Z_{r-1}$ by Theorem~\ref{b-00630}, we have $[\Delta]_{\hbar^{r-1}}^m y=[G]_{\hbar^{r-1}}$ in $\mathrm{BPS}_{\hbar^{r-1}}$. Theorem~\ref{b-00630} and~\eqref{eq-2-75} give $d_r[G]_{\hbar^{r-1}}=0$, so the derivation rule~\eqref{eq-2-78} yields $[\Delta]_{\hbar^{r-1}}^m d_ry=d_r([\Delta]_{\hbar^{r-1}}^m y)=0$. Applying $\rho_{r-1}$ gives $\rho_0([\Delta]_{\hbar^0})^m\rho_{r-1}(d_ry)=0$ in $\mathcal A_{\mathrm{Cartan}}$; cancelling the non-zero-divisor $\rho_0([\Delta]_{\hbar^0})^m$ proves $\rho_{r-1}(d_ry)=0$, so $B_r\subseteq\ker\rho_0$ and $\rho_{r-1}$ descends to $\rho_r$.~\hfill\qedsymbol
\end{proofbox}

\rthm{Theorem}{product-uniqueness-ch9-p01}{Operator mixing and renormalization}
\label{product-uniqueness-ch9-p02}%
Fix a solution of the quantum master equation, its local BV differential $\mathbf Q$, and its marked product $\star$. Given a closed completion $\mathcal O$ of $\mathcal O_0$, every closed completion $\mathcal O'$ sharing the same classical leading term $\mathcal O'_0=\mathcal O_0$ has the form
\begin{equation}\label{eq-lift-finite-freedom}
\begin{gathered}
\mathbf Q\mathcal O'=0,\quad\mathcal O'_0=\mathcal O_0
\quad\Longleftrightarrow\quad
\mathcal O'=\mathcal O+\hbar Z,\quad\mathbf QZ=0,\\
[\mathcal O']_{\mathbf Q}=[\mathcal O]_{\mathbf Q}
\quad\Longleftrightarrow\quad \hbar[Z]_{\mathbf Q}=0,
\end{gathered}
\end{equation}
where $Z$ carries the conserved charges required by the $\hbar$-shift; thus the set of completed classes over $[\mathcal O_0]_{\hbar^0}$ is the affine coset $[\mathcal O]_{\mathbf Q}+\hbar H(\mathbf Q)$ in that charge sector, and changing the completion shifts quantum products by
\begin{equation}\label{eq-product-lift-mixing}
\mathbf QA=\mathbf QB=\mathbf QZ=0
\quad\Longrightarrow\quad
[(A+\hbar Z)\star B]_{\mathbf Q}
=[A\star B]_{\mathbf Q}+\hbar[Z\star B]_{\mathbf Q}.
\end{equation}
Furthermore, any even, local, invertible $\mathbb C[[\hbar]]$-linear operator redefinition $A\mapsto A'$ equal to the identity modulo $\hbar$ and preserving the unit and physical gradings transports $(\mathbf Q,\star)$ by a unital ring isomorphism (commuting with $\mathbf Q$ whenever $\mathbf Q'=\mathbf Q$):
\begin{equation}\label{eq-product-finite-redefinition}
\begin{gathered}
\mathbf Q'A'=(\mathbf QA)',\qquad
A'\star' B'=(A\star B)',\qquad 1'=1,\\
[A]_{\mathbf Q}\longmapsto[A']_{\mathbf Q'},\qquad
[A\star B]_{\mathbf Q}\longmapsto[A'\star'B']_{\mathbf Q'}.
\end{gathered}
\end{equation}
Two admissible local product prescriptions $\star,\star'$ on the same complex $(\mathcal A_{\mathrm{loc}}[[\hbar]],\mathbf Q)$ induce the same ring structure on $H(\mathbf Q)$ if and only if their difference on every closed pair is locally $\mathbf Q$-exact:
\begin{equation}\label{eq-product-contact-equivalence}
[A\star' B]_{\mathbf Q}=[A\star B]_{\mathbf Q}
\quad\Longleftrightarrow\quad
A\star' B-A\star B\in\operatorname{im}\mathbf Q
\quad\text{in local operators}.
\end{equation}
Finally, a finite action deformation $I'[L]-I[L]$ of positive $\hbar$-order (respecting locality, symmetry, and renormalization-group flow) preserves the regularized quantum master equation if and only if it solves the Maurer--Cartan equation
\begin{equation}\label{eq-finite-action-QME}
\mathbf Q_L\bigl(I'[L]-I[L]\bigr)
+\frac12\bigl(I'[L]-I[L],I'[L]-I[L]\bigr)_L=0.
\end{equation}

\begin{proofbox}
\textbf{Proof.} Writing $\mathcal O'-\mathcal O=\hbar Z$ for $\mathcal O'_0=\mathcal O_0$, injectivity of multiplication by $\hbar$ on formal local-operator series gives
\begin{equation}\label{eq-mixing-closed}
\mathcal O'-\mathcal O=\hbar Z,\qquad
0=\mathbf Q(\mathcal O'-\mathcal O)=\hbar\mathbf QZ
\quad\Longrightarrow\quad \mathbf QZ=0,
\end{equation}
and passing to $H(\mathbf Q)$ yields~\eqref{eq-lift-finite-freedom} (retaining any $\hbar$-torsion classes obeying $\hbar[Z]_{\mathbf Q}=0$). $\mathbb C[[\hbar]]$-bilinearity of $\star$ gives~\eqref{eq-product-lift-mixing}, while transporting $(\mathbf Q,\star)$ along an invertible redefinition $A\mapsto A'$ preserves nilpotence and the Leibniz rule:
\begin{equation}\label{ch9-redefinition-proof}
\begin{gathered}
(\mathbf Q')^2A'=(\mathbf Q^2A)'=0,\qquad
(\mathbf QX)'=\mathbf Q'X',\\
\mathbf Q'(A'\star' B')
=(\mathbf Q'A')\star' B'+(-1)^{|A|}A'\star'(\mathbf Q'B'),
\end{gathered}
\end{equation}
whose inverse transports cocycles, coboundaries, and products back; \eqref{eq-product-contact-equivalence} is tautological in the quotient $H(\mathbf Q)=\ker\mathbf Q/\operatorname{im}\mathbf Q$. Lastly, subtracting the regularized quantum master equations for $I'[L]$ and $I[L]$ gives
\begin{equation}\label{ch9-action-proof}
\begin{aligned}
0={}&Q_{\mathrm{BV}}\bigl(I'[L]-I[L]\bigr)
+\hbar\Delta_L\bigl(I'[L]-I[L]\bigr)
+\bigl(I[L],I'[L]-I[L]\bigr)_L\\
&+\frac12\bigl(I'[L]-I[L],I'[L]-I[L]\bigr)_L\\
={}&\mathbf Q_L\bigl(I'[L]-I[L]\bigr)
+\frac12\bigl(I'[L]-I[L],I'[L]-I[L]\bigr)_L,
\end{aligned}
\end{equation}
which is~\eqref{eq-finite-action-QME}.~\hfill\qedsymbol
\end{proofbox}
The marked-graph prescription fixes the contact terms canonically, with any admissible change of contact prescription governed by~\eqref{eq-product-contact-equivalence}. Infinitesimally $\mathbf Q_L$-exact deformations of $I[L]$ generate quantum BV canonical transformations (integrated to finite equivalences compatible with locality and scale transport), whereas non-exact solutions of~\eqref{eq-finite-action-QME} shift physical couplings or quantization choices and are held fixed when comparing products within a given quantum theory. Finally, while multiplication on each $\mathrm{BPS}_{\hbar^{r-1}}$ is inherited from $H(Q_0)$, classical ring relations $A_0B_0=Q_0X_0$ among chosen quantum completions $\mathbf QA=\mathbf QB=0$ can acquire quantum corrections of order $\hbar$:
\begin{equation}\label{eq-product-classical-relation}
\begin{gathered}
\mathbf QA=\mathbf QB=0,\qquad A_0B_0=Q_0X_0,\\
A\star B-\mathbf QX_0=\hbar\mathcal O,\qquad
0=\hbar\mathbf Q\mathcal O\quad\Longrightarrow\quad\mathbf Q\mathcal O=0,\\
[A\star B]_{\mathbf Q}=\hbar[\mathcal O]_{\mathbf Q},
\end{gathered}
\end{equation}
where cancelling the $\hbar^0$ term and applying $\mathbf Q$ shows $\mathbf Q\mathcal O=0$; the classical relation persists unmodified in $H(\mathbf Q)$ if and only if $\hbar[\mathcal O]_{\mathbf Q}=0$.

\section{Discussion}\label{sec-conclusion}

We close with three directions.

\emph{The physical theory.} All our results are obtained in the holomorphically twisted theory. It remains to understand the same structures directly in the physical theory: the supersymmetry transformation of composite operators and, more generally, the full BV transformation that includes every symmetry together with its ghosts and auxiliary fields. This requires complete control of the counterterms and of the ultraviolet divergences of composite operators, and would give a complete account of quantum anomalies in this generalized sense and of the exact structure of local operators in supersymmetric gauge theories. It is considerably harder than the twisted computation suggests.

\emph{Lifting beyond one loop.} Our orthogonal results establish that perturbative BPS lifting in undeformed $\mathcal N=4$ Yang--Mills theory extends beyond one loop: alongside the low-level one-loop sectors ($SU(2)$ \cite{KKLP}, sixteen sectors in $SO(7)$ starting with \cite{KonishiBH}, and the BMN thresholds of $SO(9)$ and $SO(11)$ \cite[Section 6.6]{FortuityAI}) and the one-loop Pfaffian glueball family~\eqref{eq-orthogonal-d1-family}, exact rational certificates give three-loop lifts ($d_3\ne0$) in $SO(13)$ at BMN level $34$ and mixed levels $40,46,47$, and into the glueball powers $[S^6]$ in $SO(11)$, $[S^7]$ in $SO(12)$ and $SO(13)$, and $[S^8]$ in $SO(13)$. Each target survives the first two pages (Table~\ref{tab-orthogonal-d3-lifts}). For $SO(19)$, $SO(25)$ and $SO(26)$, discussion D014 of \cite{Repo} reports that the first Cartan-invisible glueball powers $[S^{10}]$ and $[S^{13}],[S^{14}]$ survive through $\mathrm{BPS}_{\hbar^4}$ and $\mathrm{BPS}_{\hbar^5}$, respectively. These lower bounds motivate the following conjecture.

\rthm{Conjecture}{conj-all-loop-lifting}{All-loop quantum lifting in $\mathcal N=4$ Yang--Mills theory}
\emph{In undeformed four-dimensional $\mathcal N=4$ Yang--Mills theory, ranging over compact gauge groups and conserved-charge sectors, every loop order $n\ge1$ produces nontrivial perturbative BPS lifts ($d_n\ne0$), and more strongly every $n$-loop supercharge $Q_n$ itself contributes nontrivially to lifting classical cohomology classes.}

Testing the $n=2$ case ($d_2\ne0$) in undeformed $\mathcal N=4$ Yang--Mills theory---where all scanned sectors to date have $d_2=0$, in contrast to the $12$ two-loop sector pairs of the $\mathcal N=1$ $SU(3)$ Leigh--Strassler deformation (Table~\ref{local-ch9-ls-lifts}) and the $SU(3)$ $\beta$-deformation---and exhibiting $\mathcal N=4$ sectors where the direct higher-loop fans $Q_{n\ge2}$ act nontrivially alongside iterated $Q_1$ transfer are immediate targets. Finally, nonperturbative lifts of perturbatively surviving classes by instantons \cite{semichiral,BK} remain a central direction: under a leading instanton differential compatible with the preserved linear superconformal Ward filtration and closed primary graviton products, source exhaustion protects the $SU(2)$ threshold $\mathcal O_0$ \cite{FortuityAI,Repo}, the level-$36$ $SU(2)$ dressed-threshold pair $(\mathcal O'_1,\mathcal O''_1)$ of \cite[Section 3.2]{KKLP}, the $G_2$ glueball cube $[S^3]$ and level-$24$ fortuitous class $(5,2,2;1,1)_9^{\mathrm{a}}$, and the $F_4$ fifth glueball $[S^5]$ \cite{Repo}, whereas fortuitous classes admitting unspanned opposite-statistics partners---such as $\mathrm{P39}$ in $SU(2)$ and the $F_1,B_1$ towers in $SU(3)$---provide concrete tests for microscopic instanton matrix elements.

\emph{Other dimensions and the gravity dual.} The circuit reduction is not tied to four dimensions: the graph integrals and their current equations exist on $\mathbb C^d$ and in holomorphic--topological theories \cite{Wang,WW}, and the iterated network reductions should evaluate the corresponding Feynman graphs. It would also be interesting to identify the counterparts of the loop corrections, of the quantum product and of lifting in the gravity dual.

\section*{Acknowledgments}\label{sec-acknowledgments}

We thank Si Li (Tsinghua University) for helpful discussions, which drew our attention to the original work of Wang \cite{Wang}. We also thank Eunwoo Lee for sustained and helpful discussions, particularly on the nilpotence of $\mathbf Q$ and the associated physical ideas. We thank Kasia Budzik for helpful discussions.

BL acknowledges support from the Peking University Global Influence High-Level Overseas Study Program for Undergraduates. KL is supported by the NSFC special fund for theoretical physics No.~12447108 and the National Key Research and Development Program of China No.~2020YFA0713000. KL acknowledges support from the Tsinghua University Short-term Visiting Scholarship for Doctoral Students. C.C. is supported by NSFC Grant No.~12575075.

AI tools were used to explore simplification routes for the Feynman graphs and to improve the algorithms and the code of the lifting tests.

\appendix

\section{Notation}\label{sec-notation}

This appendix collects the superfield of $\mathcal N=4$ Yang--Mills theory and the charge sectors and gradings used in the lifting tests.

\subsection{The \texorpdfstring{$\mathcal N=4$}{N=4} Yang--Mills superfield}\label{sec-N4-superfield}

The $\mathcal N=4$ SYM theory has the antichiral superpotential
\begin{equation}\label{b-01022}
\bar W(\gamma,0)=\frac i{3!}\varepsilon_{IJK}C_{ABC}
\delta^{AD}\delta^{BE}\delta^{CF}(\gamma^I)_D(\gamma^J)_E(\gamma^K)_F.
\end{equation}
For the twist by $Q_-^4$, the three adjoint chiral multiplets combine with the gauge fields into one superfield $C^A(\theta)$ on $\mathbb C^{2|3}$ \cite{ChangYin,BK}. The odd coordinates obey $\theta_I\bar\partial=-\bar\partial\theta_I$, with $\int d^3\theta\,\theta_1\theta_2\theta_3=1$ and $\varepsilon^{123}=\varepsilon_{123}=1$:
\begin{equation}\label{eq-A-13}
C^A(\theta)=ig\,c^A+\theta_I\delta^{AB}(\gamma^I)_B-\frac12\varepsilon^{IJK}\theta_I\theta_J(\beta_K)^A+\frac{\theta_1\theta_2\theta_3}{ig}\delta^{AB}b_B,
\end{equation}
where each $\theta_I$ implements $Q_+^I C^A=\partial_L C^A/\partial\theta_I$. With $(t_A^{\mathrm{adj}})^B{}_C=iC^B{}_{AC}$, the twisted action and tree differential take the holomorphic Chern--Simons form
\begin{equation}\label{eq-A-15}
\begin{aligned}
S_{\mathrm{tw}}
&=\int_{\mathbb C^2}\Omega\wedge
\left[\int d^3\theta\left(
\frac12\delta_{AB}C^A\bar\partial C^B
+\frac{i}{3!}C_{ABC}C^AC^BC^C\right)\right]_{(0,2)},\\
Q_0C^A&=\frac i2C^A{}_{BC}C^BC^C.
\end{aligned}
\end{equation}
Left differentiation with respect to the superfield and Berezin integration reproduce the component BV pairing and edge contraction operator:
\begin{equation}\label{eq-A-16}
\begin{aligned}
\frac{\delta}{\delta C^A(\theta)}
={}&\frac{\theta_1\theta_2\theta_3}{ig}
\frac{\partial_L}{\partial c^A}
-\frac12\varepsilon^{IJK}\theta_J\theta_K
\frac{\partial}{\partial(\gamma^I)_A}
-\theta_I\frac{\partial_L}{\partial(\beta_I)^A}
-ig\frac{\partial}{\partial b_A},\\
\frac{\delta C^B(\theta^{(2)})}{\delta C^A(\theta^{(1)})}
={}&\delta_A{}^B\prod_{I=1}^3
(\theta_I^{(1)}-\theta_I^{(2)}),
\end{aligned}
\end{equation}
\begin{equation}\label{eq-A-17}
\Delta_{vw}=-\delta^{DE}\int d^3\theta\,\frac{\delta}{\delta C_v^D(\theta)}\frac{\delta}{\delta C_w^E(\theta)} .
\end{equation}

\subsection{Charge sectors and the grading of the loop corrections}\label{sec-E-sectors}

For $\mathcal N=4$ Yang--Mills theory, we organize cochains by their letter weights and derivative counts; counting vertices and contractions determines the target sector of each loop correction and bounds the admissible obstruction orders. For the $\mathcal N=4$ letters, $\theta_I$ counts the $I$th odd-coordinate weight in $C$, and $n_\pm$ counts $\partial_{+\dot\pm}$ derivatives. For a monomial with $Y$ letters, let $e_I$ be the $I$th unit vector and write
\begin{equation}\label{eq-E-sector}
\begin{gathered}\lambda=(\theta_1,\theta_2,\theta_3;n_+,n_-),\qquad \Theta=\theta_1+\theta_2+\theta_3,\qquad \mathbf d=(1,1,1;-1,-1),\\ \theta(\gamma^I)=e_I,\qquad \theta(\beta_I)=(1,1,1)-e_I,\qquad \theta(b)=(1,1,1),\qquad \theta(c)=0.\end{gathered}
\end{equation}
Here $C^Y(\lambda)$, also written $(\lambda)_Y$, denotes the cochain sector of weight $\lambda$ and $Y$ letters, with total component degree $\Theta=\deg_\theta$:
\begin{equation}\label{eq-2-62}
\deg_\theta c=0,\quad\deg_\theta\gamma=1,\quad\deg_\theta\beta=2,\quad\deg_\theta b=3,\qquad\deg_\theta(\partial_{+\dot a}X)=\deg_\theta X.
\end{equation}
At loop order $n$, a marked product graph has $n$ cubic vertices and $2n$ physical edges, each pairing component degrees that sum to three, so degree losses add under iterated products:
\begin{equation}\label{eq-2-63}
\deg_\theta\mathcal O_n=\deg_\theta\mathcal O_0+{3n}-{6n}=\deg_\theta\mathcal O_0-3n,\qquad\mathcal O_n=0\quad\text{if }3n>\deg_\theta\mathcal O_0.
\end{equation}
Similarly, for a nonzero $Q_nX$, the $n+1$ cubic vertices and $2n+1$ edges give
\begin{equation}\label{eq-graviton-differential-degree}
\deg_\theta(Q_nX)
=\deg_\theta X+3(n+1)-3(2n+1)=\deg_\theta X-3n.
\end{equation}

The physical conserved charges are the scaling dimension $E$, the three $SU(4)_R$ Cartan charges $R_I$, and the spins $J_L,J_R$:
\begin{equation}\label{eq-E-charges}
\begin{gathered}E=\tfrac12(\Theta+Y)+n_++n_-,\qquad R_I=\theta_I-\tfrac12(\Theta-Y),\\ J_L=\tfrac12(\Theta-Y+n_++n_-),\qquad J_R=\tfrac12(n_+-n_-),\\ E=R_1+R_2+R_3+2J_L,\qquad 2(E+J_L)=2\Theta+3(n_++n_-).\end{gathered}
\end{equation}
By \eqref{eq-E-charges}, only finitely many sectors occur at each fixed level $2(E+J_L)$.

\needspace{8\baselineskip}
{\small\renewcommand{\arraystretch}{1.25}\setlength{\tabcolsep}{4pt}
\begin{xltabular}{\linewidth}{@{}l>{\raggedright\arraybackslash}Xllllll@{}}
\caption{Letters of $\mathcal N=4$ Yang--Mills theory and their charges.}\label{appE2-0020}\\
\toprule
letter & physical field & $\theta$ & $(n_+,n_-)$ & $E$ & $(R_1,R_2,R_3)$ & $J_L$ & $J_R$\\
\midrule\endfirsthead
\toprule
letter & physical field & $\theta$ & $(n_+,n_-)$ & $E$ & $(R_1,R_2,R_3)$ & $J_L$ & $J_R$\\
\midrule\endhead
$\gamma^I$ & scalar $\bar\phi_I$ & $e_I$ & $(0,0)$ & $1$ & $e_I$ & $0$ & $0$\\
$\beta_I$ & fermion $\psi^I_+$ & $(1,1,1)-e_I$ & $(0,0)$ & $\tfrac32$ & $(\tfrac12,\tfrac12,\tfrac12)-e_I$ & $\tfrac12$ & $0$\\
$b$ & field strength $f_{++}$ & $(1,1,1)$ & $(0,0)$ & $2$ & $(0,0,0)$ & $1$ & $0$\\
$\partial_{+\dot\pm}c$ & gaugino $\bar\lambda_{\dot\pm}$ & $0$ & $(1,0),(0,1)$ & $\tfrac32$ & $(\tfrac12,\tfrac12,\tfrac12)$ & $0$ & $\pm\tfrac12$\\
$\partial_{+\dot\pm}$ & covariant derivative $\nabla_{+\dot\pm}$ & $0$ & $(1,0),(0,1)$ & $1$ & $(0,0,0)$ & $\tfrac12$ & $\pm\tfrac12$\\
\bottomrule
\end{xltabular}}

\rthm{Example}{ex-threshold-charges}{Threshold charges in $SU(2)$} The $SU(2)$ threshold class $\mathcal O_0$ \cite{KKLP,ChangLinWords} has two $\gamma$ letters, five $\beta$ letters and zero derivatives:
\begin{equation*}
\begin{gathered}\lambda=(4,4,4;0,0),\quad Y=7,\\E=\frac{19}{2},\qquad R_I=\frac32,\qquad J_L=\frac52,\qquad 2(E+J_L)=24.\end{gathered}
\end{equation*}

Because the current form \eqref{eq-operatope} has degree $n$ in each dotted component, independent rescaling at fixed $\dot a\in\{\dot+,\dot-\}$ gives
\begin{equation}\label{eq-derivative-homogeneity}
\mathcal I_\Gamma
\bigl(\lambda_{v,+\dot a}\mapsto \kappa\lambda_{v,+\dot a};\,
z_e^{+\dot a}\mapsto \kappa^{-1}z_e^{+\dot a}\bigr)
=\kappa^n\,\mathcal I_\Gamma(\lambda;z),
\qquad \kappa>0.
\end{equation}
The $\lambda$ powers supply output derivatives while the $z$ powers remove input derivatives, so $Q_n$ adds $n$ derivatives in each dotted direction \cite[Equations (3.44)--(3.45)]{semichiral}. Since $Q_0$ preserves both derivative totals, we may choose homogeneous contractions $(i_r,\pi_r,h_r)$ in \eqref{eq-page-transfer}.

\rthm{Proposition}{appE2-0024}{Sector shifts and charge increments} \emph{For $n\ge0$ and $r\ge1$,}
\begin{equation}\label{eq-E-grading}
Q_n:C^Y(\lambda)\to C^{Y+1-n}(\lambda-n\mathbf d),\qquad d_r:\mathrm{BPS}_{\hbar^{r-1}}^Y(\lambda)\to \mathrm{BPS}_{\hbar^{r-1}}^{Y+1-r}(\lambda-r\mathbf d),
\end{equation}
\emph{and both maps carry $(\Delta E;\Delta R_I;\Delta J_L,\Delta J_R)=(\tfrac12;\tfrac12,\tfrac12,\tfrac12;-\tfrac12,0)$.}

\begin{proofbox}
\textbf{Proof.} Counting vertices and edges gives the letter and flavour shifts
\begin{equation*}
\Delta Y=3(n+1)-2(2n+1)=1-n,\qquad\theta_I(Q_nX)-\theta_I(X)=(n+1)-(2n+1)=-n,
\end{equation*}
while \eqref{eq-derivative-homogeneity} supplies $n$ derivatives in each dotted direction. By induction from $\mathbf Q^{(0)}=\mathbf Q$, the homogeneous contractions and transferred coefficients obey
\begin{equation}\label{local-ch8-transfer-grading}
\begin{gathered}
h_r:\mathrm{BPS}_{\hbar^{r-1}}^Y(\lambda)\longrightarrow \mathrm{BPS}_{\hbar^{r-1}}^{Y+r-1}(\lambda+r\mathbf d),\\
[\hbar^m]\mathbf Q^{(r)}:
\mathrm{BPS}_{\hbar^{r-1}}^Y(\lambda)\longrightarrow \mathrm{BPS}_{\hbar^{r-1}}^{Y+1-r-m}\bigl(\lambda-(r+m)\mathbf d\bigr).
\end{gathered}
\end{equation}
Indeed, if a term of $[\hbar^m]\mathbf Q^{(r+1)}$ contains $\ell$ perturbation factors of total order $m+1$ and $\ell-1$ copies of $h_r$, then
\begin{equation}\label{local-ch8-transfer-grading-calculation}
\begin{aligned}
\Delta Y&=\ell(1-r)-(m+1)+(\ell-1)(r-1)=-r-m,\\
\Delta\lambda&=-(\ell r+m+1)\mathbf d+(\ell-1)r\mathbf d=-(r+m+1)\mathbf d.
\end{aligned}
\end{equation}
Setting $m=0$ for $d_{r+1}$ and inserting the shifts into \eqref{eq-E-charges} gives
\begin{equation*}
\Delta\Theta=-3n,\qquad\Delta(n_++n_-)=2n,\qquad(\Delta E;\Delta R_I;\Delta J_L,\Delta J_R)=\left(\frac12;\frac12,\frac12,\frac12;-\frac12,0\right).
\end{equation*}
\par\vspace{-\medskipamount}\noindent\hspace*{\fill}\qedsymbol
\end{proofbox}

Although the sectors $(\lambda+k\mathbf d)_{Y+k}$ ($k\in\mathbb Z$) share identical physical charges, the cochain grading separates distinct loop orders and bounds the active differentials on any sector.

\rthm{Corollary}{b-00758}{Obstruction and $d_r$-exactness bounds} \emph{A nonvanishing obstruction $d_r[\mathcal O_0]_{\hbar^{r-1}}\ne0$ requires $r\le\min_I\theta_I$, and $d_r$-exactness $[\mathcal O_0]_{\hbar^{r-1}}\in\operatorname{im}d_r$ requires $r\le\min(n_+,n_-)$. In particular, no nonzero class with zero input derivatives is $d_r$-exact for $r\ge1$.}

\begin{proofbox}
\textbf{Proof.} Nonnegativity of the target weights in \eqref{eq-E-grading} requires $\theta_I-r\ge0$ ($\Longleftrightarrow r\le\min_I\theta_I$) and, for a source mapping into $(\lambda)_Y$, $n_\pm-r\ge0$ ($\Longleftrightarrow r\le\min(n_+,n_-)$).~\hfill\qedsymbol
\end{proofbox}

\section{The twisted complex and the physical \texorpdfstring{$Q$}{Q}-cohomology}\label{app-twist-cohomology}

This appendix proves the identification \eqref{eq-twist-cohomology} of Section~\ref{canonical-changes-of-variables}, in the notation of that section, and gives its form in perturbation theory.

\rthm{Proposition}{prop-twist-cohomology}{The twisted complex and the BPS operators} \emph{Consider local operators invariant under constant gauge transformations and independent of the undifferentiated ghost $c_{\mathrm{gh}}^A$. In each finite-dimensional $2(E+J_L)$ eigenspace, the isomorphisms \eqref{eq-twist-cohomology} read}
\begin{equation*}
H\bigl((S_{\mathrm{tw}},\,\cdot\,)\bigr)
\;\cong\;H\bigl((S_{\mathrm{BV}},\,\cdot\,)\bigr)
\;\cong\;H\Bigl(Q_-\,;\ \frac{\text{gauge-invariant local operators}}{\text{equations of motion}}\Bigr).
\end{equation*}
\emph{The cohomology of the total differential, which contains the gauge BRST transformation, the equations of motion and $Q_-$, is thus the same as the cohomology of $Q_-$ on the BRST cohomology of the physical theory.}

\begin{proofbox}
\textbf{Proof.} The canonical transformations generated by $\Psi$ and $\Psi'$ are changes of variables, and by \eqref{eq-T-representative} every cocycle of $S_{\mathrm{BV,new}}=S_{\mathrm{tw}}+S_{\mathcal T}$ has a representative that depends only on the retained fields $\mathcal S$ and their antifields; this gives the first isomorphism. For the second, write
\begin{equation*}
S_{\mathrm{BV}}=S_{\mathrm{phys,BRST}}+S_Q,\quad
S_{\mathrm{phys,BRST}}=S_{\mathrm{phys}}+\int d^4x\,(s\Phi^{\mathcal A})\Phi^\ddagger_{\mathcal A},\quad
S_Q=\int d^4x\,(Q_-\Phi^{\mathcal A})\Phi^\ddagger_{\mathcal A},
\end{equation*}
and $D_0=(S_{\mathrm{phys,BRST}},\,\cdot\,)$, $D_Q=(S_Q,\,\cdot\,)$. Let $\mathrm{gh}$ be the ghost number and $J_L=J_{+-}$ the Cartan generator of the rotations of the undotted index, as in \eqref{eq-E-charges}; the supercharge $Q_-$ has $J_L=-\tfrac12$. Then
\begin{equation*}
\underbrace{D_0}_{(J_L,\mathrm{gh})=(0,1)}{}^{2}=0,\qquad
\underbrace{D_0D_Q+D_QD_0}_{(J_L,\mathrm{gh})=(-\frac12,1)}=0,\qquad
\underbrace{D_Q}_{(J_L,\mathrm{gh})=(-\frac12,0)}{}^{2}=0,
\end{equation*}
the three components of $(S_{\mathrm{BV}},S_{\mathrm{BV}})=0$ of fixed $J_L$, so the local operators form a double complex.\footnotemark{} Its first page is the cohomology of $D_0$, the local BRST cohomology of the gauge theory: on this relative complex it is concentrated in ghost number $\mathrm{gh}=0$, where it is the space of gauge-invariant local operators modulo the equations of motion \cite{BBH}. The differential induced on it by $D_Q$ is $Q_-$. The higher differentials $d_r$, $r\ge2$, change $\mathrm{gh}$ by $1-r\ne0$ and therefore vanish, which gives the second isomorphism.~\hfill\qedsymbol
\end{proofbox}
\footnotetext{\label{fn-weight}Both differentials preserve $2(E+J_L)$, with $E$ the scaling dimension and dimensionful couplings counted with their dimensions. The $J_L$ filtration is bounded in each finite-dimensional weight space, so its spectral sequence converges.}

Proposition~\ref{prop-twist-cohomology} is the classical statement. Within a perturbative quantization of the physical theory that preserves the symmetries below, one can compare the total quantum differential with the quantum supercharge on physical operators. Give the supercharge a constant ghost $\epsilon$ with $(J_L,\mathrm{gh})=(\tfrac12,1)$, so that $S_{\mathrm{BV}}(\epsilon)=S_{\mathrm{phys,BRST}}+\epsilon S_Q$ has $(J_L,\mathrm{gh})=(0,0)$; the physical theory is the part of order $\epsilon^0$, and the twist is $\epsilon=1$. A perturbative quantization that preserves $\mathrm{gh}$ and $J_L$ gives, at $\epsilon=1$, a differential on local operators of the form
\begin{equation*}
\begin{gathered}
\mathbf D=\sum_{k\ge0}D_k,\qquad
D_k:\ (J_L,\mathrm{gh})\longmapsto(J_L-\tfrac k2,\,\mathrm{gh}+1-k),\qquad
\mathbf D^2=0,\\
D_0=(S_{\mathrm{phys,BRST}},\,\cdot\,)+O(\hbar),\qquad D_1=(S_Q,\,\cdot\,)+O(\hbar),
\end{gathered}
\end{equation*}
with $D_k=O(\hbar)$ for $k\ge2$. Here $D_0$ is the quantum BRST differential of the physical theory, with the loop corrections to the gauge transformations and to the equations of motion of composite operators; $D_1$ is the quantum supercharge, with its anomalous terms; and the $D_k$ with $k\ge2$ record that the quantum supercharge squares to zero only up to BRST-exact terms, $D_1^2=-D_0D_2-D_2D_0$.

\rthm{Proposition}{prop-twist-cohomology-quantum}{Quantum BPS operators} \emph{Let $V$ consist of local operators invariant under constant gauge transformations and independent of the undifferentiated ghost $c_{\mathrm{gh}}^A$. Let $\mathbf D=\sum_{k\ge0}D_k$ be a $\mathbb C[[\hbar]]$-linear differential on $V[[\hbar]]$ of the form above that preserves $2(E+J_L)$. Assume the eigenspaces of $2(E+J_L)$ in $V$ are finite dimensional (footnote~\ref{fn-weight}).}

(i) \emph{The cohomology of $D_0$ is concentrated in ghost number zero, and the space $P=H^0(D_0)$ of quantum physical operators is a free $\mathbb C[[\hbar]]$-module with $P/\hbar P$ the classical space of gauge-invariant local operators modulo the equations of motion.}

(ii) \emph{$D_1$ induces a differential $Q_\hbar$ on $P$, the quantum supercharge acting on physical operators.}

(iii) \emph{The complexes $(V[[\hbar]],\mathbf D)$ and $(P,Q_\hbar)$ are homotopy equivalent over $\mathbb C[[\hbar]]$. In particular}
\begin{equation*}
H\bigl(V[[\hbar]],\mathbf D\bigr)\;\cong\;H\bigl(Q_\hbar\,;\,H(D_0)\bigr)
\end{equation*}
\emph{as $\mathbb C[[\hbar]]$-modules, torsion included: a class is lifted at the same order in $\hbar$ on both sides.}

\begin{proofbox}
\textbf{Proof.} (i) At each value of $2(E+J_L)$, $(V[[\hbar]],D_0)$ is a bounded complex of free $\mathbb C[[\hbar]]$-modules of finite rank whose reduction modulo $\hbar$ is the classical complex of $D_0$, with cohomology in ghost number zero only (proof of Proposition~\ref{prop-twist-cohomology}). The universal coefficient sequence
\begin{equation*}
0\to H^{\mathrm{gh}}(D_0)/\hbar H^{\mathrm{gh}}(D_0)\to H^{\mathrm{gh}}\bigl(D_0|_{\hbar=0}\bigr)\to\ker\bigl(\hbar:H^{\mathrm{gh}+1}(D_0)\to H^{\mathrm{gh}+1}(D_0)\bigr)\to0
\end{equation*}
gives $H^{\mathrm{gh}}(D_0)=\hbar H^{\mathrm{gh}}(D_0)$ for $\mathrm{gh}\ne0$, hence $H^{\mathrm{gh}}(D_0)=0$ by Nakayama's lemma, and then shows that $H^0(D_0)$ has no $\hbar$-torsion and reduces to the classical cohomology.

(ii) The components of $\mathbf D^2=0$ that change $J_L$ by $-\tfrac12$ and by $-1$ are $D_0D_1+D_1D_0=0$ and $D_1^2+D_0D_2+D_2D_0=0$.

(iii) Since $P$ is free, the complex $(V[[\hbar]],D_0)$ splits: there are $\mathbb C[[\hbar]]$-linear maps $\iota:P\to V[[\hbar]]$, $\pi:V[[\hbar]]\to P$ and a homotopy $h$ of $(J_L,\mathrm{gh})=(0,-1)$ with $\pi\iota=1$ and $\iota\pi=1-D_0h-hD_0$. Treat $\delta=\sum_{k\ge1}D_k$ as a perturbation of $D_0$. It lowers $J_L$, which is bounded at each value of $2(E+J_L)$, so $\delta h$ is nilpotent and the homological perturbation lemma \cite[Lemma 2.4]{Crainic} gives a homotopy equivalence between $(V[[\hbar]],\mathbf D)$ and $P$ with the differential
\begin{equation*}
\pi\delta(1+h\delta)^{-1}\iota=\pi\Bigl(\delta-\delta h\delta+\delta h\delta h\delta-\cdots\Bigr)\iota .
\end{equation*}
A term with the factors $D_{a_1},\ldots,D_{a_n}$ has $(J_L,\mathrm{gh})=(-\tfrac A2,1-A)$ with $A=a_1+\cdots+a_n\ge n$. Because $P$ lies in ghost number zero, only $A=1$ contributes, that is, the single term $\pi D_1\iota=Q_\hbar$.~\hfill\qedsymbol
\end{proofbox}

Under the hypotheses of Proposition~\ref{prop-twist-cohomology-quantum}, the total differential $\mathbf D$ and the quantum supercharge on physical operators have the same cohomology. This compares two complexes within the chosen quantization of the physical theory. The differential $\mathbf Q$ of \eqref{eq-1-1}, however, is obtained by quantizing the classical twisted action $S_{\mathrm{tw}}$ directly. Identifying its complex with $(V[[\hbar]],\mathbf D)$ requires a quantum comparison compatible with local operator renormalization. Independence of gauge fixing and regularization within a fixed BV quantization \cite{Costello} does not by itself construct this comparison. We assume this identification when interpreting the results as physical quantum BPS cohomology, as stated in Section~\ref{sec-brackets}.

\section{The spanning-tree solution of the circuit}\label{sec-B-spanning}

The reductions of Section~\ref{sec-B-local-reduction} do not use the global solution of the circuit. We recall it here for comparison with earlier work. A formalism already exists for computing all integrals of the form \eqref{eq-operatope} \cite[Theorem 4 and Corollary 2]{Wang}, which solves the discrete Laplace equation on the circuit exactly and expresses all edge currents in terms of the external currents $\lambda$ and the resistances $t_e$. The expanded formula has one term per spanning tree. At fixed resistances one can instead solve the reduced Laplacian system for the vertex potentials.

To compute the edge currents from the spanning trees, first solve the conservation law on each spanning tree $T$: deleting $e\in T$ gives its current as the source sum on the component containing $e(0)$. Then sum over the spanning trees with the weights in \eqref{eq-1-24}; the result automatically satisfies the loop voltage equation and solves the full network \cite{Wang},
\begin{equation}\label{eq-1-24}
\begin{aligned}
y^T_{e,+\dot a}
&=\begin{cases}
\displaystyle\sum_{\substack{v\text{ connected to }e(0)\\
\text{in }T\setminus\{e\}}}\lambda_{v,+\dot a},&e\in T,\\[3pt]
0,&e\notin T,
\end{cases}\\[4pt]
y_{e,+\dot a}(t)
&=\frac{\displaystyle\sum_{\substack{T\\\mathrm{spanning\ trees}}}
\left(\prod_{h\in I\setminus T}t_h\right)y^T_{e,+\dot a}}
{\displaystyle\sum_{\substack{T\\\mathrm{spanning\ trees}}}\prod_{h\in I\setminus T}t_h}.
\end{aligned}
\end{equation}

A common rescaling of the resistances does not affect the formula, so we may impose $\sum_{e\in I}t_e=1$. Then the integral is just the Jacobian of the change of variables from the $y$'s to the independent $t$'s. The moves of Section~\ref{sec-B-local-reduction} replace this global solution by local changes of the resistances, in which the spanning-tree denominator of \eqref{eq-1-24} cancels.

\section{Covariant derivatives in the superpotential}\label{app-covariant-derivatives}

The superpotential in \eqref{eq-Psi} may also contain superspace covariant derivatives. For the standard off-shell chiral and vector multiplets, any local gauge-invariant Lorentz-scalar chiral or antichiral density obeys
\begin{equation}\label{eq-derivative-chirality}
\bar\nabla_{\dot a}W=0,\qquad \nabla_a\bar W=0.
\end{equation}

\rthm{Proposition}{prop-d-chiral-cohomology}{Covariant chiral cohomology} \emph{For the standard off-shell chiral and vector multiplets, a compact gauge group and linearly transforming matter, the local gauge-invariant Lorentz-scalar densities satisfying \eqref{eq-derivative-chirality} admit the decomposition}
\begin{equation}\label{eq-spinor-chiral-decomposition}
\begin{aligned}
W&=W(\Phi,W_a)+\frac14\bar\nabla^2K,\\
\bar W&=\bar W(\bar\Phi,\bar W_{\dot a})+\frac14\nabla^2K^\dagger,
\end{aligned}
\end{equation}
\emph{where the ordinary superpotentials depend only on undifferentiated matter and field-strength superfields, while $K$ is a gauge-invariant local scalar that may contain superspace covariant derivatives.} \emph{The decomposition holds coefficientwise in formal power series, with finitely many derivatives at each order.}

\begin{proofbox}
\textbf{Proof.}
Brandt proved the decomposition for the linearized multiplets in \cite[Eq.~(6.8) and Appendix A]{Brandt1993}. The off-shell multiplets used here have the required decomposition \cite[Section 10 and Appendix C.2]{Brandt1996}. Applying the covariant form of Brandt's theorem \cite[Appendix B, Eqs.~(B.32)--(B.33)]{Brandt2002} to a ghost-independent scalar gives the first line of \eqref{eq-spinor-chiral-decomposition}. Complex conjugation gives the second.~\hfill\qedsymbol
\end{proofbox}

Field-strength factors generated by commutators of superspace covariant derivatives are absorbed into the ordinary superpotential $W(\Phi,W_a)$ in \eqref{eq-spinor-chiral-decomposition}.

\rthm{Proposition}{prop-d-projector-removal}{Removal of the chiral projection} \emph{The projector terms in \eqref{eq-spinor-chiral-decomposition} assemble into a full-superspace D-term with $\delta K=K+K^\dagger$, which is removed in the BV complex by the canonical antifield translation generated by}
\begin{equation}\label{eq-spinor-dterm-generator}
\delta\Psi=\frac18\int d^4x\,
\nabla_+\bar\nabla^2\delta K\Big|.
\end{equation}

\begin{proofbox}
\textbf{Proof.}
Using $\nabla^2=2\nabla_-\nabla_+$ and superspace integration by parts with vanishing spacetime boundary terms gives
\begin{equation}\label{eq-spinor-projector-dterm}
\begin{aligned}
\delta S_{\mathrm{phys}}
&=\frac12\int d^4x\,
\nabla_-\nabla_+\left(\frac14\bar\nabla^2K\right)\Big|
+\frac14\int d^4x\,
\bar\nabla^2\left(\frac14\nabla^2K^\dagger\right)\Big|\\
&=\frac18\int d^4x\,
\nabla_-\nabla_+\bar\nabla^2(K+K^\dagger)\Big|\\
&=Q_-\delta\Psi=(s+Q_-)\delta\Psi,
\qquad s\delta\Psi=0.
\end{aligned}
\end{equation}
With Euler--Lagrange functional derivatives and the sign convention of \eqref{eq-B-2},
\begin{equation}\label{eq-spinor-canonical-removal}
\begin{aligned}
\Phi^\ddagger_{\mathcal A,\mathrm{new}}
&=\Phi^\ddagger_{\mathcal A,\mathrm{old}}
+\frac{\delta_L\delta\Psi}{\delta\Phi^{\mathcal A}},\\
S_{\mathrm{BV,new}}
&=\left.S_{\mathrm{phys}}\right|_{\delta K=0}
+\delta S_{\mathrm{phys}}
+\int d^4x\,((s+Q_-)\Phi^{\mathcal A})
\Phi^\ddagger_{\mathcal A,\mathrm{new}}
-(s+Q_-)\delta\Psi\\
&=\left.S_{\mathrm{phys}}\right|_{\delta K=0}
+\int d^4x\,((s+Q_-)\Phi^{\mathcal A})
\Phi^\ddagger_{\mathcal A,\mathrm{new}}.
\end{aligned}
\end{equation}
Because $\delta\Psi$ depends solely on fields and their derivatives and is independent of antifields, the translation is triangular with identity diagonal, has unit Berezinian, and satisfies $\Delta\delta\Psi=0$.~\hfill\qedsymbol
\end{proofbox}

\rthm{Proposition}{prop-d-unchanged-twist}{Twisted operations} \emph{Fix the ordinary superpotential in \eqref{eq-spinor-chiral-decomposition}, the canonical insertions, and the quantization prescription of Section~\ref{sec-brackets}.} \emph{The projector sector leaves the canonical $H(\bar\partial)$ representatives, the loop operators $Q_n$, and the quantum product $\star$ unchanged.}

\begin{proofbox}
\textbf{Proof.}
The chiral F-term remains $Q_-$-exact by \eqref{eq-Psi}, so after the translation \eqref{eq-spinor-canonical-removal} the twist retains only the ordinary antichiral term. Eliminating the contractible pairs of Table~\ref{b-00904} yields the unchanged interaction density
\begin{equation}\label{eq-derivative-interaction}
\begin{gathered}
\mathcal I=-\frac g2C^A{}_{BC}b_Ac^Bc^C
-igc^A(t_A)^j{}_i\gamma_j\beta^i
+\bar W(\gamma,w),\\
w_{\dot a}^A=\frac12\partial_{+\dot a}c^A.
\end{gathered}
\end{equation}
The canonical $H(\bar\partial)$ representatives remain the holomorphic zero-form letters $c,b,\gamma,\beta$ and their holomorphic derivatives, while their dictionary to untwisted physical fields shifts by
\begin{equation}\label{eq-spinor-representative-shift}
\delta\beta^{(0)i}
=\frac{\delta_L\delta\Psi}{\delta\bar{\mathcal F}_i},
\qquad
\delta b_A^{(0)}
=-\frac i2\frac{\delta_L\delta\Psi}{\delta\lambda_-^A}.
\end{equation}
For fixed canonical insertions $\mathcal O,A,B$, the free BV pairing, gauge fixing, and interaction satisfy
\begin{equation}\label{eq-spinor-unchanged-operations}
\begin{gathered}
\delta S_{\mathrm{free}}=\delta\mathcal I=0,
\qquad \delta K_t=\delta P_{\varepsilon,L}=0,\\
\delta(Q_n\mathcal O)=0\quad(n\ge0),
\qquad \delta(A\star B)=0.
\end{gathered}
\end{equation}
Because the projector vertices are removed prior to the Wick expansion, the form-degree count for $Q_n\mathcal O$ remains
\begin{equation}\label{eq-spinor-degree-count}
\underbrace{|I|-1}_{\text{propagators}}
+\underbrace{2}_{\text{heat kernel}}
=\underbrace{2(|V|-1)}_{|V|-1\ \text{integrated points}},
\qquad |V|=n+2,\quad |I|=2n+1,
\end{equation}
so the classification of Section~\ref{sec-restrictions} applies verbatim for $n\ge1$: fans for interactions of degree at most three, and fans or spiders for an ordinary superpotential of arbitrary degree.~\hfill\qedsymbol
\end{proofbox}

\section{Star--triangle reduction with a source}\label{app-sourced-star}

\begin{figure}[ht!]
\centering
\includegraphics[width=\linewidth]{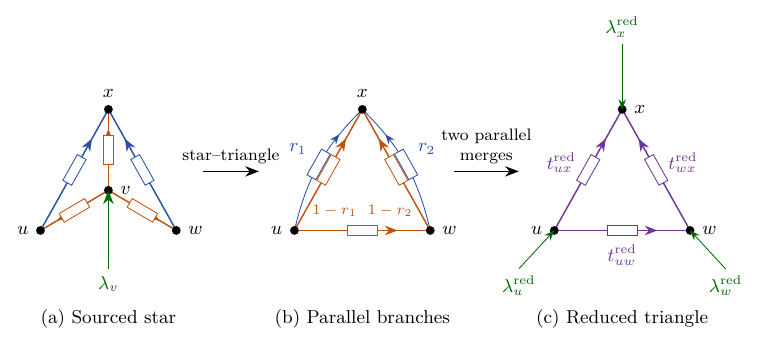}
\caption{Star--triangle reduction with source $\lambda_v$: the source is redistributed to $u,w,x$ by \eqref{local-B-sourced-current-map}, and the original $ux,wx$ branches merge with the new triangle edges.}
\label{fig-sourced-star}
\end{figure}

Theorem~\ref{b-01159} gave the star--triangle move at a source-free degree-three vertex; when the eliminated vertex $v$ carries an external current source $\lambda_v$, the transformation redistributes $\lambda_v$ across its three neighbours with resistance-dependent weights before merging the parallel edges (Figure~\ref{fig-sourced-star}).

\rthm{Theorem}{thm-sourced-star}{Star--triangle with a source} \emph{In a Laman graph, let $v$ have degree three and neighbours $u,w,x$, with $ux,wx$ present and $uw$ absent, ordered as in Theorem~\ref{b-01159}. After the star--triangle change and the two parallel merges, let $r_1,r_2\in(0,1)$ be the current fractions of the original edges $ux,wx$, and set}
\begin{equation}\label{eq-sourced-Sigma}
\Sigma=t_{uw}^{\mathrm{red}}
+\frac{t_{ux}^{\mathrm{red}}}{1-r_1}
+\frac{t_{wx}^{\mathrm{red}}}{1-r_2}
\end{equation}
\emph{for the sum of the triangle resistances prior to merging. The reduced sources and original currents are}
\begin{equation}\label{local-B-sourced-current-map}
\begin{aligned}
\lambda_u^{\mathrm{red}}
&=\lambda_u+\frac{t_{wx}^{\mathrm{red}}}{(1-r_2)\Sigma}\lambda_v,&
\lambda_w^{\mathrm{red}}
&=\lambda_w+\frac{t_{ux}^{\mathrm{red}}}{(1-r_1)\Sigma}\lambda_v,\\
\lambda_x^{\mathrm{red}}
&=\lambda_x+\frac{t_{uw}^{\mathrm{red}}}{\Sigma}\lambda_v,\\
y_{vu}
&=\frac{t_{wx}^{\mathrm{red}}}{(1-r_2)\Sigma}\lambda_v
-y_{uw}^{\mathrm{red}}-(1-r_1)y_{ux}^{\mathrm{red}},\\
y_{vw}
&=\frac{t_{ux}^{\mathrm{red}}}{(1-r_1)\Sigma}\lambda_v
+y_{uw}^{\mathrm{red}}-(1-r_2)y_{wx}^{\mathrm{red}},\\
y_{vx}
&=\frac{t_{uw}^{\mathrm{red}}}{\Sigma}\lambda_v
+(1-r_1)y_{ux}^{\mathrm{red}}+(1-r_2)y_{wx}^{\mathrm{red}},\\
y_{ux}&=r_1y_{ux}^{\mathrm{red}},&y_{wx}&=r_2y_{wx}^{\mathrm{red}}.
\end{aligned}
\end{equation}
\emph{Choosing a spanning tree $T^{\mathrm{red}}$ containing $ux$ and $wx$, the $n$-loop kernel is the full current-area pullback}
\begin{equation}\label{local-B-sourced-operator}
\begin{aligned}
\mathcal I_\Gamma(\lambda;z)
={}&-\int_{(0,1)^2\times\mathbb{RP}_{>0}^{2n-2}}
e^{-\sum_{e\in I\,\mathrm{(edges)}}y_{e,+\dot a}z_e^{+\dot a}}\\
&\quad\times[-d(r_1y_{ux,+\dot+}^{\mathrm{red}})
\wedge d(r_1y_{ux,+\dot-}^{\mathrm{red}})]\\
&\quad\wedge[-d(r_2y_{wx,+\dot+}^{\mathrm{red}})
\wedge d(r_2y_{wx,+\dot-}^{\mathrm{red}})]
\wedge\bigwedge_{\substack{e\in I^{\mathrm{red}}\setminus T^{\mathrm{red}}\\e\ne uw}}
(-d^2y_e^{\mathrm{red}}),
\end{aligned}
\end{equation}
\emph{where $d$ varies $(r_1,r_2)$ and the positive reduced resistance ratios at fixed original sources, with reduced currents evaluated at the resistance-dependent sources in \eqref{local-B-sourced-current-map}.}

\begin{proofbox}
\textbf{Proof.} The subset count in Theorem~\ref{b-01159} proves that the reduced graph is Laman. The star--triangle change divides $\lambda_v$ in proportion to the incident star conductances $(1/t_{vu},1/t_{vw},1/t_{vx})$, whose normalized values equal the unmerged opposite triangle resistances $(S_v/t_{vu},S_v/t_{vw},S_v/t_{vx})$ divided by their sum $\Sigma$, and parallel merging supplies the factors $1-r_1,1-r_2$ and the current maps \eqref{local-B-sourced-current-map}. Replacing $ux,wx$ in $T^{\mathrm{red}}$ by $vu,vw,vx$ gives a spanning tree of $\Gamma$ whose chords are $ux,wx$ together with the retained chords other than $uw$; their area wedge is the integrand in \eqref{local-B-sourced-operator}, with the overall minus sign from orientation reversal.~\hfill\qedsymbol
\end{proofbox}

For higher-degree superpotentials, the output count \eqref{eq-1-23} permits multiple source vertices inside a single interaction tree: iterating Theorem~\ref{thm-sourced-star} then yields rational currents and current-area densities whose source transfers and reverse maps require only polynomially many arithmetic operations in the graph size at fixed resistance ratios.

\end{document}